\documentclass[11pt,a4paper]{article}
\pdfoutput=1
\usepackage{jheppub}

\usepackage{amsmath,epsfig}
\usepackage{amssymb,amsfonts}
\usepackage{latexsym}
\usepackage[latin1]{inputenc}
\usepackage{slashed}
\usepackage{empheq}
\numberwithin{equation}{section}
\usepackage{dsfont}
\usepackage{cancel}
\usepackage{overpic}
\usepackage{subcaption}
\usepackage{subeqnarray}
\usepackage{xcolor}

\usepackage{graphicx}
\usepackage{longtable}
\usepackage{color}
\usepackage{bbm}

\newcommand{\exclude}[1]{}
\def\a#1{\alpha_{#1}}

\def\beq{\begin{equation}}
\def\eeq{\end{equation}}
\def\be{\begin{equation}}
\def\ee{\end{equation}}
\def\bea{\begin{eqnarray}}
\def\eea{\end{eqnarray}}
\def\bal{\begin{align}}
\def\eal{\end{align}}

\def\2b2[#1,#2][#3,#4]{\left( \begin{array}{cc} #1 & #2 \\ #3 & #4 \end{array}
\right)}
\def\3b3[#1,#2,#3][#4,#5,#6][#7,#8,#9]{\left( \begin{array}{ccc} #1 & #2 #3 \\
#4 & #5 & #6\\#7&#8&#9\end{array} \right)}

\newcommand\fverb{\setbox\pippobox=\hbox\bgroup\verb}
\newcommand\fverbdo{\egroup\medskip\noindent%
                        \fbox{\unhbox\pippobox}\ }
\newcommand\fverbit{\egroup\item[\fbox{\unhbox\pippobox}]}

\newcommand{\bear}{\begin{eqnarray}}

\newcommand{\eear}{\end{eqnarray}}
\newcommand{\de}{\partial}
\newcommand{\bsea}{\begin{subeqnarray}}
\newcommand{\esea}{\end{subeqnarray}}
\newbox\pippobox

\def\f{\varphi}

\def\6{\partial}
\def\a{\alpha}

\def\sp{\;\;\;,\;\;\;}

\def\sq
\def\a{\alpha}

\title{Holographic RG flows on curved manifolds and quantum phase transitions}

\author[\natural]{J.~K.~Ghosh,}
\author[\flat, \natural]{E.~Kiritsis,}
\author[\natural]{F.~Nitti,}
\author[\natural]{L.~T.~Witkowski}

\affiliation[\natural]{\href{http://www.apc.univ-paris7.fr}{APC, AstroParticule et Cosmologie}, Universit\'e Paris Diderot, CNRS/IN2P3, CEA/IRFU, \\
Observatoire de Paris, Sorbonne Paris Cit\'e,\\
 10, rue Alice Domon et L\'eonie Duquet, 75205 Paris
Cedex 13, France}
\affiliation[\flat]{\href{http://hep.physics.uoc.gr}{Crete Center for Theoretical Physics},
Department of Physics,\\
University of Crete, 71003 Heraklion, Greece}

\preprint{CCTP-2017-7\\
\hphantom{AAAAAAAAAAAAAAAAAAAAAAAAAAAAAAAAAAAAAAAAlll} ITCP-IPP 2017/18}

\abstract{Holographic RG flows dual to QFTs on maximally symmetric curved manifolds (dS$_d$,
  AdS$_d$, and $S^d$) are considered in the framework of Einstein-dilaton gravity in $d+1$ dimensions. A general dilaton potential is used and the flows are driven by a scalar relevant operator.
  The general properties of such flows are analyzed and the UV and IR asymptotics computed.
 New RG flows can appear at  finite curvature which do not have a zero
  curvature counterpart. The so-called
  `bouncing' flows, where the $\beta$-function has a branch cut at which it changes sign, are found to persist at finite curvature. Novel  quantum first-order phase transitions are found, triggered
  by a variation in the $d$-dimensional curvature in theories allowing
  multiple ground states.
}

\begin{document}
\maketitle

\section{Introduction, summary of results and outlook}

Renormalization group (RG) flows in Quantum Field Theory (QFT) are usually studied in flat space.
There are, however, many reasons to consider QFTs on curved manifolds
and study the associated RG flows.

One reason is that curved manifolds are considered in order to render QFTs well defined or well controlled in the IR, by taming IR divergences. There are many facets of this idea, going back to \cite{cw}, and to \cite{kk} for a similar approach of regulating IR divergences in string theory. On the holographic side of QFTs, this is the role played by global AdS space. There, the QFT lives on $R\times S^d$, where the spatial part is a sphere.

Curved manifolds also provide IR modifications to supersymmetric QFTs in a way that supersymmetric indices or other supersymmetric observables are well-defined, \cite{Komar, Marte}. Moreover, they provide control parameters on which supersymmetric observables can depend upon.
Non-trivial backgrounds in general can be useful in deriving non-trivial results like exact $\beta$-functions, \cite{NSVZ}. Curved manifolds (spheres) were used to regulate otherwise singular holographic solutions in \cite{buchel1}. Holographic flows with de Sitter slices were discussed in \cite{buchel2}.

Further objects of interest are partition functions on curved manifolds. In particular, the partition function on spheres  was argued to serve as an analogue of the c-functions in odd dimensions. The case of three dimensions is well known, \cite{F1,1103.1181} but from holographic arguments the case can be made also for other dimensions, \cite{1006.1263,1011.5819}.
The dynamics of QFTs on curved manifolds may have interesting and different structure from that on flat manifolds, especially in the case of QFTs on AdS manifolds, \cite{ABTY}.

Cosmology is another important context where quantum  field
theories on curved space-times may be considered. There is a vast  body of work in this context.  In
connection to the renormalization group flow aspect, for example, the non-perturbative RG group on de Sitter backgrounds  was
studied for large-$N$ scalar field theories in \cite{Tsamis:1992xa, Tsamis:1994ca, Tsamis:1996qm,Ramsey:1997qc,Burgess:2009bs,1105.4539,1306.3846}.

There is a folk theorem which says that RG flows of QFTs on curved
manifolds are very similar to those on flat manifolds. The argument
for this  is that $\beta$-functions are determined by UV/short-distance divergences and the short distance structure of a given QFT is independent of its curvature.
Although the leading intuition of such statements is basically correct, the folk theorem fails on several grounds. Indeed, the leading UV divergences are independent of curvature. However, subleading ones do depend on curvature.
We will see this clearly and in a controlled fashion in this
paper, although it is well known to experts,
\cite{HSke,skenderislec,skenderisham}. A further observation is that already for
CFTs, curvature is a source of breaking of scale invariance via the
conformal anomaly, \cite{Duff}. For general QFTs driven by relevant
couplings, the $\beta$ functions do in general depend on curvature,
\cite{Osborn}. The same is true for the vacuum expectation values of
operators. More generally, one may expect that curvature becomes very
important in the IR and this expectation is, in general, correct.

\vspace{0.3cm}
In this work we will study the RG behavior of QFTs on maximally
symmetric curved space-times using the framework of holography.
The holographic correspondence, \cite{9711200, 9802109, 9802150}, provides a map between QFT and gravity/string-theories in higher dimensions, at least in the limit of large $N$. In this context, the holographic dimension serves as an effective RG scale in the dual QFT, thus geometrizing the notion of RG flow. In essence, RG flows can be understood as bulk evolution in the holographic dimensions \cite{9810126, 9903190, 9904017, 9912012, 0105276, 0404176, 0702088, Papadimitriou:2007sj, 0707.1324, 0707.1349,1010.1264,1010.4036,1106.4826,1112.3356,1205.6205,1310.0858,1401.0888,exotic}, \cite{1006.1263}.

In this framework, there is additional motivation for studying RG flows in the presence of curvature.
\begin{itemize}
\item Vacuum expectation values depend crucially on curvature. In perturbation theory, vevs are trivial and therefore they do not affect the RG running. In holographic QFTs however, they are non-trivial generically, they do depend on curvature, and they do affect the RG flows.
In this work we will show many non-trivial modifications, driven by
curvature,  of (holographic) RG flows. These modifications may be
so significant  that they can trigger  quantum phase
transitions between different ground states of the theory. For example, in \cite{1108.6070} quantum phase transitions were observed for a supersymmetric field theory on $\mathbbm{R} \times S^3$. In this work we show the presence of such a quantum phase transition in an explicit example for a field theory on dS$_4$.\footnote{Phase transitions triggered by a change in space-time curvature in CFTs on de Sitter space \emph{at finite temperature} were addressed in \cite{1007.3996}.}

\item Holography provides a means of calculating the induced action for the background metric in QFT. Integrating out a QFT coupled to a background metric provides a Schwinger functional for the metric that can be turned into an effective action for the expectation value of the stress tensor. This is the starting point for many cosmology setups, like that of Starobinsky, \cite{staro}, and its generalizations, \cite{kcosmo}.
    In the case of a CFT, this effective action is rather simple as the only scale is the UV cutoff, and there are only a finite number (depending on the space-time dimension) of curvature-related terms that can appear.
For example in four dimensions one obtains a cosmological constant at
order ${\cal O}(\Lambda^4)$, an Einstein term at order ${\cal
  O}(\Lambda^2)$ and the conformal anomaly at order ${\cal
  O}(\log\Lambda^2)$. The curvature squared term is scheme dependent
 due to the form of the conformal anomaly. All other terms are suppressed by the cutoff.

In more general QFTs, the situation is more complex as now the QFT has more dimensionful parameters, that can enter the effective action for the metric/curvature. Such effective actions are largely unexplored, as computing them in perturbation theory is a difficult task.
Here, in the context of holographic theories we will provide a concrete framework to compute them. More detailed results will appear in a forthcoming publication \cite{Ftheorem}.

In particular, we will be working at constant curvature ${\cal R}$, and the effective action we will compute is a function of ${\cal R}$. In this computation, derivatives of the curvature are neglected and therefore the effective action is valid for manifolds whose curvature scale is sufficiently slowly varying. In that sense it is similar to the DBI action for gauge fields in string theory. This approach is complementary to the one followed in e.g.~\cite{1401.0888}, where RG flows were studied in theories where the curvature tensor is small but generic (i.e.~the space-time is not maximally symmetric).

\item In the Hartle-Hawking proposal for the wave-function
  of the universe, \cite{HH}, such a wave-function is determined by a
  gravitational path integral. It has been proposed in \cite{HH2,HHH} that holography can be used to define such a path integral and generate the relevant wave-functions. Our results provide the spatial curvature dependence of such wave-functions.

\item In \cite{self} a new approach was proposed using holography in order to implement the self-tuning idea for the cosmological constant.
In order to further address solutions for the universe brane with constant (positive or negative) curvature one must use for the bulk the type of solutions we are studying in this paper. Their application to the self-tuning problem will be discussed in a forthcoming publication.
\end{itemize}
In the following we summarize our setup and results.

\subsection{Summary of results}
We use a two-derivative $(d + 1)$-dimensional model described by the action:
\begin{align}
\label{eq:actionintro} S[g, \f]= \int du \, d^dx \, \sqrt{-g} \left(R^{(g)} - \frac{1}{2} \partial_a \f \partial^a \f - V(\f) \right)  \, .
\end{align}
We consider solutions with a scalar field profile and a metric ansatz
given by:
\begin{align} \label{eq:flowintro}
\f = \f(u), \qquad ds^2 = du^2 + e^{2 A(u)} \zeta_{\mu \nu} dx^{\mu} dx^{\nu} \, .
\end{align}
We will study solutions where $\zeta_{\mu \nu}$ is a fixed
(i.e.~$u$-independent) metric on a maximally symmetric $d$-dimensional
space-time  with positive or negative curvature. With our work,  we
extend to these space-times the systematic analysis of flat RG flows
presented in  \cite{exotic}. Via gauge/gravity duality, the ansatz (\ref{eq:flowintro}) describes
RG flows of field theories defined on manifolds with constant positive curvature
($S^d$ or dS$_d$, depending on the signature\footnote{The form of the solution as well as most
of our final results do not depend on the space-time signature.}) or constant
negative curvature ($H^d$, AdS$_d$).

The scalar field profile $\f(u)$ is interpreted as the running
coupling associated to a scalar operator $\mathcal{O}$ of the field
theory. We will focus our attention on scalar potentials that display extrema with negative cosmological constant, which can serve
as a UV/IR fixed points for the dual field theory. More specifically,
we will consider geometries which contain an asymptotic boundary
region, where the scalar field reaches the maximum of the potential and
the full metric asymptotes to (the boundary region of) $d+1$-dimensional Anti-de Sitter space-time.

 From
the dual field theory standpoint, these flows describe a UV CFT which
is defined on a curved space-time and deformed by a relevant
operator. In this language, the two parameters defining the UV
theory are  $R^{\textrm{uv}}$ (the scalar curvature of
the space-time where and the UV CFT is  defined) and the leading
coefficient in the scalar field near-boundary expansion, $\f_-$ (the
coefficient of the relevant deformation). They appear in the
near-boundary asymptotic expansion as:
\be \label{boundaryas}
ds^2 \simeq du^2 + e^{-2u/\ell} \gamma^{\textrm{uv}}_{\mu\nu}dx^\mu dx^\nu,
\quad \f \simeq \f_- \, \ell^{(d-\Delta)} \, e^{(d-\Delta)u/\ell}, \quad u \to -\infty.
\ee
where $\gamma^{\textrm{uv}}$ is a metric whose constant curvature is $R^{\textrm{uv}}$.
Both parameters $R^{\textrm{uv}}$ and $\f_-$  break
conformal invariance explicitly.   One of our  main goals will be to
characterize the interplay between these two different kinds of deformation. The
asymptotic analysis of the solution (\ref{eq:flowintro}) will also define precisely what
exactly is the curvature of the space-time seen by the UV CFT.

In the rest of this subsection we give a short  summary and discussion
of the results which are derived and presented in the main body of the paper.

\begin{figure}[t]
\centering
\begin{overpic}
[width=0.65\textwidth]{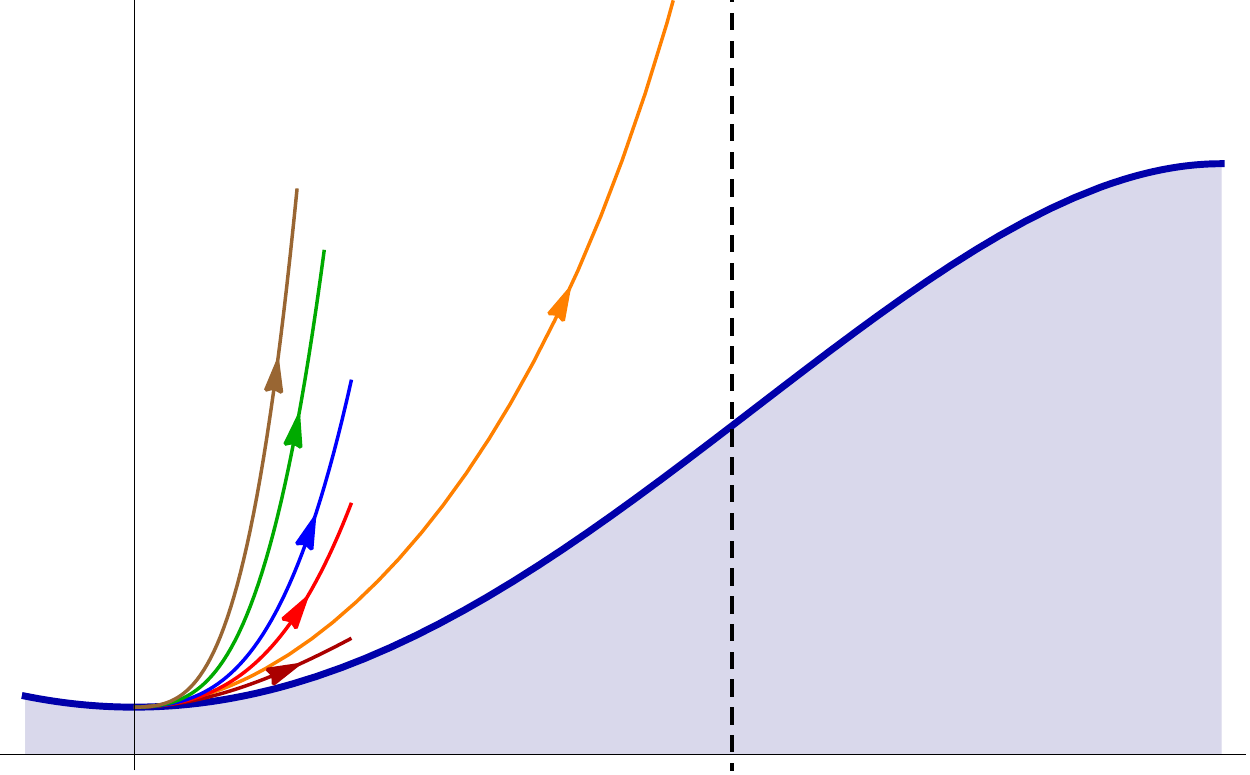}
\put(4,7){UV}
\put(60,3){$\f_0$}
\put(60,58){IR}
\put(99,3){$\f$}
\put(-2,58){$W(\f)$}
\put(26.8,34){$W_{C, \mathcal{R}}(\f)$}
\put(71.5,32){$\sqrt{-\tfrac{4(d-1) V(\f)}{d}}$}
\end{overpic}
\caption{Family of solutions $W_{C, \mathcal{R}}(\f)$ in the vicinity
  of a UV fixed point located at a maximum of the potential. For a
  field theory with a given UV value of the source and fixed
  $R^{\textrm{uv}}$ only a finite number of these solutions (here only
  one) can be completed into a flow reaching an IR end point at
  $\f_0$. The shaded region below the blue curve is not accessible for
$R^{\textrm{uv}}\geq 0$.}
\label{fig:flowsintro}
\end{figure}
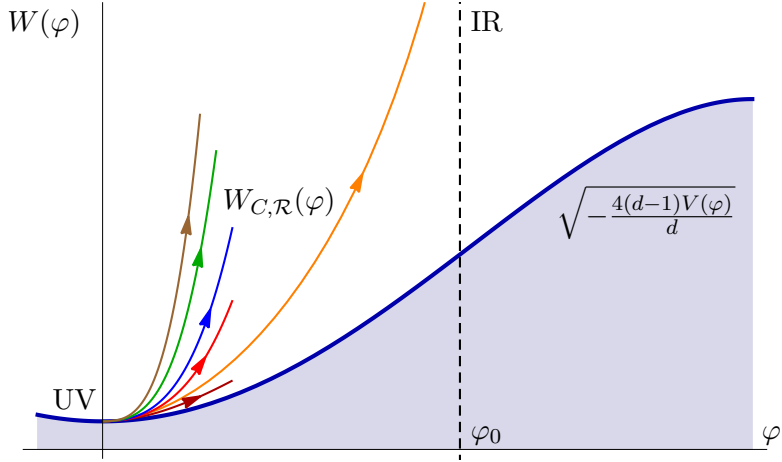

\vspace{0.3cm}
\noindent \textbf{First order formalism (Section \ref{sec:1storder}).}
The connection between the bulk geometry and RG flow in the dual field
theory is made by using a first order formalism adapted to the case
where the induced metric on constant $u$ slices has constant curvature.\footnote{A similar first order
  formalism was developed in \cite{SkenderisTownsend} by defining complex
  superpotentials.  The latter version of the first order formalism is
equivalent to ours, as we show in Appendix \ref{complex}.} One defines  two independent  scalar
functions (superpotentials) $W(\f)$ and $S(\f)$ which satisfy:
\begin{align}
W(\f(u)) \equiv - 2(d-1) \frac{dA(u)}{du} \, , \qquad S(\f(u)) \equiv \frac{d\f(u)}{du} \, ,
\end{align}
and which are solutions of a set of non-linear first order differential
equations \eqref{eq:EOM7} and \eqref{eq:EOM8}.
Assuming the scale
factor plays the same role of energy scale as in the flat case, the
holographic $\beta$-function can be expressed in terms of $W(\f)$ and $S(\f)$:
\begin{align}
\beta(\f) \equiv \frac{d \f}{d A} = -2(d-1)\frac{S(\f)}{W(\f)} \, .
\label{4}\end{align}
The space of regular solutions for $W(\f)$ and $S(\f)$ coincides with the
space of possible RG flows up to a choice of two initial conditions:
(1) The value of the deformation parameter in the UV, $\f_-$, which is
encoded in the leading scalar field asymptotics in equation (\ref{boundaryas}).
It is called the source and corresponds to the UV coupling constant of the relevant operator dual to the scalar.
 (2) The UV value of the scalar curvature $R^{\textrm{uv}}$ of the manifold on which the field theory is defined. Therefore, by classifying all solutions $W(\f)$, $S(\f)$ to the equations  \eqref{eq:EOM7} and \eqref{eq:EOM8} for a given potential $V(\f)$, we can characterize all possible RG flows corresponding to a given bulk theory.

What we find is as follows:

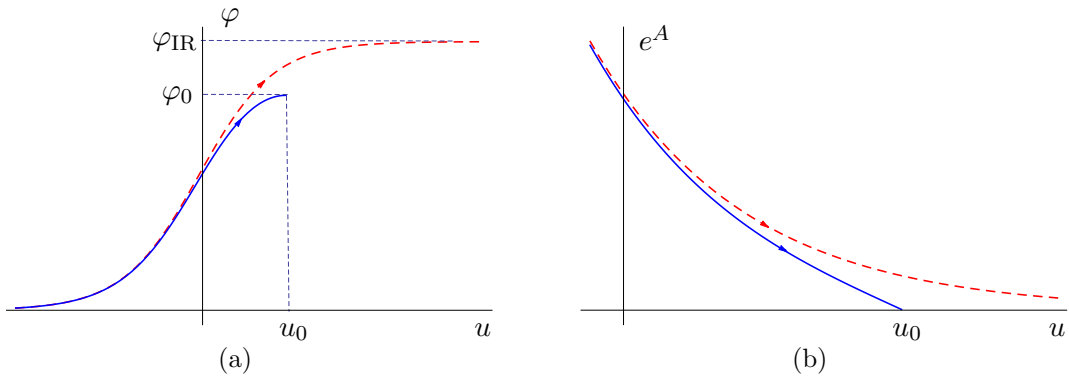
\begin{figure}[t]
\centering
\begin{subfigure}{.5\textwidth}
 \centering
   \begin{overpic}[width=0.85\textwidth,tics=10]{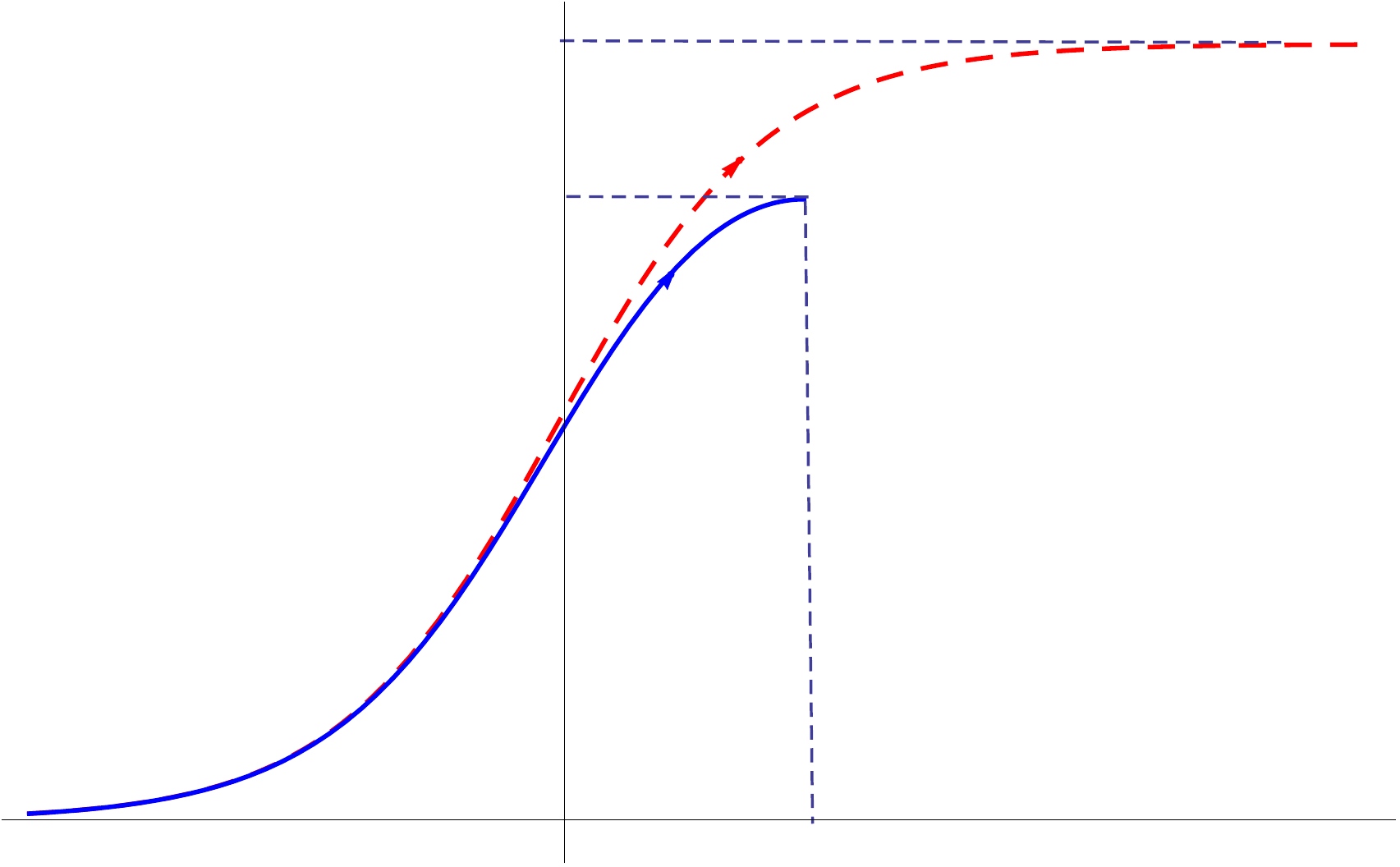}
\put (44,63) {$\f$} \put (30,58) {$\f_{\textrm{IR}}$} \put (32,47) {$\f_{0}$} \put (96,-2) {$u$} \put (56,-2) {$u_0$}
\end{overpic}
 \caption{\hphantom{A}}
  \label{positiveRflowphi}
\end{subfigure}%
\begin{subfigure}{.5\textwidth}
  \centering
 \begin{overpic}[width=0.85\textwidth,tics=10]{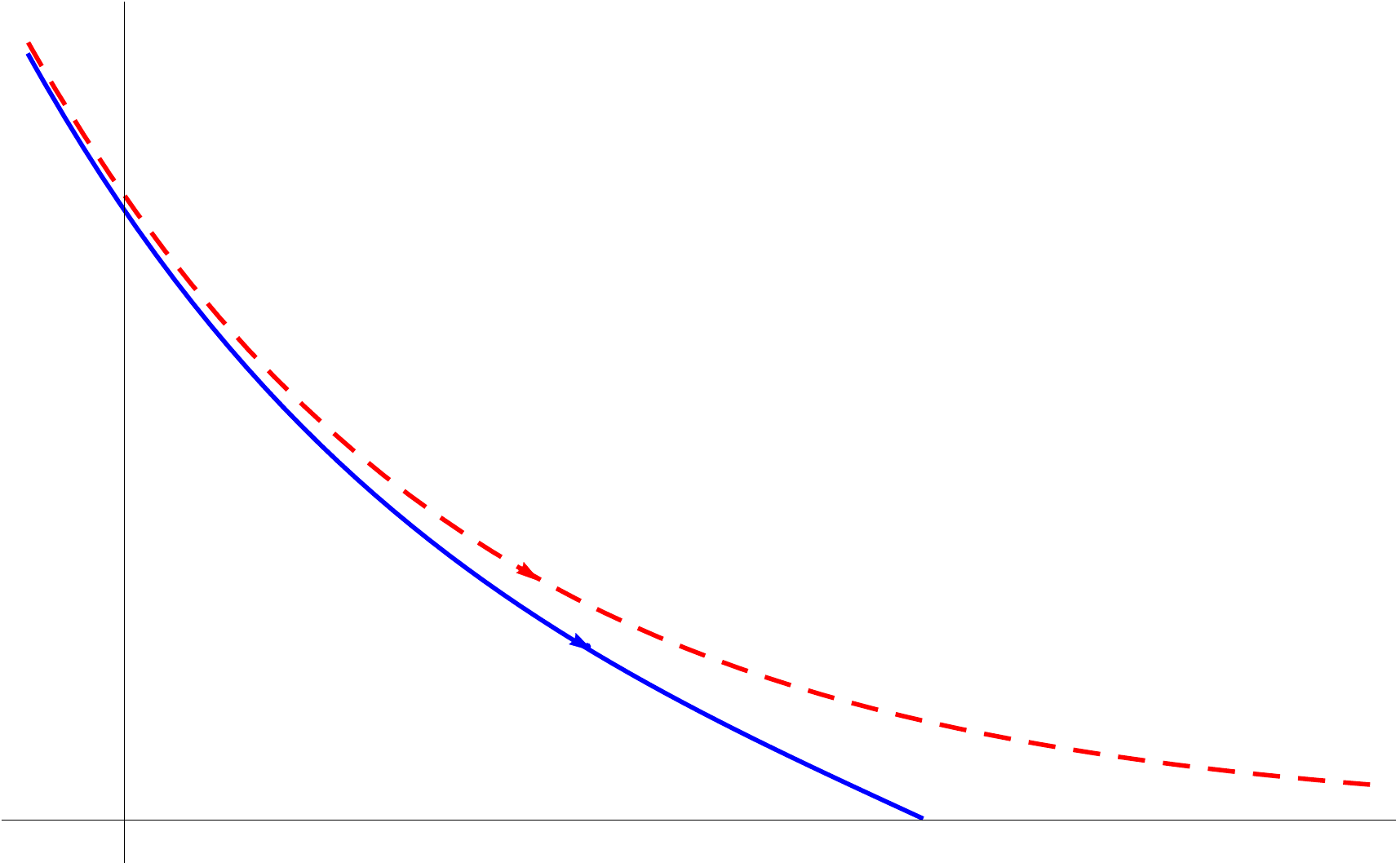}
\put (12,57) {$e^A$} \put (96,-2) {$u$} \put (64,-2) {$u_0$}
\end{overpic}
\caption{\hphantom{A}}
  \label{positiveRflowA}
\end{subfigure}
\caption{The solid lines show the scalar field (left) and scale factor
  (right) profiles of a positive curvature RG flow geometry, from the
  UV ($u\to -\infty$, $\f\to 0$) to the IR endpoint ($u=u_0$, $\f =
  \f_0$). The dashed lines represent the solutions with zero
  curvature, extending all the way to $u\to +\infty$ and to the IR fixed point at $\f=\f_{\textrm{IR}}$.}
\label{positiveRflow}
\end{figure}

\vspace{0.3cm}
\noindent \textbf{UV fixed points (Section \ref{sec:asymp}).} It is a
well-known result from holography  that, when the boundary theory is
in flat space,  UV fixed points are associated
with extrema of the potential $V$. This persists at finite boundary
curvature ($R^{\textrm{uv}}\neq 0$).
\begin{itemize}

\item \textbf{Maxima of $V$:} For maxima of $V$, we find solutions
  describing a relevant deformation away from a UV fixed
  point. Such solutions come as
  \emph{two-parameter families} $W_{C, \mathcal{R}}(\f)$, $S_{C,
    \mathcal{R}}(\f)$, where  $C$  and $\mathcal{R}$ are dimensionless
  parameters related to the vev of the
  deforming operator and to the UV
  value of the scalar curvature $R^{\textrm{uv}}$, respectively. At most a finite subset of these solutions can be
  extended to globally regular solutions corresponding to RG
  flows. This is shown schematically in figure \ref{fig:flowsintro}.

\item \textbf{Minima of $V$:} Flows away from a minimum of $V$ also
  exist  and are driven by the vev of an irrelevant operator
  \cite{exotic}. There is only
  one free parameter $\mathcal{R}$ related to $R^{\textrm{uv}}$.  These solutions  are all \emph{badly singular} in generic theories and can only be extended to globally regular solutions in special cases.
\end{itemize}

\noindent \textbf{IR limit (Section \ref{IRasymp}).} The behavior in the innermost region of the
geometry (far from the UV boundary) depends on the sign of the
curvature $R^{\textrm{uv}}$ of the boundary metric. Recall that, in the flat case, the IR endpoint of a
  holographic RG flow corresponds
to the scalar field asymptoting to a minimum of the bulk potential. In
that case, in the far IR region, the bulk geometry asymptotes to
AdS${}_{d+1}$ in the interior.
\begin{itemize}
\item \textbf{Field theory on manifolds with $R^{\textrm{uv}}>0$ (Section \ref{positivecurvatureflows})}:
  Here, non-singular flows stop at  an IR endpoint, where $e^A
  \rightarrow 0$, $\f$ takes on a finite value $\f_0$, and
  $S(\f)=\dot{\f}=0$.

At the endpoint, the function $W$ diverges as $W \sim |\f -
  \f_0|^{-1/2}$. This  behavior is  shown schematically in figures
  \ref{fig:flowsintro} (superpotential $W$) and \ref{positiveRflow}
  (scalar field and scale factor).
The bulk geometry is regular and  becomes
  approximately  AdS$_{d+1}$ when approaching the IR endpoint. An
  important remark is that, when the field theory curvature is non-vanishing, the IR endpoint {\em cannot}
  be located at a minimum of the scalar potential.

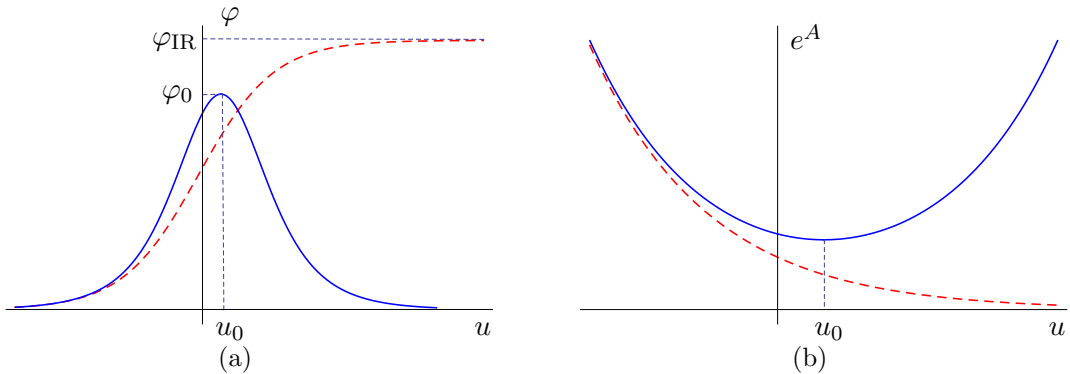
\begin{figure}[t]
\centering
\begin{subfigure}{.5\textwidth}
 \centering
   \begin{overpic}[width=0.85\textwidth,tics=10]{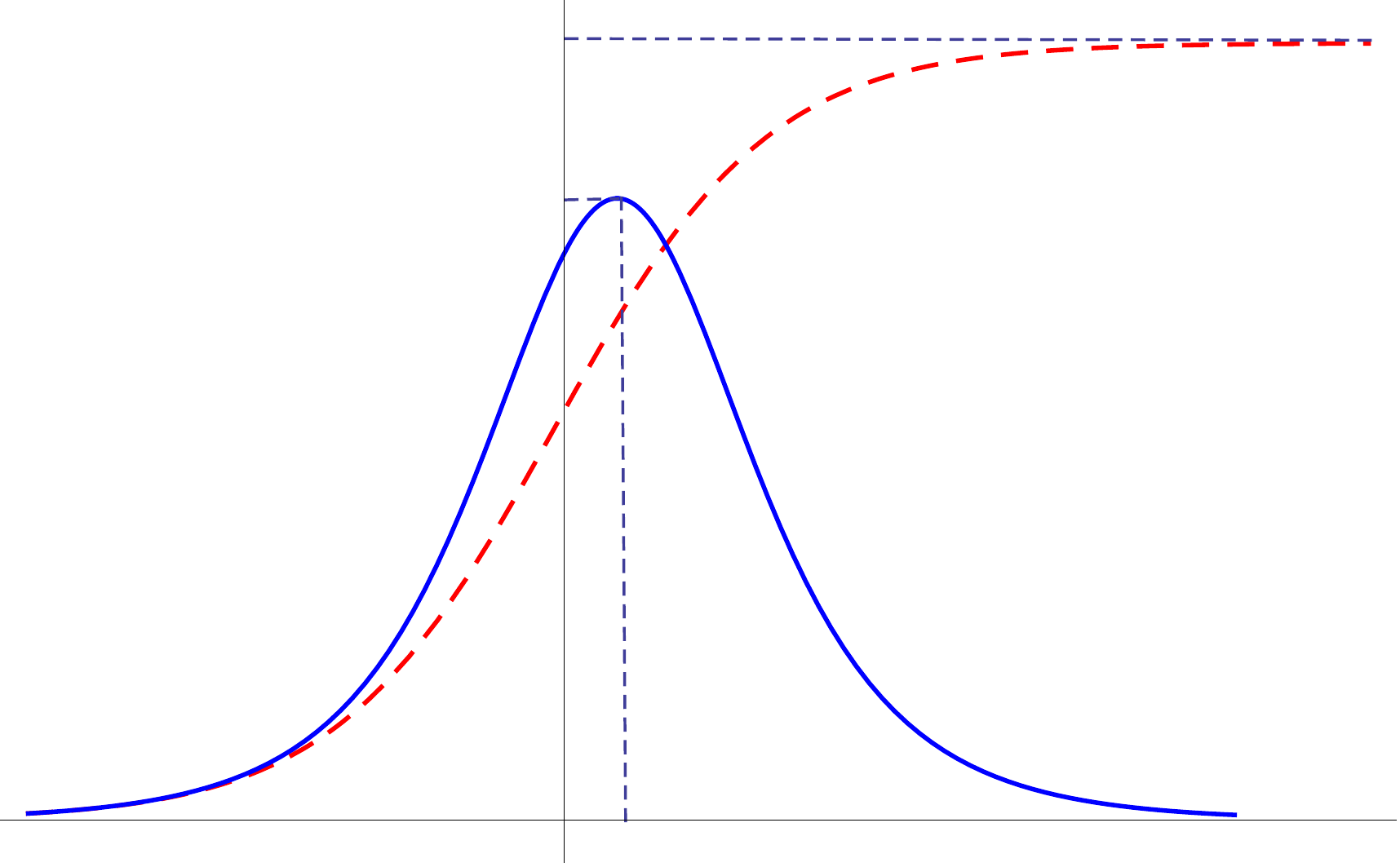}
\put (44,63) {$\f$} \put (30,58) {$\f_{\textrm{IR}}$} \put (32,47) {$\f_{0}$} \put (96,-2) {$u$} \put (43,-2) {$u_0$}
\end{overpic}
 \caption{\hphantom{A}}
  \label{negativeRflowphi}
\end{subfigure}%
\begin{subfigure}{.5\textwidth}
  \centering
 \begin{overpic}[width=0.85\textwidth,tics=10]{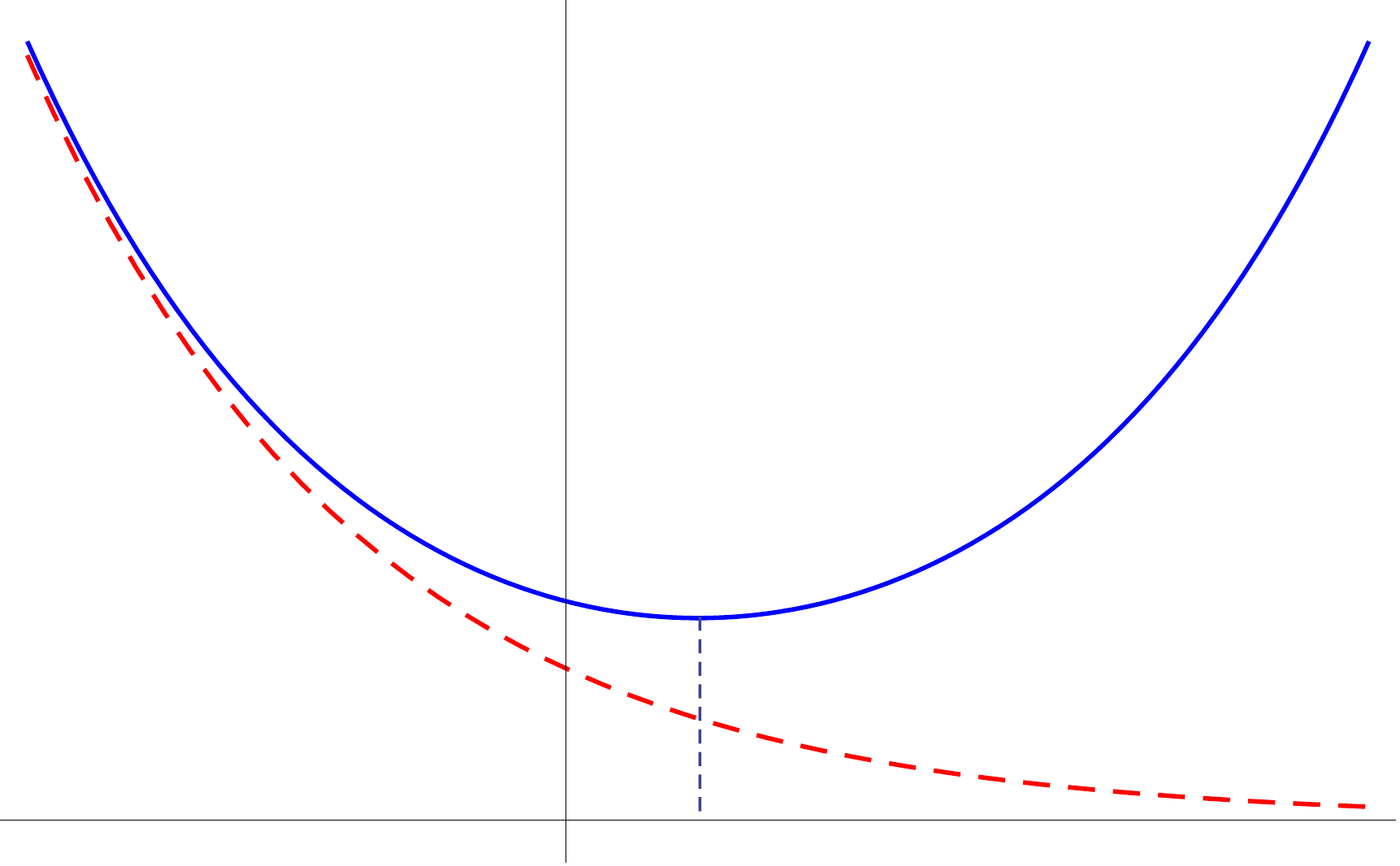}
\put (43,57) {$e^A$} \put (96,-2) {$u$} \put (48,-2) {$u_0$}
\end{overpic}
\caption{\hphantom{A}}
  \label{negativeRflowA}
\end{subfigure}
\caption{The solid lines show the scalar field (left) and scale factor (right) profiles of a
   negative curvature RG flow geometry, from the  left boundary  ($u\to -\infty$,
   $\f\to 0$) to the turning point ($u=u_0$, $\f = \f_0$), to the
   right boundary  ($u\to +\infty$,
   $\f\to 0$). The solution is symmetric around $u_0$.   The dashed lines
   represent the zero-curvature solution interpolating from the UV
   boundary to the IR fixed point at $\f=\f_{IR}$, and featuring a
   monotonic scale factor.}
\label{negativeRflow}
\end{figure}

\item \textbf{Field theory on manifold with $R^{\textrm{uv}} <0$ (\ref{negativecurvatureflows})}: In
  this case,  the flow eventually reaches  a turning point $\f_0$ where both $A(u)$
  and $\f(u)$ invert their direction, but  the value of $e^A$ remains
  finite. These points are characterized by  $S(\f_0) = W(\f_0)
  = 0$.   As $u$ increases past the turning point,  the geometry
  connects back to the boundary of AdS${}_{d+1}$. Therefore, in this case,
  there is no IR endpoint, but rather an IR ``throat'' connecting two
  boundary regions. This behavior is shown schematically  in Figure
  \ref{negativeRflow}.
This situation was already discussed in
  \cite{maldamaoz} in the case of pure gravity. As it was pointed out
  there, the two UV regions are part of the same AdS${}_{d+1}$ boundary,
  and are connected via the AdS${}_d$ boundary of the codimension-one
  slices.
\end{itemize}

\vspace{0.3cm}
\noindent \textbf{Complete non-singular RG flows (Section
  \ref{sec:examples}).}
\vspace{0.1cm}

\noindent In general, any  point $\f_0$  can be an IR end point ($R^{\textrm{uv}}>0$)
or turning point ($R^{\textrm{uv}}<0$), as long as $\f_0$ is
not an extremum of $V$. However, for a given curvature $R^{\textrm{uv}}$, and
for  every value $\f_0$, there is a unique solution corresponding to a
flow ending at $\f_0$ and connecting to the UV boundary.
Such points, for $R^{\textrm{uv}}>0$, are the fixed points of the flow, where the associated $\beta$-function, defined in (\ref{4}), vanishes as discussed in section \ref{sec:betafunc}

We construct  several examples of curved holographic RG flow solutions
that start from a UV fixed point at a maximum of the potential,  and
have a regular interior where the geometry displays the general features
described in the previous paragraphs.

The value  $\f_{0}$ (be it an endpoint or a turning point) for a given regular flow is determined by the values of
$R^{\textrm{uv}}$ and $\f_-$ governing the leading UV asymptotics in
equation (\ref{boundaryas}). More precisely, the value $\f_{0}$ is
completely determined by the  dimensionless combination
  ${\mathcal R} \equiv R^{\textrm{uv}} |\f_-|^{2/(d-\Delta)}$, where $\Delta$ is the
  dimension of the dual operator corresponding to $\f$.


In particular, the following observations hold for generic
potentials and either sign of the curvature:  \emph{Increasing} $|R^{\textrm{uv}}|$ while keeping the UV
source $\f_-$ fixed causes the IR endpoint (or turning
point) $\f_{0}$ to move closer to the starting point of the flow at the
UV maximum. Conversely, \emph{decreasing} $|R^{\textrm{uv}}|$ while keeping the the UV
source $\f_-$ fixed  causes $\f_{0}$ to move away from the starting point of the flows at the UV maximum. For $R^{\textrm{uv}} \rightarrow 0$ the IR endpoint (or turning point) approaches a minimum of $V$.

\begin{figure}[t]
 \centering
\begin{overpic}
[width=0.45\textwidth]{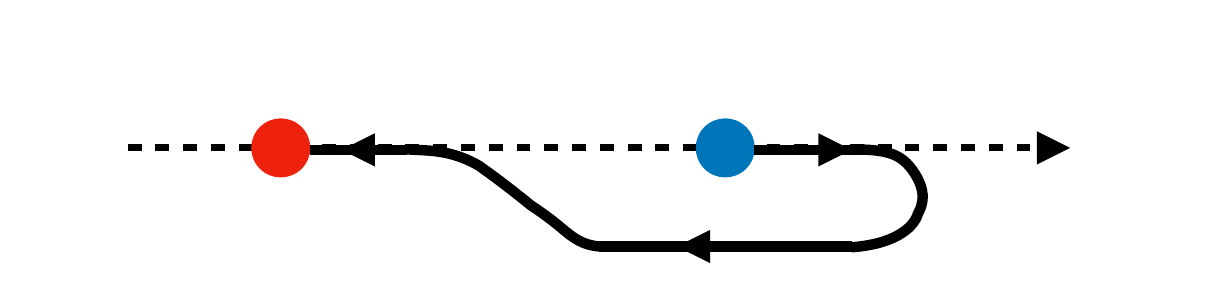}
\put (85,17) {$\f$} \put (55,17) {UV} \put (20,17) {IR}
\end{overpic}
\caption{Schematic structure of a field theory with an RG flow
    exhibiting a ``bounce''.}
\label{fig:bounceschematic-intro}
\end{figure}

In the simplest RG flow solutions, the scalar field is monotonic
between the UV fixed point and the IR critical point $\f_0$, which
typically is reached before the (would-be flat) IR fixed point (as in
figure \ref{positiveRflowphi}).  This is, however,
not always the case.  For example, already in the flat case it was
shown in  \cite{exotic}  that
holographic RG flows may exhibit phenomena which seem exotic from the
point of view of QFT perturbation theory. For one, holographic RG flows
may \emph{bounce}, i.e.~the flow may reverse direction in coupling
constant space (see fig.~\ref{fig:bounceschematic-intro}) without stopping.

In this work we find that flows exhibiting bounces  persist for field
theories on curved manifolds. Interestingly, for field theories on
AdS${}_4$ we find that bounces are generic features of RG flows if the
UV operator dimension is $3<\Delta<4$. This is unlike the
situation in the flat or positively curved case, where this does not
occur generically.

\vspace{0.3cm}
\noindent \textbf{Quantum Phase Transitions (Section \ref{QPT}).}
\vspace{0.1cm}

\noindent A field theory on a curved manifold, defined by a value of the UV source
$\f_-$  and of the UV curvature $R^{\textrm{uv}}$ may display several
saddle points, which in the holographic dual framework correspond to different
RG flow geometries. The question then remains which one of the ground states is the true vacuum of the theory. This can be determined by calculating the free energies of the flows corresponding to the various saddle points: the solution with the lowest free energy is the true vacuum.

If we vary the value of $R^{\textrm{uv}}$, the RG flows deform and the
vevs and free energies of the various vacua change. Interestingly,
under a variation of $R^{\textrm{uv}}$ the identification of the true
vacuum may change, i.e.~the system may exhibit  phase transitions. This corresponds to a {\em quantum} phase transition as our system is at zero temperature.\footnote{A richer structure may be obtained at finite temperature, but we will not pursue this here.} The control parameter in this case is the curvature $R^{\textrm{uv}}$.

\begin{figure}[t]
 \centering
\begin{overpic}
[width=0.45\textwidth]{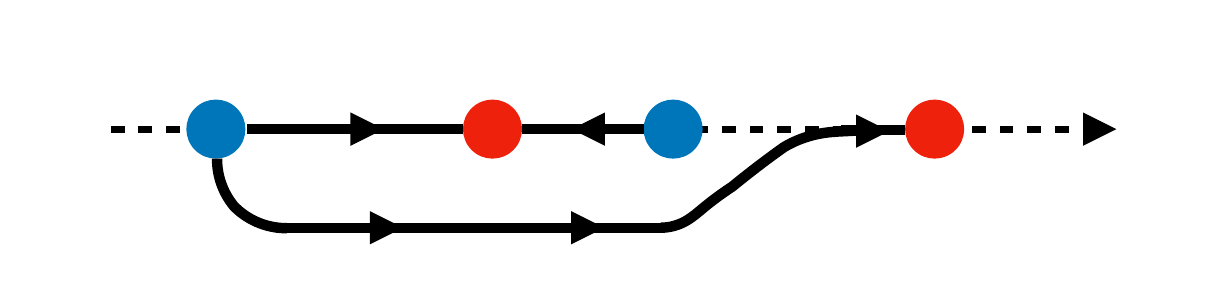}
\put (89,17) {$\f$} \put (13.5, 17) {UV${}_1$} \put (37, 17) {IR${}_1$} \put (51.5, 17) {UV${}_2$} \put (74, 17) {IR${}_2$}
\end{overpic}
\caption{Schematic structure of a field theory which presents multiple
  RG-flows: In particular, there are two flows starting at the fixed
  point UV$_1$, one going to the closer IR fixed point IR$_1$, the
  second skipping IR${}_1$ and ending at  IR${}_2$.}
\label{fig:skipschematic}
\end{figure}
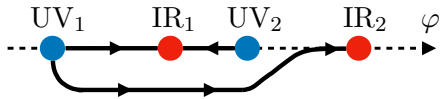

We find further evidence that curvature-driven first order quantum phase transitions {\em do} arise in a
holographic framework. Previously, this was observed holographically for supersymmetric field theories on $\mathbbm{R} \times S^3$ in \cite{1108.6070}. In Section \ref{QPT} we present an explicit
example that exhibits this phenomenon\footnote{A somewhat similar
  phenomenon, although one cannot talk about a proper quantum phase
  transition, was already observed since the very early days of holography in
  \cite{witten-thermo} for $N=4$ SYM: the flat-space theory is in the
  deconfined phase at arbitrarily low temperature, whereas the theory
  on $\mathbbm{R} \times S^3$ is in the confined phase up to a finite
 temperature set by the curvature scale}. The  example we discuss
here is based on a potential which was already studied in \cite{exotic} in the
flat case. Its main feature is to allow two regular flows in the
zero-curvature case, one connecting two neighboring (UV and IR) fixed
points, the second {\em skipping} the closest available IR fixed point
and ending at a fixed point at larger field value. This is represented
schematically in fig.~\ref{fig:skipschematic}.

In \cite{exotic}, it was
found that, for the theory in flat space, the preferred ground state
is the skipping solution. Turning on a small positive curvature, the
skipping and non-skipping solutions persist, their free-energy
difference decreases up to a critical value ${\mathcal R}_c$ where it
changes sign: for ${\mathcal R} >  {\mathcal R}_c$ the dominant saddle
point is now the shorter (non-skipping) flow. Finally,  skipping flows
disappear beyond a finite  value larger than ${\mathcal R}_c$ (see
figure \ref{phasetransition}).

This is the typical structure of a first order phase transition, where
a stable phase becomes unstable then disappears, as a function of a
control parameter. Usually these kinds of phase transitions are driven
by temperature (e.g.~Hawking-Page transition for global
AdS-Schwarzschild black holes). Here it takes place at zero temperature
and is driven by one of the ``couplings'' (the curvature is a
non-normalizable  deformation in the AdS/CFT language) of the UV theory.

\subsection{Open questions and outlook}

From the point of view of  field theory, this work offers a new
(holographic) insight on RG flows on curved manifolds. In particular,
we have seen that positive curvature flows end at points of maximal
symmetry.  From the gravity side, this symmetry is just the full set of AdS
isometries. However, it is unclear how this symmetry is realized on the field
theory side, since (unlike in the case of flat IR fixed points) it does
not obviously reduce to conformal transformations (or rather conformal
isometries) on the field theory coordinates. It would be interesting
to investigate this further.

One important question is whether  the solutions discussed in
this work are perturbatively stable. In the case of flat sections, it
was shown in \cite{exotic} that, regardless of the details of the bulk
geometries, solutions which reach an IR fixed point are stable under
small perturbations. This includes bouncing solutions, as the bounce
does not introduce any particular features in the fluctuation
equations which may trigger instabilities. In the curved case the
situation is less clear. To reach the same conclusion one would have
to develop a complete fluctuation analysis around the ansatz
(\ref{eq:flowintro}). This is beyond the scope of this
work. Nevertheless, in section \ref{sec:stability} we present a
general discussion which allows to draw some partial conclusions and
suggests that perturbative stability is preserved when we introduce
a non-vanishing curvature.

Another important question concerns the boundary conditions and stability for
solutions in which the field theory is on a negatively curved
space. When the field theory space-time is AdS$_d$, it turns out that
the solution is not single valued on the
boundary, but it  must contain a defect along the portion  of
the  AdS$_{d+1}$ boundary which corresponds to the AdS$_d$ boundary
(see the discussion in section 4.4). This can be avoided by quotienting
the AdS$_d$ slices by a finite group, but
as argued in \cite{maldamaoz}, this may introduce perturbative
instabilities. It would be interesting to understand these issues in
more detail.

Another interesting application of our analysis is in the context of $3$-dimensional field theories, and the associated
$F$-theorem \cite{1103.1181}. The original definition of the $F$-theorem is in terms of the free energy $F=-\ln Z_{S^3}$ of a field theory on $S^3$. However, it is an entanglement entropy \cite{1006.1263,1011.5819} (in particular the renormalized entanglement entropy of \cite{1202.2070}) rather than the free energy which gives rise to a suitable $F$-function, which decreases monotonically from UV to IR \cite{1202.5650}. An interesting open question is whether a suitable $F$-function can be defined using RG flows in curved space-times. In holography, this question was addressed in \cite{Taylor} in the large curvature limit and it was shown that the free energy on $S^3$ does in general not give rise to a suitable $F$-function. We will extend this analysis in an upcoming publication \cite{Ftheorem} and describe how a monotonic $F$-function can be recovered from holographic RG flows on $S^3$.

As we have mentioned earlier, one of the original motivations for this
work was to find curved space-time solutions (eventually
time-dependent) in the self-tuning framework developed in
\cite{self}. In that case, the interesting solutions would be two
distinct RG flow solutions (one containing the UV, the other
containing an IR region) glued by a domain wall in the transverse
direction. The study of the junction conditions in \cite{self} will
then be extended to the curved case, and will give an indication of
the dynamics of the domain wall away from the flat ground state
solution.

Finally, it would be interesting to study examples of Einstein-dilaton
gravity where the flat space-time field theory is confining, as those
described in \cite{0707.1349},  and
investigate the phase structure (and possible phase transitions) in
the presence of curvature (see also \cite{1007.3996} for a related discussion
of holographic confining theories on space-times with non-zero curvature).

\section{Holographic space-times with curved slicing and RG flows}

Our object of study is an Einstein-scalar theory in $(d+1)$-dimensions
with Euclidean or Lorentzian metric (in the latter case we use the
convention $(-,+,+, \ldots, +)$). In the Lorentzian case, it is  described by the following two-derivative action:
\begin{align}
\label{eq:action} S[g, \f]= \int du \, d^ dx \, \sqrt{|g|} \left(R^{(g)} - \frac{1}{2} \partial_a \f \partial^a \f - V(\f) \right) + S_{GHY} \, ,
\end{align}
where we also included the Gibbons-Hawking-York term $S_{GHY}$. The
Euclidean action is defined by setting $S_E = - S$ and changing the
metric to positive signature.

In this work we will be interested in boundary field theories defined
on curved maximally symmetric $d$-dimensional space-times. Then,
without loss of generality, we can employ domain wall coordinates and
choose the following ansatz for $\f$ and the $(d+1)$-dimensional
metric (for both Euclidean and Lorentzian signatures):
\begin{align}
\label{eq:metric} \f = \f(u) \, , \qquad ds^2 = du^2 + e^{2 A(u)} \zeta_{\mu \nu} dx^{\mu} dx^{\nu} \, .
\end{align}
Here, $A(u)$ is a scale factor that depends on the coordinate $u$ only, while $\zeta_{\mu \nu}$ is a metric describing a $d$-dimensional maximally symmetric space-time. As a consequence of maximal symmetry we have\footnote{In the case of a foliation with positive curvature slices, these can be given by $d$-dimensional de Sitter space or a $d$-sphere. In the rest of the paper we will mainly refer to the case of the sphere, keeping in mind that the results will also hold for de Sitter.}
\begin{align}
\label{eq:Rzeta}
R^{(\zeta)}_{\mu \nu} = \kappa \zeta_{\mu \nu} \, , \quad R^{(\zeta)} = d \kappa \, , \quad \textrm{with} \quad \kappa = \left\{
  \begin{array}{c l}
   \hphantom{-} \frac{(d-1)}{\alpha^2} & \quad \textrm{dS}_d~~{\rm or} ~~S^d\\
   0 & \quad \mathcal{M}_d \\
- \frac{(d-1)}{\alpha^2} & \quad \textrm{AdS}_d\\
  \end{array} \right. \, \ ,
\end{align}
where $\alpha$ is the curvature length scale and $\mathcal{M}_d$ denotes $d$-dimensional Minkowski space. We also define the induced metric $\gamma_{\mu \nu}$ on a $d$-dimensional slice at constant $u$ which is given by
\begin{align}
\label{eq:indmet}
\gamma_{\mu \nu} \equiv e^{2 A(u)} \zeta_{\mu \nu} \, .
\end{align}

In the following, we will also adhere to the following shorthand notation. Derivatives with respect to $u$ will be denoted by a dot while derivatives with respect to $\f$ will be denoted by a prime, i.e.:
\begin{align}
\dot{f}(u) \equiv \frac{d f(u)}{du} \, , \qquad g'(\f) \equiv \frac{d g(\f)}{d \f} \, .
\end{align}

Varying the action \eqref{eq:action} with respect to the metric and the scalar $\f$ gives rise to the equations of motion:
\begin{align}
\label{eq:EOM1} 2(d-1) \ddot{A} + \dot{\f}^2 + \frac{2}{d} e^{-2A} R^{(\zeta)} &=0 \, , \\
\label{eq:EOM2} d(d-1) \dot{A}^2 - \frac{1}{2} \dot{\f}^2 + V - e^{-2A} R^{(\zeta)} &=0 \, , \\
\label{eq:EOM3} \ddot{\f} +d \dot{A} \dot{\f} - V' &= 0 \, .
\end{align}
These equations are the same for both Lorentzian and Euclidean
signatures, so all our results will hold for both cases (unless stated explicitly).

Holographic RG flows are in one-to-one correspondence with regular
solutions to the equations of motion
\eqref{eq:EOM1}--\eqref{eq:EOM3}. Hence, in the following we will be
interested in solutions to these equations for various choices of the
potential $V(\f)$.

To be specific, we will assume that $V(\f)$  has at least one
  maximum, where $V$ takes a negative value. This ensures that there exists a
conformal fixed point, and a family of asymptotically AdS
solutions which correspond to deforming the theory away from the fixed
point by a relevant operator.

 In addition, $V(\f)$ may have other
maxima and/or minima representing distinct UV or IR fixed points for
the dual CFT. The aim of the next subsection is to study the solutions
{\em at} the fixed points in the presence of curvature, but with the
relevant deformation set to zero.

\subsection{Conformal fixed points}
\label{sec:AdSfixedpoints}
A conformal fixed point corresponds to a solution of
\eqref{eq:EOM1}--\eqref{eq:EOM3} with $\f=\textrm{const.}$,
i.e.~$\dot{\f}=0=\ddot{\f}$.  Such solutions are
associated with extrema $\f_{\textrm{ext}}$ of the potential, at which
$V'(\f_{\textrm{ext}})=0$.

In this case, the solution
(\ref{eq:metric}) is always the space-time AdS${}_{d+1}$, written in different coordinate
systems, regardless of the curvature of the $d$-dimensional slices
with metric $\zeta_{\mu \nu}$. Indeed,
AdS${}_{d+1}$ admits $d$-dimensional de Sitter (or sphere, in the Euclidean version), Minkowski or AdS slicings of the form
(\ref{eq:metric}), with (see e.g.~\cite{KarchRandall}):
\begin{align}
\label{eq:AdSScaleFactor}
e^{A(u)} = \left\{
  \begin{array}{l l l}
   \displaystyle{\frac{\ell}{\alpha} \sinh \left(-\frac{u+c}{\ell}\right)}, & \qquad
   -\infty < u \leq -c,  &
   \qquad \textrm{dS}_d \textrm{ or } S^d\\
& \\
   \displaystyle{\exp\left(- \frac{u+c}{\ell}\right)}, &  \qquad -\infty < u <
   +\infty, &  \qquad \mathcal{M}_d \\
& \\
 \displaystyle{
   \frac{\ell}{\alpha} \cosh
   \left(\frac{u+c}{\ell}\right)}, & \qquad -\infty < u <
   +\infty, &  \qquad \textrm{AdS}_d \\
  \end{array} \right. \, \ .
\end{align}
Here we introduced the AdS${}_{d+1}$ length $\ell$ which is defined via
$V(\f_{\textrm{ext}})=-\frac{d(d-1)}{\ell^2}$, while the length scale
$\alpha$ was introduced in \eqref{eq:Rzeta}. We also chose the
boundary of AdS${}_{d+1}$ to be located at $u \rightarrow
-\infty$. The parameter $c$ is an integration constant.

 Although the bulk space-time is the same, the asymptotic boundary is different in
the three cases. This   leads to inequivalent boundary
theories. As the $d$-dimensional boundaries of the space-times
(\ref{eq:AdSScaleFactor}) are all conformally equivalent, at the fixed
point  the effect of curvature is completely encoded in the conformal
anomaly. This will change when we consider RG-flows and introduce an
explicit breaking of conformal invariance.

In the case of Minkowski slices, $c$ can be removed by a conformal rescaling of the boundary metric and we can set $c=0$. However, for dS and AdS slices the constant $c$ contains physical information and cannot be removed. In particular, $c$ determines the curvature of the $d$-dimensional slices.
To see this, first note that the length scale $\alpha$ is unphysical and can be chosen freely. It is an artifact of splitting the induced metric $\gamma_{\mu \nu}$ on the $d$-dimensional slices into a scale factor $e^{A(u)}$ and $\zeta_{\mu \nu}$. This can be seen e.g.~by evaluating $R^{(\gamma)}$, i.e.~the scalar curvature associated with the induced metric $\gamma_{\mu \nu}$ on the $d$-dimensional slices, which is given by
\begin{align}
R^{(\gamma)} = e^{-2A(u)} R^{(\zeta)} \, .
\end{align}
For a conformal fixed point we hence find
\begin{align}
R^{(\gamma)} = \left\{ \begin{array}{c l}
  \displaystyle{\hphantom{-}  \frac{d(d-1)}{\ell^2} \sinh^{-2}
    \frac{u+c}{\ell}} & \quad \textrm{dS}_d \textrm{ or } S^d
  \\
& \\
\displaystyle{ -\frac{d(d-1)}{\ell^2} \cosh^{-2} \frac{u+c}{\ell}} & \quad \textrm{AdS}_d \\
  \end{array} \right. \, ,
\end{align}
i.e.~the parameter $\alpha$ has dropped out from the expressions. Instead, it is the parameter $c$ which contains information about the curvature.

An important quantity will be the scalar curvature $R^{\textrm{uv}}$,
which we define as the scalar curvature associated  with the UV limit
of the (rescaled) induced metric $\gamma^{\textrm{uv}}_{\mu \nu} \equiv \lim_{u
  \rightarrow -\infty} e^{2u/ \ell} \gamma_{\mu \nu}$.  Indeed,
  the metric $\gamma^{\textrm{uv}}_{\mu\nu}$ is the leading boundary data
  appearing in the Fefferman-Graham expansion near the boundary, and
  in holography it is interpreted as the metric of the dual  field
  theory. The associated curvature  $R^{\textrm{uv}}$  is given by:
\begin{align}
\label{eq:RUVdef}
R^{\textrm{uv}} = \lim_{u \rightarrow -\infty} e^{-2u/ \ell} R^{(\gamma)} \, ,
\end{align}
For the case of a conformal fixed point we  find
\begin{align}
R^{\textrm{uv}} = \pm \frac{4d(d-1) }{\ell^2} \, e^{2c/\ell} \, ,
\end{align}
where the positive (negative) sign is appropriate for dS (AdS) curvature.

For later use, it will be convenient to expand the scale factor \eqref{eq:AdSScaleFactor} in the vicinity of the boundary $u \rightarrow -\infty$. Restricting attention to the cases of $S^d$/AdS${}_d$ slicings we obtain
\begin{align}
\nonumber A(u) &= \ln \Big( \frac{\ell}{2 \alpha} \Big) - \frac{u+c}{\ell} \mp e^{2(u+c)/ \ell} + \mathcal{O}(e^{4u/ \ell}) \\
\label{eq:AdSScaleFactorExpansion} &=\frac{1}{2} \ln \Big(\frac{R^{(\zeta)}}{R^{\textrm{uv}} } \Big) - \frac{u}{\ell} - \frac{\ell^2 R^{\textrm{uv}}}{4d(d-1)} \, e^{2u/ \ell} + \mathcal{O}(e^{4u/ \ell})
\end{align}
where the last line is valid for both $S^d$/AdS${}_d$ slicings.

For most practical purposes it will be convenient to set $R^{(\zeta)} = R^{\textrm{uv}}$, which we can always achieve by choosing $\alpha= \tfrac{\ell}{2} \, e^{-c/\ell}$ for the arbitrary parameter $\alpha$. This is the convention that we will adopt throughout this paper. Note that in this case the constant term in \eqref{eq:AdSScaleFactorExpansion} vanishes.
This also holds for space-times which only asymptote to AdS${}_{d+1}$
near the boundary: We can always ensure that $R^{(\zeta)} =
R^{\textrm{uv}}$ for an asymptotically AdS space-time by setting the
constant term in the near-boundary expansion of $A(u)$ to zero.

\vspace{0.3cm}
\noindent{\bf Summary.} A solution of the form (\ref{eq:metric}) in which $\f$
is fixed to an extremum of the scalar
potential is dual to a CFT living on a manifold whose curvature can be
arbitrary, and depends on the choice of the leading term in the
boundary asymptotics (\ref{eq:AdSScaleFactorExpansion}).
As a final remark, the definition \eqref{eq:RUVdef} of
$R^{\textrm{uv}}$ is also valid for asymptotically AdS${}_{d+1}$
space-times, i.e.~space-times where $A(u)$ approaches the form
\eqref{eq:AdSScaleFactor} at the boundary $u \rightarrow - \infty$,
but may depart from that form in the interior.

\subsection{The first order formalism} \label{sec:1storder}
We now turn our attention to solutions where the scalar field has a
non-trivial dependence on the coordinate $u$.
Our  goal is to interpret  solutions to the equations \eqref{eq:EOM1}--\eqref{eq:EOM3} in terms of RG flows. 

To this end it will be convenient to rewrite the second-order Einstein
equations as a set of first-order equations, which will allow an
interpretation as gradient RG flows. This is locally always possible,
except at special points where $\dot{\f}=0$, which we will later refer
to as bounces, as previously observed in \cite{exotic}. Given a
solution,  as long as
$\dot{\f}(u)\neq 0$, we can invert the relation between $u$ and
$\f(u)$ and define the following {\em scalar} functions of $\f$:
We define,
\begin{align}
\label{eq:defWc} W(\f) & \equiv -2 (d-1) \dot{A} \, , \\
\label{eq:defSc} S(\f) & \equiv \dot{\f} \, , \\
\label{eq:defTc}  T(\f) & \equiv e^{-2A} R^{(\zeta)} = R^{(\gamma)} \, .
\end{align}
where the expressions on the right hand side are evaluated at
$u=u(\f)$.

In terms of the functions defined above, the equations of motion \eqref{eq:EOM1} --\eqref{eq:EOM3} become
\begin{align}
\label{eq:EOM4} S^2 - SW' + \frac{2}{d} T &=0 \, , \\
\label{eq:EOM5} \frac{d}{2(d-1)} W^2 -S^2 -2 T +2V &=0 \, , \\
\label{eq:EOM6} SS' - \frac{d}{2(d-1)} SW - V' &= 0 \, ,
\end{align}
which are  coordinate independent, first-order non-linear
differential equations in field space. In the flat case $T=0$
  and we recover the usual superpotential equation for $W$ by setting
  $S = W'$. Thus, the difference between $S$ and $W'$ is a measure of
  the curvature.

Note that  equations (\ref{eq:EOM4}-\ref{eq:EOM6}) are algebraic in $T$. We can hence partially solve this system by eliminating $T$ so that we are left with the following two equations
\begin{align}
\label{eq:EOM7} \frac{d}{2(d-1)} W^2 + (d-1) S^2 -d S W' + 2V &=0 \, , \\
\label{eq:EOM8} SS' - \frac{d}{2(d-1)} SW - V' &= 0 \, .
\end{align}
Next,  we can solve the second equation \eqref{eq:EOM8} for $W$ algebraically. Substituting into the first equation \eqref{eq:EOM7} we obtain
\be
 dS^3S''-{d\over 2}S^4  -S^2 (S')^2-{d\over d-1}S^2V +{(d+2)}SS'V' -dS^2V''-(V')^2 =0
\label{eq:EOM9}
\ee
This is a second order equation in $S$ and its integration requires
two integration constants. As we will see in the next section,  one
integration constant will be related to the vev of the perturbing
operator, while the other integration constant will be related to the UV curvature.

In the following we will work both with the full set of equations
\eqref{eq:EOM4}--\eqref{eq:EOM6}, with the two equations
\eqref{eq:EOM7}--\eqref{eq:EOM8} or with \eqref{eq:EOM9} choosing
whichever is more convenient.

A few important properties of the functions $W$, $S$ and $T$ are discussed below
(see appendix \ref{properties} for more details):
\begin{enumerate}
\item At a generic point in field space, there exist two branches of
  solutions, with opposite signs of $S$ and $W'$. In each branch,  $S$
  and $W'$ have the same sign.
\item We define the {\em critical curve} $B(\f)$ as:
\be \label{criticalB}
B(\f)=\sqrt{-\frac{4(d-1)}{d}V(\f)}
\ee
In the flat case,  $W(\f)$ has to satisfy $|W(\f)| \geq B(\f)$, and
equality can be reached only where $W'=0$.
For  positive non-zero curvature, the bound is stricter, and the
critical curve cannot be reached, except in the UV where $T \rightarrow 0$. On the other hand, for negative
curvature $W$ can cross the critical curve.
\item Curvature invariants are finite as long as $S(\f)$ and $V(\f)$
  are both finite. On the other hand a divergent $W(\f)$  does
  not necessarily imply that the solution is singular.
\item Points where $S \to 0$ (i.e.~$\f(u)$ has an extremum) correspond to points of enhanced symmetry:
  In fact, around these points the metric is approximately maximally
  symmetric, i.e.
\be\label{maxsym}
S(\f_*) = 0 \quad \Rightarrow \quad R_{ab} =  {V(\f_*) \over d-1}  g_{ab}.
\ee
This has to be
  confronted with generic points, where $R^{(g)}_{\mu\nu} = k g_{\mu\nu}$,
  $R^{(g)}_{uu} = k' g_{uu}$, with $k'-k = S^2/2$.
\end{enumerate}

\section{Perturbative analysis near extrema of the potential}
\label{sec:asymp}
 We will now examine solutions of our system in the vicinity of extremal points of $V$. We then proceed to determining the near-boundary geometry. Without loss of generality we take the extremum to be at $\f=0$. It will then be sufficient to consider the potential
\begin{align}
\label{eq:Vextremum} V = - \frac{d(d-1)}{\ell^2} - \frac{m^2}{2} \f^2
+ \mathcal{O}(\f^3) \, ,
\end{align}
where we will choose $m^2 >0$ for maxima and $m^2 <0$ for minima. In the following we will solve equations \eqref{eq:EOM4}--\eqref{eq:EOM6} for $W(\f)$, $S(\f)$ and $T(\f)$ near $\f=0$. The relevant calculations are performed in appendix \ref{app:smallRexp}. Here we present and discuss the results.

\subsection{Expansion near maxima of the potential}
\label{sec:maximaofV}
We work in an expansion in $\f$ about the maximum at $\f=0$. Like in the
case of zero boundary curvature discussed in e.g.~\cite{exotic}, there
are two branches of solutions to equations \eqref{eq:EOM4}--\eqref{eq:EOM6}, and we will distinguish them by the subscripts $(+)$ and $(-)$. The $(-)$ solutions are:
\begin{align}
\nonumber W_{-}(\f) & = \frac{1}{\ell} \left[2(d-1) + \frac{\Delta_-}{2} \f^2 + \mathcal{O}(\f^3) \right] + \frac{\mathcal{R}}{d \ell} \, |\f|^{\frac{2}{\Delta_-}} \ [1+ \mathcal{O}(\f) + \mathcal{O}(\mathcal{R})] \\
\label{eq:Wmsol} & \hphantom{=} \,  + \frac{C}{\ell} \, |\f|^{\frac{d}{\Delta_-}} \ [1+ \mathcal{O}(\f)+ \mathcal{O}(C) + \mathcal{O}(\mathcal{R})] \, , \\
 \nonumber \\
\label{eq:Smsol} S_{-}(\f) & = \frac{\Delta_-}{\ell} \f \ [1+ \mathcal{O}(\f)] + \frac{Cd}{\Delta_- \ell} \, |\f|^{\frac{d}{\Delta_-}-1} \ [1+ \mathcal{O}(\f) + \mathcal{O}(C)] \, , \\
\nonumber & \hphantom{=} \,  + \frac{1}{\ell} \mathcal{O}\left( \mathcal{R} |\f|^{\frac{2}{\Delta_-}+1} \right) + \frac{1}{\ell} \mathcal{O}\left(\mathcal{R} C |\f|^{\frac{2+d}{\Delta_-}-1} \right) \\
\nonumber \\
\label{eq:Tmsol} T_{-}(\f) &= \ell^{-2} \, \mathcal{R} \, |\f|^{\frac{2}{\Delta_-}} [1+ \mathcal{O}(\f) + \mathcal{O}(C) + \mathcal{O}(\mathcal{R})] \, ,
\end{align}
where $C$ and ${\mathcal R}$ are integration constants, and we have
defined:
\be
\label{eq:Deltadef} \Delta_{\pm}  = \frac{1}{2}\left( d \pm \sqrt{d^2-  4 m^2 \ell^2} \right) \, \qquad \textrm{with} \quad 0 < m^2 < \frac{d^2}{4 \ell^2} \,
\ee
The $(+)$ solution is given by:
\begin{align}
\label{eq:Wpsol} W_{+}(\f) & = \frac{1}{\ell} \left[2(d-1) + \frac{\Delta_+}{2} \f^2 + \mathcal{O}(\f^3) \right] + \frac{\mathcal{R}}{d \ell} \,
|\f|^{\frac{2}{\Delta_+}} \ [1+ \mathcal{O}(\f) +
\mathcal{O}(\mathcal{R})] \, , \\
\nonumber \\
\label{eq:Spsol} S_{+}(\f) & = \frac{\Delta_+}{\ell} \f \ [1+
\mathcal{O}(\f)] + \mathcal{O}\left( \mathcal{R}
  |\f_-|^{\frac{2}{\Delta_+}+1} \right) \, ,\\
\nonumber \\
\label{eq:Tpsol} T_{+}(\f) &=\ell^{-2} \, \mathcal{R} \, |\f|^{\frac{2}{\Delta_+}} [1+ \mathcal{O}(\f) + \mathcal{O}(\mathcal{R})] \, .
\end{align}
The  above expressions describe two continuous families of solutions,
whose structure is a {\em universal} analytic expansion in integer
powers of $\f$, plus a series of non-analytic, subleading terms which, in principle, depend on two
(dimensionless) integration constants $C$ and ${\cal R}$.
Note that the $(-)$-branch of solutions depends on two
integration constants $C$ and $\mathcal{R}$, while only the
integration constant $\mathcal{R}$ appears in the solutions of the
$(+)$-branch.  The notation $\mathcal{O}(C)$ and
$\mathcal{O}(\mathcal{R})$ does not imply that $C$ or $\mathcal{R}$
have to be small. Rather, this is shorthand to indicate that these terms will be accompanied by higher powers of $\f$ thus justifying their omission.\footnote{For example, the solution for $W$ on the $(-)$-branch can be written schematically as the following triple expansion:
\begin{align}
W_{-}(\f) = \frac{1}{\ell} \sum_{l=0}^{\infty} \sum_{m=0}^{\infty} \sum_{n=0}^{\infty} A_{l,m,n} \left(C \, |\f|^{d/\Delta_-} \right)^l \, \left(\mathcal{R} \, |\f|^{2/\Delta_-} \right)^m \, \f^n
\end{align}
}

The solutions \eqref{eq:Wmsol}--\eqref{eq:Tmsol} generalize the
near-extremum superpotential solutions that arise in the flat case,
where only the integration constant $C$ was present. Indeed, setting
${\mathcal R}=0$ we recover the flat result for $W(\f)$ and we also
find $S = W'$ and $T=0$. This indicates that ${\mathcal R}=0$ is
strictly related to the curvature of the $d$-dimensional metric, a
fact that will be confirmed explicitly below.

Given our results for $W$, $S$ and $T$, we are now in a position to solve for $\f(u)$ and $A(u)$. For the $(-)$-branch we solve \eqref{eq:defSc} and  \eqref{eq:defWc} subject to \eqref{eq:Smsol} and \eqref{eq:Wmsol} to obtain
\begin{align}
\label{eq:phimsol} \f(u) &= \f_- \ell^{\Delta_-}e^{\Delta_-u / \ell} \left[ 1+ \mathcal{O} \left(\mathcal{R} |\f_-|^{2/\Delta_-} e^{2u/\ell} \right) + \ldots \right] \\
\nonumber & \hphantom{=} \ + \frac{C d \, |\f_-|^{\Delta_+ / \Delta_-}}{\Delta_-(d-2 \Delta_-)} \, \ell^{\Delta_+} e^{\Delta_+ u \ell} \left[ 1+ \mathcal{O} \left(\mathcal{R}|\f_-|^{2/\Delta_-} e^{2u/\ell} \right) + \ldots \right] + \ldots \, , \\
\label{eq:Amsol} A(u) &= \bar{A}_- -\frac{u}{\ell} - \frac{\f_-^2 \, \ell^{2 \Delta_-}}{8(d-1)} e^{2\Delta_- u / \ell}  -\frac{\mathcal{R}|\f_-|^{2/\Delta_-} \, \ell^2}{4d(d-1)} e^{2u/\ell} \\
\nonumber & \hphantom{=} \ - \frac{\Delta_+ C |\f_-|^{d/\Delta_-} \, \ell^d}{d(d-1)(d-2 \Delta_-)}e^{du/\ell} +\ldots \, .
\end{align}
Here, we introduced the integration constants $\f_-$ and $\bar{A}_-$. We can repeat the analysis for the $(+)$-branch of solutions by solving \eqref{eq:defSc} and  \eqref{eq:defWc} subject to \eqref{eq:Spsol} and \eqref{eq:Wpsol}. We find:
\begin{align}
\label{eq:phipsol} \f(u) &= \f_+ \ell^{\Delta_+}e^{\Delta_+ u / \ell} \left[ 1+ \mathcal{O} \left(\mathcal{R} |\f_+|^{2/\Delta_+} e^{2u/\ell} \right) + \ldots \right] + \ldots \, , \\
\label{eq:Apsol} A(u) &= \bar{A}_+ -\frac{u}{\ell} - \frac{\f_+^2 \, \ell^{2 \Delta_+}}{8(d-1)} e^{2\Delta_+ u / \ell}  -\frac{\mathcal{R}|\f_+|^{2/\Delta_+} \, \ell^2}{4d(d-1)} e^{2u/\ell} +\ldots \, ,
\end{align}
where we introduced the integration constants $\f_+$ and $\bar{A}_+$. A few comments are in order.
\begin{itemize}
\item As our solutions for $W$, $S$ and $T$ are only valid for small $\f$, the above results are the leading terms in $\f(u)$ and $A(u)$ for $u \rightarrow -\infty$.
\item For both the $(+)$ and $(-)$-branch, the result for $A(u)$ exhibits the behavior expected for a scale factor in the near-boundary region of an AdS$_{d+1}$ space-time with length scale $\ell$, as can be seen by comparing with \eqref{eq:AdSScaleFactorExpansion}. As explained in section \ref{sec:AdSfixedpoints}, for asymptotically AdS space-times we can always choose a metric ansatz such that $R^{(\zeta)}=R^{\textrm{uv}}$. This amounts to setting $\bar{A}_{\pm}=0$.
\item
For the $(-)$-branch of solutions, we identify $\f_-$ as the source for the scalar operator $\mathcal{O}$ in the boundary field theory associated with $\f$. The vacuum expectation value of $\mathcal{O}$ depends on $C$ and is given by
\begin{align}
\langle \mathcal{O} \rangle_- = \frac{Cd}{\Delta_-} \, |\f_-|^{\Delta_+ / \Delta_-} \, .
\end{align}
\item
For the $(+)$-branch of solutions, the bulk field $\f$ is also associated with a scalar operator $\mathcal{O}$ in the boundary field theory. However, in this case the source is identically zero, yet there is a non-zero vev given by
\begin{align}
\langle \mathcal{O} \rangle_+ = (2\Delta_+-d) \, \f_+ \, .
\end{align}
There is also an associated moduli space of vevs, as $\f_+$ is arbitrary, being an integration constant of the first order flow equation.

\item We can learn even more by comparing the two expressions for $A(u)$ in \eqref{eq:Amsol} and \eqref{eq:AdSScaleFactorExpansion}. Matching the coefficients of $e^{2u/\ell}$ implies
\begin{align}
\mathcal{R} \equiv \left\{ \begin{array}{c l}
  R^{\textrm{uv}} |\f_-|^{-2/\Delta_-} & \qquad (-)\textrm{-branch} \\
  R^{\textrm{uv}} |\f_+|^{-2/\Delta_+} & \qquad (+)\textrm{-branch} \\
  \end{array} \right. \, .
\label{r}
\end{align}
Thus, the integration constant $\mathcal{R}$ is related to the curvature $R^{\textrm{uv}}$ of the manifold on which the UV QFT is defined.
\item Interestingly, the $(-)$ and $(+)$-branches of solutions are not completely unrelated. One can check that given a solution of the $(-)$-branch, we can arrive at a solution on the $(+)$-branch by performing the following rescaling:
\begin{align}
\f_- \rightarrow 0 \, ,\quad  C\to +\infty\, , \quad C |\f_-|^{\frac{\Delta_+}{\Delta_-}} = \textrm{const.} \, , \qquad \mathcal{R}|\f_-|^{\frac{2}{\Delta_-}}= \textrm{const.} \,
\end{align}
To be specific, under this rescaling the solutions in
\eqref{eq:phimsol} and  \eqref{eq:Amsol} can be brought into the form
of \eqref{eq:phipsol} and  \eqref{eq:Apsol}. This gives rise to
another interpretation of the $(+)$-branch of solutions. In
particular, for fixed $R^{\textrm{uv}}$ a $W_+$ solution is the upper
envelope of the family of $W_-$ solutions parameterized by $C$. This
is similar to the flat case \cite{Papadimitriou:2007sj,exotic}.
\end{itemize}
Overall, the above findings imply that, as in the flat case,
\emph{maxima of the potential are associated with UV fixed
  points}. The bulk space-time asymptotes to AdS${}_{d+1}$ and
reaching the maximum of the potential is equivalent to reaching the
boundary. Moving away from the boundary corresponds to a flow leaving
the UV. Flows corresponding to solutions on the $(-)$-branch are
driven by the existence of a non-zero source $\f_-$ for the perturbing
operator $\mathcal{O}$. Flows corresponding to solutions on the
$(+)$-branch are driven purely by a non-zero vev for $\mathcal{O}$,
as the source vanishes identically.

Finally, notice  that as expected near a  UV
fixed point, the curvature terms proportional to ${\mathcal R}$ only
enter as subleading corrections in $W$, $S$ and
the bulk solution \eqref{eq:phimsol}--\eqref{eq:Apsol}.

\subsection{Expansion near minima of the potential}
\label{sec:minimaofV}

In the following, we will display solutions for $W(\f)$, $S(\f)$ and $T(\f)$ corresponding to flows that either leave or arrive at minima of the potential. We will again consider the potential \eqref{eq:Vextremum}, but now we have $m^2<0$. In the following, we present the results, while detailed calculations can be found in appendix \ref{app:smallRexp}. Again, we find that there exist $(+)$ and $(-)$-branches in the space of solutions, which we will discuss in turn.

\subsubsection*{$(+)$-branch}
For the $(+)$-branch the expansions around a minimum of $V$ take the same form as in the vicinity of a maximum. Therefore, we  obtain:
\begin{align}
\label{eq:Wpsol2} W_{+}(\f) & = \frac{1}{\ell} \left[2(d-1) + \frac{\Delta_+}{2} \f^2 + \mathcal{O}(\f^3) \right] \\
\nonumber & \hphantom{=} \,  + \frac{\mathcal{R}}{d \ell} \, |\f|^{\frac{2}{\Delta_+}} \ [1+ \mathcal{O}(\f) + \mathcal{O}(\mathcal{R})] \, , \\
\label{eq:Spsol2} S_{+}(\f) & = \frac{\Delta_{+}}{\ell} \f \ [1+ \mathcal{O}(\f)] + \frac{1}{\ell} \mathcal{O}\left( \mathcal{R} |\f|^{\frac{2}{\Delta_+}+1} \right) \, , \\
\label{eq:Tpsol2} T_{+}(\f) &= \ell^{-2} \, \mathcal{R} \, |\f|^{\frac{2}{\Delta_+}} [1+ \mathcal{O}(\f) + \mathcal{O}(\mathcal{R})] \, ,
\end{align}
with $\Delta_{+}$ defined as before in \eqref{eq:Deltadef}, but now we have that $m^2 < 0$. The integration constant $\mathcal{R}$ is continuous and gives rise to a family of solutions. As before, we can integrate to obtain:
\begin{align}
\label{eq:phipsol2} \f(u) &= \f_+ \ell^{\Delta_+}e^{\Delta_+ u / \ell} \left[ 1+ \mathcal{O} \left(\mathcal{R} |\f_+|^{2/\Delta_+} e^{2u/\ell} \right) + \ldots \right] + \ldots \, , \\
\label{eq:Apsol2} A(u) &= \bar{A}_+ -\frac{u}{\ell} - \frac{\f_+^2 \, \ell^{2 \Delta_+}}{8(d-1)} e^{2\Delta_+ u / \ell}  -\frac{\mathcal{R}|\f_+|^{2/\Delta_+} \, \ell^2}{4d(d-1)} e^{2u/\ell} +\ldots \, ,
\end{align}
This solution has the following interpretation:
\begin{itemize}
\item Recall that the expansions of $W$, $S$ and $T$ are valid only for small $\f$. As $\Delta_+ > d$ at a minimum, it follows from \eqref{eq:phipsol2} that small $\f$ requires $u \rightarrow - \infty$. Using \eqref{eq:Apsol2}, this in turn implies that $e^{A(u)} \rightarrow \infty$ when approaching the minimum of the potential. This is the behavior expected when approaching a UV fixed point.
\item In the boundary QFT, the bulk field $\f$ will be associated with an operator $\mathcal{O}$. However, the absence of a term $\sim e^{\Delta_- u / \ell}$ in \eqref{eq:phipsol2} implies that the source of this operator vanishes. On the other hand, there is a non-zero vev given by
\begin{align}
\langle \mathcal{O} \rangle_+ = (2 \Delta_+-d) \f_+ \, .
\end{align}
The solution \eqref{eq:phipsol2} is to be interpreted as a flow purely driven by a vev.
\item Matching the expression \eqref{eq:Apsol2} with the near-boundary expansion \eqref{eq:AdSScaleFactorExpansion} we again find that $\mathcal{R}$ is related to the curvature $R^{\textrm{uv}}$ of the background manifold of the UV QFT:
\begin{align}
\mathcal{R} = R^{\textrm{uv}} |\f_+|^{-2/ \Delta_+} \, .
\end{align}
\end{itemize}
To summarize, as in the flat case,  for solutions of the $(+)$-branch
\emph{minima of the potential correspond to UV fixed points}. Flows
from such UV fixed points are not driven by a source for the
perturbing operator $\mathcal{O}$, but rather  by its vev $\langle
\mathcal{O} \rangle_+$. This corresponds to a spontaneous breaking of
conformal invariance.

As in the flat case, these solutions are generically singular in the
IR, because unlike the $(-)$-branch solutions departing from a UV {\em
  maximum}, they have no continuous adjustable parameter (beside the UV data
$R^{\textrm{uv}}$ )  which one may vary to select a solution with
regular interior geometry. In the $(-)$-branch on the other hand, this
role is played by the extra integration constant $C$. Nevertheless,
one can construct  specific examples where regular solutions of the
$(+)$ type exist \cite{exotic}.

\subsubsection*{$(-)$-branch}
Interestingly, we will need to distinguish between the two cases where the boundary field theory is defined on a curved manifold ($R^{\textrm{uv}} \neq 0$) and a flat manifold ($R^{\textrm{uv}} = 0$). A key result is that for $R^{\textrm{uv}} \neq 0$ the $(-)$-branch of solutions does not exist. A proof (at the level of the functions $\f(u)$ and $A(u)$) can be found in appendix \ref{app:Aphiminima}. More specifically, there are no solutions in the $(-)$-branch corresponding to flows that either leave or arrive at a minimum of $V$ for $R^{\textrm{uv}} \neq 0$. Such a solution only exists if $R^{\textrm{uv}} = 0$ and we recover the result from \cite{exotic}, where RG flows for QFTs on flat manifolds were studied.

For $R^{\textrm{uv}} = 0$ we have that $T_{-}(\f)=0$ identically. In addition, \eqref{eq:EOM1} implies that $S(\f)= W'(\f)$ and the solution is completely specified by $W(\f)$. Our findings can be summarized as follows:
\begin{align}
R^{\textrm{uv}} \neq 0: \qquad & \textrm{No solution.} \\
\label{eq:Wmsol2} R^{\textrm{uv}} = 0: \qquad & W_-(\f) = \frac{1}{\ell} \left[2(d-1) + \frac{\Delta_-}{2} \f^2 + \mathcal{O}(\f^3) \right] \ .
\end{align}
For the case $R^{\textrm{uv}} = 0$ we can determine $\f(u)$ and $A(u)$ by integrating \eqref{eq:defWc}, \eqref{eq:defSc} subject to \eqref{eq:Wmsol2} and $S=W'$. One obtains
\begin{align}
\label{eq:phisol3} \f(u) &= \f_- \, \ell^{\Delta_-} \, e^{\Delta_- u / \ell} + \ldots \, , \\
\label{eq:Asol3}A(u) &= \bar{A} - \frac{u}{\ell} - \frac{\f_-^2}{8 (d-1)} \, \ell^{2 \Delta_-} \, e^{2 \Delta_- u / \ell} + \ldots \, ,
\end{align}
with $\bar{A}$ and $\f_-$ integration constants. We recover the
known result that for $R^{\textrm{uv}} = 0$ \emph{minima of the
  potential correspond to IR fixed points}, as $\f$ small now
requires $u \rightarrow + \infty$ (as $\Delta_-<0$ at a minimum) which
in turn implies $e^{A(u)} \rightarrow 0$.
On the other hand, no such IR limit  exist in the presence of
curvature, unlike around a UV fixed point, where curvature only added
subleading corrections.  One is then led to  wonder what
is the fate of curved  RG flows in the interior where curvature
becomes the dominant driving parameter. This is the subject of the next section.

\section{The geometry in the interior} \label{IRasymp}

After having analyzed the behavior close to the UV boundary, in this
section we analyze the geometry in the interior. In particular, we
will be interested in the way the space-time can ``end'' in a regular
way, i.e.~where the scale factor shrinks to zero but the bulk
curvature invariants are finite. These are the curved analogs of flat IR
fixed points.

Unlike what happens close to the AdS boundary, where curvature leads only to
subleading corrections to the near-boundary asymptotics, in the
interior curvature can drastically change the geometry with respect to
the flat case. As we will see, for positive curvature, the RG flow
reaches an end before the (would-be) IR fixed point of the flat theory; for
negative curvature, on the other hand, both the scalar field and the
scale factor turn around and the flow reaches the boundary on both
sides.  Non-zero curvature can also deform the geometry near {\em bounces} (points
where the scalar field inverts its flow direction \cite{exotic} but
the scale factor is still monotonically decreasing).

In this analysis, a key role is played by the critical points in the 1st order equations. These are points where
$\dot{\phi} = 0$. Before discussing the non-zero curvature case, we
briefly review the analysis of critical points for zero curvature.

In the flat case critical points correspond
to points along the flow with vanishing $W'$. Extrema of $W$ in the
interior of the geometry
belong to two classes: regular IR fixed points, and bounces \cite{exotic}.
\begin{itemize}
\item {\bf Flat IR fixed points.} A regular IR endpoint of the geometry is attained, in the flat case,
when the flow asymptotes to a minimum (which we denote by $\f_{IR}$)
of the scalar potential. In this case,  $\dot{\f} \to 0$ and the geometry in the
interior is  asymptotically AdS${}_{d+1}$,
with
\be
e^A(u) \simeq  e^{-u/\ell_{IR}},  \qquad \f(u) \to \f_{IR},
\qquad u \to +\infty
\ee
In the language of the superpotential, as $\f\to \f_{IR}$, $W\to
2(d-1)/\ell_{IR}$ and $W'\to 0$, and
the (flat) holographic $\beta$-function vanishes:
\be
\beta(\f) = - 2(d-1) {W' \over W} \to 0 \, , \qquad \f \to \f_{IR}.
\ee
The condition $W'(\f_{IR})=0$ by itself does not imply that the flow has
reached a fixed point: for that, we also need $W''(\f)$ to be finite, and
this only happens when $V'(\f_{IR}) =0$, i.e.~$\f_{IR}$ is also a
minimum of the bulk potential as we have assumed above.
If the latter condition is not met, then the geometry features a bounce, as discussed below.

\item {\bf Flat Bounces.} A generic extremum of $W$, i.e.~a
point $\f_B$ in field space such that $W'(\f_B) = 0$ but $V'(\f_B) \neq 0$, corresponds to a
{\em bounce}, i.e.~a point where the flow inverts its direction
\cite{exotic}. The superpotential becomes singular because $\f$ is not a
good coordinate around such points. The flow however can be  continued by
gluing to another branch, where $\dot{\f}$ has the opposite
sign. Close to a bounce, the superpotential behaves as:
\be
W(\f) \simeq W_b \pm c (\f_B - \f)^{3/2}+ O\Big((\f_B- \f)^2\Big),
\ee
where $c$ is a constant and we have supposed that the flow reaches the
bounce from below ($\f< \f_B$). The two signs correspond to the two
branches of the superpotential along the RG flow before and after the
bounce, which can be glued at $\f=\f_B$ giving rise to a regular
solution  $\f(u), A(u)$, in which  all curvature
invariants are finite. The holographic $\beta$ function still vanishes at $\f_B$,  but it  becomes multivalued:
\be
\beta(\f) \simeq \mp{3(d-1)\over W_b} \sqrt{\f_B- \f} +
O\Big((\f_B - \f)\Big),
\ee
\end{itemize}

It is natural to ask what happens to  IR fixed points and bounces when the
boundary theory lives on a curved space-time. In the maximally symmetric  case
analyzed in this paper, the condition $\dot{\f} = 0$ is now the
vanishing of the function $S(\f)$, which in the flat case coincides
with $W'$.  The
analysis of the possible ways in which $S$ can vanish is  presented in
Appendix \ref{App:interior}. As it is shown there,
 the asymptotic behavior of the functions $W$, $S$ and $T$
always takes the general form of an expansion in half-integer powers
of  $x\equiv (\f_0 -
\f)$ (which we assume to be positive, for simplicity):
\begin{align}
S(x)&=\sqrt{x} \left(S_0+S_1 \sqrt{x}+S_2 x +\cdots\right)  \label{Sbounce} \ ,\\
W(x)&=\frac{1}{\sqrt{x}}\left(W_0+W_1 \sqrt{x}+W_2 x+\cdots \right)\label{Wbounce} \ ,\\
T(x)&=\frac{1}{x}\left(T_0+T_1\sqrt{x}+T_2 x+\cdots  \right)\label{Tbounce} \ ,
\end{align}
On the other hand, we have assumed that the potential has a regular expansion around $\f_0$:
\be \label{Vbounce}
V(x)=V_0+ V_1 x + V_2 x^2 + \cdots\ .
\ee
Depending on the sign of the curvature and on the values of the the
coefficients in (\ref{Sbounce}-\ref{Tbounce}), the solution around $\f
= \f_0$ can be   of be three possible types:
\begin{enumerate}
\item Fixed points (positive curvature only)
\item Reflection points (negative  curvature only)
\item Bounces (both signs of the curvature)
\end{enumerate}
We proceed to discuss each one in turn.

\subsection{Positive curvature flows: IR endpoints}  \label{positivecurvatureflows}
Let us suppose that $\f_0$ is such that $S(\f_0)=0$, and the leading
coefficients $S_0$, $W_0$ and $T_0$ are all non-vanishing (this corresponds to
case (b) in Appendix \ref{App:interior}). Then,   $T
= R^{(\zeta)} e^{-2A}$ diverges at $\f_0$,  implying that the
scale factor  $e^A$ shrinks to zero size. However, this  does not
imply a singularity but as we will see below, it represents a
coordinate horizon (or a regular end of space in the Euclidean signature).

To leading order in $\f_0-\f$, we have:
\be \label{caseb}
S(\f) \simeq S_0  (\f_0 - \f)^{1/2}, \qquad W(\f) \simeq {W_0  \over
  (\f_0 - \f)^{1/2}},
\ee
\be \label{Tcaseb}
 T  \simeq {T_0 \over  (\f_0 - \f)}
\ee
where the coefficients are fixed by the superpotential equations to be:
\be\label{coeff}
S_0^2=\frac{2 V_1}{d+1}, \qquad W_0= (d-1) S_0, \qquad T_0=\frac{d(d-1)
}{4(d+1)}\, S_0^2
\ee
We see that this solution, obtained assuming $\f < \f_0$, makes sense
only for $V_1>0$, i.e. $V'(\f_0) < 0$ (cfr.~equation \eqref{Vbounce}).  Similarly,  it is easy to show
that, for $V'(\f_0)>0$  we have to reach $\f_0$ from above. In both
cases, equation (\ref{Tcaseb}) implies $T>0$. From the
definition (\ref{eq:defTc}), this in turn implies that such behavior
can occur only for {\em positive curvature}.

With  expressions (\ref{caseb}-\ref{Tcaseb}) for $W$, $S$ and $T$,
we can integrate equations (\ref{eq:defWc}-\ref{eq:defSc}) order by order in $(\f_0 -
\f)$ to find the expressions for
the scale factor $A(u)$ and the scalar field $\f(u)$. We  define  the
``end of space''   point $u_0$ where
\be
\f(u_0) = \f_0, \qquad e^{A(u_0)} = 0 ,
\ee
To lowest order, one finds:
\begin{equation}
\f=\f_0 - \frac{S_0^2}{4}(u-u_0)^2 +  O((u-u_0)^3) \, , \qquad  A(u)= \ln (u-u_0) +
A_0 + O(u-u_0). \label{IRG2}
\end{equation}
The parameter $A_0$ is an integration constant for equation
(\ref{eq:defWc}), but it is determined algebraically by the asymptotic form of the superpotential $T(\f)$ close to $\f_0$:
  \begin{equation}
  T\equiv R^{(\zeta)} e^{-2 A(u)} \simeq \frac{T_0}{\f_0-\f}
  \end{equation}
Inserting the expansions (\ref{IRG2}), and using the relations
(\ref{coeff}) and the fact that $R^{(\zeta)} = d(d-1)/\alpha^2$,  one easily finds:
  \begin{equation} \label{Azero}
  A_0=-\log \alpha
  \end{equation}
Interestingly, it is precisely (and only) for this value of $A_0$ that
there is no  singularity at $u_0$. Recall that the metric is given by:
\begin{equation}
 ds^2=du^2+e^{2 A(u)} ds_{\textrm{dS}_{\alpha}}^2 = du^2+\alpha^2 e^{2 A(u)} ds_{\textrm{dS}_{\alpha=1}}^2 \, ,
 \end{equation}
 where $ds_{\textrm{dS}_{\alpha}}^2$ is the de Sitter metric with radius $\alpha$.
 Near $u\rightarrow u_0$, this becomes,
 \begin{equation}
 ds^2\simeq du^2+\alpha^2 e^{2 A_0}(u-u_0)^2ds_{\textrm{dS}_{\alpha=1}}^2.
 \end{equation}
 For $A_0=-\log \alpha$,  the four-dimensional part of the metric
 vanishes exactly like it does at the ``end of space'' in the dS
 slicing of AdS${}_{d+1}$, as can be seen from equation
 (\ref{eq:AdSScaleFactor}) by expanding the scale factor around the
 point $u=-c$.  Equivalently, if we go to Euclidean signature,
 unit-curvature dS$_d$ becomes the unit sphere $S_d$, and for
 $A_0=-\log \alpha$ the  metric close to $u_0$ is approximately:
\be
ds^2_E = d\rho^2 + \rho^2 dS_d^2 , \qquad \rho \equiv |u-u_0|
\ee
i.e. the metric around the origin of Euclidean $d+1$-space in
spherical coordinates. In the Minkowski version, $u=u_0$ is a coordinate
singularity (it is a horizon).



Notice that the qualitative behavior of the solution  is different,
and  the IR endpoint discussed here does not exist, if $V'(\f_0)=0$:
indeed, in this case $T_0=0$. Thus, solutions with positive slice
curvature cannot reach a would-be flat IR fixed point $\f_{IR}$, which
corresponds to a minimum of the scalar potential. This is in agreement
with the result found in section 3, that curved RG flows cannot end at
minima of the potential.

From the above analysis  we can infer the general structure of
a positive curvature RG flow in a model with a bulk potential having
several maxima and minima. Consider  a {\em flat} RG flow connecting a
UV maximum at $\f = \f_{UV}$  to an IR minimum at $\f=\f_{IR}$. In the region
$\f_{UV}<\f< \f_{IR}$, we have $V'(\f) < 0$ by construction. Thus,  adding
positive curvature forces the flow to stop at an intermediate point
$\f_{UV}< \f_0 < \f_{IR}$. This behavior is illustrated in Figure
\ref{positiveRflow}.

The position of the  endpoint in field space is determined
by equations (\ref{coeff}) and by the value of the UV data, i.e.~the
source term $\f_-$ in the scalar field UV asymptotics, and the boundary
curvature $R^{\textrm{uv}}$. This can be seen as follows: from the IR
equations, choosing the endpoint $\f_0$ determines all the expansion
coefficients in the superpotentials, (\ref{coeff}), through the values
of $V(\f)$ and its derivatives at $\f_0$. On the other hand, the
integration constant $A_0$ in the scale factor is fixed by the
curvature parameter by equation (\ref{Azero}). Since all the
integration constants are fixed,  if we  integrate the
equations towards the UV we will find a given value of the UV source
and vev parameters,  $\f_-$ and $\f_+$. Turning it around, for a
given set of UV data $\f_-$ and $R^{\textrm{uv}}$, there exist a single regular solution
with a specific vev  $\f_+(\f_-, R^{\textrm{uv}})$ and a specific endpoint
$\f_0(\f_-, R^{\textrm{uv}})$.

 As we will see in the next section, and as expected on general
 grounds, the position of the endpoint actually depends only on the
 dimensionless combination ${\mathcal R} \equiv R^{\textrm{uv}}
 (\f_-)^{-{2/\Delta_-}}$. Indeed, from the UV expansions
 (\ref{eq:Wmsol}-\ref{eq:Tmsol}) it is clear that the superpotentials
 know only about the integration constant ${\mathcal R}$, and do not depend
 separately on the source $\f_-$. Thus it is the combination
 ${\mathcal R}$ which determines the endpoint. As ${\cal R}$ ranges from zero to infinity, the endpoint moves from $\f_{IR}$ to $\f_{UV}$.

Finally, the IR endpoint discussed in this section, constitutes
  a special case of the general result we present as property 5 of
  superpotentials in Appendix \ref{properties}, namely that around a point where $S=0$, the bulk metric is
 approximately  maximally symmetric. In other words, the IR fixed
 point is a point of enhanced symmetry. This is similar to the flat
 case, where at an IR fixed point, Poincar\'e symmetry is upgraded to
 conformal invariance.  Here, from the boundary  point of view, the invariance  is conformal symmetry written in
  ``unusual'' coordinates.

\subsection{Negative curvature flows: AdS throat}
\label{negativecurvatureflows}
The endpoint behavior  presented in the previous section requires
positive curvature slices. If the  curvature is negative, then we must
consider $(a)$ in
Appendix \ref{App:interior}. This  generically corresponds to a
bounce (which will be discussed in the next section)  {\em except} in
the special case $W_0=W_1=0$ (referred to as case (c) in
Appendix \ref{App:interior}):
\bea
&& S_0^2 = 2V_1, \quad W_0 = W_1 =0, \quad T_0 = T_1 = 0 \\
&& W_2 = {4V_0\over d S_0}, \qquad T_2 = V_0.
\eea
At leading order the superpotentials are (assuming $\f<\f_0$):
\be
S\simeq S_0 (\f_0-\f)^{1/2}+\ldots, \quad W \simeq W_2 (\f_0-\f)^{1/2}+\ldots, \quad
T \simeq T_2 +\ldots
\ee
This solution is possible  only for negative slice curvature, since $T= e^{-2A}R^{(\zeta)}$  has
the same sign as the potential $V(\f_0)$, which we assumed to be negative definite.  Notice that now $W$ vanishes at $\f_0$,
and $T= e^{-2A}R^{(\zeta)}$ remains finite. This implies that
$\dot{A}=0$ and  $A$ is
finite  at $\f_0$, i.e. the scale factor has a
turning point.

By integrating the first order flow equations, one can show that this solution corresponds to the following geometry for $u \rightarrow u_0$:
\bea
\label{eq:AdSslice}
&& \f(u)\simeq \f_0 - \frac{S_0^2}{4}(u-u_0)^2  , \\
&& ds^2 \simeq  du^2 + e^{2A_0}\left(1 + {(u-u_0)^2 \over \ell_{IR}^2}\right) \,
ds_{\textrm{AdS}_d, \alpha}^2,
\eea
where $A_0$ is an integration constant, $ds_{\textrm{AdS}_d, \alpha}^2$ is a metric on AdS${}_d$ with length
scale $\alpha$ and $\ell_{IR}^2 = d(d-1)/|V(\f_0)|$.
 The behavior around
 $u_0$ matches that of  space AdS${}_{d+1}$, with length scale
 $\ell_{IR}$, sliced with AdS${}_d$ hypersurfaces, as seen in (\ref{eq:AdSScaleFactor}).

The conditions $\dot{A}(u_0) = \dot{\f}(u_0) = 0$ imply that we can
continue the  geometry for $u > u_0$ by gluing its reflected image
around $u_0$. The resulting geometry is regular  since the metric and
scalar field, as well as their derivatives, are all continuous across $u_0$.  If the solution  starts at a UV
fixed point reached as  $u\to -\infty$,  the resulting doubled geometry
will reach a ``second'' UV  boundary again as $u\to +\infty$,
connected by a throat where the scale factor reaches a minimum, as
shown in figure \ref{negativeRflow}.

This geometry
does not really have two disconnected boundaries: rather, the two
boundaries at $u=\pm \infty$ are connected through the boundary of the
lower-dimensional AdS$_{d}$ slices. This was discussed in
\cite{maldamaoz} in the case of the  AdS$_{d+1}$ geometry written in
AdS$_d$ slices, and we will return to this point in subsection
\ref{sec-coordinates}.

As was the case for the endpoint in the positive curvature case, the
turning point $\f_0$ is completely determined by the UV data $\f_-$
and $R^{\textrm{uv}}$.

\subsection{Bounces} \label{bounces}

The generic situation of such a point $\f_B$ with vanishing $S$ and
finite $W$  corresponds to $W_0 = 0$ but $W_1\neq0
$  (case (a) in  Appendix
\ref{App:interior}). To leading order, the superpotentials behave as
(assuming $(\f <  \f_B$):
\be \label{bounce1}
S\simeq S_0(\f_B-\f)^{1/2} +\ldots , \quad W \simeq W_1 + W_2
(\f_B-\f)^{1/2} +\ldots, \quad T \simeq T_2 + \ldots
\ee
where
\be\label{bounce2}
S_0^2 = 2V_1, \quad  T_2 = V_0 + {d W_1^2  \over 4(d-1)},
\ee
and $W_1$ is arbitrary ($W_1=0$ corresponds to the special case of the
AdS throat analyzed in the previous subsection).
 In this case, the scalar field has a turning point, but the scale
 factor has a non-vanishing derivative.  Since $V_0<0$,  $T_2$ does
not have a definite sign and therefore this solution is allowed for both positive
and negative curvature.

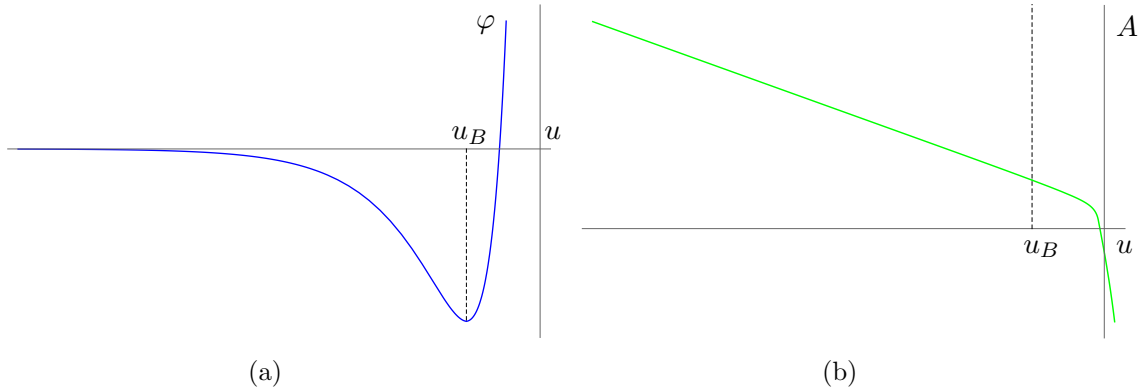
\begin{figure}[t]
\centering
\begin{subfigure}{.5\textwidth}
 \centering
   \begin{overpic}[width=0.95\textwidth,tics=10]{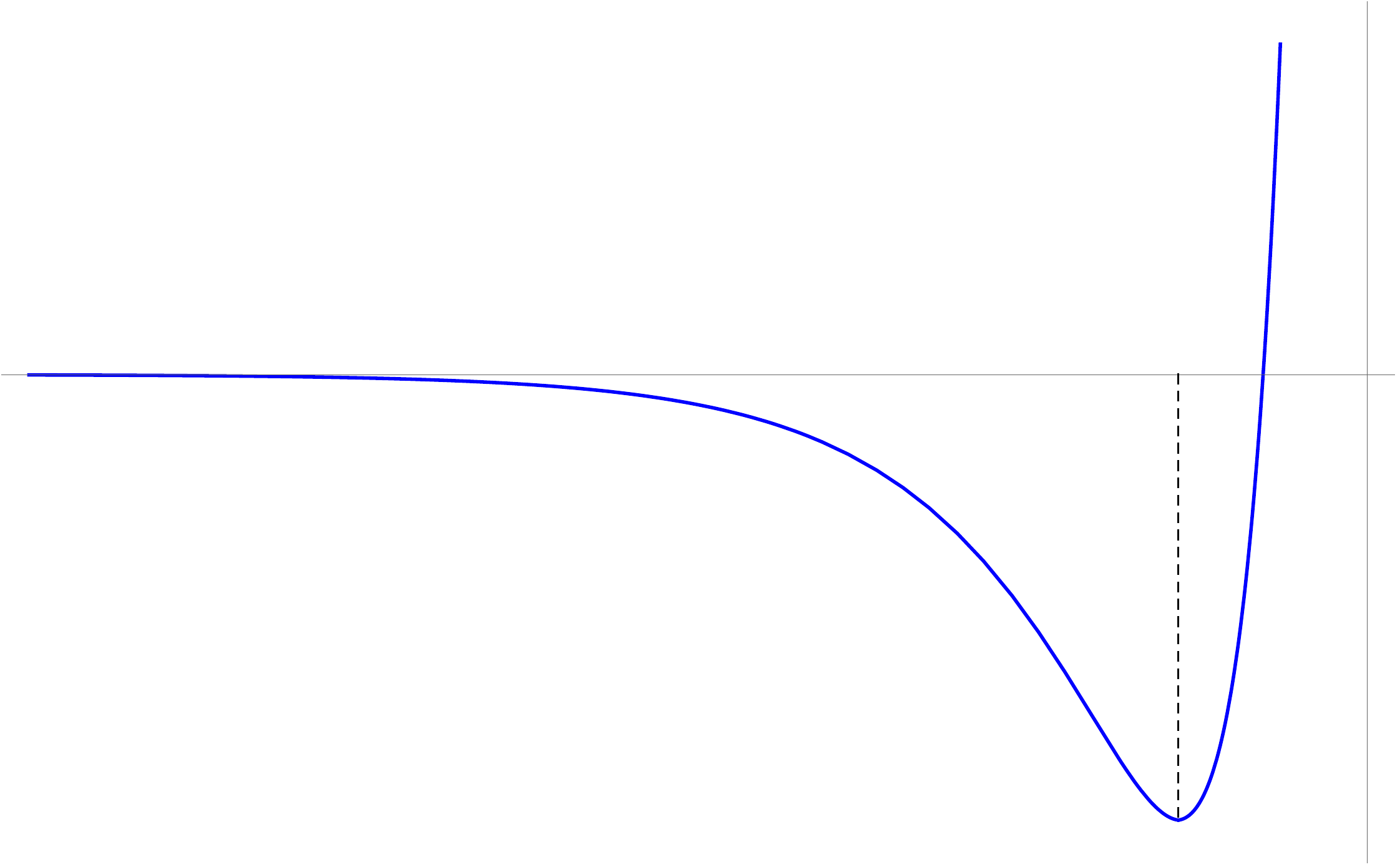}
\put (86,57.5) {$\f$} \put (98.5,37) {$u$} \put (81.5,37) {$u_B$}
\end{overpic}
 \caption{\hphantom{A}}
  \label{phibouncefig}
\end{subfigure}%
\begin{subfigure}{.5\textwidth}
  \centering
 \begin{overpic}[width=0.95\textwidth,tics=10]{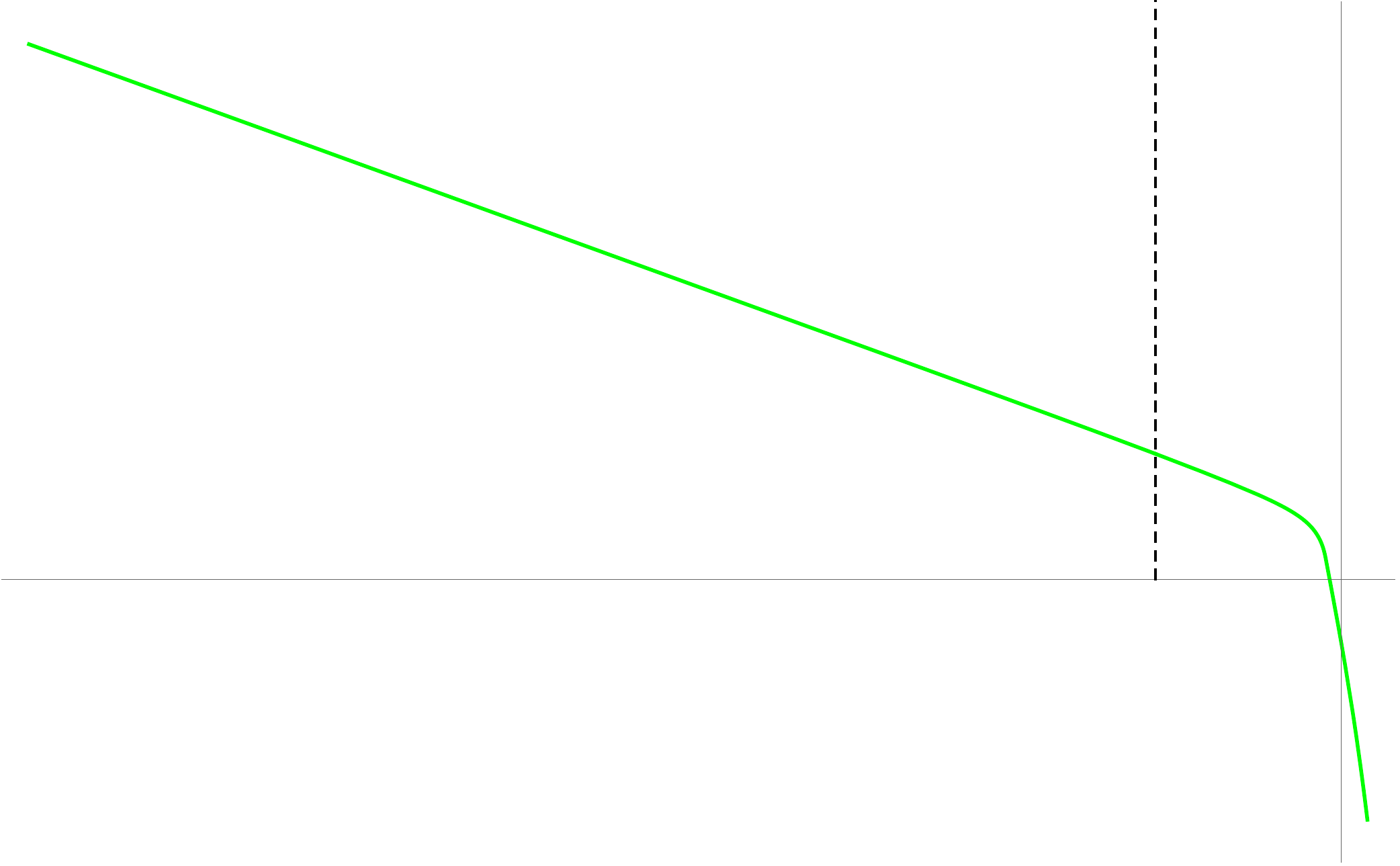}
\put (98,56) {$A$} \put (98,16) {$u$} \put (81,16) {$u_B$}
\end{overpic}
\caption{\hphantom{A}}
  \label{Abounce}
\end{subfigure}
\caption{$\f$ and $A$ vs.~$u$ for an example exhibiting a bounce at $u=u_B$.}
\label{fig:bounceschematic}
\end{figure}

The solution (\ref{bounce2}) actually describes two branches, with
$S_0=\pm \sqrt{V_1}$, corresponding to the two branches of the
superpotential equations discussed in section \ref{sec:1storder}: at a
bounce, these two branches can be glued, giving rise to a regular
geometry, exactly as in the flat case \cite{exotic}. Integrating the
superpotential equations we find the bulk solution  close to $\f_B$ in
terms of the $u$ coordinate (where $u=u_B$ is the bounce point):
\begin{equation} \label{bounce3}
\f(u)\simeq\f_B-\frac{S_0^2}{4}\left(u - u_B\right)^2 +\ldots \quad A(u)=A_B-\frac{W_1}{2(d-1)}(u-u_B) +\ldots
\end{equation}
This solution is regular both for $u< u_B$ ($S(\f)>0$) and $u> u_B$
($S(\f>0$). The superpotentials are singular because at $\f_B$ because
$\f$ ceases to be a good coordinate at $\f_B$, but the geometry
(\ref{bounce3}) is smooth.  The typical scalar field profile is illustrated in Fig.~\ref{phibouncefig}. Like in the flat case, we refer to this as a bouncing flow.

The fact that the  geometry is regular at the bounce can be also be
seen from the fact that all the curvature invariants can be written in terms of
the potential $V(\f)$ and the function $S(\f)$ only. As the behavior of $S(\f)$ is regular near the
bounce, all the curvature invariants are finite, as one can see from
the expressions in Appendix \ref{curvature inv}. 

We can relate the value of the scale factor at the bounce by
evaluating  the relation $T = R^{(\zeta)} e^{-2A}$ at $u=u_B$, using
equations (\ref{bounce1}) and (\ref{bounce3}):
\begin{equation}
T_2= R^{(\zeta)} e^{-2 A_B}. \label{defTbounce}
\end{equation}
Using the expression of $T_2$ from Eq. \eqref{bounce2}, we find a
 relation between  $A_B$, $W_1$ and $V(\f_B)$:
 \begin{equation}
A_B=-\ln(\alpha)-\frac{1}{2}\ln \Bigg{\lvert}  \left(\frac{W_1}{2(d-1)} \right)^2-\frac{1}{\ell_B^2}  \Bigg{\lvert} \ \label{AB}
\end{equation}
where we have defined
\be
V(\f_B)=-\frac{d(d-1)}{\ell_B^2}\;.
\ee
As $S=0$ at the bounce,  the bulk geometry is approximately maximally
symmetric, as shown in Appendix \ref{properties}.

\subsection{The geometry of curved RG flows} \label{sec-coordinates}

In this section we clarify some aspects about the global geometry of
the RG flow solutions discussed above. In particular, we are
interested in the geometry close to the boundary and in the deep
interior (the endpoint of the flow).

For simplicity, we will
consider  explicitly the situation where the scalar field does not run:
such solutions are just AdS$_{d+1}$ space-time written in an unusual
coordinate system, and the analysis can be performed
analytically. However,  all the qualitative conclusions carry on to
non-trivial RG flow solutions both close to the boundary and close to
the flow endpoint, where the geometry is approximately AdS$_{d+1}$.

We start by recalling some facts about various coordinate systems in
AdS$_{d+1}$.  The ``flat'' slicing  coordinates  $(z,t,\vec{x})$
cover  the Poincar\'e patch of  AdS$_{d+1}$, and in these coordinates the metric is given by:
\be \label{geo1}
ds^2  = {\ell^2 \over z^2 }\left( dz^2 - dt^2 + \sum_{i=1}^{d-1}dx^i dx^i
\right),  \qquad z>0.
\ee
Global coordinates $(\rho,\psi,\Omega_i)$,   on
AdS$_{d+1}$ are related to the Poincar\'e coordinates by:
\bea
&& \cosh\rho = {\left[\Delta^2 + (2\ell t)^2\right]^{1/2} \over 2\ell
  z}, \qquad \sin \psi = {2 \ell t \over \left[\Delta^2 + (2\ell
      t)^2\right]^{1/2}} \label{geo2} \\
&& \Omega_i = {2 \ell x_i \over \left[\Delta^2 + (2\ell
      t)^2 -  (2\ell  z)^2\right]^{1/2}}, \qquad i = 1\ldots d-1,  \label{geo3}\\
&& \Delta \equiv z^2 +\ell^2 + |\vec{x}|^2 -t^2 . \nonumber
\eea
In global coordinates the metric reads:
\be\label{geo1b}
ds^2 = \ell^2 \left[d\rho^2 -\cosh^2 \rho d\psi^2 + \sinh^2\rho d\Omega^2\right].
\ee
The AdS boundary is located at $\rho\to +\infty$ in global
coordinates, and it is reached as $z\to 0$ (with the $(t,x_i)$ coordinates
fixed) in the Poincar\'e patch.  In addition, a single  point on the
boundary (corresponding to $\psi = 0, \Omega_i = 0$ in global coordinates) can
also be reached as $z \to +\infty$. This point also corresponds to spatial
infinity ($|\vec{x}| \to +\infty$) on any fixed-$z$ slice.  The
Poincar\'e horizon is reached as $z\to \infty$, $t\to \pm \infty$, $z = \pm
t + \textrm{subleading terms}$, and corresponds to $\rho$ finite
(e.g.~$\vec{x}=0$ on the Poincar\'e horizon is located at $\rho = 0,
\psi= \pi/2, \Omega_i = 0$). For a detailed discussion of how the
$AdS$ boundary is described in Poincar\'e coordinates,  the
reader is referred  to \cite{Bayona}. Finally, recall that one can go
to conformal global coordinates by changing variables to $\tan \Theta
= \sinh\rho$, which leads to:
\be \label{globalconformal}
ds^2 = {\ell^2 \over \cos^2\Theta}\left[d\Theta^2 - d\psi^2 + \sin^2\Theta d\Omega^2\right].
\ee
In these coordinates the AdS boundary is at $\Theta = \pi/2$ and it is manifestly conformal to $R_{\psi}\times S^{d-1}$.
\subsubsection*{Positive curvature}

We now consider the positive curvature slicing of AdS$_{d+1}$,
\be \label{geo4}
ds^2  = du^2 + \ell^2 \sinh^2 {u\over \ell} \left[{-d\tau^2 + \delta_{ij}dx^i
  dx^j\over  \tau^2}\right], \qquad \tau \in (-\infty, 0), \quad u\in   (-\infty, 0).
\ee
This metric corresponds to the scale factor given in
(\ref{eq:AdSScaleFactor}) for the dS${}_d$ case, where for
simplicity we have chosen $c=0$. For the positive
curvature submanifolds at fixed $u$, we have chosen to write the
metric as  the cosmological patch of de Sitter space $dS_{d}$ in
conformal time.
We emphasize that for pure AdS this is a choice of coordinates, whereas
in the case of RG flows the metric ansatz below corresponds to a
different solution from flat-section RG flows.

\begin{figure}[t]
\begin{center}
\includegraphics[width=8cm]{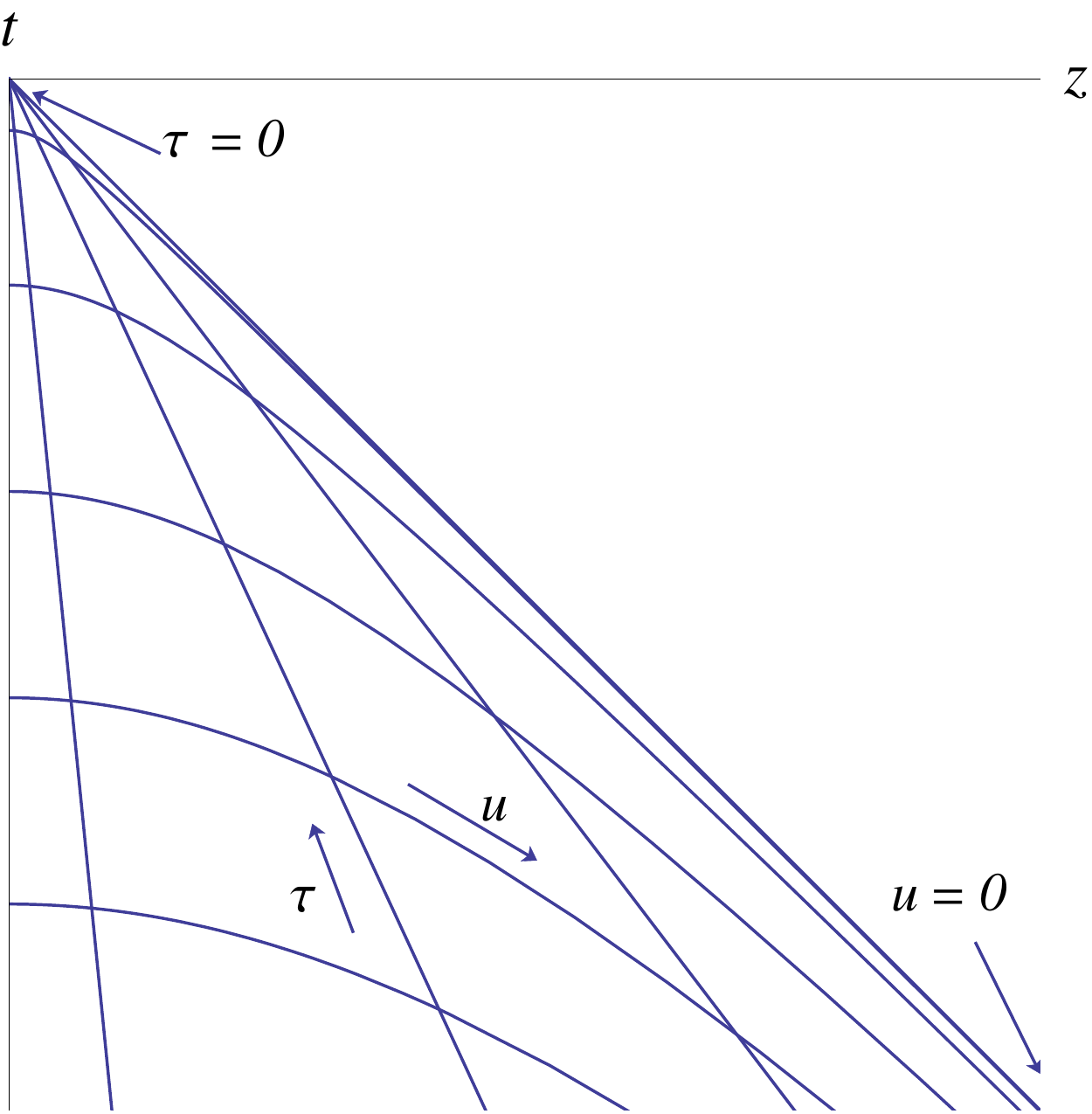}
\caption{Relation between Poincar\'e coordinates $(t,z)$ and
  dS-slicing coordinates $(\tau,u)$. Constant $u$ curves are half straight
  lines all ending at the origin $(\tau\to 0^-)$; Constant $\tau$
  curves are branches of hyperbolas ending at $u=0$ (null infinity on the
  $z=-t$ line). The boundary $z=0$ corresponds to $u\to -\infty$.}
\label{dSPoinc}
\end{center}
\end{figure}

One can check that the coordinate transformation
between the forms (\ref{geo4}) and (\ref{geo1}) of the metric is given by:
\bea
&& t = - \tau  \coth {u \over \ell}, \qquad z = {\tau \over \sinh {u
    \over \ell}},  \label{geo5}\\
&& u = -\ell \cosh^{-1}\left(-{t\over z}\right), \qquad \tau = - (t^2-z^2)^{1/2}.  \label{geo6}
\eea
As is clear from these equations, the dS coordinates only cover
a wedge of the Poincar\'e patch, corresponding to  $t \in (-\infty,
0)$ and $0< z \leq |t|$. It is also instructive  to go to hyperbolic coordinates:
\be \label{geo6b}
t = \tau \cosh\theta, \quad z = \tau \sinh\theta , \quad \cosh {u\over
  \ell} = -\coth \theta .
\ee
The metric in these coordinates reads:
\be\label{geo6c}
ds^2  = {\ell^2\over \sinh^2\theta} \left[ d\theta^2 + {-d\tau^2 + \delta_{ij}dx^i
  dx^j\over  \tau^2}\right], \quad \tau \in (-\infty, 0), \quad
\theta\in   (-\infty, 0).
\ee
with the boundary located at $\theta =0$.  The curves $\tau = const$
are branches of hyperboles, whereas the curves $u=const$ are straight
lines through the origin which can be parametrized by  $\theta$ (see
figure \ref{dSPoinc}).

We can identify the following  limits:
\begin{itemize}
\item As $u \to -\infty$, $(\theta=0)$ the two time coordinates become
  equal, $t=\tau$,  and
  $z \to 0$. This corresponds to going to the boundary.
\item The limit $\tau \to 0^-$ , i.e.~the far
  future in dS, also corresponds to  the
  boundary $z = 0$, as well as the future end of the $t$ half-line,
  $t=0$.
\item The ``endpoint'' $u\to 0$,  $(\theta=-\infty)$  corresponds to the line  $z=|t|$, with both
  $t\to\infty$ and $t\to-\infty$. This is the past  horizon of
  Poincar\'e AdS.
\end{itemize}
The same remarks  apply to the same regions in the RG flow
geometry.

\begin{figure}[t]
\begin{center}
\includegraphics[width=10cm]{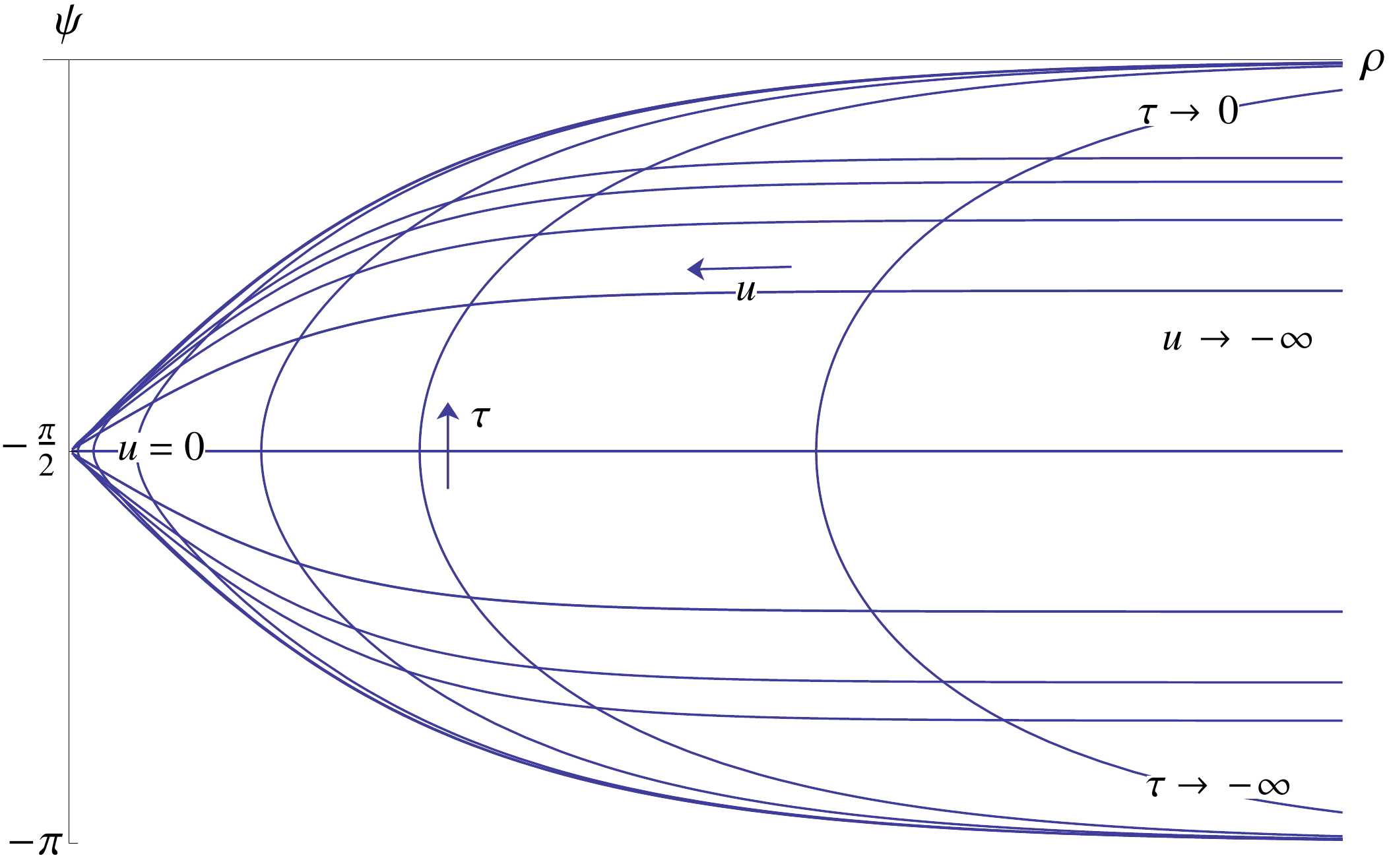}
\caption{Embedding of the dS patch in global coordinates. The flow endpoint
  $u=0$ corresponds to the point $\rho=0, \psi=-\pi/2$ in global
  coordinates. the $AdS$ boundary is at $\rho=+\infty$ and it is
  reached along $u$ as  $u\to -\infty$, and along $\tau$ both as
  $\tau\to -\infty$ and as $\tau \to 0$.}
\label{dSGlobal}
\end{center}
\end{figure}

\begin{figure}[h!]
\begin{center}
\includegraphics[width=7cm]{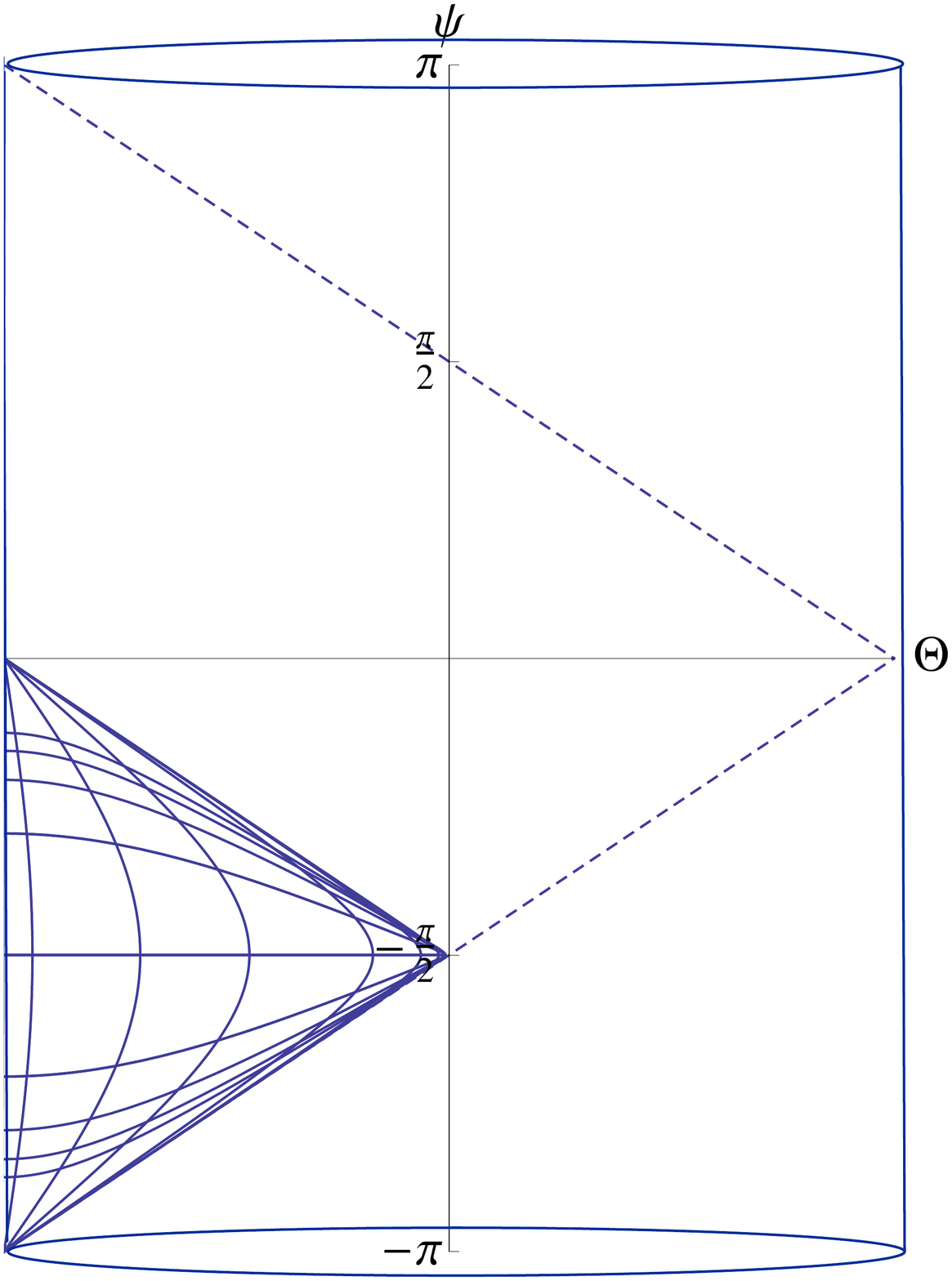}
\caption{Embedding of the dS patch in global conformal
  coordinates, $\tan\Theta = \sinh \rho$, where each
  point is a $d-1$ sphere ``filled'' by $\Theta$. The boundary is at $\Theta= \pi/2$. The
  dashed lines correspond to the Poincar\'e patch embedded in global
  conformal coordinates. The flow endpoint  $u=0$ is situated on the Poincar\'e horizon.}
\label{dSPoincGlobal}
\end{center}
\end{figure}

To have a better understanding of the ``endpoint'' $u \to
0$, it is convenient to see what it corresponds to in global coordinates.  Using equations (\ref{geo2}) and (\ref{geo6b}) we  find the
transformation between global and dS-slicing coordinates, which at $\vec{x}
= 0$ reads:
\bea
&& \cosh\rho = {\left[\left(\ell^2 -\tau^2\right)^2 + 4\tau^2 \ell^2
    \cosh^2\theta\right]^{1/2} \over  2\ell\tau \sinh\theta}, \label{geo7}\\
&& \nonumber \\
&&\sin \psi = - {2\ell \tau \cosh\theta \over \left[ \left(\ell^2 -\tau^2\right)^2 + 4\tau^2 \ell^2
    \cosh^2\theta \right]^{1/2}}. \label{geo8}
\eea
In the IR limit $\theta \to -\infty$, and for any  $\tau$ fixed, we
reach the point $\rho = 0, \psi = -\pi/2$, which is a point in the
interior of AdS space-time.  On the other hand, the
far  past and future are on the boundary at $\psi = -\pi$ and $\psi=
0$ respectively. This embedding is illustrated in figure
\ref{dSGlobal}.  Figure \ref{dSPoincGlobal} represents the conformal
diagram of the same
geometry,  where the AdS boundary is now at $\Theta= \pi/2$.

\subsubsection*{Negative curvature}

We now repeat the analysis for the case negative curvature.  We write
the metric of AdS$_{d+1}$ using the scale factor from equation
(\ref{eq:AdSScaleFactor}) in the negative curvature case,
\begin{align} \label{geo9}
ds^2  &= du^2 + \ell^2 \cosh^2 {u\over \ell} \left[{d\xi^2 - dt^2 + \eta_{ij}dx^i
  dx^j\over  \xi^2}\right], \\
\nonumber  \xi &\in (0,+\infty), \quad u\in
(-\infty, +\infty), \quad i=1, \ldots, d-2.
\end{align}
We have set the constant $c=0$ for simplicity, and written the AdS$_d$
slice metric in Poincar\'e coordinates.

As we discussed in the previous subsections, the metric
(\ref{geo9})
seems to have two distinct boundaries, at $u = \pm \infty$. Below we
will clarify the meaning of the boundary structure, and argue that the
boundary conditions at the two endpoints must be identified.

\begin{figure}[t]
\begin{center}
\includegraphics[width=5cm]{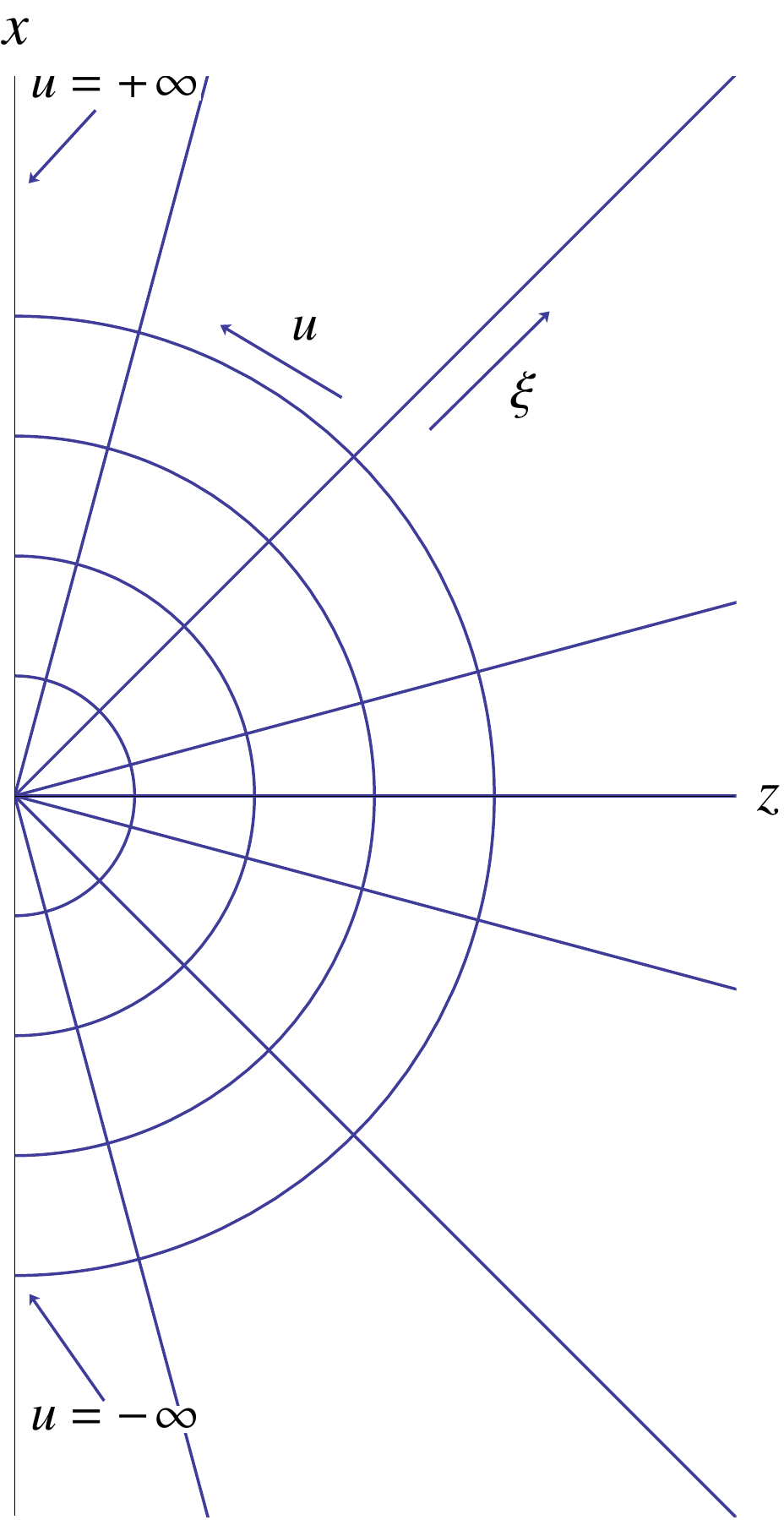}
\caption{Relation between Poincar\'e coordinates $(x,z)$ and
  AdS-slicing coordinates $(\xi,u)$. Constant $u$ curves are half straight
  lines all ending at the origin $(\xi\to 0^-)$; Constant $\xi$
  curves are semicircle joining the two halves of the  boundary at $u=\pm\infty$.}
\label{AdSPoinc}
\end{center}
\end{figure}

First, let us write the coordinate transformation that brings the
Poincar\'e AdS metric (\ref{geo1}) to the form (\ref{geo9}). To this
end, we
single out one spatial  coordinate $x \equiv x^1$ which will be traded for
the AdS$_d$ radial coordinate $\xi$ (the other coordinates $x^2\ldots
x^{d-1}$ being spectators). Then, the appropriate transformation is:
\bea
&& x = \xi  \tanh {u \over \ell}, \qquad z = {\xi\over \cosh {u
    \over \ell}},  \label{geo10}\\
&& u = \ell \sinh^{-1}\left({x\over z}\right), \qquad \xi = (x^2+z^2)^{1/2}.  \label{geo11}
\eea
It is instructive to use polar coordinates:
\be \label{polar}
x = \xi \sin\theta, \quad z = \xi \cos\theta, \quad \sinh {u\over
  \ell} = \tan\theta.
\ee
We notice the following features:
\begin{itemize}
\item
The lines of constant $\theta$ are straight lines in the half-plane
$(x,z>0)$, reaching the origin as $\xi \to 0$ (Poincar\'e boundary of the
lower-dimensional AdS$_d$ slice) (see  figure \ref{AdSPoinc}).
\item
The Poincar\'e boundary $z=0, x\in (-\infty,+\infty)$ is split in two
parts: $x>0$ corresponds to $u\to +\infty$ ($\theta = \pi/2$), whereas
$x<0$ is reached as $u\to -\infty$ ($\theta = -\pi/2$).
\item The lines of constant $\xi$ are circles joining the two
  half-boundaries from $\theta= -\pi/2$ to $\theta = \pi/2$.
\item The two halves of the boundary are connected through the surface
  $\xi=0$ (which is on the boundary for any value of $u$ (or
  $\theta$)). This surface corresponds to the boundary of the {\em
    lower dimensional} AdS space.
\end{itemize}

Therefore, the two boundaries at $u=\pm \infty$ are not really
disconnected, but they are connected through the lower-dimensional
AdS boundary. Because of this, the geometry really has only one boundary, as
was already noted in  \cite{maldamaoz} in the case of the pure AdS
geometry. This fact  carries over to the our RG flow
solutions, however there is a subtlety due to the non-trivial scalar
field profile.  In our case, the scalar field is constant on each
radial slice. In the coordinates (\ref{polar}),  this  is a function
$\f(\theta)$, which does not vanish at the origin $\xi=0$. Thus, the
solution can only be defined up to a cut-off $\xi = \epsilon$, but  it
cannot be extended to a regular function all the way to the {\em
  slice} AdS boundary $\xi=0$. The full boundary is thus the two
half lines $\theta = \pm\pi/2$ starting at $x=\pm\epsilon$, connected
by the half circle $\xi = \epsilon$ with $-\pi/2 <\theta <\pi/2$. The
fact that one cannot take the limit $\epsilon\to 0$ in a regular way
suggests that, on the field theory side,  the theory lives on an $AdS$ space with a
defect (or a source) on the boundary. The precise nature of such a
defect is a matter of speculation, and we will not investigate it
further here.

There is another setup that is described by the AdS-sliced metrics we consider. As already suggested in \cite{maldamaoz}, we may orbifold the AdS slices by discrete Fuchsian groups in order to make them have finite volume. For AdS$_2$ for example one obtains genus $g\geq 2$ Riemann surfaces with negative curvature.
In such a case, the full bulk space-time has now two disconnected boundaries.
There are possible instabilities in that case studied for the full AdS geometry in \cite{maldamaoz}.
Our results are applicable to cases where such instabilities do not
arise.

Notice that regularity of the RG flows implies that the
solution must be symmetric around $\theta=0$, since at that point both
the derivative of the  scale factor and the scalar field
vanish. Thus, the UV boundary conditions on $x<0$ must be the
same as for $x>0$.

Finally, notice that  knowledge of  the bulk solution automatically
imposes boundary conditions for the fields in the lower dimensional theory on
AdS$_{d}$.

\subsection{Comments on stability}
\label{sec:stability}
To conclude this section we briefly comment on the issue whether the solutions we
are discussing are stable under small perturbations.  Here we will
not perform a complete stability analysis, which is technically involved
and  is beyond the scope of this paper. Nevertheless, we can  draw
some hints about stability from general
considerations and previous experience with the same issue in
flat-slicing solutions. This will not be conclusive, but it will lay the ground for a more thorough future
analysis. We start with some preliminary considerations, which we will then
illustrate explicitly in the case of tensor fluctuations.

In $(d+1)$-dimensional Einstein-scalar theories where the metric ansatz
respects $d$-dimensional maximal symmetry, physical fluctuations can
be decomposed into
tensor, vector and the scalar components with respect to the
$d$-dimensional constant curvature metric $\zeta_{\mu\nu}$. Linearized
diffeomorphisms play the role of gauge transformations, and one can
show that the gauge-invariant propagating fluctuations reduce to a
single propagating, gauge-invariant  scalar mode, as well as a single
tensor mode. The latter is obtained by perturbing the background
slice metric by
a symmetric transverse traceless 2-tensor,
\be\label{stab1}
\zeta_{\mu\nu} \to \zeta_{\mu\nu} + h_{\mu\nu}, \quad \nabla^{(\zeta),
  \mu}h_{\mu\nu} =0 = \zeta^{\mu\nu}h_{\mu\nu} \, .
\ee
The definition of the single gauge-invariant scalar fluctuation
$\xi(u,x^\mu)$ is much more complex, as it involves the background
functions $A(u)$ and $\varphi(u)$.

The question of
stability can be phrased in terms of the positivity of
eigenvalues $\lambda = m^2$
of suitable Hamiltonians for a set of
Schr\"odinger-like problems (one for each independent bulk
perturbation). The Hamiltonian is the operator governing the radial
evolution.  The corresponding eigenvalues $m^2$ represent the mass
spectrum in a Kaluza-Klein decomposition in terms of $d$-dimensional
modes, and stability requires that they are above a certain minimal
value (which is zero in flat space). Absence of
ghosts on the other hand is automatically guaranteed by the fact that
all sources satisfy the null-energy condition.

It is known (see e.g.~\cite{exotic} for a self-contained
 discussion) that, in scalar-tensor models of type (\ref{eq:action}),
 asymptotically AdS solutions with {\em flat} constant-$u$ induced  metric,
 are stable under small perturbations, at least  when the
Schr\"odinger problem is well defined without imposing extra boundary
conditions in the IR. This condition is satisfied by a large class of bulk
potentials, as long as they do not grow too fast at large $\varphi$, and
it certainly applies when there is an AdS fixed point in the IR. In
the latter case, one can show that the spectrum of $m^2$ for both
tensor and scalar fluctuations  is continuous and starting at
zero (excluded).  This holds regardless of the details of the
bulk geometry, and it applies in particular to geometries which
include a bounce \cite{exotic}. In other words, in the flat case the
presence of a bounce does neither destabilize the tensor nor the scalar sector.

Let us now move on to the curved case. Intuitively, when we  put the theory on
a positively curved manifold,  we may  expect to gap a spectrum which was
gapless in the flat theory. 
We will see that this expectation is reflected in a modification of the
IR asymptotic behavior of the effective Schr\"odinger potential for the tensor
perturbations. Furthermore, experience with the flat-slicing case
suggests that the tensor and scalar Schr\"odinger potentials obey the same
UV and IR asymptotics, which is universally governed by the
asymptotically AdS form of the metric. Therefore, we
expect that a similar IR modification will also gap the scalar
fluctuations. The presence of a bounce is not expected to
introduce specific stability problems, since none  were there in the
flat case to start with.

Let us now discuss the negatively curved case. 
One might
expect that in this case the spectra become discrete, since the bulk
is enclosed by two UV AdS boundaries which behave like infinite
potential walls. However  the situations is
complicated by the fact that, as discussed in section \ref{sec-coordinates}, one
needs to regulate the  boundary on both sides,  in order to obtain a
consistent asymptotic behavior on the slice-AdS boundary. This will
again make the spectrum discrete, but the eigenvalues will crucially
depend on the boundary conditions one imposes at the regulated
boundary.

To illustrate some of the previous points we now present a more quantitative
analysis for the tensor perturbations, which turn out to be quite tractable. The  linearized second order equation governing the tensor fluctuations
is universal, i.e.~independent of the fields which source the
background. It only involves the scale factor, and in our ansatz it
reads (see e.g.~\cite{KarchRandall}):
\be\label{stab2}
\left[-\de_u^2 - d (\de_u A) \de_u - \left(\nabla^{(\zeta),\mu}\nabla^{(\zeta)}_\mu - {2K}\right) \right] h_{\mu\nu}(u,x^\mu) = 0.
\ee
where $K = \alpha^{-2}$ for positive curvature, and $K=-\alpha^{-2}$
for negative curvature.  One then decomposes this equation  in terms of $d$-dimensional modes
satisfying the appropriate linear equation for a tensor of mass $m^2$,
plus a radial equation for the profile,
\be\label{stab3}
h_{\mu\nu} = h(u)\epsilon_{\mu\nu}(x),  \quad \left(\nabla^{(\zeta), \mu}\nabla^{(\zeta)}_\mu \pm {2\over
    \alpha^2}\right) \epsilon_{\mu\nu} = m^2 \epsilon_{\mu\nu}  \, ,
\ee
\be\label{stab4}
\left[-{d^2\over du^2} - d {d A \over du} {d\over du} \right] h(u) =
m^2 h(u) \, .
\ee
Equation (\ref{stab3})  reduces to the linearized Einstein equation
for $m^2=0$. Thus, in this language, $m^2$ corresponds to the Pauli-Fierz
mass parameter.

One then performs the following transformation on both the radial coordinate
and the radial profile $h(u)$,
\be \label{stab5}
du = e^{A(r)}dr, \qquad \psi(r) = e^{{(d-1)\over 2}A(r)} h(r),
\ee
to obtain the Schr\"odinger-like equation
\be\label{stab6}
\left[-{d^2 \over dr^2} + V(r) \right]\psi(r) = m^2 \psi(r), \qquad
  V(r) = {(d-1)^2 \over 4}\left({d A \over dr}\right)^2 +{(d-1) \over
    2}  {d^2 A \over dr^2}.
\ee
We now discuss the positive and negative curvature cases separately.
\begin{itemize}
\item {\bf Positive curvature:}
In this case $r\to 0$ as $u\to -\infty$ (UV boundary) and $r\to
+\infty$ as $u\to u_c$ (IR endpoint).  One then finds
\be\label{stab7}
A(r) \to \left\{\begin{array}{cl} -\log {r\over \ell} &\quad r\to 0 \\
    & \\
    -{r\over \alpha} & \quad r \to +\infty \end{array}\right. \,,
\quad V(r) \to \left\{\begin{array}{cl} {d^2 -1 \over 4r^2} &\quad
    r\to 0 \\ & \\
    {(d-1)^2\over 4\alpha^2} & \quad r \to +\infty \end{array}\right.
\ee
In the UV, as in the zero-curvature case, the potential diverges as
$1/r^2$.  In the IR,  it now asymptotes a positive constant instead
of tending to zero as in the flat case. The asymptotics (\ref{stab7}) imply that  the continuous
spectrum is gapped by $(d-1)^2/4\alpha^2$. With this
asymptotic behavior one can show following \cite{exotic} that the Hamiltonian
operator (\ref{stab6}) is strictly positive, which implies
$m^2>0$. This however is not enough, because a massive, transverse traceless
symmetric tensor must in addition satisfy the Higuchi bound
\cite{higuchi,derham},
\be\label{stab8}
m^2 \geq {(d-2) \over \alpha^2}
\ee
 Since $(d-1)^2/4 \geq (d-2)$ for any $d$, the modes in the
continuous spectrum  {\em do} satisfy the Higuchi bound
\eqref{stab8}.  It is suggestive that the gap in the continuous
spectrum is sufficient to satisfy the bound, although we cannot exclude the
existence of discrete eigenmodes with masses in the range $0< m^2 <
(d-2)/\alpha^2$. This issue can only be settled with a detailed
 study of the full solution, which most likely has to be performed
 numerically.

\item {\bf Negative curvature:}
In this case the Schr\"odinger problem is in a finite size box $0<r<L$
whose endpoints are the images of the two boundaries at  $u \to \pm
\infty$. The potential has the following asymptotic behavior
\be
 V(r) \to \left\{\begin{array}{cl} {d^2 -1 \over 4r^2} &\quad
    r\to 0 \\ & \\
     {d^2 -1 \over 4(L-r)^2} & \quad r \to L \end{array}\right. \, .
\ee
Taken at face value, the Hamiltonian operator is positive and all
modes respect the unitarity bound  for symmetric tensors in AdS$_d$,
which is just $m^2>0$ (there is no equivalent of the Higuchi bound in
AdS). However, as we have mentioned above, we must consider cutting
off the two boundaries at finite $u$. The spectrum then depends on the
boundary conditions.  Therefore, without a more complete theory of how these
solutions arise  and what imposes the boundary conditions, it is
premature to draw definite conclusion about the positivity of the
spectrum.\footnote{One can expect  that if the boundary
  condition is imposed by some physical well-behaved source (e.g. a
  lower-dimensional brane with positive energy) no instability should
  arise from the boundary conditions. For example, if we impose vanishing of the
wave-function at the endpoint (which is the extension of the
normalizability condition in the full range of $r$, and corresponds
to an infinitely massive extended source at the boundary) then
positivity of $m^2$  still holds.}
\end{itemize}

The analysis above strongly hints at the fact that no unstable tensor
modes are present. However, obtaining the equation for the scalar perturbation is much more
complicated: one expects to obtain
an equation of a similar form as (\ref{stab6}), in which  the
coefficients will involve not only the scale factor but also the scalar-field profile. Nevertheless we
may expect (as experience with the flat case suggests) that  the behavior  of the Schr\"odinger potential close to
the UV boundary and,  in the positive curvature case, to the IR
endpoint to be similar to the one observed for tensor modes. This in turn
would imply that  the scalar spectrum is positive (with a gapped
continuous component).  We leave these questions for future
investigation.

We conclude this section with a comment about ``bouncing'' solutions,
where the scalar field is non-monotonic.
Notice that nothing special happens to the Schr\"odinger  potential
(\ref{stab6}) at points corresponding to bounces: the scale factor
there is finite, as well as its first and second derivatives, and no
special features arise.
One can argue that the same holds for scalar
perturbations, even without knowing the precise form of the scalar
perturbation  equation.  This is known to be the case for the
flat-slicing geometries, in which indeed the
bounce does not destabilize the fluctuations \cite{exotic}. Adding
curvature  will most likely only modify the perturbation equation  by
further mass-like terms, as it is the case for the tensor equation
(\ref{stab2}). This will not dramatically change the fact that the
presence of the bounce does not significantly affect stability.

\section{Examples of  complete RG flows} \label{sec:examples}
In this section, we will display solutions corresponding to full RG flows for different choices of potentials. The examples are chosen to illustrate various properties of RG flows of theories on curved manifolds described in the previous sections.  The flows will originate from a UV fixed point at a maximum of the potential, which we choose to locate at $\f=0$ for convenience, and will end at an IR point $\f_{0}$, which does not need to be an extremum of the potential. We will distinguish between flows where $\f$ changes monotonically between the UV and IR points and situations where the flow in $\f$ changes direction, an effect referred to as a bounce. In addition, we will describe how the flows are deformed by changes to UV data such as the dimensionless curvature $\mathcal{R}$.

\subsection{Generic flows}
\label{sec:generic}
We will refer to flows as generic if they exhibit the following two properties: they originate at a maximum of the potential and end in the region between this maximum and the nearest minimum. In addition, $\f$ changes monotonically along the flow from UV to IR. We call such solutions `generic' as they arise for generic potentials as long as they possess at least one maximum.

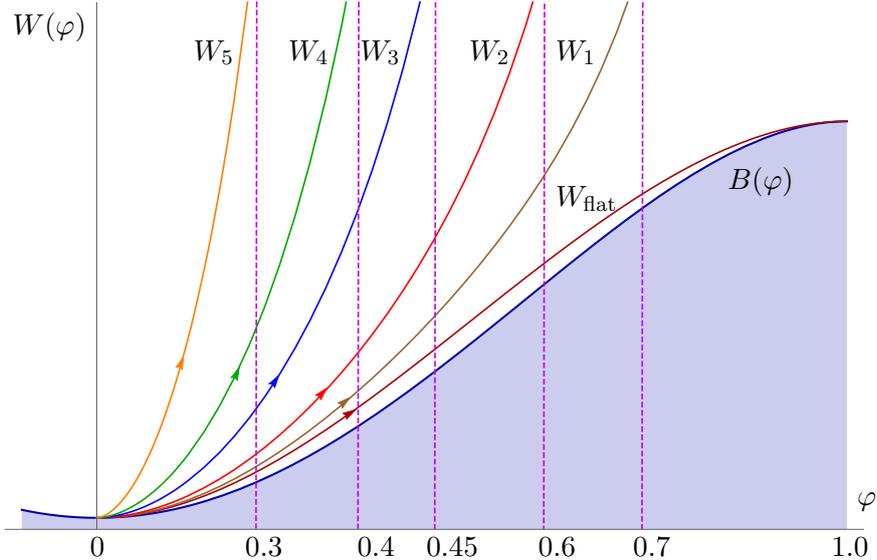
\begin{figure}[t]
\centering
\begin{overpic}
[width=0.75\textwidth]{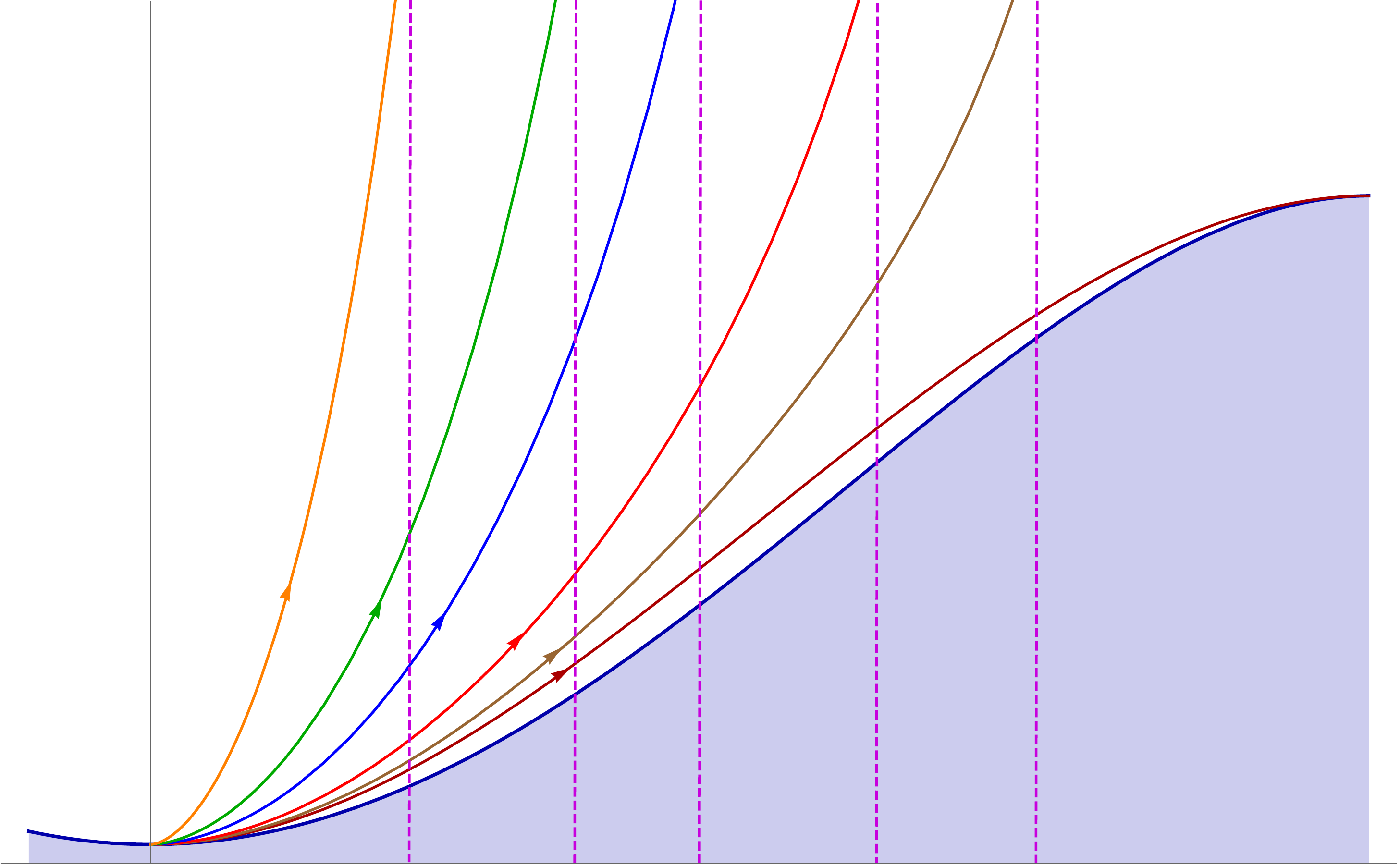}
\put(10,-3){0}
\put(73,-3){0.7}
\put(62,-3){0.6}
\put(49,-3){0.45}
\put(41,-3){0.4}
\put(28,-3){0.3}
\put(96,-3){1.0}
\put(99,3){$\f$}
\put(1,58){$W(\f)$}
\put(22,55){$W_5$}
\put(33,55){$W_4$}
\put(41.3,55){$W_3$}
\put(54,55){$W_2$}
\put(64,55){$W_1$}
\put(64,38){$W_{\text{flat}}$}
\put(84,40){$B(\f)$}
\end{overpic}
\caption{Solutions for $W(\f)$ with $\mathcal{R} \geq 0$ and for the
  potential \protect\eqref{generic} with $\Delta_-=1.2$. The five
  solutions $W_i(\f)$ with $i=1, \ldots, 5$ differ in the value of
  their IR endpoint $\f_0$. The critical curve is defined as $B(\f)=\sqrt{-3 V(\f)}$. In the case of zero and positive curvature, the superpotential $W(\f)$ cannot enter the are below the critical curve, which is depicted as the shaded region.}
\label{fig1examples}
\end{figure}

To illustrate generic flows, a potential exhibiting a maximum and at
least one minimum will be sufficient.\footnote{In fact, the existence
  of a minimum at a finite value of $\f$ is not strictly necessary,
  but will be useful for illustrating further properties of these
  flows. Similar conclusions apply also to potentials that extend to
  infinity in field space, which in the flat case were also discussed in \cite{exotic}.} For this purpose  we  consider the following quadratic-quartic potential:
\begin{equation}
\label{generic}
V(\f)=-\frac{d(d-1)}{\ell^2}-\frac{m^2}{2}\f^2+\lambda\f^4  \, .
\end{equation}
This  potential has one maximum at $\f_{\textrm{max}}=0$. We will also find it convenient to choose $\lambda=m^2/4$ such that the minima occur at $\f_{\textrm{min}} = \pm 1$. We then proceed to studying RG flows by solving \eqref{eq:EOM7}--\eqref{eq:EOM8} numerically for $W(\f)$ and $S(\f)$. In practice, it is easiest to specify boundary conditions for $W$ and $S$ at or close to the IR end point $\f_0$. The relevant boundary conditions for RG flows are described in sections \ref{positivecurvatureflows} and \ref{negativecurvatureflows} for the two cases of a QFT  on a sphere or~AdS. Given the symmetry of the setup, we restrict our attention to flows that end in the region $\f_0 \in [0,1]$.

\subsubsection*{QFT on S${}^d$: $\mathcal{R} > 0$}
In Fig.~\ref{fig1examples} we exhibit solutions for the superpotential $W(\varphi)$ corresponding to generic RG flows for a bulk potential given by \eqref{generic} and for $\mathcal{R} > 0$. To be specific, we have set $\Delta_-=1.2$, but our observations will hold more generally.

\begin{itemize}
\item The main result is that for every value of $\f_0$ between
  $\f_{\textrm{max}}=0$ and $\f_{\textrm{min}}=1$ there exists a
  unique solution to the superpotential equations (\ref{eq:EOM4}-\ref{eq:EOM6}) corresponding to an RG flow originating from the UV fixed point at  $\f_{\textrm{max}}=0$ and ending at $\f_0$. For illustration, in Fig.~\ref{fig1examples} we have chosen to display five such solutions labelled by $W_i(\f)$ with $i=1, \ldots, 5$, that differ in the value of the IR endpoint $\f_0$. However, there is no solution for a flow with $\mathcal{R} > 0$ with an endpoint at $\f_0=\f_{\textrm{min}}$ exactly. In contrast, such a flow arises for $\mathcal{R}=0$ and we have included the corresponding superpotential $W_{\textrm{flat}}$ in the figure.
\item Note that the solutions $W_i(\f)$ diverge when approaching their corresponding IR end points. The divergence is of the form $\sim (\f_0-\f)^{-1/2}$ as expected for a RG flow with $\mathcal{R} > 0$ (see sec.~\ref{positivecurvatureflows}). Recall that this divergence does not imply a singularity in the bulk geometry.
\item Returning to Fig.~\ref{fig1examples}, in the vicinity of the UV fixed point the solutions are described by the family of solutions collectively denoted by $W_{-}(\f)$ in section \ref{sec:asymp}. These solutions depend on the two continuous parameters $\mathcal{R}$ and $C$. This is consistent with the existence of a unique solution corresponding to a RG flow for every $\f_0$. Picking a solution with the correct IR behavior for a RG flow fixes one combination of the two parameters. The remaining freedom is then equivalent to the choice of IR end point $\f_0$.

\begin{figure}[t]
\centering
\begin{subfigure}{.5\textwidth}
 \centering
   \begin{overpic}[width=1.0\textwidth,tics=10]{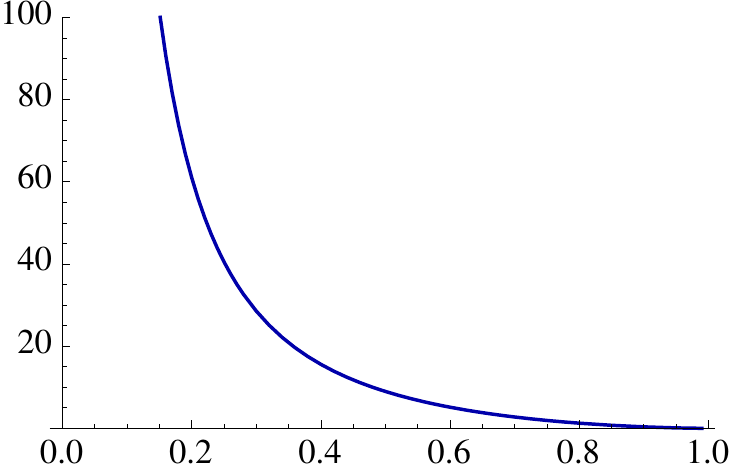}
\put (93,12) {$\f_0$} \put (13,61) {$\mathcal{R}$}
\end{overpic}
 \caption{\hphantom{A}}
  \label{fig:Rvsphi0a}
\end{subfigure}%
\begin{subfigure}{.5\textwidth}
  \centering
 \begin{overpic}[width=1.0\textwidth,tics=10]{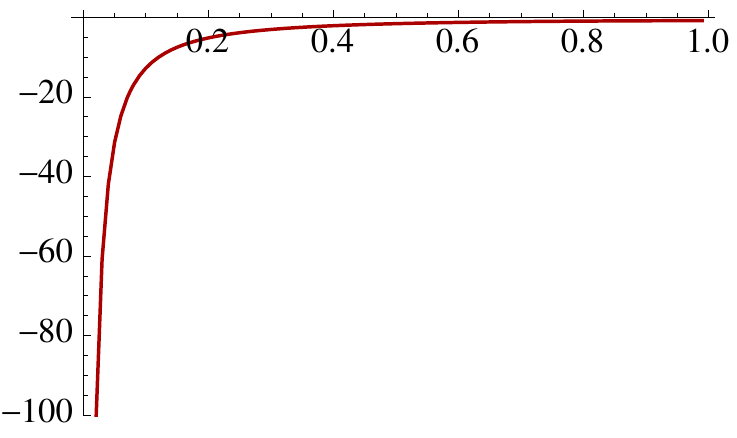}
\put (93,45) {$\f_0$} \put (18,3) {$C$}
\end{overpic}
\caption{\hphantom{A}}
  \label{fig:Cvsphi0a}
\end{subfigure}
\caption{QFT on S${}^d$: Dimensionless curvature $\mathcal{R}$ and dimensionless vev $C$ as a function of IR endpoint $\f_0$  for the
  potential \protect\eqref{generic} with $\Delta_-=1.2$. Both these quantities diverge as $\f_0$ approaches the UV fixed point at $\f=0$.}
\label{fig:RCvsphi0a}
\end{figure}

 \item Given a numerical solution, we can extract the corresponding
   values of $\mathcal{R}$ and $C$ explicitly by fitting the UV region
   with the asymptotics (\ref{eq:Wmsol}-\ref{eq:Tmsol}).    An
   interesting observation is that there exists an inverse relation
   between $\mathcal{R}$ and $\f_0$, i.e.~flows with endpoints closer
   to the UV fixed points exhibit larger values of
   $\mathcal{R}$. Further,  $\mathcal{R}$ diverges when
   $\f_0$ approaches the UV fixed point. On the other hand, when the IR endpoint approaches $\f_{\textrm{min}}$ the value of $\mathcal{R}$ asymptotes to zero. In fact, for the potential considered here the evolution of $\mathcal{R}$ with $\f_0$ is monotonic and displayed in Fig.~\ref{fig:Rvsphi0a}.
\item For completeness, we also exhibit $C$ as a function of $\f_0$ in
  Fig.~\ref{fig:Cvsphi0a}. We find that $C$ is negative, but its
  behavior is otherwise similar to that of $\mathcal{R}$. In
  particular, its absolute value increases with decreasing $\f_0$ and
  diverges when $\f_0 \rightarrow 0$. However, as $\f_0 \rightarrow
  \f_{\textrm{min}}$ we find  numerically that $C \neq 0$ .
 \item Given a solution $W_i(\f)$, we can also determine the
  corresponding scale factor $e^{A(u)}$ and scalar field profile
  $\f(u)$. This introduces the remaining integration
  constant $\f_-$ into the solution. In figure \ref{fig:invscale} we
  plot the  scale factors $e^{A(u)}$ and the scalar field profile $\f(u)$ corresponding to
  the solutions $W_1(\f)$, $W_2(\f)$ and $W_3(\f)$ in figure \ref{fig1examples}. The scale factor $e^{A(u)}$ shrinks to zero at a finite value of $u$, in agreement with our
  identification of $\f_0$ with an IR endpoint. The scalar field arrives at the IR value $\f_0$ at that point.

\end{itemize}

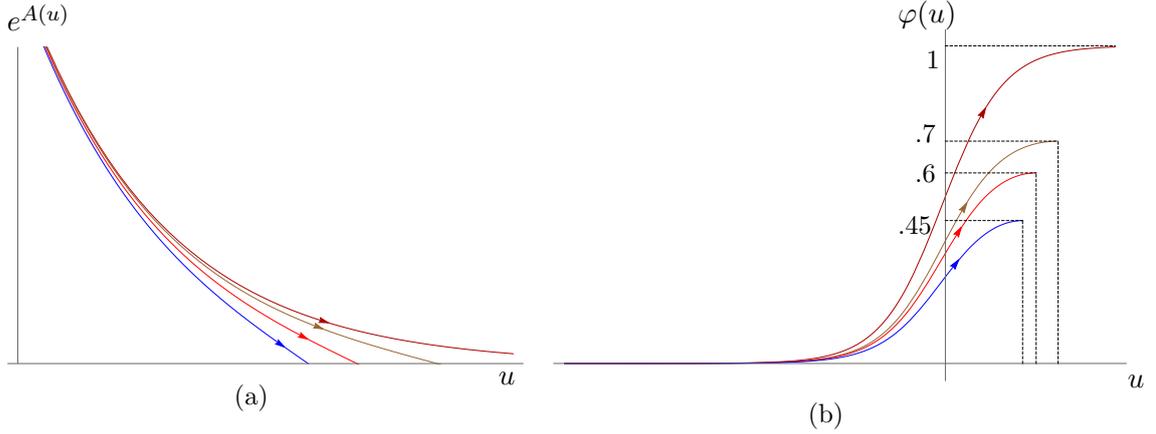
\begin{figure}[t]
\begin{subfigure}{.5\textwidth}
 \centering
   \begin{overpic}[width=.9\textwidth]{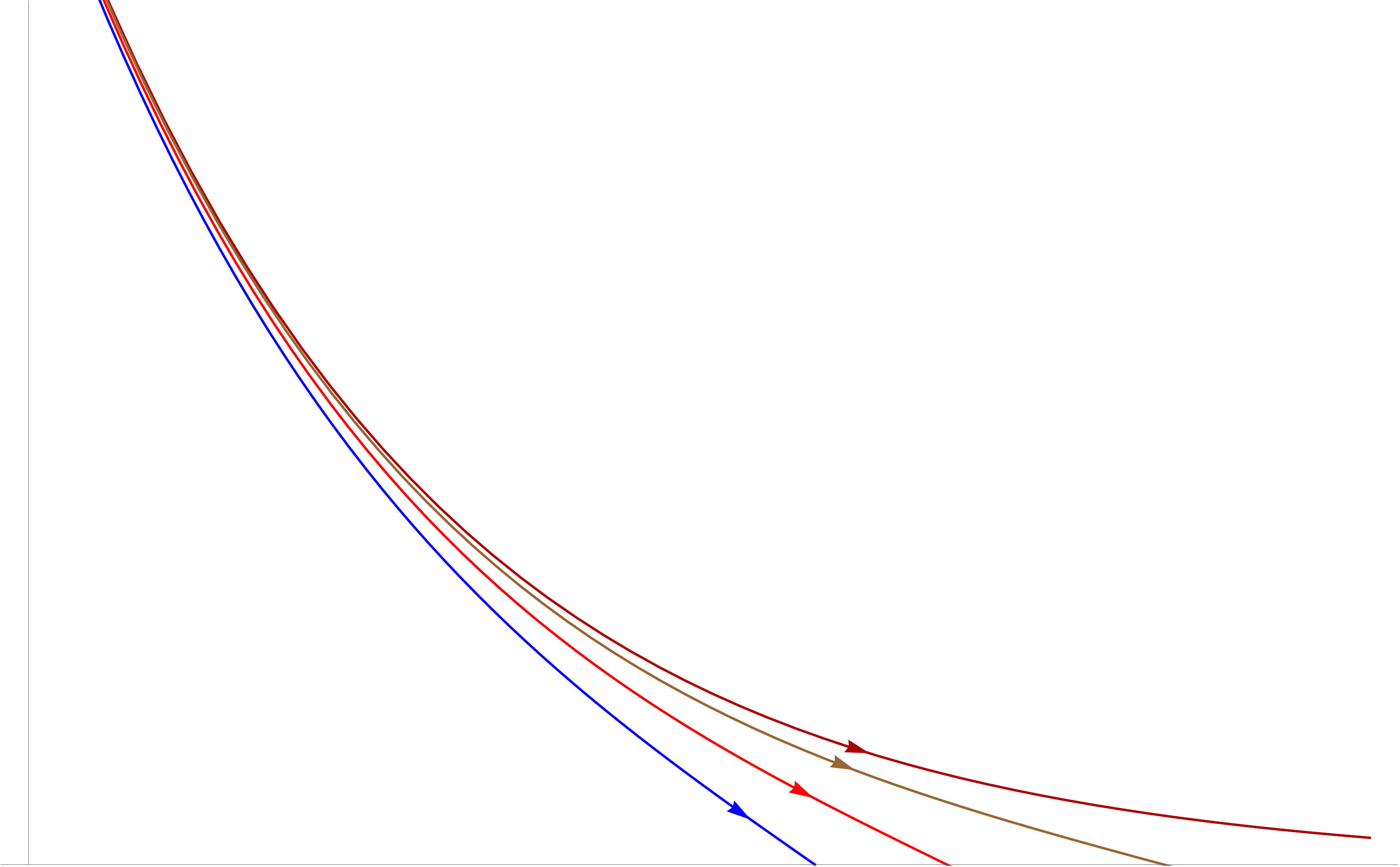}
   \put(95,-4){$u$}
    \put(0,65){$e^{A(u)}$}
\end{overpic}
 \caption{\hphantom{A}}
\end{subfigure}%
\begin{subfigure}{.5\textwidth}
 \begin{overpic}[width=1.0\textwidth]{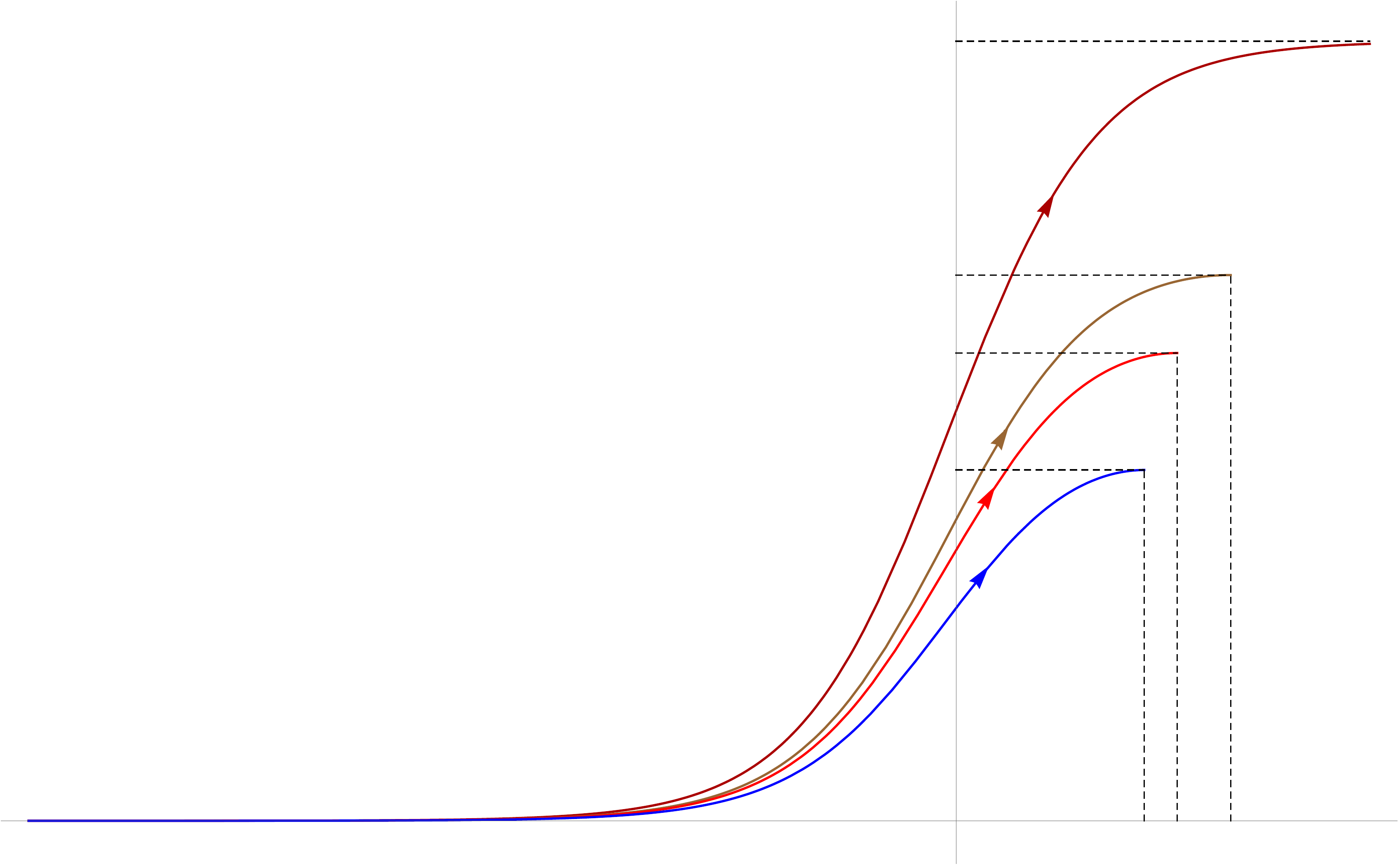}
 \put(100,-1){$u$}
 \put(60,63){$\f(u)$}
  \put(65,55){\small{1}}
   \put(63,42){\small{.7}}
     \put(63,35){\small{.6}}
         \put(60,26){\small{.45}}
\end{overpic}
 \caption{\hphantom{A}}
\end{subfigure}
\caption{ (a): Scale factor $e^{A}$ as a function of $u$ for the flows denoted by 1,2,3 in Fig.~\protect\ref{fig1examples}. For the flat case, the scale factor is depicted as the dark red curve and only shrinks to zero at $u\to \infty$. For the positive curvature cases, the scale factor $e^{A}$ shrinks to zero at finite values of $u$. (b): Scalar field $\f(u)$ as a function of $u$. For the flat case, the scalar field arrives at the minimum of the potential at $\f=1$ at the end of the flow. Here it is denoted by the darker red line. For the curved cases, the scalar field does not reach the minimum. The figure shows the scalar field profile for flows denoted by 1,2,3 in Fig.~\protect\ref{fig1examples}.   } \label{fig:invscale}
\end{figure}

The above findings can be summarized as follows. Consider a field
theory with fixed source $\f_-$ on a background with fixed curvature
$R^{\textrm{uv}}$, such that $\mathcal{R} = R^{\textrm{uv}}
|\f_-|^{-2/ \Delta_-}$ is fixed. 
The
flow will end at a value $\f_0$, which is determined by
 $\mathcal{R}$. The larger the value of $\mathcal{R}$, the closer
this end point $\f_0$ will be to $\f_{\textrm{max}}$. This is in agreement with QFT intuition.
 The remaining
parameter $C$ is then determined by the requirement that the solution
is has the correct (regular) behavior at the endpoint.

\subsubsection*{QFT on AdS${}_d$: $\mathcal{R} < 0$}
We repeat the above analysis for the case with $\mathcal{R} < 0$. We
will find that many properties observed for positive $\mathcal{R}$
also hold for negative $\mathcal{R}$. In this case however the IR
endpoint at $\f_0$ is replaced by an IR {\em turning point} for both $\f$ and $A$ (see section \ref{negativecurvatureflows}).
 There are also other important differences which we will highlight. To be specific, we will again
work with the potential given by \eqref{generic}. In the following, we collect our observations. It will be convenient to distinguish the cases $\Delta_- < 1$ and $\Delta_- >1$.

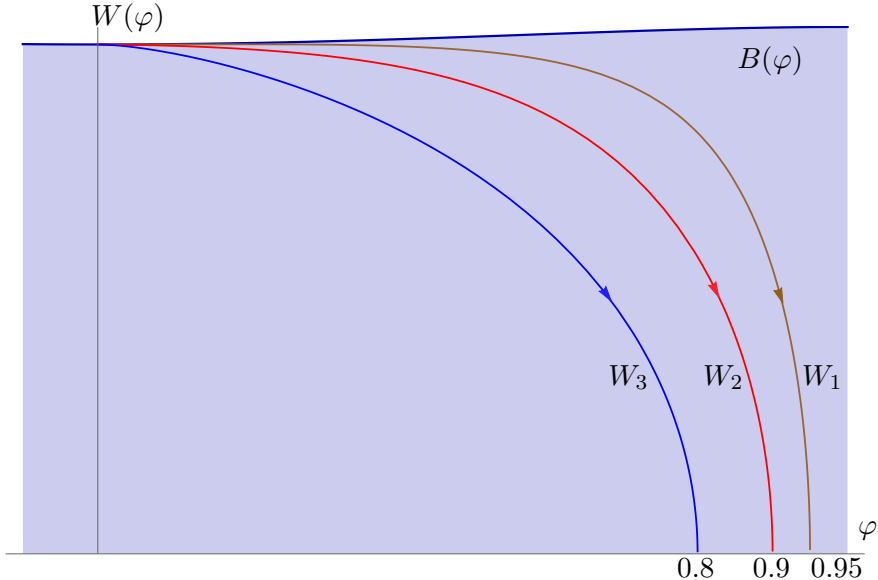
\begin{figure}[t]
\centering
\begin{overpic}
[width=0.75\textwidth]{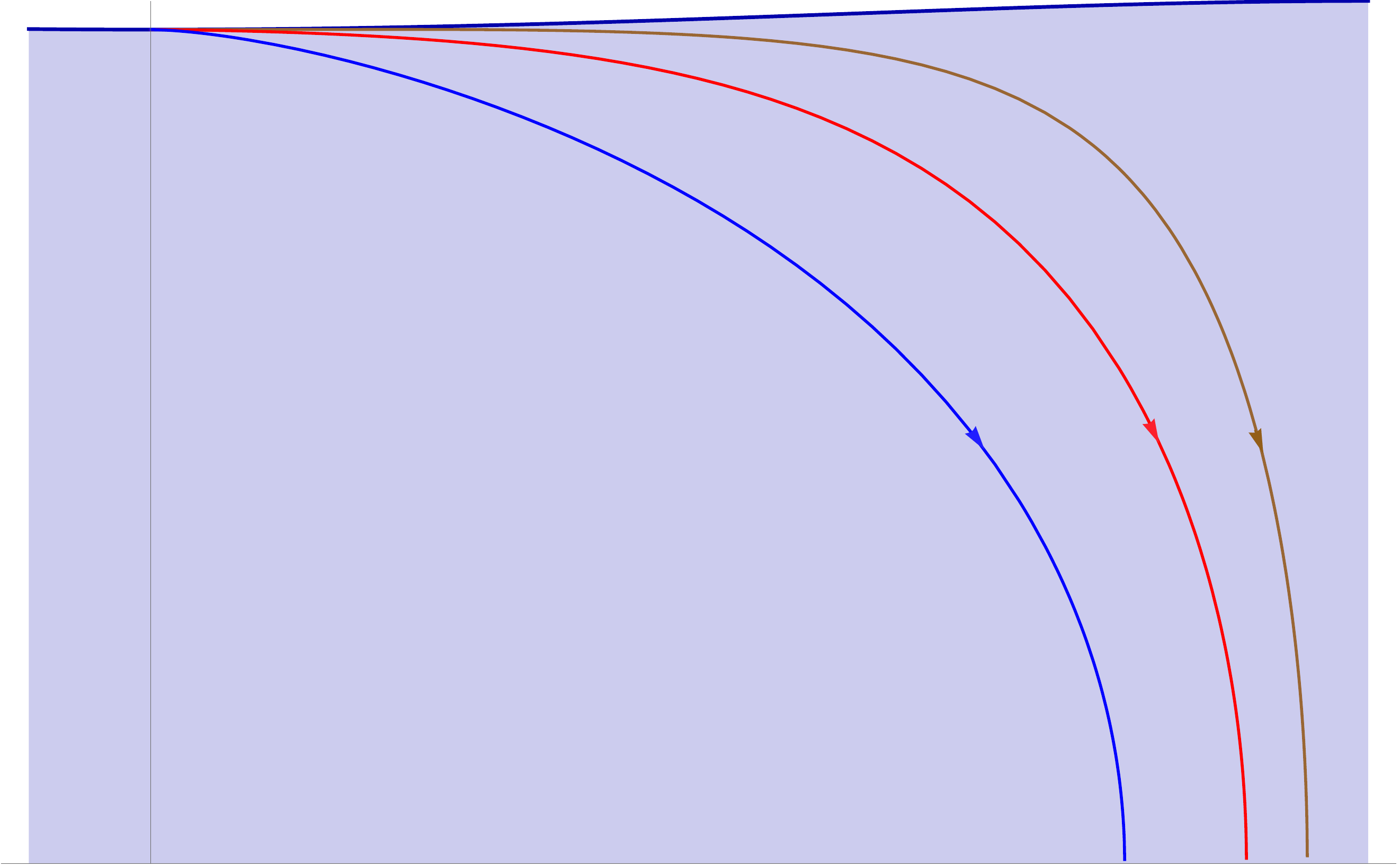}
\put(86.75,-2.5){0.9}
\put(93.5,-2.5){0.95}
\put(78,-2.5){0.8}
\put(99,2.5){$\f$}
\put(10,62){$W(\f)$}
\put(70,20){$W_3$}
\put(81,20){$W_2$}
\put(92.5,20){$W_1$}
\put(85,57){$B(\f)$}
\end{overpic}
\caption{Solutions for $W(\f)$ with $\mathcal{R} < 0$ and for the
  potential \protect\eqref{generic} with $\Delta_-=1.2$. The three
  solutions $W_i(\f)$ with $i=1, 2,3$ have different IR turning points
  $\f_0$. For negative curvature case, the superpotential $W(\f)$ can go into the area below the critical curve $B(\f)=\sqrt{-3 V(\f)}$ which is the shaded region.}\label{WAdS}
\end{figure}

\begin{figure}[t]
\begin{subfigure}{.5\textwidth}
 \centering
   \begin{overpic}[width=.9\textwidth]{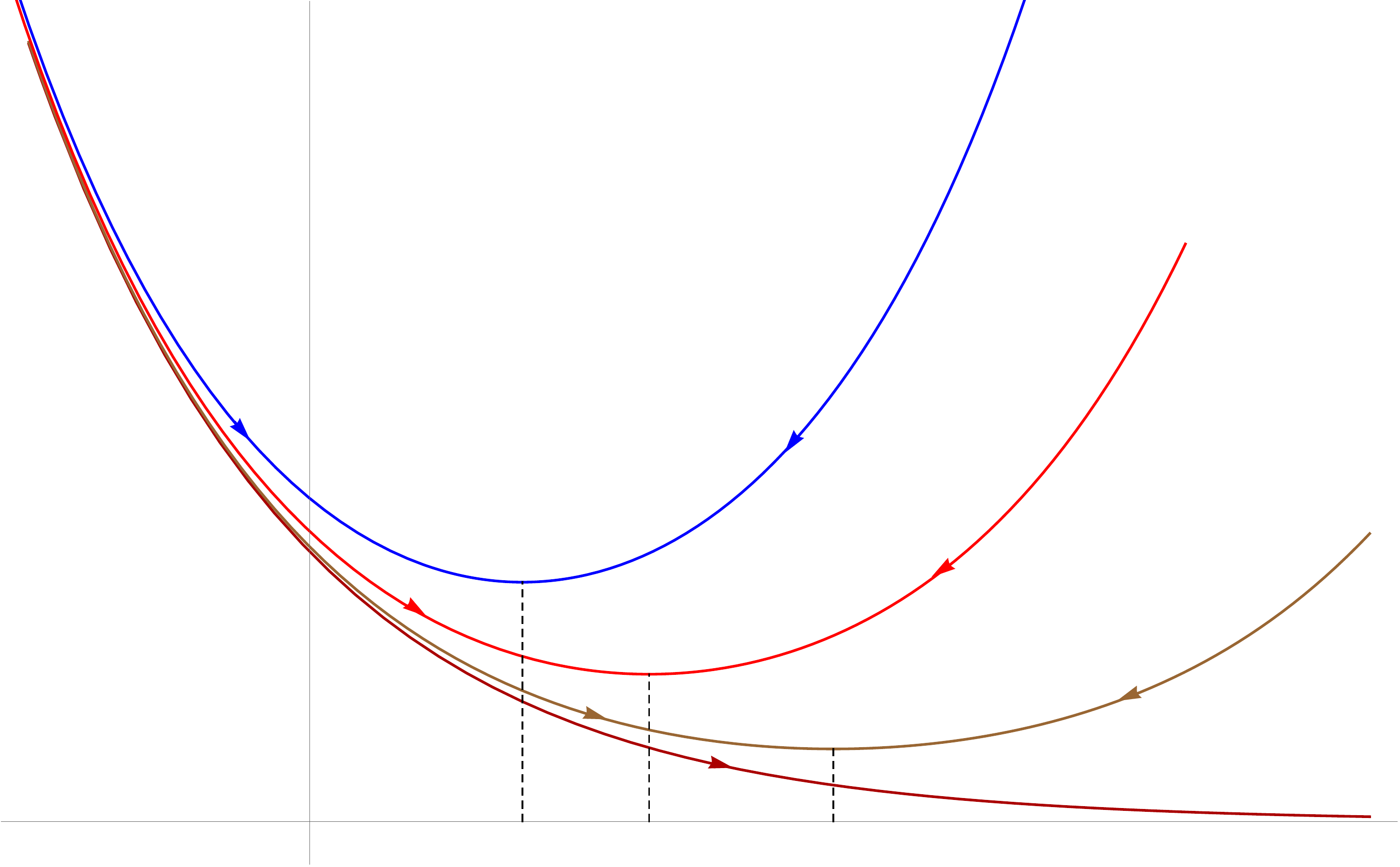}
   \put(95,-1){$u$}
     \put(20,63){$e^{A(u)}$}
\end{overpic}
 \caption{\hphantom{A}}
\end{subfigure}%
\begin{subfigure}{.5\textwidth}
 \begin{overpic}[width=.9\textwidth]{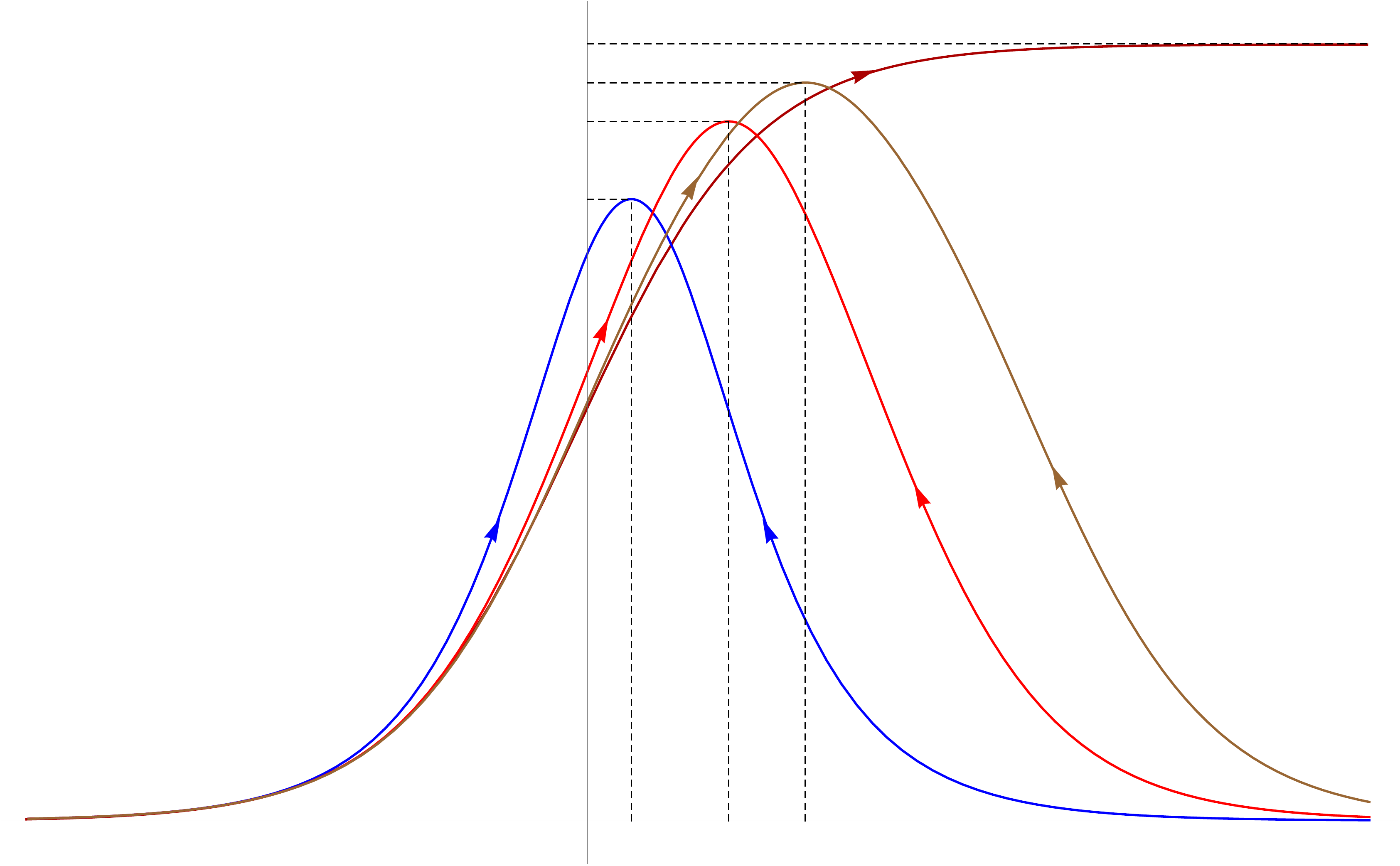}
 \put(95,-1){$u$}
    \put(30,65){$\f(u)$}
    \put(40,58){\tiny{1}}
      \put(36,55){\tiny{.95}}
       \put(38,52){\tiny{.9}}
        \put(38,46){\tiny{.8}}
\end{overpic}
 \caption{\hphantom{A}}
\end{subfigure}
\caption{(a): $e^{A(u)}$ vs.~$u$ and (b) $\f(u)$ vs.~$u$ corresponding to the flows in Fig.~\protect\ref{WAdS}. The coloring is the same as in Fig.~\protect\ref{WAdS}. The dark red curve corresponds to the solution for the flat case. The vertical lines represent the reflection points. } \label{TAdS}
\end{figure}

\begin{figure}[t]
\centering
\begin{subfigure}{.5\textwidth}
 \centering
   \begin{overpic}[width=0.9\textwidth,tics=10]{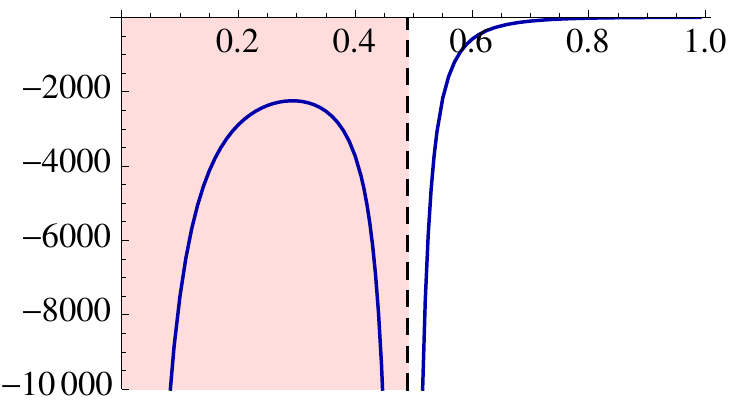}
\put (93,44) {$\f_0$} \put (52,59) {$\f_c$} \put (15,-5) {$\mathcal{R}$}
\end{overpic}
 \caption{\hphantom{A}}
  \label{fig:Rvsphi0c}
\end{subfigure}%
\begin{subfigure}{.5\textwidth}
  \centering
 \begin{overpic}[width=0.9\textwidth,tics=10]{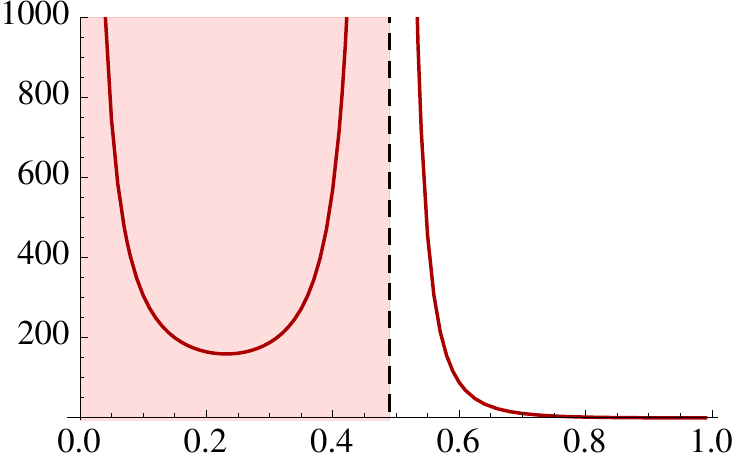}
\put (93,12) {$\f_0$} \put (49,63) {$\f_c$} \put (17,54) {$C$}
\end{overpic}
\caption{\hphantom{A}}
  \label{fig:Cvsphi0c}
\end{subfigure}
\caption{QFT on AdS${}_d$: dimensionless curvature $\mathcal{R}= R^{\textrm{uv}} |\f_-|^{-2 / \Delta_-}$ and dimensionless vev $C = \tfrac{\Delta_-}{d}\langle \mathcal{O} \rangle |\f_-|^{-\Delta_+/\Delta_-}$ vs.~$\f_0$ for the potential \protect\eqref{generic} with $\Delta_-=1.2$. Flows with turning points in the rose-colored region leave the UV fixed point at $\f=0$ to the left before bouncing and finally ending at positive $\f_0$. Flows with turning points in the white region are direct: They leave the UV fixed point at $\f=0$ to the right and do not exhibit a reversal of direction. The flow with turning point $\f_c$ on the border between the bouncing/non-bouncing regime corresponds to a theory with vanishing source $\f_-$. As a result, both $\mathcal{R}$ and $C$ diverge at this point.}
\label{fig:RCvsphi0c}
\end{figure}
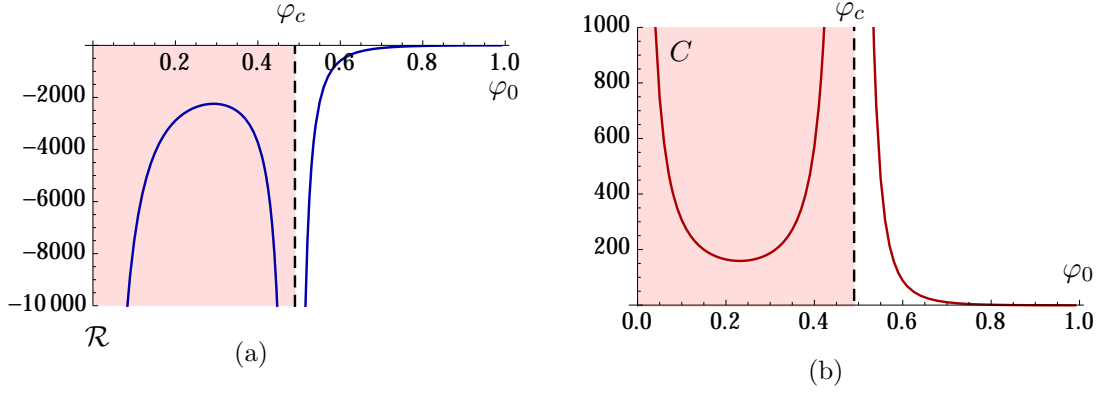

\begin{itemize}
\item For $\Delta_- < 1$ there exists a unique solution to equations
   (\ref{eq:EOM4}-\ref{eq:EOM6})  corresponding
  to a {\em monotonic}  RG flow for every value of IR turning point $\f_0 \in [0,\f_{\textrm{min}}]$. This is identical to what we have found for flows with $\mathcal{R} > 0$, and for any $\Delta_-$.
\item For $\Delta_- > 1$ an RG flow solution exists for every value of
  IR turning point $\f_0 \in [0,\f_{\textrm{min}}]$, but
  these are   not necessarily monotonic. In particular, we find that
  generic flows can only have $\f_0 \in [\f_c,
  \f_{\textrm{min}}]$, where $\f_c > \f_{\textrm{max}} $ is determined by the specific  choice
  of potential. On the other hand,  flows with $\f_0
  \in [ 0, \f_c]$ exist, but are non-monotonic: the scalar field
  $\f(u)$ starts from $0$ towards {\em negative} values, before
  turning around, crossing $\f=0$ again and reaching a value $\f_0 \in
  [0, \f_c]$. We will refer to such solutions as \emph{bouncing} flows
  and they will be discussed in more detail in section
  \ref{sec:bounceexample}. Note that these bouncing flows are dual to
  a different class of theories than the monotonic flows, since the UV
  source must have a different sign in the two cases.

\item In Fig.~\ref{WAdS} we display three solutions for $W(\f)$ with
  $\mathcal{R}<0$. The results are obtained for the potential
  \eqref{generic} with $\Delta_-=1.2$. The three solutions labelled by
  $W_1$, $W_2$ and $W_3$ differ in the value of their corresponding IR
  turning point $\f_0$. Note that $W(\f)$ goes to zero at $\f_0$,
  with infinite slope as expected from the analytical results in
  sec.~\ref{negativecurvatureflows}. The inverse scale factor
  $e^{A(u)}$ reaches a minimum at the loci where $W$ vanishes (see
  Fig.~\ref{TAdS}). This is due to the fact that $A$ reaches a minimum
  at the turning point.
\item Returning to figure \ref{WAdS}, in the vicinity of the UV fixed point the solutions are again described by the family of solutions collectively denoted by $W_{-}(\f)$ in section \ref{sec:asymp}. We can extract the constants $\mathcal{R}$ and $C$ for generic flows. As in the positive curvature case, we find that the absolute value $|\mathcal{R}|$ increases monotonically with a decrease in the value of the IR (turning) point $\f_0$. For $\Delta_- < 1$ we find that the increase in $|\mathcal{R}|$ continues until $\f_0 \rightarrow 0$ at which point $|\mathcal{R}|$ diverges. This changes for
$\Delta_- > 1$. In this case $|\mathcal{R}|$ already diverges when $\f_0 \rightarrow \f_c$ for some positive value $\f_c$. This is shown in Fig.~\ref{fig:Rvsphi0c}, where we display $\mathcal{R}$ as a function of $\f_0$ for the potential \eqref{generic} with $\Delta_-=1.2$. In this example we find that $\f_c \approx 0.49$. The reason for this behavior is that the flow ending at $\f_c$ corresponds to a theory with vanishing source $\f_-$, which in turn implies that $\mathcal{R}= R^{\textrm{uv}} |\f_-|^{-2 / \Delta_-}$ diverges.
The rose-colored area in Fig.~\ref{fig:Rvsphi0c} corresponds to $\f < \f_c$, which is the region where direct flows cannot arrive. Flows with turning points in this region leave the UV fixed point at $\f=0$ to the left before reversing direction. We will discuss such bouncing flows in more detail in section \ref{sec:bounceexample}.

\item For completeness, we display $C$ as a function of $\f_0$ in Fig.~\ref{fig:Cvsphi0c}. The results are again obtained for the potential \eqref{generic} with $\Delta_-=1.2$. For $\f_0 \rightarrow \f_{\textrm{min}}$ the parameter $C$ takes some finite value. The value of $C$ then increases monotonically with decreasing $\f_0$ until $C$ diverges for $\f_0 \rightarrow \f_c$. This can be understood as the divergence of $C = \tfrac{\Delta_-}{d}\langle \mathcal{O} \rangle |\f_-|^{-\Delta_+/\Delta_-}$ for $\f_- \rightarrow 0$. The rose-colored region again corresponds to the area $\f< \f_c$ where direct flows cannot end.
\end{itemize}


To summarize, generic flows for $\mathcal{R} <0$ exhibit many phenomena already observed for $\mathcal{R} >0$. In particular, the inverse relation between $|\mathcal{R}|$ and $\f_0$ persists. A new phenomenon is the appearance of bouncing solutions for $\Delta_->1$. We will discuss these solutions in detail in the following section.

\begin{figure}[t]
\centering
\begin{overpic}
[width=0.75\textwidth]{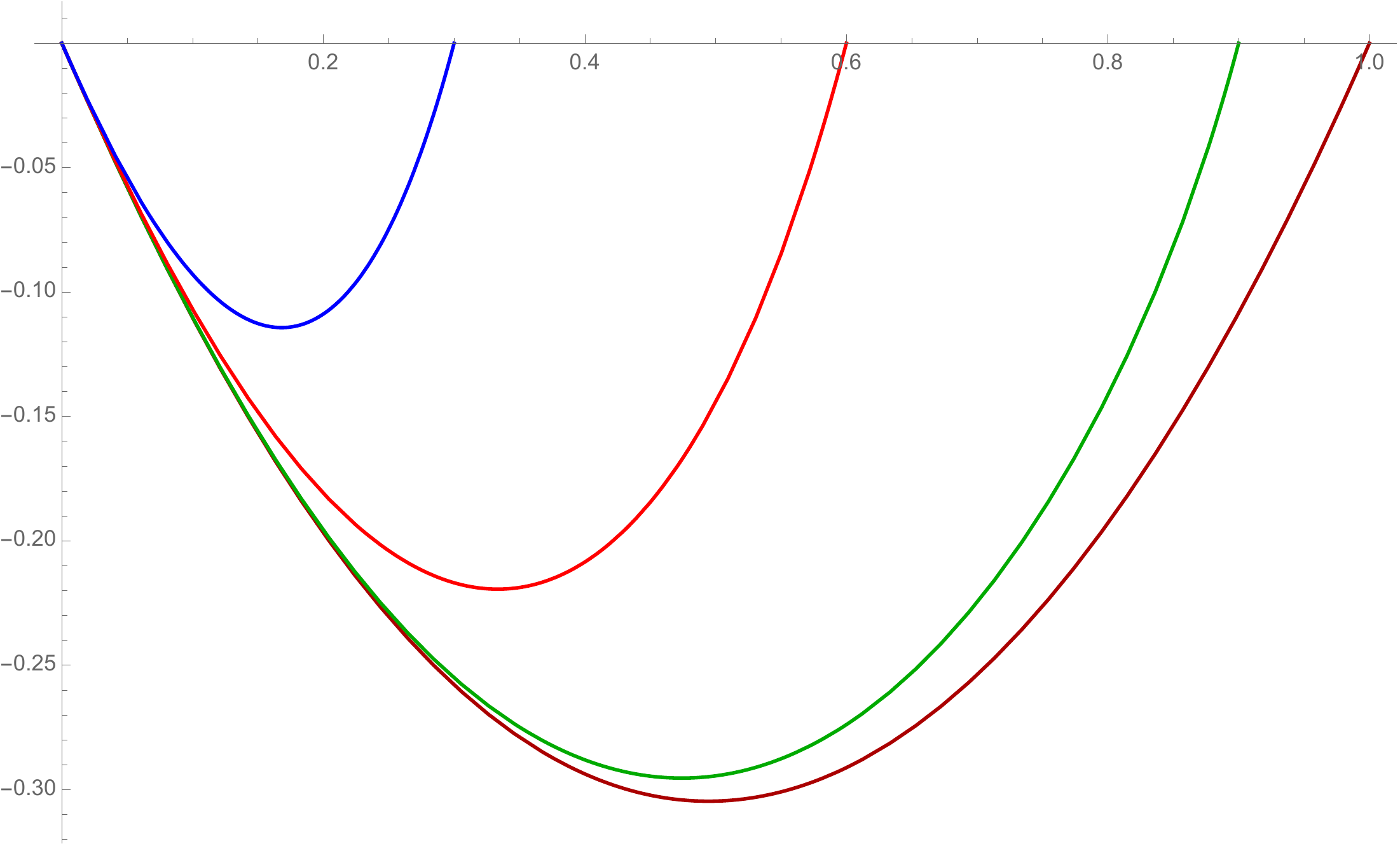}
\put(99,52){$\f$}
\put(-5,54){$\beta(\f)$}
\put(28,40){$\beta_3$}
\put(58,40){$\beta_2$}
\put(78,40){$\beta_1$}
\put(93,40){$\beta_{\textrm{flat}}$}
\end{overpic}
\caption{Solutions for $\beta(\f)= -2(d-1) \tfrac{S(\f)}{W(\f)}$ with $\mathcal{R} \geq 0$ and for the
  potential \protect\eqref{generic} with $\Delta_-=1.2$. The five
  solutions $\beta_i(\f)$ with $i=1, \ldots, 5$ differ in the value of
  their IR endpoint $\f_0$.}
\label{fig:beta}
\end{figure}
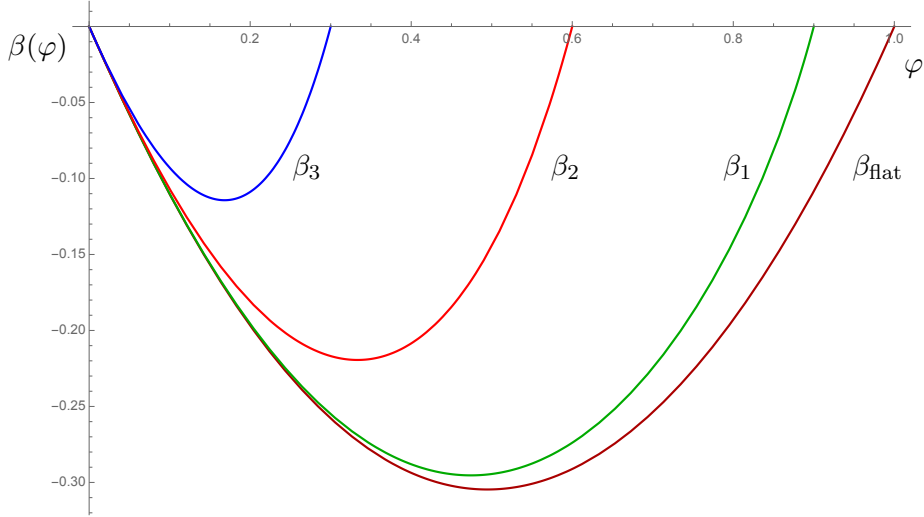

\subsection{The holographic $\beta$-function}
\label{sec:betafunc}
Having discussed generic flow solutions for $W(\f)$ and $S(\f)$, let us now turn to the holographic $\beta$-function given by
\begin{align}
\beta(\f) \equiv \frac{d \f}{d A} = -2(d-1)\frac{S(\f)}{W(\f)} \, .
\end{align}
In Fig.~\ref{fig:beta} we plot examples of the holographic $\beta$-function for generic flows in the potential \eqref{generic} with $\Delta_-=1.2$. The dark red curve is $\beta_{\textrm{flat}}(\f) = -2(d-1) W'(\f)/W(\f)$ for a flat flow. The remaining curves $\beta_i(\f)$ with $i=1,2,3$ correspond to RG flow solutions with $\mathcal{R} = 0.09$,  $1.29$, $7.14$ respectively. Note that the $\beta$-functions show the same asymptotic behavior in the UV, i.e.
\begin{align}
\label{eq:betaUV} \beta(\f) = - \Delta_- \f + \ldots \,+{\cal O}\left({\cal R}~|\f|^{1+{2\over \Delta_{-}}}\right) ,
\end{align}
where we have indicated the first non-trivial curvature correction at small curvature.
Moving away from the UV fixed point at $\f=0$, the $\beta$-functions begin to change due to the curvature. The departure from $\beta_{\textrm{flat}}$ is the more pronounced, the higher the value of $\mathcal{R}$. Most notably, depending on the value of $\mathcal{R}$, the various $\beta$-functions vanish at different values of $\f_0$. 
Using results from sections \ref{sec:minimaofV} and \ref{positivecurvatureflows}, we find that
the $\beta$-functions exhibit the following asymptotic form near their respective IR end-points:
\begin{align}
\mathcal{R} = 0:& \qquad \beta_{\textrm{flat}} = \Delta_-^{\textrm{min}} (\f_\textrm{min} -\f) + \mathcal{O} \big((\f_\textrm{min} -\f)^{2} \big) \, , \\
\mathcal{R} > 0:& \qquad \beta_i = -2 (\f_{0,i} -\f) + \mathcal{O} \big((\f_{0,i} -\f)^{3/2} \big)  \, , \quad \textrm{for} \quad i=1,2,3 \, ,
\end{align}
where $\f_{0,i}$ are the respective end-points of the flows. 
The difference in IR behavior between flat flows and flows with $\mathcal{R} > 0$ can be understood as follows. For flat flows,  IR fixed-points are determined by the potential, as they correspond to minima of $V$. Correspondingly, the vanishing of $\beta_{\textrm{flat}}$ at the fixed point is set by $\Delta_-^{\textrm{min}}$, which depends on the potential at the minimum. In contrast, for flows with $\mathcal{R} > 0$, it is the existence of non-zero curvature which cuts off the flows in the IR, rather than a property of the potential. As a result, the behavior of $\beta_i$ in the IR does not depend on the behavior of the potential at the IR end-point. In fact, the behavior $\beta_i = -2 (\f_{0,i} -\f)$ for $\mathcal{R} > 0$ is universal regardless of $V$ and $\mathcal{R}$. The effect of non-zero curvature $\mathcal{R}$ is to fix the value of the IR end point $\f_{0,i}$.

Last, we  comment on flows with $\mathcal{R} <0$. The behavior of $\beta(\f)$ in the UV is again given by \eqref{eq:betaUV}. However, at the turning point $\f_0$ the $\beta$-function does not vanish, but remains finite. Using our expressions from sec.~\ref{negativecurvatureflows} one finds
\begin{align}
\beta(\f_0) = d(d-1) \frac{|V' (\f_0)|}{V(\f_0)} \, .
\end{align}
This indicates that the turning points $\f_0$ in the negative curvature case are not fixed points of an RG flow.

\subsection{Bouncing Solutions}
\label{sec:bounceexample}

Bouncing solutions are made possible by the fact that at a generic
point $W(\f)$ and $S(\f)$ are typically multivalued. We have already
observed in the previous section that bounces are generic for negative
curvature. Here we present a few more examples.

\subsubsection*{QFT on $S^d$: $\mathcal{R}>0$}
\begin{figure}[t]
 \centering
\begin{overpic}
[width=0.5\textwidth]{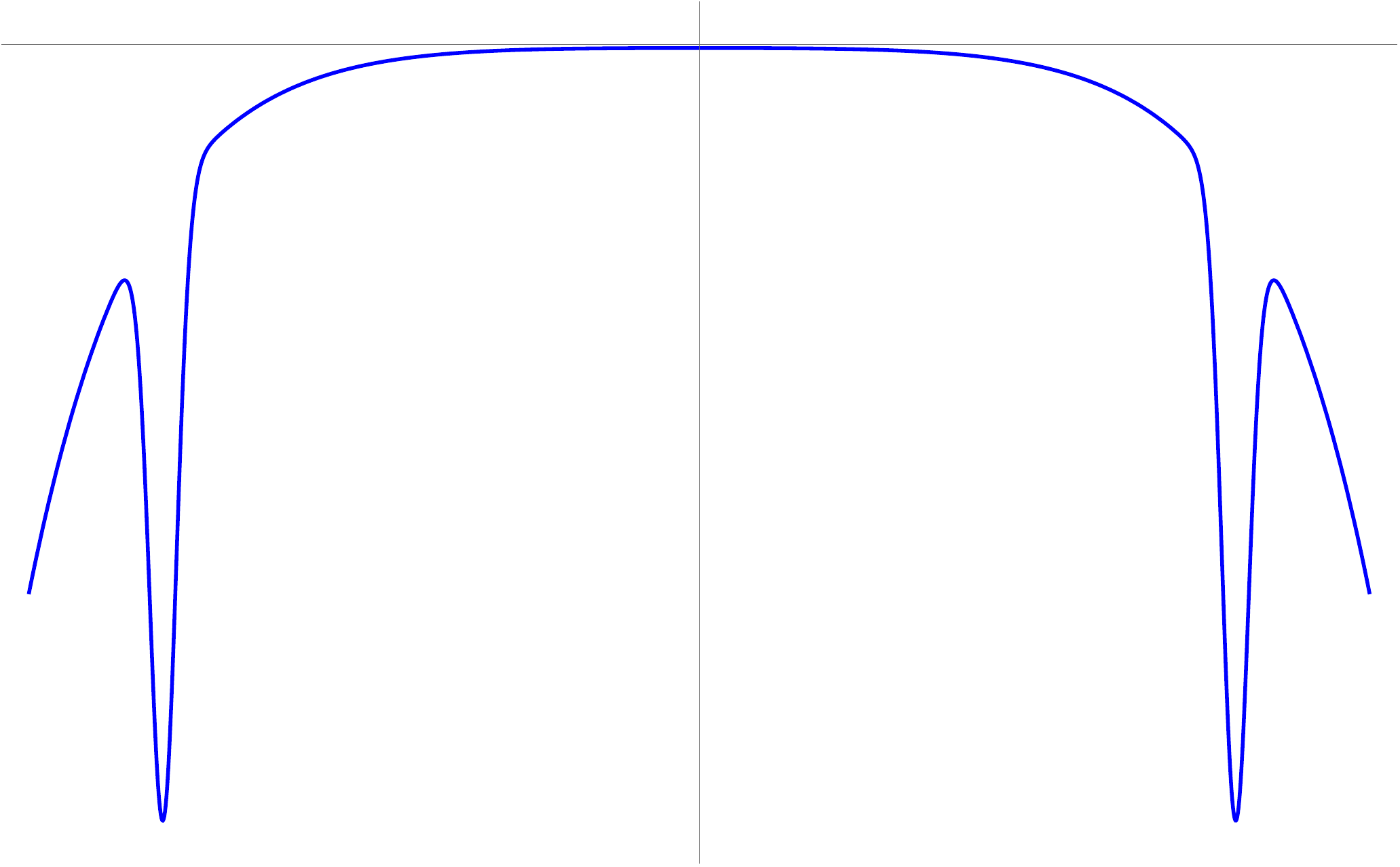}
\put(52,2){$V(\f)$}
\put(52,53){$0$}
\put(96,54){$\f$}
\end{overpic}
\caption{Plot of the potential \protect\eqref{eq:bounceex} with  parameters $\Delta_+=2.91$,
  $\sigma=0.01$, $c=2000$, $\Gamma= 2 / \sqrt{3}$, $\f_*=4$.}
\label{Vgauss}
\end{figure}

We now consider the  potential:
\begin{align}
\label{eq:bounceex} V= & -\frac{\Delta_+(d-\Delta_+)}{2 \ell^2} \f^2+ \frac{d(d-1)}{\ell^2} \left(\frac{\Gamma}{2} \f^2 - \cosh (\Gamma \f) \right) \\
\nonumber & - \frac{|c|}{\ell^2} \left( e^{- \frac{(\f-\f_*)^2}{2 \sigma}} + e^{- \frac{(\f+\f_*)^2}{2 \sigma}} + e^{- \frac{\f_*^2}{2 \sigma}} \left(\frac{1}{\sigma} - \frac{\f_*^2}{\sigma^2} \right) \f^2 - 2 e^{- \frac{\f_*^2}{2 \sigma}} \right) \, .
\end{align}
which depends on the parameters $\Delta_+$, $\Gamma$, $\f_*$ and
$c$. The potential is an inverted parabola superimposed on an inverted
cosh. In addition, there are two inverted Gaussians with peaks at $\pm
\f_*$ and whose height and width can be adjusted using $c$ and
$\sigma$. The potential thus has a maximum at $\f=0$. In addition, we
choose the parameters $\Gamma$, $c$ and $\sigma$ such that there are
also two minima close to $\pm \f_*$. The parameter $\Delta_+$ is
consistent with the definition of $\Delta_\pm$ in \eqref{eq:Deltadef},
i.e.~it is given by \eqref{eq:Deltadef} with $m^2 =
V''|_{\f=0}$. A plot of this potential is shown in Fig.~\ref{Vgauss} for parameters $\Delta_+=2.91$,
  $\sigma=0.01$, $c=2000$, $\Gamma= 2 / \sqrt{3}$, $\f_*=4$. This potential is constructed with the intention to give rise to bouncing solutions. The two Gaussians introduce two steep features in the potential, which increase the likelihood of bounces.

\begin{figure}[t]
 \centering
\begin{overpic}
[width=0.75\textwidth]{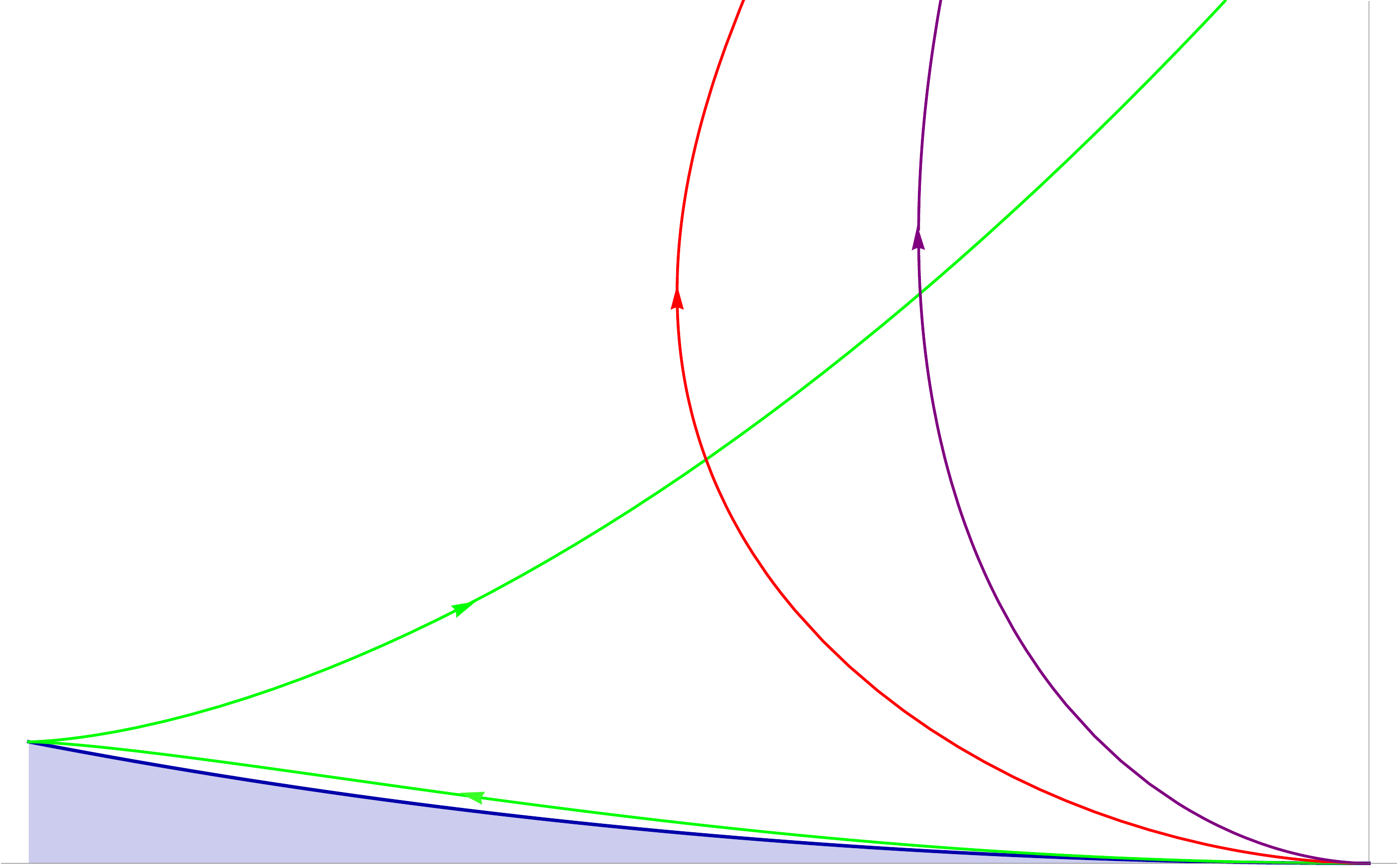}
\put(-1,2){$\f$}
\put(99,59){$W(\f)$}
\put(51,50){$W_1$}
\put(67,50){$W_2$}
\put(40,18){$W_{\text{flat}}$}
\put(6,3){$B(\f)$}
\end{overpic}
\caption{Solutions for $W(\f)$ for flows exhibiting a single bounce
  for the potential \protect\eqref{eq:bounceex} with $\Delta_+=2.91$,
  $\sigma=0.01$, $c=2000$, $\Gamma= 2 / \sqrt{3}$, $\f_*=4$. $W_{1,2}$
  are solutions with $\mathcal{R}_2> \mathcal{R}_1 >0$, while $W_{\textrm{flat}}$ is
  the solution in the same potential with $\mathcal{R}=0$. The shaded area is the region below the critical curve $B(\f)=\sqrt{-3 V(\f)}$.}
\label{WbouncedS}
\end{figure}
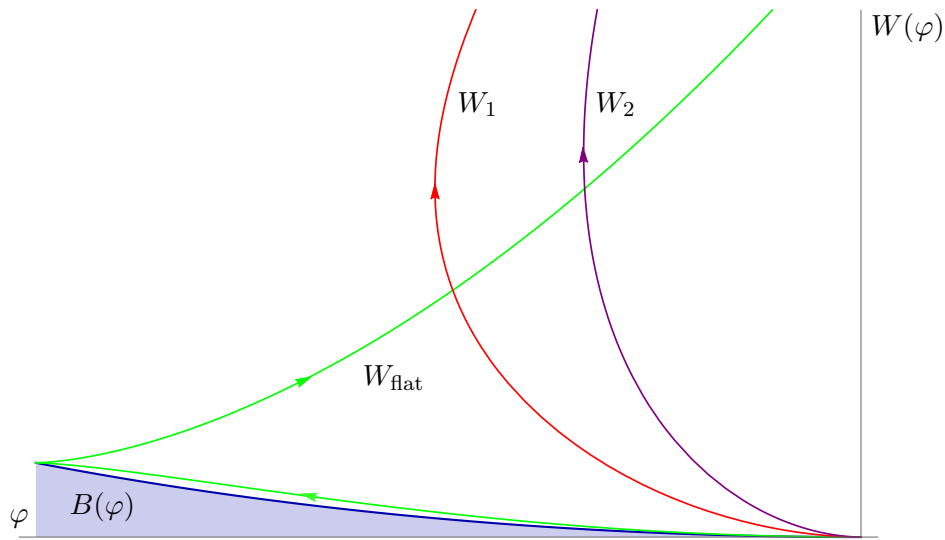

Fig.~\ref{WbouncedS} shows two bouncing solutions $W_1(\f)$ and
$W_2(\f)$ with $\mathcal{R} >0$. For comparison, we also plot
$W_{\textrm{flat}}$, which is the bouncing solution in the same
potential for $\mathcal{R}=0$.  The flows in Fig.~\ref{WbouncedS} originate from a UV fixed point at a maximum of $V$ at $\f_{\textrm{max}}=0$, departing towards negative values of $\f$. The flows then exhibit a bounce, reverse direction and eventually flow to the regime with $\f>0$. Depending on the precise form of the potential for $\f>0$, the flows may bounce again (even repeatedly), diverge, or approach an IR end point.

For $\mathcal{R}=0$  the bounce locus is the critical curve, i.e.~$W_{\textrm{flat}}(\f_B)=B(\f_B)$. In addition $W_{\textrm{flat}}'\sim \pm (\f-\f_B)^{1/2}$ in the vicinity of a bounce. As shown in \cite{exotic}, these two properties are shared by all bouncing solutions for $\mathcal{R}=0$.

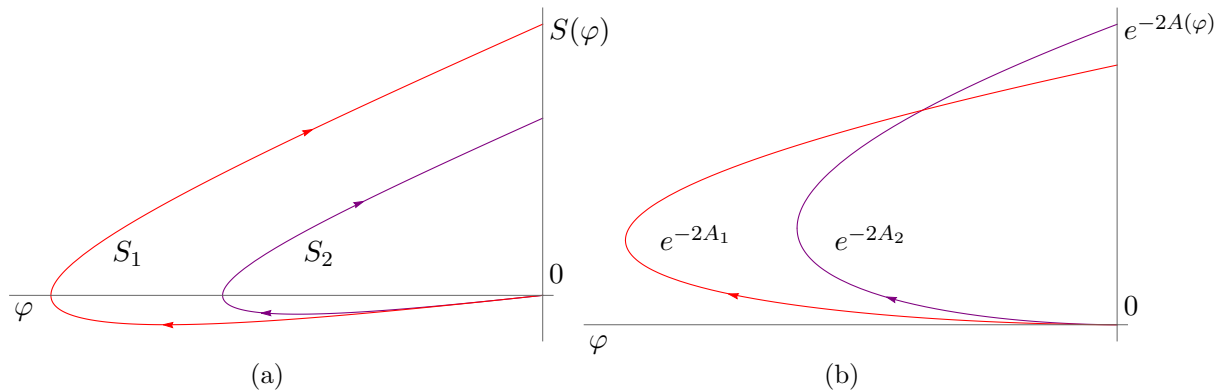
\begin{figure}[t]
\centering
\begin{subfigure}{.5\textwidth}
 \centering
   \begin{overpic}[width=0.95\textwidth,tics=10]{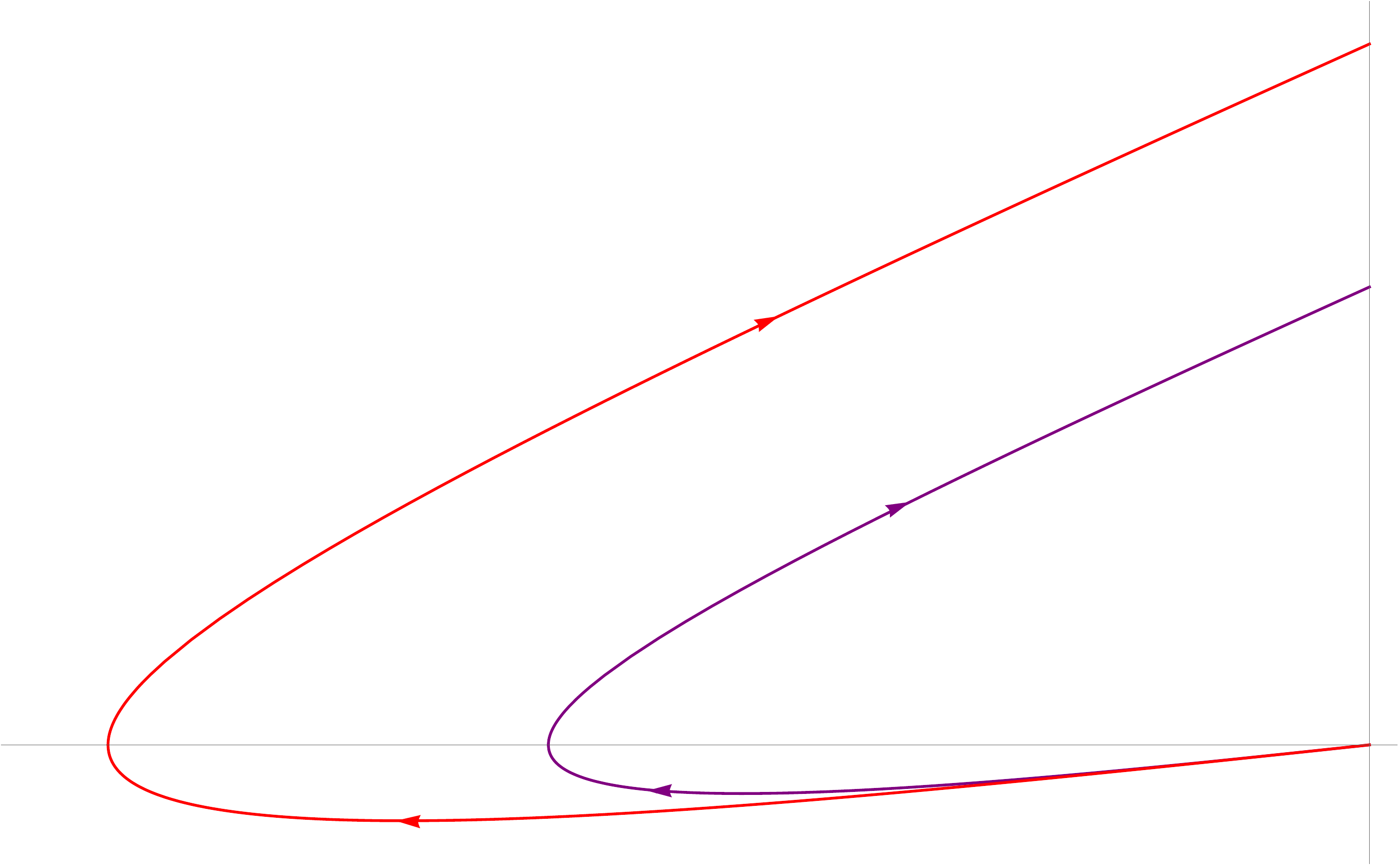}
\put (99,56) {$S(\f)$} \put (19,15) {$S_1$} \put (54,15) {$S_2$} \put (1,5) {$\f$} \put(99,11){$0$}
\end{overpic}
 \caption{\hphantom{A}}
  \label{fig:Sbounce}
\end{subfigure}%
\begin{subfigure}{.5\textwidth}
  \centering
 \begin{overpic}[width=0.95\textwidth,tics=10]{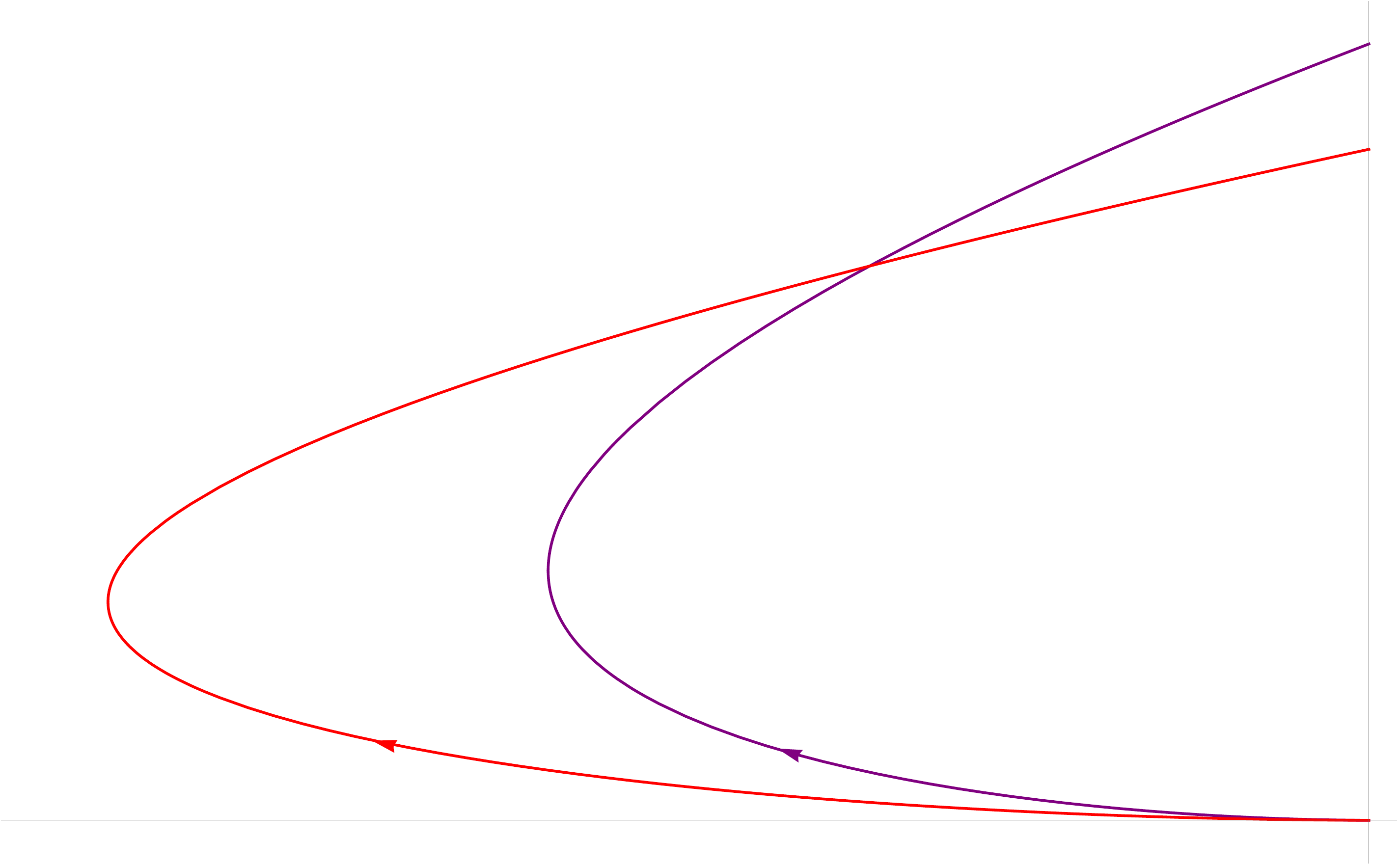}
\put (99,56) {$e^{-2A(\f)}$} \put (14,17) {$e^{-2A_1}$} \put (46,17) {$e^{-2A_2}$} \put (1,-1) {$\f$} \put(99,5){$0$}
\end{overpic}
\caption{\hphantom{A}}
  \label{fig:TdSbounce}
\end{subfigure}
\caption{Plots of $S_{1,2}(\f)$ and $e^{-2A_{1,2}(\f)}$ corresponding to the two solutions $W_{1,2}(\f)$ in figure \protect\ref{WbouncedS}. The solutions are obtained for the potential \protect\eqref{eq:bounceex} with $\Delta_+=2.91$, $\sigma=0.01$, $c=2000$, $\Gamma= 2 / \sqrt{3}$, $\f_*=4$.}
\label{fig:STbounce}
\end{figure}

For $\mathcal{R}>0$ we can make the following observations:
\begin{itemize}
\item For $\mathcal{R}>0$ the bounce locus does not lie on the
  critical curve (\ref{criticalB}). Instead we find that $W(\f_B) > B(\f_B)$ for $\mathcal{R}>0$ (see Fig.~\ref{WbouncedS}). Another interesting observation is that $W' \sim \pm (\f-\f_B)^{-1/2}$ in the vicinity of a bounce for $\mathcal{R} \neq 0$. Thus, while $W'$ changes sign at the bounce, it also diverges at this point.
\item We can also determine the values $\mathcal{R}_1$ and $\mathcal{R}_2$ corresponding to the two flows $W_1$ and $W_2$. Here we have $\mathcal{R}_2 > \mathcal{R}_1$. Studying further bouncing flows with $\mathcal{R} \neq 0$ the following behavior emerges: the larger the value of $|\mathcal{R}|$, the closer bounces occur to the UV fixed point.
\item For completeness, we also plot the functions $S$ and the inverse scale factor $e^{-2A}$ corresponding to the two solutions $W_1$ and $W_2$. This is shown in figures \ref{fig:Sbounce} and \ref{fig:TdSbounce} respectively. Note that $e^{-2A}$ takes finite values throughout, including at the bouncing locus $\f_B$. The bulk geometry is thus perfectly regular along the flow including the bounces.
\end{itemize}

We can summarize our findings as follows. Consider a field theory with
fixed UV source $\f_-$ and defined on a manifold with UV curvature
$R^{\textrm{uv}}$. One observation of this section is that for a
suitable potential $V$, solutions corresponding to RG flows with bounces
exist. Decreasing  $R^{\textrm{uv}}$ (and thus
$\mathcal{R}=R^{\textrm{uv}} |\f_-|^{-2/\Delta_-}$), the bounce  moves
further away from the UV fixed point and approaches the critical curve
$B(\f)$. For $R^{\textrm{uv}} \rightarrow 0$ the bounce eventually
reaches the critical curve. On the contrary, increasing $R^{\textrm{uv}}$ moves the bounce  towards the UV fixed point.

\subsubsection*{QFT on AdS$_d$: $\mathcal{R}<0$}
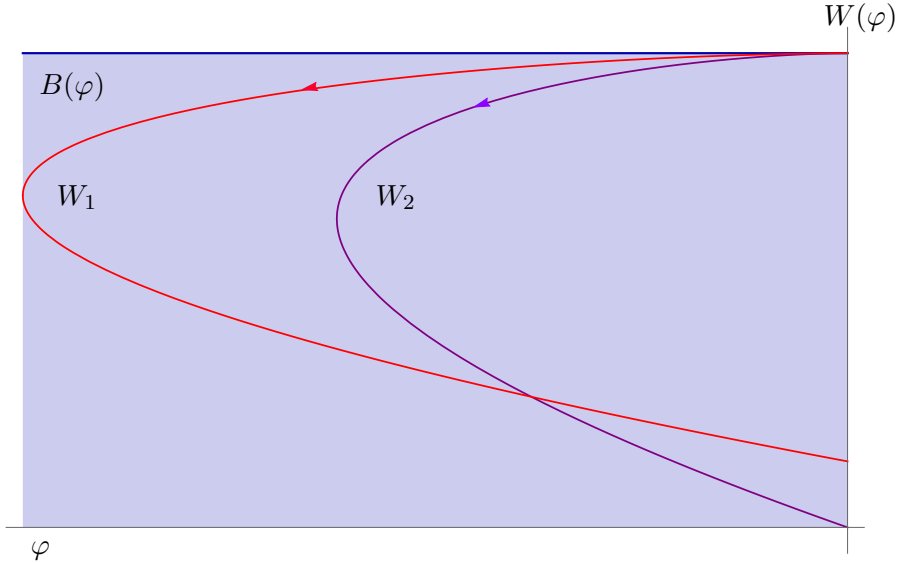
\begin{figure}[t]
 \centering
\begin{overpic}
[width=0.75\textwidth]{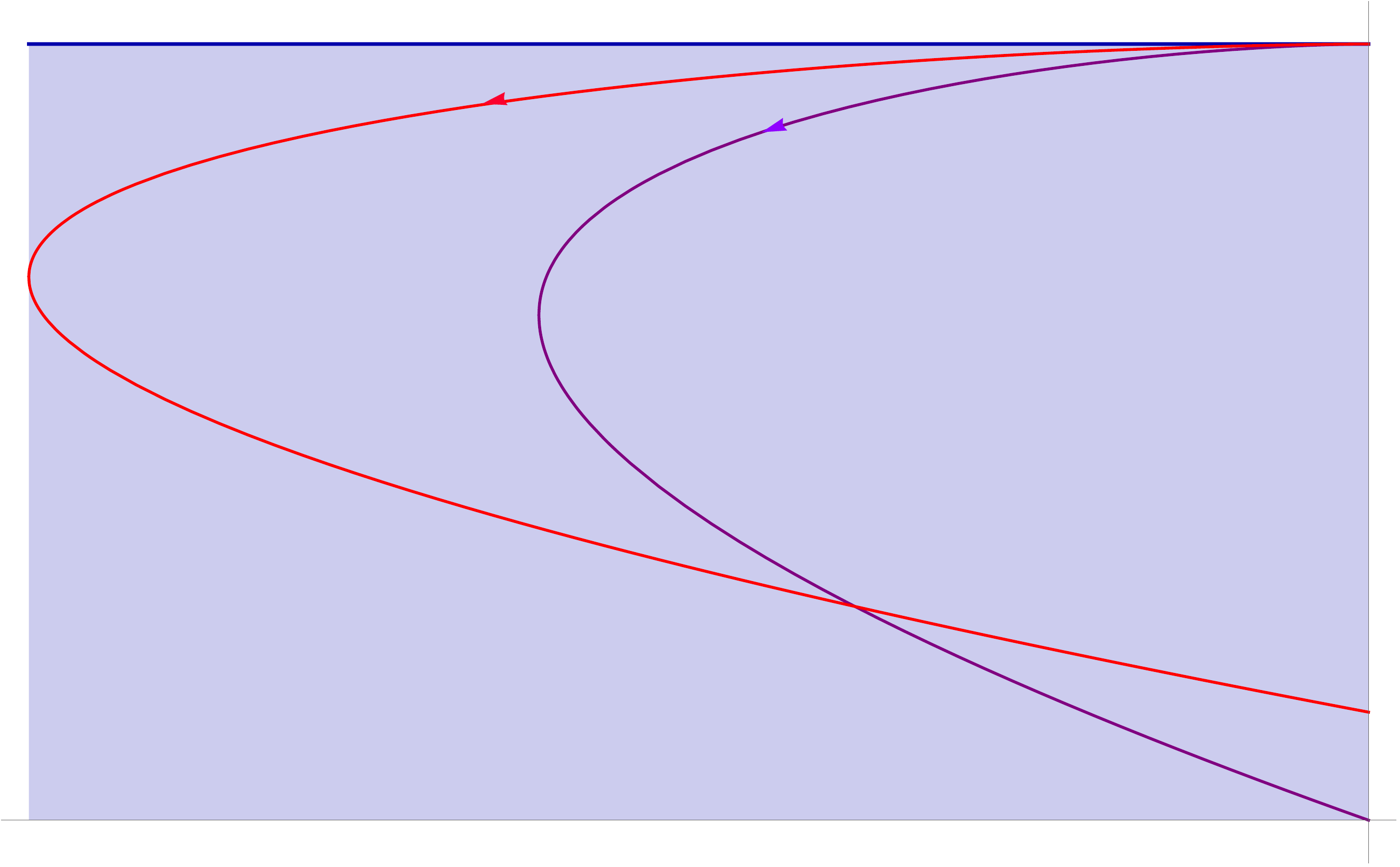}
\put(3,0.25){$\f$}
\put(95,62){$W(\f)$}
\put(6,41){$W_1$}
\put(43,41){$W_2$}
\put(4,54){$B(\f)$}
\end{overpic}
\caption{Plot of solutions for $W(\f)$ for flows exhibiting a single
  bounce for the potential \protect\eqref{eq:bounceex} with
  $\Delta_+=2.91$, $\sigma=0.01$, $c=2000$, $\Gamma= 2 / \sqrt{3}$,
  $\f_*=4$. $W_{1,2}$ are solutions with $\mathcal{R} <0$, and
  $|\mathcal{R}_2 | > |\mathcal{R}_1|$.}
\label{WbounceAdS}
\end{figure}

Bouncing solutions also exist for $\mathcal{R}<0$, i.e.~for field theories defined on negatively curved manifolds. Our observations from above also hold here: Bounce loci $\f_B$ do not occur on the critical curve, but now we have $W(\f_B) < B(\f_B)$ for $\mathcal{R}<0$ (see Fig.~\ref{WbounceAdS}). Similarly, increasing $|\mathcal{R}|$ moves $\f_B$ towards the UV fixed point. For example, for the two flows $W_1$ and $W_2$ in Fig.~\ref{WbounceAdS} we have $\mathcal{R}_2 > \mathcal{R}_1$.

An important difference to the case $\mathcal{R}>0$ is that for $\mathcal{R}<0$ bouncing solutions are generic. Let us explain what we mean by this. For $\mathcal{R}>0$ bouncing solutions may exist in suitable potentials, but given a generic potential, one does not expect the existence of bouncing solutions. This is to be contrasted with the situation for $\mathcal{R}<0$. There any potential with at least one maximum with $\Delta_- >1$ will allow for bouncing solutions.

This can be understood analytically from the results of appendix
\ref{sec:largeR}. There we study solutions for flows in purely
quadratic potentials, where we restricted our focus to flows that end or turn
close to the UV fixed point where they originated from. The main observation of relevance here is as follows: we find that
for $\mathcal{R}<0$ such flows will always bounce if $\Delta_->1$. Now
note that sufficiently close to a maximum the potential can always be
approximated by a quadratic function, and the results from appendix \ref{sec:largeR} will hold. Thus we expect that for $\Delta_->1$ flows with turning points sufficiently close to this maximum will necessarily bounce. This is indeed what we observe numerically.

Thus, for $\mathcal{R}<0$ bounces can occur in very simple potentials,
like the quadratic-quartic potential in \eqref{generic} and which we
will discuss in the following. However, now we will consider this
potential for the range $\f \in [-\f_{\textrm{min}},
\f_{\textrm{min}}] =[-1,1]$. As we have seen in section \ref{sec:generic}, for
negative curvature and $\Delta_->1$ there exists a critical value
$\f_c$ for the IR turning point below which the solution goes through
a bounce before reaching the turning point. The critical value $\f_c$
depends on the details of the potential: for example,
a numerical analysis shows that, for $\Delta_-=1.2$,   $\f_c=0.49$, while
for $\Delta_-=1.4$ $\f_c=0.63$.

We will now examine how the quantities $\mathcal{R}$ and $C$ vary as the endpoint of the flow $\f_0$ is varied from $\f_{\textrm{max}}=0$ to $\f_{\textrm{min}}=1$. For $\Delta_-=1.2$ this has already been discussed for the non-bouncing flows with $\f_0 \gtrsim 0.49$. We will now complement these plots with the results for bouncing flows, i.e.~for $|\f_0| \lesssim 0.49$. This is shown in fig.~\ref{fig:RCvsphi0c}.

Interestingly, in addition to the usual divergence in $\mathcal{R}$ and $C$ for $\f_0 \rightarrow \f_{\textrm{max}}$ we find that both $\mathcal{R}$ and $C$ also diverge when $\f_0 \rightarrow \f_c$, i.e.~when $\f_0$ approaches the boundary between bouncing and non-bouncing flows. This can be understood as follows. Non-bouncing flows with turning points in $\f_0 > \f_c$ correspond to flows that leave the UV fixed point to the right, and as a consequence have $\f_- > 0$. Bouncing flows with $\f_0 < \f_c$ leave the UV fixed point to the left before turning around, and thus have $\f_- < 0$. Thus, when $\f_0$ passes through $\f_c$ the source has to change sign. Note that in terms of the physical UV curvature $R^{\textrm{uv}}$ and the vev $\langle \mathcal{O} \rangle$ the two quantities $\mathcal{R}$ and $C$ can be written as:
\begin{align}
\mathcal{R} = \frac{R^{\textrm{uv}}}{|\f_-|^{2 / \Delta_-}} \, , \qquad C = \frac{\Delta_-}{d} \frac{\langle \mathcal{O} \rangle}{|\f_-|^{\Delta_+ / \Delta_-}} \, ,
\end{align}
The divergence for $\f_0 \rightarrow \f_c$ can thus be interpreted as follows. To change sign the source has to pass through zero. As long as $R^{\textrm{uv}}$ and $\langle \mathcal{O} \rangle$ remain finite, letting $\f_- \rightarrow 0$ makes both $\mathcal{R}$ and $C$ diverge.

\begin{figure}[t]
 \centering
\begin{overpic}
[width=0.45\textwidth]{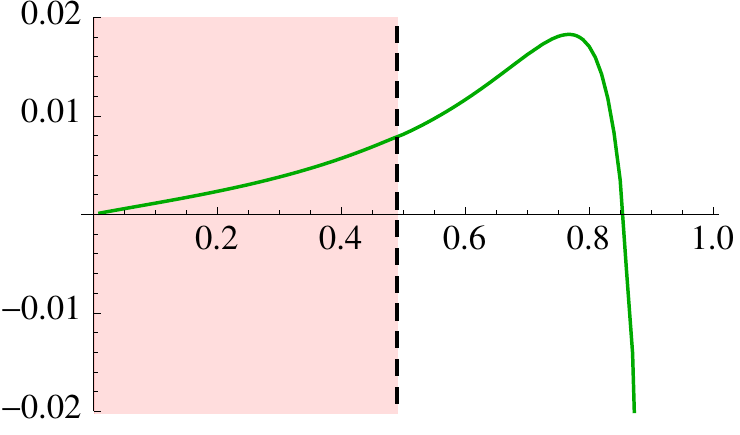}
\put (93,35) {$\f_0$} \put (50,60) {$\f_c$} \put (17,50) {$\frac{C}{\mathcal{R}^{\Delta_+/2}}$}
\end{overpic}
\caption{QFT on AdS${}_d$: $C/ \mathcal{R}^{\Delta_+/2}$ vs.~$\f_0$ for the potential \protect\eqref{generic} with $\Delta_-=1.2$. Flows with turning points in the rose-colored region exhibit a bounce.}
\label{fig:CoverRvsphi0}
\end{figure}

This interpretation can be checked by plotting the quantity
\begin{align}
\chi \equiv \frac{\langle \mathcal{O} \rangle}{(R^{\textrm{uv}})^{\Delta_+/2}} = \frac{d}{\Delta_-} \frac{C}{\mathcal{R}^{\Delta_+/2}} \, ,
\end{align}
which is manifestly independent of $\f_-$ and should thus be finite for $\f_0 \rightarrow \f_c$. This is indeed what we observe. In fig.~\ref{fig:CoverRvsphi0} we plot $\tfrac{\Delta_- \chi}{d}$ vs.~$\f_0$ for the potential \eqref{generic} with $\Delta_-=1.2$. We find that $\chi$ is finite and continuous when $\f_0$ passes through $\f_c$.

\section{A holographic quantum phase transition driven by curvature} \label{QPT}
In the framework of holography there may be multiple RG flows that originate from a single UV fixed point but flow to different IR end points. From the field theory point of view this corresponds to a theory that exhibits multiple vacua, with a one-to-one correspondence between the number of flows and the number of vacua.

This behavior was recently described in \cite{exotic} for holographic RG flows of field theories on flat manifolds. There it was observed that in addition to a flow from a maximum to the nearest minimum, there may also exist flows that \emph{skip} the nearest minimum to end at the next minimum.  We find that this situation persists if we add non-zero curvature: At a fixed curvature and UV source there may be multiple RG flow emanating from the same UV fixed point, but terminating at different IR end points. The different flows are distinguished by the subleading term which determines the vev of the dual operator.

The various flows correspond to different saddle points of the action and are hence in one-to-one correspondence with vacua of the field theory. In the case of multiple flows the question then arises which of these ground states is the true vacuum. In the dual gravitational picture this is equivalent to the questions which saddle point is dominant in the gravitational path integral. This can be answered by comparing the free energies associated with the various vacua  and identifying the ground state with the lowest free energy as the true vacuum.

If we vary the value of $R^{\textrm{uv}}$ the RG flows deform and the
vevs and free energies of the various vacua change. Interestingly,
under a variation of $R^{\textrm{uv}}$ the identification of the true
vacuum may change, i.e.~the system may exhibit a phase transition. This corresponds to a {\em quantum} phase transition as our system is at zero temperature. The control parameter in this case is the curvature $R^{\textrm{uv}}$.

In the following, we review the calculation of the free energy for a given RG flow. We then turn to a specific example based on a suitably chosen potential exhibiting multiple RG flows originating from the same UV fixed point. For this example we study the evolution of the free energies for the various flows under a change of curvature. Ultimately, we will observe a change of sign of the free energy difference between the different saddle points, indicating the existence of a phase transition driven by curvature.

\subsection{The free energy}
\label{sec:Fgeneral}
The free energy is given by the Euclidean on-shell action on the
gravitational side. This can be determined by starting from the action
\eqref{eq:action} and evaluating it on a given solution. One arrives
at the expression\footnote{As explained in section \ref{sec:AdSfixedpoints} we choose
  to set $R^{(\zeta)} = R^{\textrm{uv}}$. This can be done by a
  suitable choice for the integration constant entering $A(u)$.}:
  \begin{equation}
     S_{\text{{on-shell}}}=M_P^{d-1} \int d^dx \sqrt{-\zeta}\left(  \left[ e^{d A} W \right]_{\textrm{UV}} +\frac{2  R^{\textrm{uv}}}{d}\int_{\textrm{UV}}^{\textrm{IR}} du \, e^{(d-2)A(u)} \right) , \label{Sonshellu}
     \end{equation}
where the subscript ${\textrm{UV}}$ denotes a quantity to be evaluated at the UV boundary. We also reinstated the Planck mass $M_P$. Whenever we write $S_{\text{{on-shell}}}$ this corresponds to the on-shell action for a theory on a manifold with Lorentzian signature. We can obtain the corresponding Euclidean action by simply multiplying with $(-1)$.

In the following, let us briefly sketch how \eqref{Sonshellu} can be derived. The starting point is the action \eqref{eq:action}. First, rewrite the bulk curvature scalar $R^{(g)}$ in terms of $\dot{\f}^2$ and the potential $V$ as shown in appendix \ref{curvature inv}. In the next step insert for the potential $V$ using the equation of motion \eqref{eq:EOM2}. An important contribution comes from the Gibbons-Hawking-York boundary term $S_{GHY}$ which we have left implicit so far. This is given by
\begin{align}
S_{GHY} = 2 M_P^{d-1} \left[\int d^d x \, \sqrt{-\gamma} \, K \right]_{\textrm{UV}} = -2 d  M_P^{d-1}  \int d^dx \sqrt{-\zeta} \left[e^{dA} \dot{A} \right]_{\textrm{UV}} \, ,
\end{align}
where $K=-d \dot{A}_{\textrm{UV}}$ is the extrinsic curvature of the boundary. Here $\gamma_{\mu\nu}$ is the induced metric defined in \eqref{eq:indmet}. Putting everything together we arrive at \eqref{Sonshellu}. The second term in \eqref{Sonshellu} is only present for field theories on curved manifolds, i.e.~when $R^{\textrm{uv}} \neq 0$. For field theories on flat space-time only the first term in \eqref{Sonshellu} contributes.

The on-shell action is divergent and the divergences are originating from the UV boundary at $u \rightarrow 0$, which we choose to coincide with $\f=0$. To regulate we define a UV cut-off as $\f=\f_{UV}$. The divergences can then be removed by the addition of suitable counter-terms. The number of required counter-terms depends on the number of space-time dimensions. For example for $d=4$ the required counter-terms are \cite{Papadimitriou:2007sj}:
     \begin{align}
S_{ct}^{(0)} &=-M_P^{d-1} \int_{\textrm{UV}}d^dx \sqrt{-\gamma} \, W_{0,ct}(\f) \  ,  \label{ct0}\\
S_{ct}^{(1)} &= \hphantom{-} M_P^{d-1} \int_{\textrm{UV}} d^d x\sqrt{-\gamma} \, R^{(\gamma)} \, U_{ct}(\f)\ , \label{ct1} \\
S_{ct}^{(2)} &=  -M_P^{d-1} \int_{\textrm{UV}} d^d x\sqrt{-\gamma} \, (R^{(\gamma)})^2 \frac{\ell^3}{48 \Delta_-} \log(\f_{UV}) \ . \label{ct2}
     \end{align}
$S_{ct}^{(n)}$ is the counter-term required at the order $(R^{(\gamma)})^n$. Also note that the counter-terms are manifestly covariant. The functions $W_{0,ct}(\f)$ and $U_{ct}(\f)$ satisfy the following equations \cite{1401.0888,Papadimitriou:2007sj} :
\begin{align}
& \frac{d}{4(d-1)}W_{0,ct}^2(\f)-\frac{1}{2}(W_{0,ct}'(\f))^2=-V ,\\
& W_{0,ct}' (\f)U'_{ct}(\f)-\frac{d-2}{2(d-1)} W_{0,ct}(\f) U_{ct}(\f)=1 .
\end{align}
Near $\f=0$, we can find these functions as a series in $\f$, \cite{Myers,Kraus,HaroSkenderisSolodukhin} :
\begin{align}
& W_{0,ct}(\f)\simeq \frac{2(d-1)}{\ell}+\frac{\Delta_- }{2\ell}\f^2+ \frac{C_{ct}}{\ell} |\f |^{d/\Delta_-}+\cdots , \label{W0ct} \\
& U_{ct}(\f) \simeq \frac{\ell}{2-d}+\frac{d \ell^3}{2-d} B_{ct} |\f_{-}|^{2/\Delta_-} |\f |^{(d-2)/\Delta_-}+\cdots . \label{Uct}
\end{align}
Here $C_{ct}$ and $B_{ct}$ are the integration constants for the counter-terms. A particular choice for $C_{ct}$ and $B_{ct}$ corresponds to a renormalization scheme. Here and in the following we will choose $C_{ct}=B_{ct}=0$ which we will refer to as  ``minimal subtraction".

\begin{figure}[t]
\centering
\begin{overpic}
[width=0.6\textwidth]{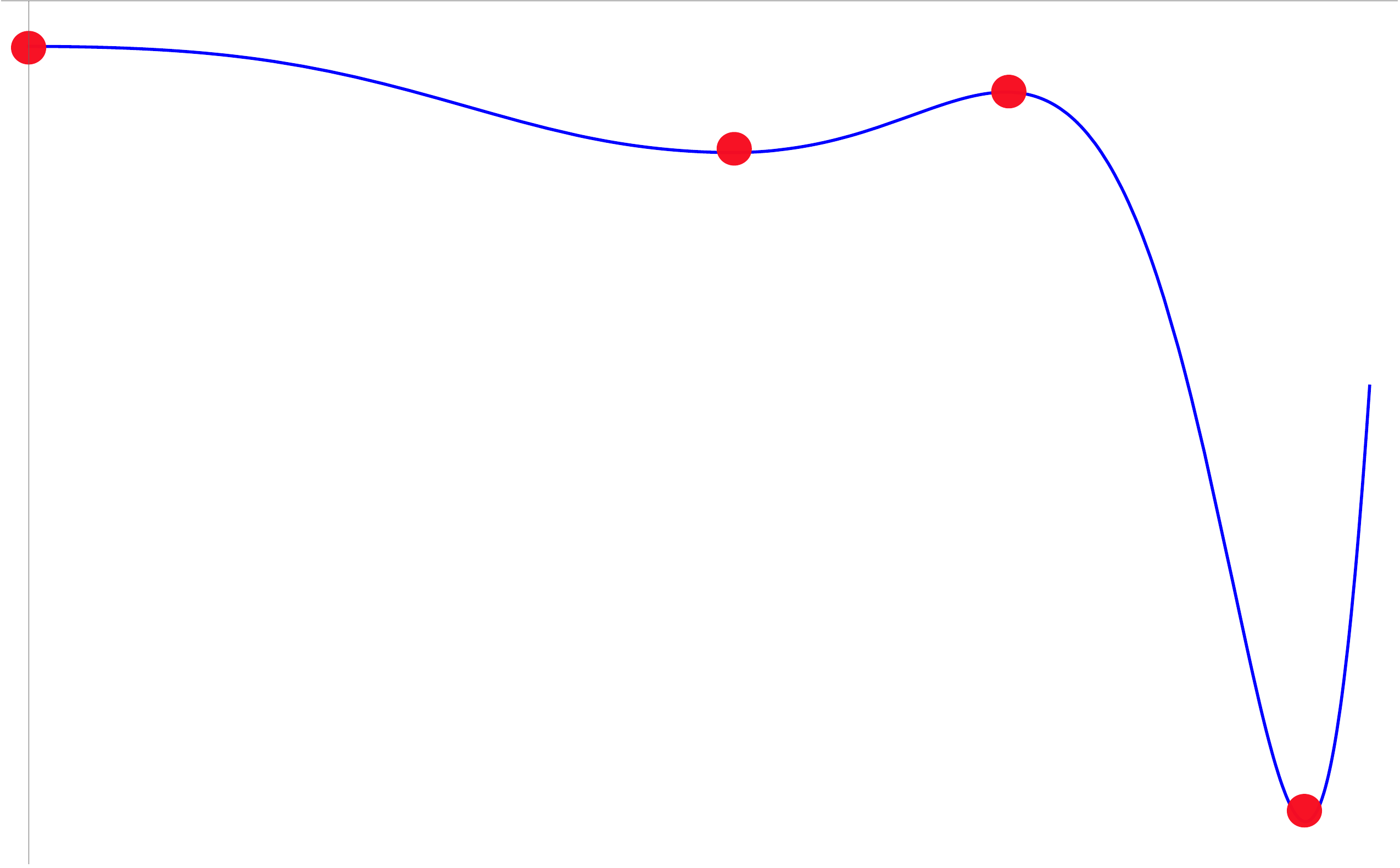}
\put (0,65) {V($\f$)}
\put (3,53) {UV$_1$}
\put (50,45) {IR$_1$}
\put (69,49) {UV$_2$}
\put (91,-1.5) {IR$_2$}
\put (96,58) {$\f$}
\end{overpic}
\caption{Plot of the degree-12 potential which allows skipping solutions. This potential has several extrema which are denoted as UV$_1$, UV$_2$, IR$_1$ and IR$_2$. }
  \label{skipnew}
\end{figure}
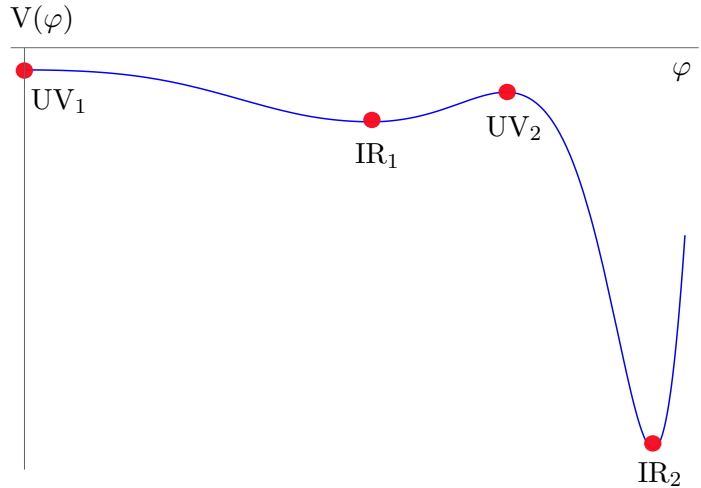

We can now write down an expression for the free energy $F$. It is given by the renormalized Euclidean on-shell action $S_E^{\textrm{ren}}$, i.e.~the Euclidean on-shell action including the counter-terms. In case we are working in a space-time with a Lorentzian signature, a Wick rotation is needed to pass to Euclidean signature. In this circumstances the on-shell action picks up a minus sign. We thus have
\begin{align}
F = S_E^{\textrm{ren}} \, , \qquad \textrm{with} \qquad S_E^{\textrm{ren}} =
-S_{\textrm{on-shell}}^{\textrm{ren}} \, .
\end{align}

\subsection{Skipping flows and a quantum phase transition driven by curvature}
To illustrate  the  concept of skipping flows,  we set $d=4$ and consider a
potential which is a polynomial of degree 12, whose explicit form is
not essential and can
be found in  \cite{exotic}.  A plot of (part of) this potential is shown in figure \ref{skipnew}. It has several extrema, denoted by UV$_1$, UV$_2$, IR$_1$ and IR$_2$   In this section, we will determine the possible RG flows as we increase the dimensionless UV curvature parameter, $\mathcal{R}$.

\subsubsection*{Flat case: $\mathcal{R}=0$}
\begin{figure}[t]
\centering
\begin{overpic}
[scale=.4]{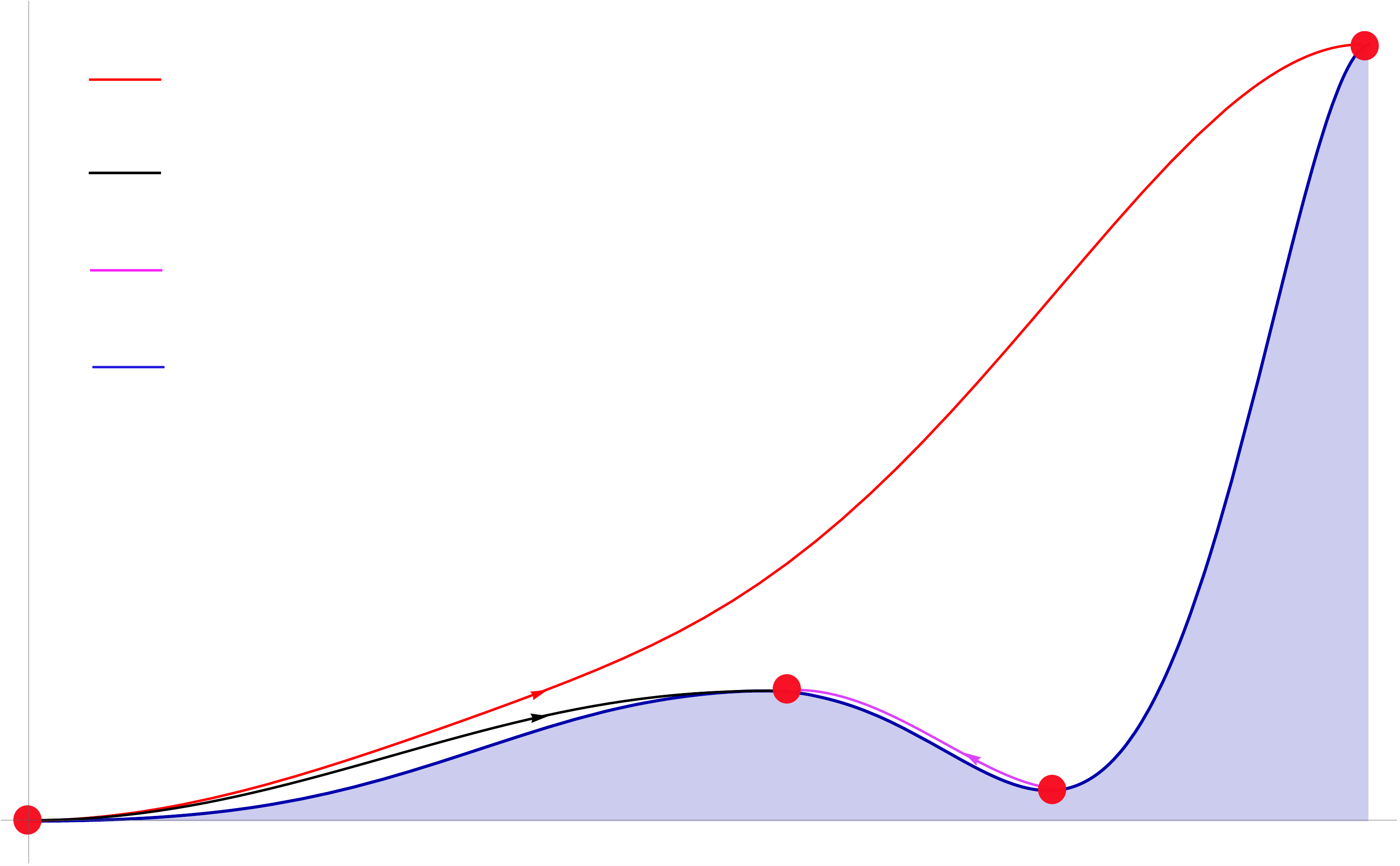}
\put(100,2){$\f$}
\put(1,62){$W(\f)$}
\put(3,5){UV$_1$}
\put(55,9){IR$_1$}
\put(72,7){UV$_2$}
\put(95,60){IR$_2$}
\put(12,55){$W_{12}(\f)$}
\put(12,48.5){$W_{11}(\f)$}
\put(12,41.5){$W_{21}(\f)$}
\put(12,34.5){$ B(\f)=\sqrt{-3V(\f)}$}
\end{overpic}
\caption{The plot represents the superpotentials corresponding to the
  three regular  holographic RG flow solutions arising from the
  potential in figure \protect\ref{skipnew}. Two of them are standard interpolating from UV$_1$ to
  IR$_1$ and from UV$_2$ to IR$_1$. The third  one skips IR$_1$ and
  ends up at IR$_2$. The shaded region is the forbidden region below
  the critical curve $B(\f)$. The arrows represent the direction of
  the flow from the UV to the IR.}\label{skipflatnew}
\end{figure}
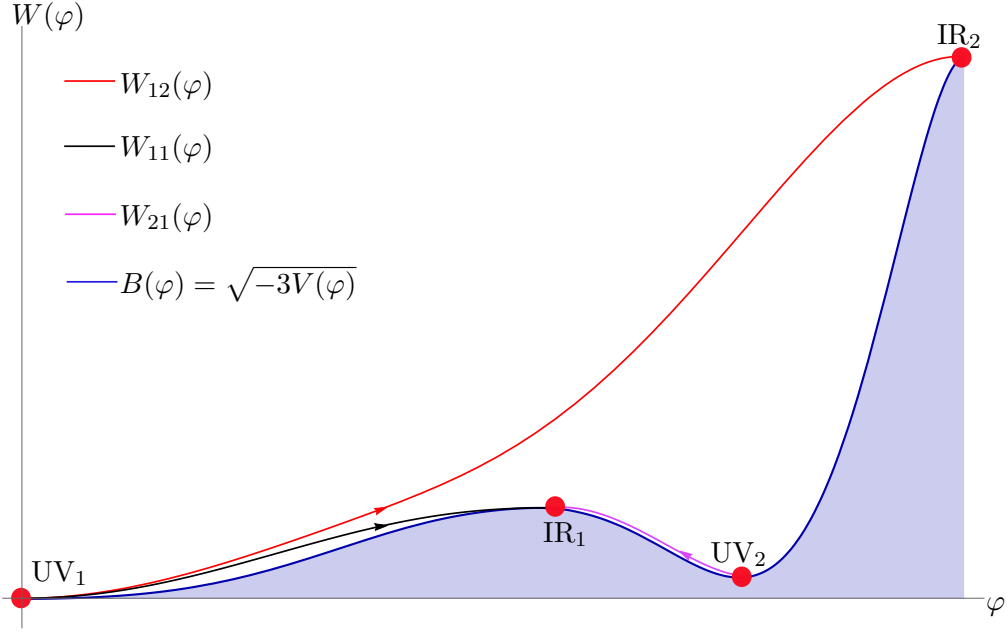

We first consider the RG flow solutions for the QFT in flat space-time.
The corresponding solutions for $W(\f)$ are shown in
Fig.~\ref{skipflatnew}. In this case, there are three RG flows in
total, two originating from UV$_1$ and one from UV$_2$. We will focus on the flows emanating from UV$_1$.  Among the two RG flows emanating from the UV$_1$, the one denoted by $W_{11}(\f)$ interpolates between UV$_1$ and the nearest minimum to the right IR$_1$. The second flow denoted $W_{12}(\f)$ skips
the fixed point IR$_1$ and ends at IR$_2$.

To decide which solution is the dominant saddle point in the path
integral, we need to compare their corresponding free energies at
fixed boundary condition $\f_-$. As described in \cite{exotic} the flow with the higher
value of vev parameter $C$ is the dominant saddle point: this is the
skipping flow, as the solution with larger $C$ has a larger $W(\f)$ in
the UV. Consequently, the skipping flow represents the true ground-state of the dual QFT in flat space-time.

Although our focus will be on flows starting at UV$_1$, for completeness we
briefly discuss the flow from UV$_{2}$, also depicted in
Fig.~\ref{skipflatnew}.  There is a single  flow which starts from
UV$_{2}$, denoted as $W_{21}(\f)$, and it interpolates between the
UV$_{2}$ and IR$_{1}$. There is no solution starting
from UV$_{2}$ and ending at IR$_{2}$ as there can be at most one flow that
can end at a given IR point from a given direction. This is because regular
solutions in the IR do not admit small deformations \cite{exotic}.  Here the
fixed point IR$_{2}$ is already ``taken'' by the skipping flow $W_{12}(\f)$.

We now  turn on positive curvature on the $d$-dimensional slice and see how this
affects the various RG flows solutions presented above.

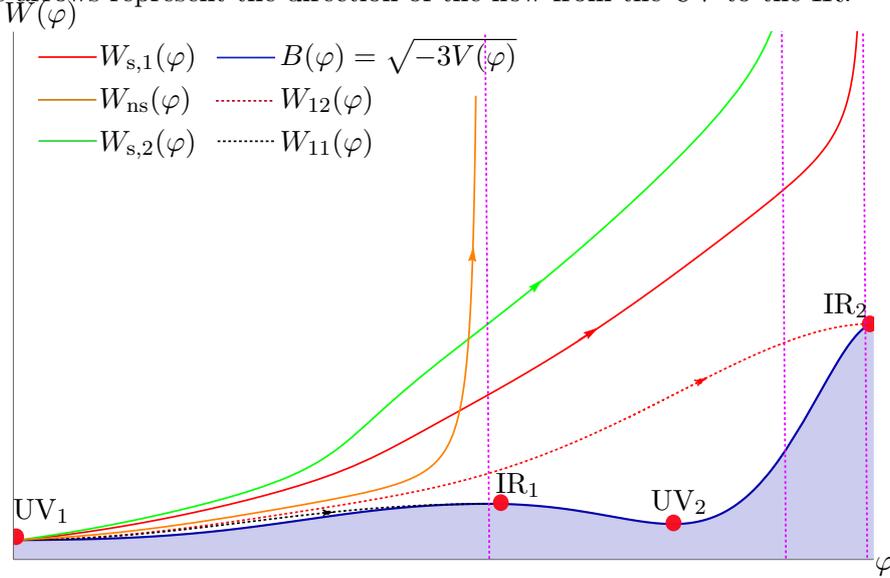
\begin{figure}[t]
\centering
\begin{overpic}
[scale=.4]{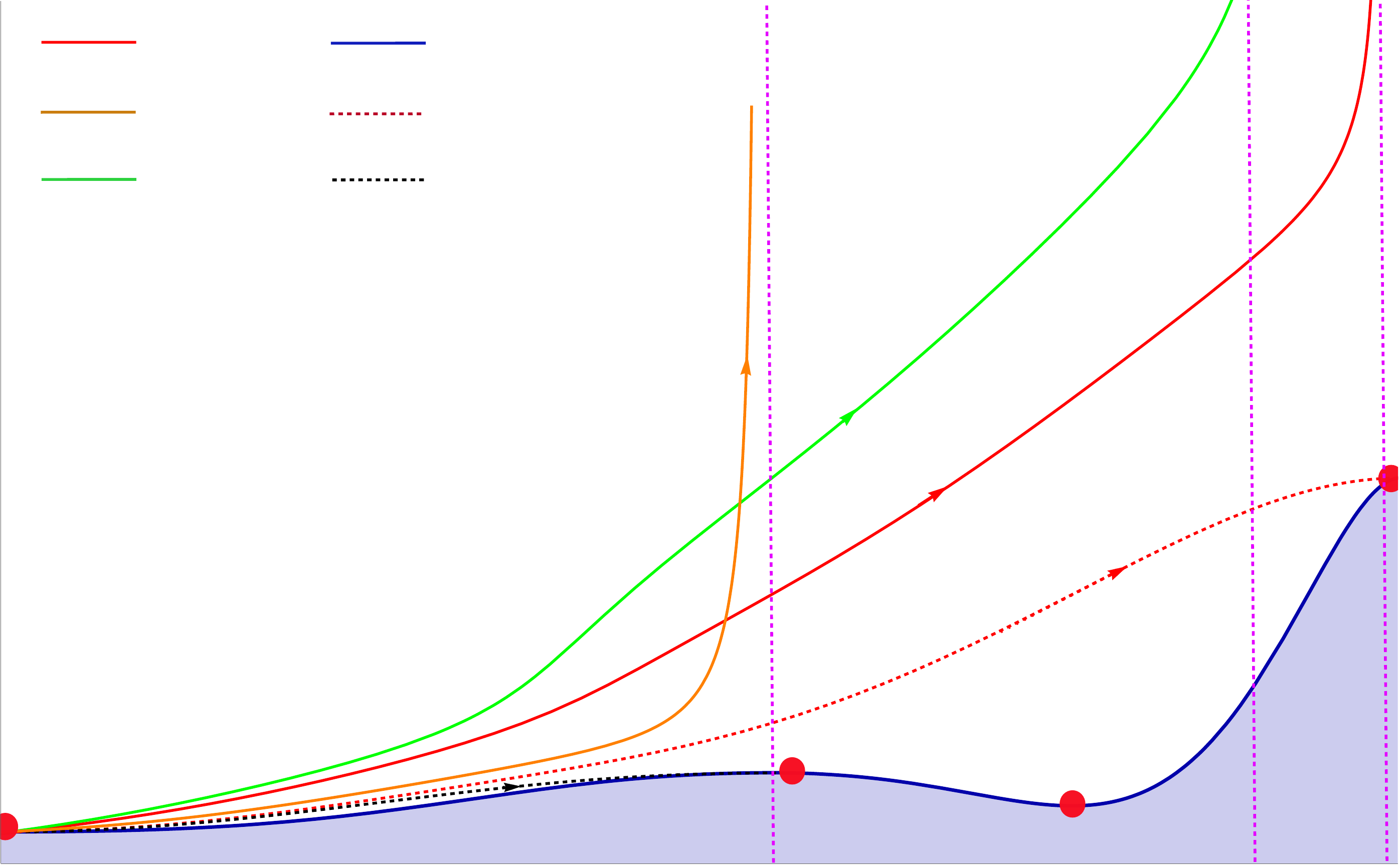}
\put(100,-1){$\f$}
\put(-1,63){$W(\f)$}
\put(0,5){UV$_1$}
\put(56,8){IR$_1$}
\put(74,6){UV$_2$}
\put(94,29){IR$_2$}
\put(10,58){$W_{\text{s},1}(\f)$}
\put(10,53){$W_{\text{ns}}(\f)$}
\put(10,48){$W_{\text{s},2}(\f)$}
\put(31,58){$B(\f)=\sqrt{-3V(\f)}$}
\put(31,53){$W_{12}(\f)$}
\put(31,48){$W_{11}(\f)$}
\end{overpic}
\caption{The solid lines represent the superpotential $W(\f)$
  corresponding to the three different solutions starting from UV$_1$
  which exist at  small positive curvature. Two of them (red and green
  curves) are skipping
  flows  and the third one (orange curve) is non-skipping. For
  comparison, we also show the flat RG flows (dashed curves)}\label{skipcurvedsmall}
\end{figure}

\subsubsection*{Finite curvature $\mathcal{R} > 0$; Flows from UV$_1$}
Turning on a  small positive curvature, there are now three flows (instead of the two at $\mathcal{R}=0$)  emanating from UV$_1$ and shown in the Fig.~\ref{skipcurvedsmall}. Two of them are skipping the nearest possible IR region (between
$\f=0$ and IR$_1$) and end up near IR$_2$. These two are denoted as
$W_{\text{s},1}(\f)$ and  $W_{\text{s},2}(\f)$ and represented  as red
and green curves respectively in the Fig.~\ref{skipcurvedsmall}.
The third flow is the non-skipping flow and denoted as
$W_{\text{ns}}(\f)$. It starts from UV$_1$ and ends up in the region
between $\f=0$ and IR$_1$. The two  flows $W_{\text{ns}}(\f)$  and $W_{\text{s},1}(\f)$ follow the general pattern seen in section 6. They
are deformations of the corresponding flat flows (represented as the
dotted black and red curves in Fig.~\ref{skipcurvedsmall}) and end slightly
before the respective IR fixed points. The flow
$W_{\text{s},2}(\f)$ on the other hand is a new branch which only
exists for non-zero curvature.

\begin{figure}[h]
\centering
\begin{overpic}
[scale=.4]{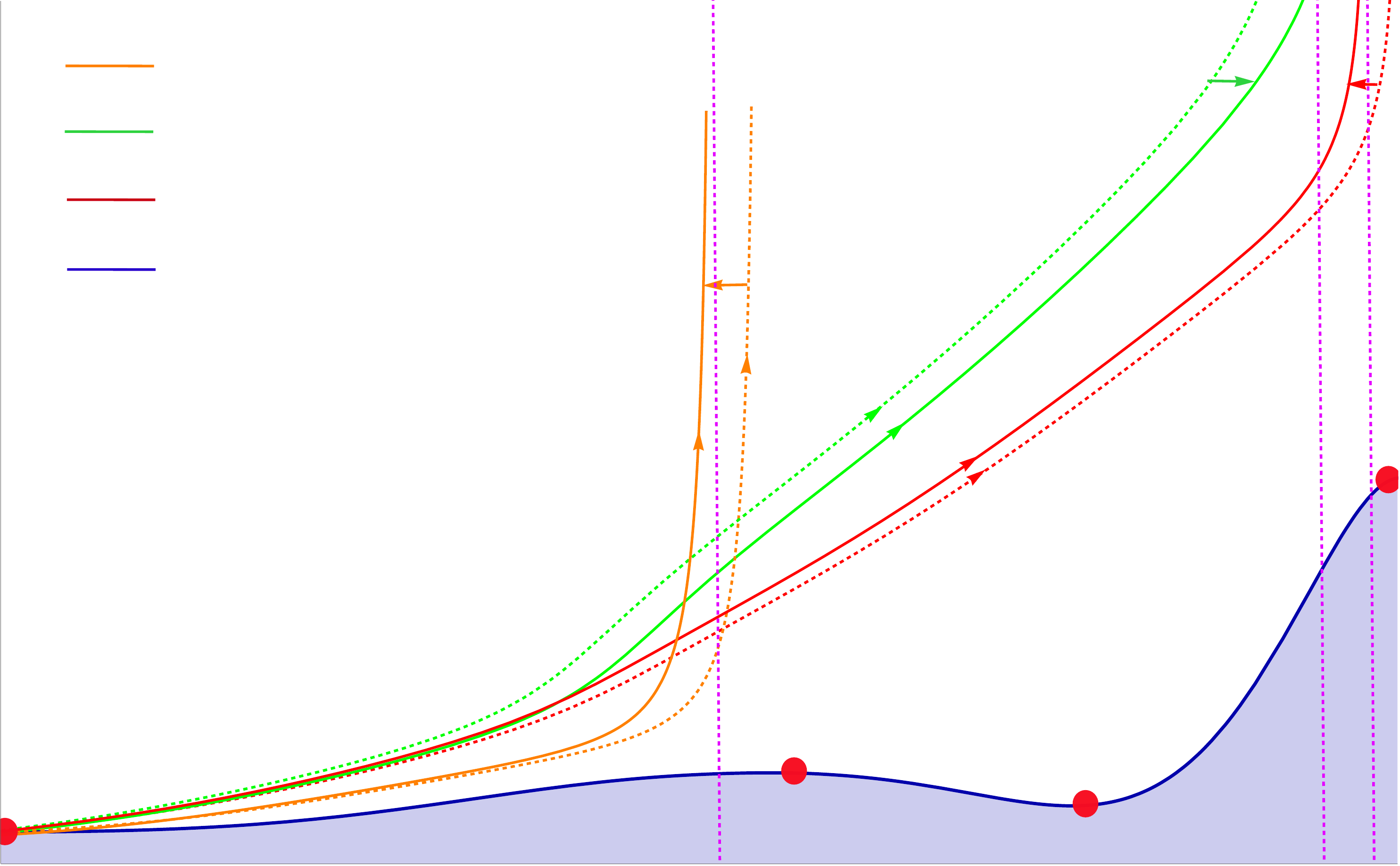}
\put(100,-.5){$\f$}
\put(-.5,62){$W(\f)$}
\put(0,5){UV$_1$}
\put(56,8){IR$_1$}
\put(74,6){UV$_2$}
\put(94,29){IR$_2$}
\put(12,56){$W_{\text{ns}}(\f)$}
\put(12,51.5){$W_{\text{s},2}(\f)$}
\put(12,46.5){$W_{\text{s},1}(\f)$}
\put(12,41.5){$B(\f)=\sqrt{-3V(\f)}$}
\end{overpic}
\caption{As the curvature is increased, IR end points of the skipping flows move toward each other. The non-skipping IR end point moves closer to UV$_1$. }\label{skipincreased}
\end{figure}
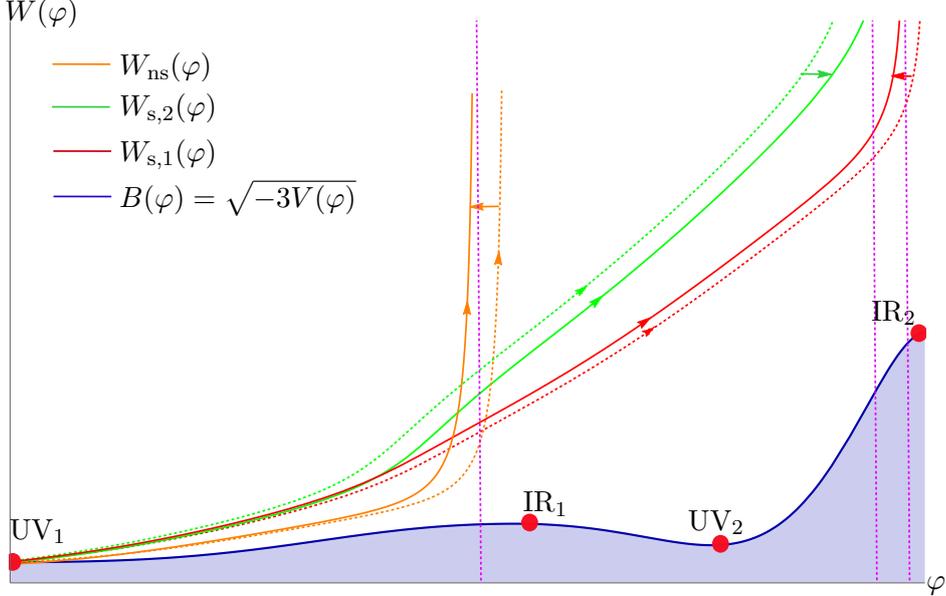

As $\mathcal{R}$ is increased, the flow are deformed  as shown in Fig.~\ref{skipincreased}.
The IR end points of the skipping flows $W_{\text{s},2}(\f)$ and
$W_{\text{s},1}(\f)$, move towards each other, up to a critical value
of $\mathcal{R}$ where the two solutions merge. Above this value, the
skipping solution disappears and only the non-skipping flow is left.  The IR end point of the
non-skipping flow  $W_{\text{ns}}(\f)$ moves  to the left
from the IR$_1$ and it approaches the UV fixed point point UV$_{1}$ as $\mathcal{R}$ is increased, as expected in the
generic case.

\subsubsection*{Finite curvature: $\mathcal{R} > 0$; Flows from UV$_2$}
\begin{figure}[t]
\centering
\begin{overpic}
[scale=.5]{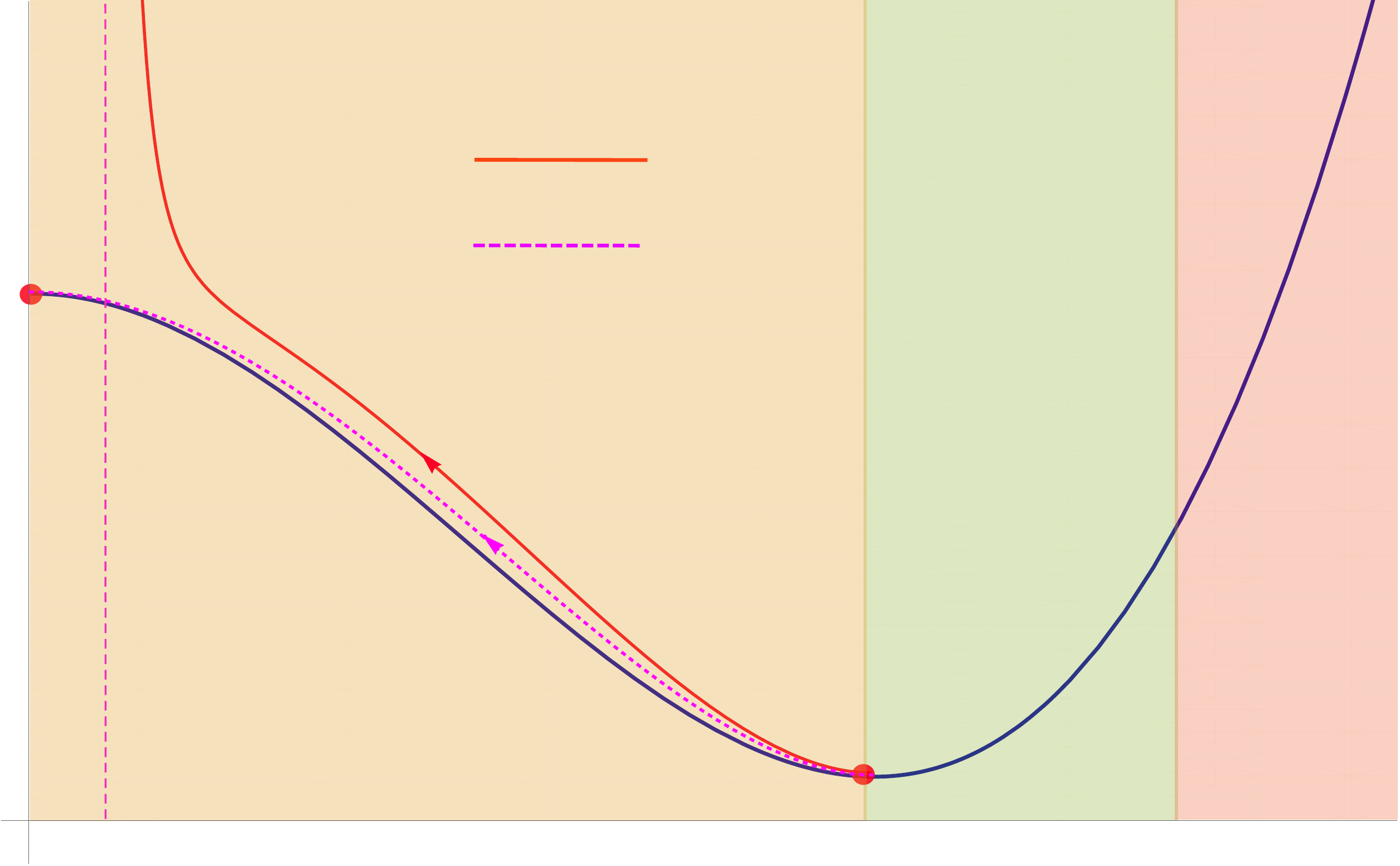}
\put(101,2.5){$\f$}
\put(2.25,59){$W(\f)$}
\put(59,8){UV$_2$}
\put(2,37){IR$_1$}
\put(20,55){Non-bouncing}
\put(20,52){$\f_-<0$}
\put(68,55){Non-}
\put(68,52){bouncing}
\put(68,49){$\f_->0$}
\put(84.25,59){Bouncing}
\put(84.25,56){$\f_-<0$}
\put(86,26){$B(\f)$}
\put(47,49){$W_{2nb,n}(\f)$}
\put(47,43){$W_{21}(\f)$}
\end{overpic}
\caption{Superpotential $W_{2nb,n}(\f)$ for a non-bouncing flows starting from UV$_2$. In this case the source is negative. This type of flow comes from a curvature deformation of the flat case flow $W_{21}$ and exists for all values of the curvature.}\label{UVIIneg}
\end{figure}

\begin{figure}[t]
\centering
\begin{overpic}
[scale=.42]{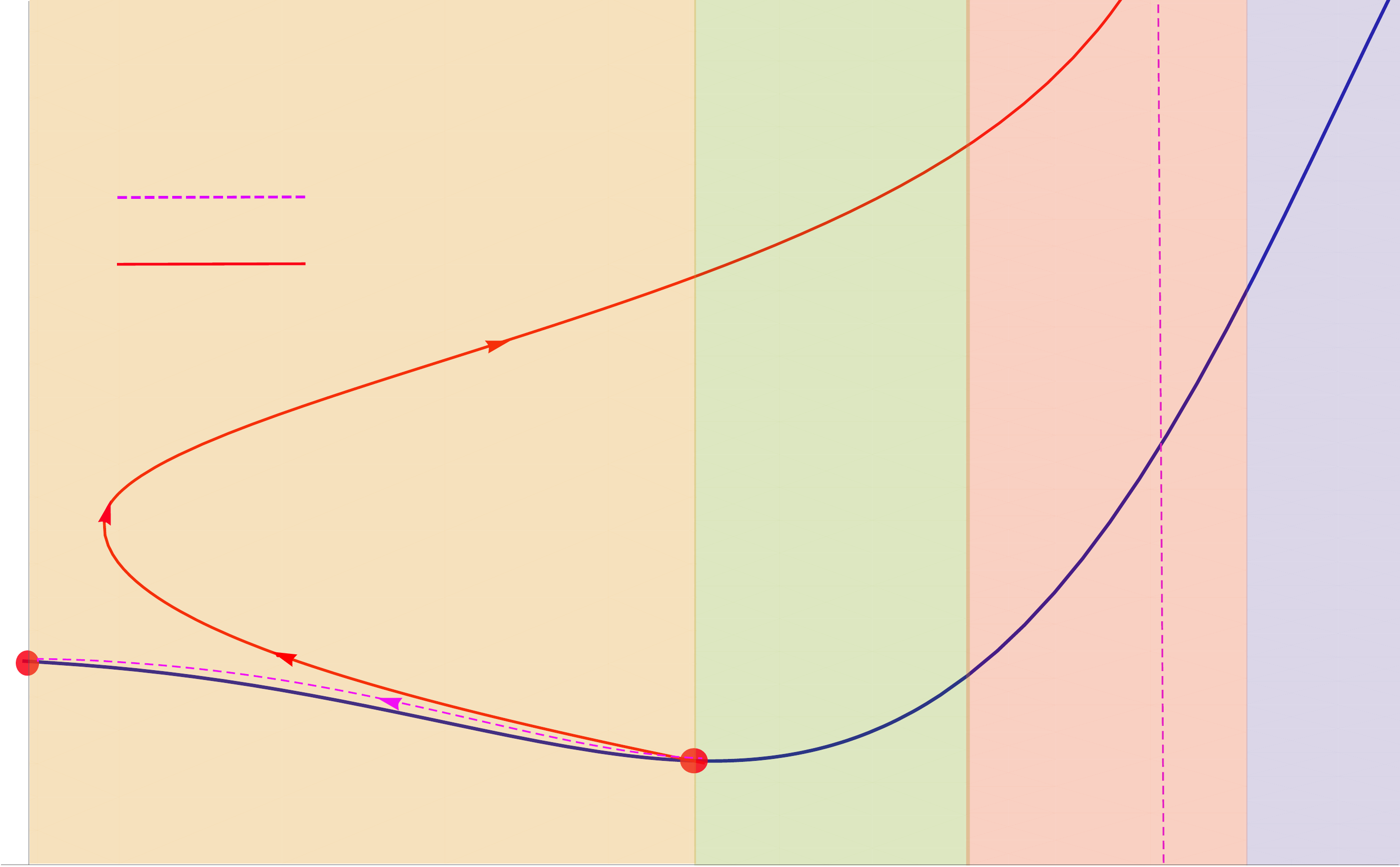}
\put(101,0){$\f$}
\put(3,58.5){$W(\f)$}
\put(48,4){UV$_2$}
\put(2,11){IR$_1$}
\put(20,55){Non-bouncing}
\put(20,52){$\f_-<0$}
\put(51,55){Non-}
\put(51,52){bouncing}
\put(51,49){$\f_->0$}
\put(70,40){Bouncing}
\put(70,37){$\f_-<0$}
\put(70,12){$B(\f)$}
\put(24,42){$W_{2b,n}(\f)$}
\put(24,47){$W_{21}(\f)$ }
\end{overpic}
\caption{Superpotential $W_{2b,n}(\f)$ for a bouncing flows starting
  from UV$_2$. For bouncing flows, the
  source is negative. This type of flow also comes from a curvature
  deformation of the flat case flow $W_{21}$ (dotted line) and exist
  for all values of the curvature.}\label{UVIIbneg}
\end{figure}
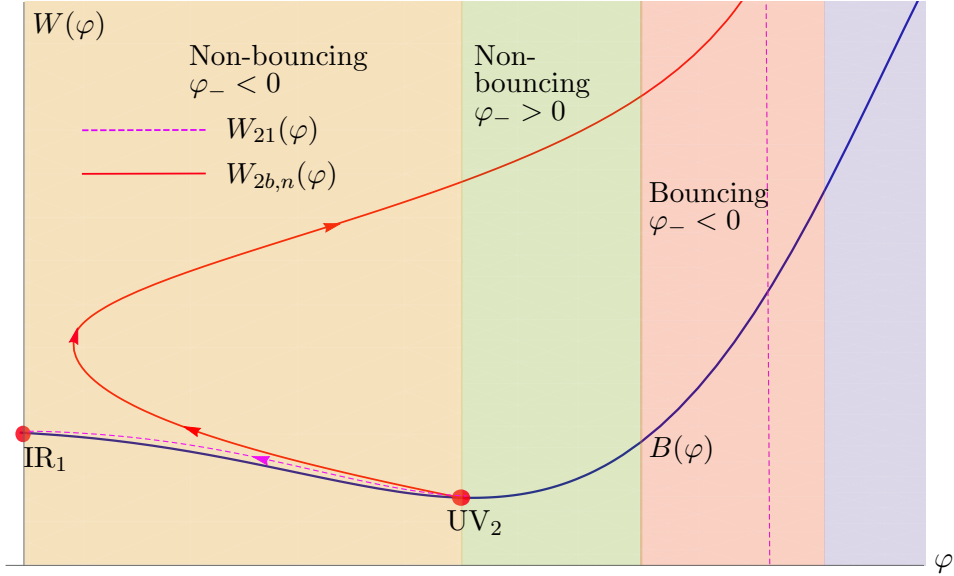

For completeness, we will also describe  the flows starting from
UV$_2$ as we increase the curvature. In the flat case, there was only
one such flow, with a negative source (see $W_{21}(\f)$ in
Fig.~\ref{skipflatnew}). If the curvature is slightly
increased, this flow splits into two. One starts from UV$_2$  and
ends in the region between IR$_1$ and UV$_2$. This is shown as the
red curve  W$_{2nb,n}(\f)$ in Fig.~\ref{UVIIneg}. The other flow,
represented  as the red curve W$_{2b,n}(\f)$ in Fig.~\ref{UVIIbneg},
starts form UV$_2$, {\em bounces} in the  region between IR$_1$ and UV$_2$ and ends  in the region $\f > \text{UV}_2$. These two flows, namely the non-bouncing flow and the bouncing flow with negative source, exist for all values of curvature.

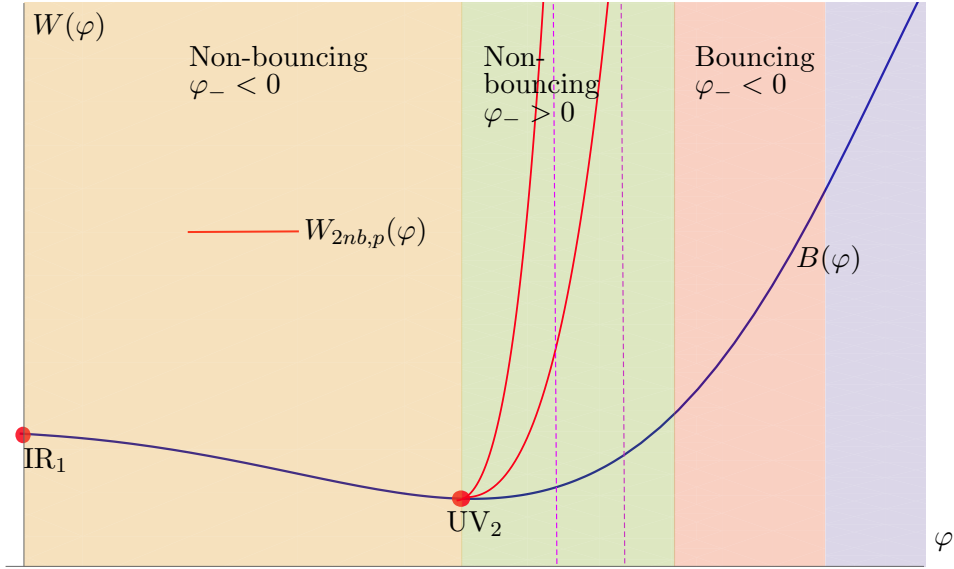
\begin{figure}[t]
\centering
\begin{overpic}
[scale=.42]{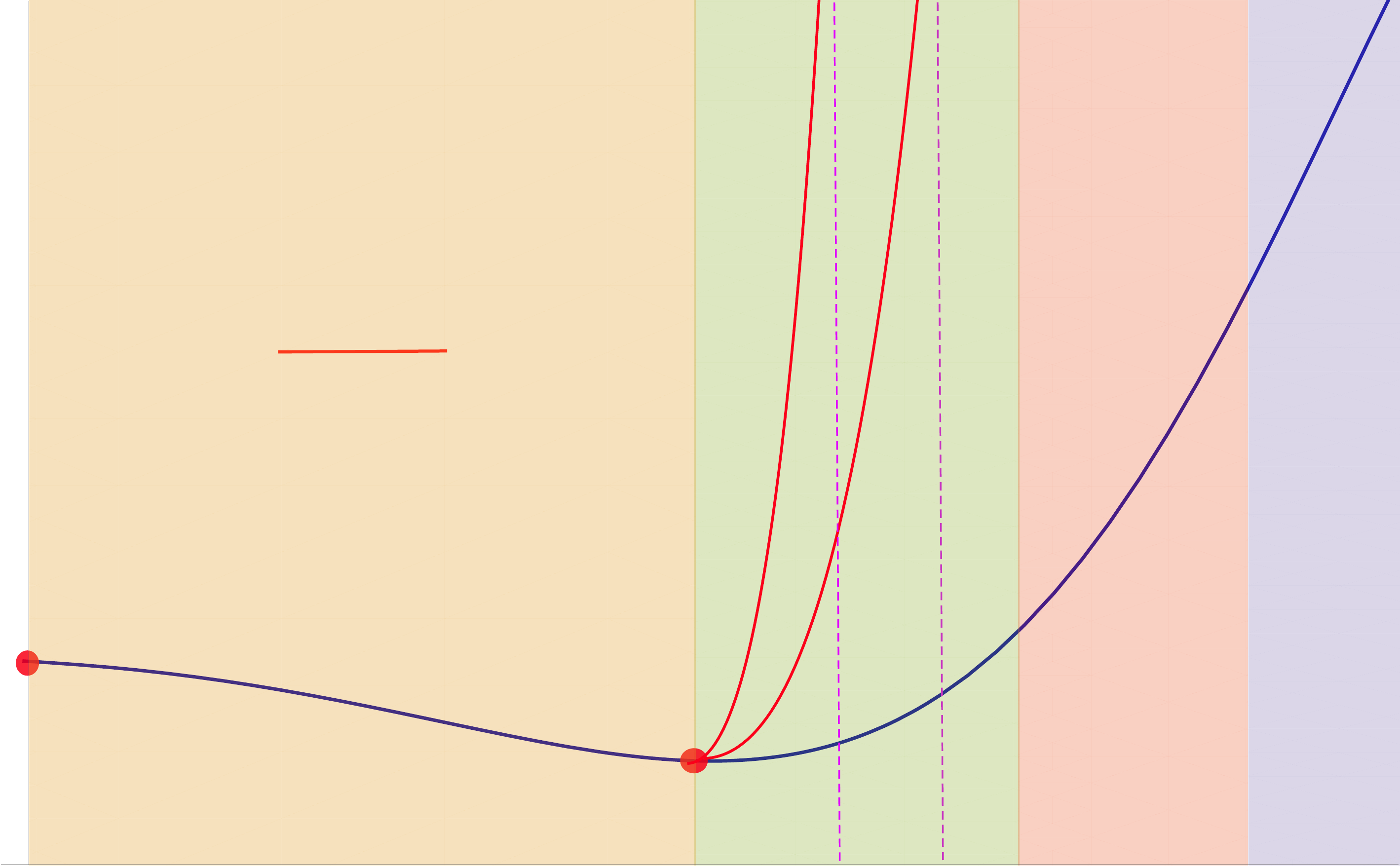}
\put(101,2.5){$\f$}
\put(3,58.5){$W(\f)$}
\put(48,4){UV$_2$}
\put(2,11){IR$_1$}
\put(20,55){Non-bouncing}
\put(20,52){$\f_-<0$}
\put(52,55){Non-}
\put(52,52){bouncing}
\put(52,49){$\f_->0$}
\put(75,55){Bouncing}
\put(75,52){$\f_-<0$}
\put(86,33){$B(\f)$}
\put(32.5,36){$W_{2nb,p}(\f)$}
\end{overpic}
\caption{Two non-bouncing flows $W_{2nb,p}(\f)$ from UV${}_2$ with identical value of $\mathcal{R}$. Interestingly, flows of this type only exist for $\mathcal{R} > \mathcal{R}_{\textrm{min}}$, i.e.~when the dimensionless curvature is larger than some minimal value. As a result, such flows do not come from a curvature deformation of a $\mathcal{R}=0$ flow. }\label{UVIIpos}
\end{figure}

 Beyond a certain value of the  curvature,  two new solutions appear
 that emanate from UV$_2$ in addition to the bouncing and the
 non-bouncing flows. These have   a positive source and have no
 $\mathcal{R}=0$ counterpart.  They end  in  region $\f >
 \text{UV}_2$. They are shown as the red curves $W_{2nb,p}(\f)$
 in Fig.~\ref{UVIIpos}.

\subsubsection*{Summary and observations}
\begin{figure}[t]
\begin{overpic}
[scale=.48]{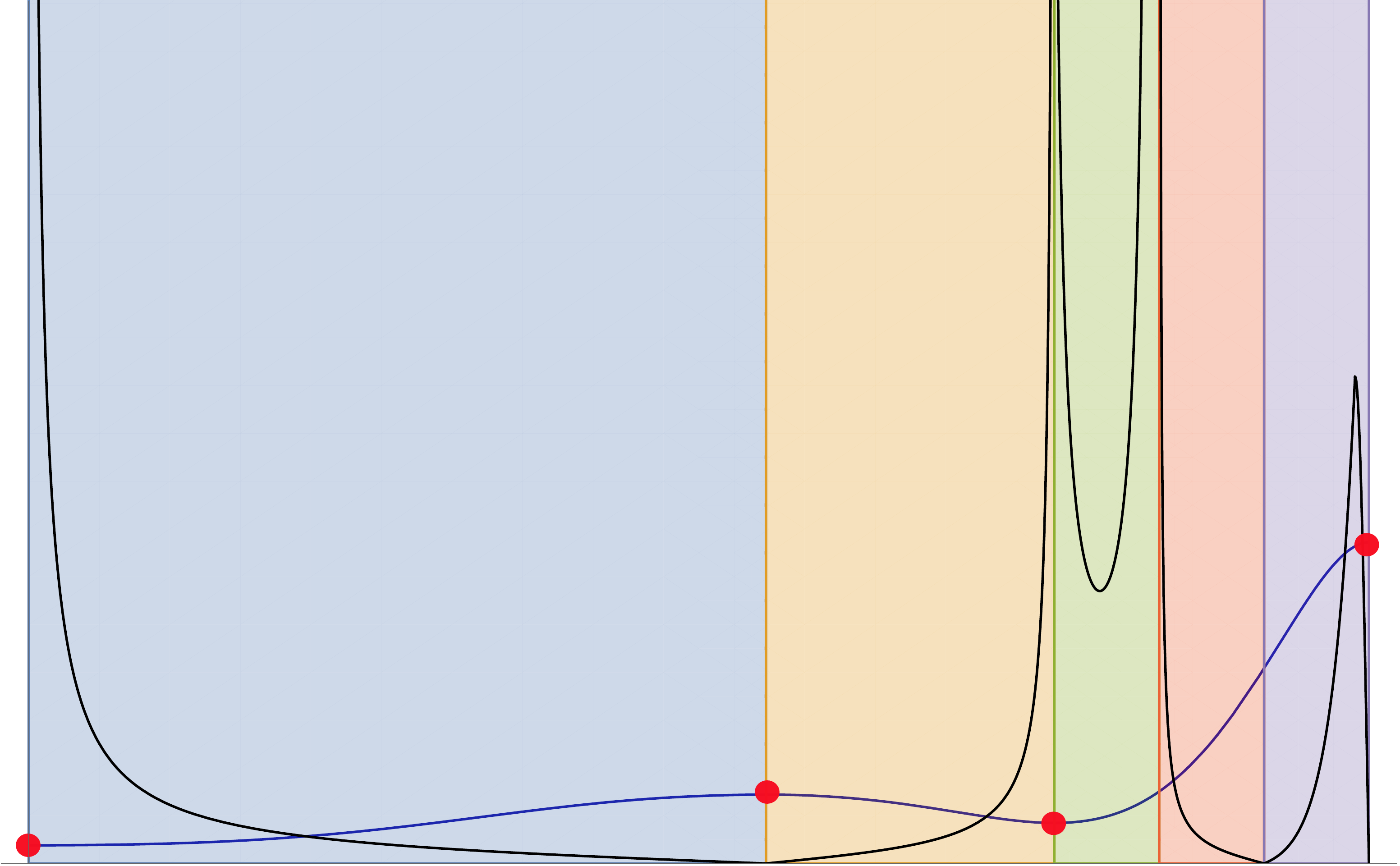}
\put(100,-1.3){$\f_0$}
\put(82,-1.3){$\f_*$}
\put(89.5,-1.3){$\f_!$}
\put(0,63){$\mathcal{R}$}
\put(2.5,3){UV$_1$}
\put(55.5,6.5){IR$_1$}
\put(74,4.5){UV$_2$}
\put(92,24){IR$_2$}
\put(40,5.5){$B(\f)$}
\put(20,60){\tiny{Non-skipping}}
\put(20,58){\tiny{$\f_->0$}}
\put(20,56){\tiny{From UV$_1$}}
\put(58,60){\tiny{Non-bounce}}
\put(58,58){\tiny{$\f_-<0$}}
\put(58,56){\tiny{From UV$_2$}}
\put(77,60){\tiny{Non-}}
\put(76,58){\tiny{bounce}}
\put(75.75,56){\tiny{$\f_->0$}}
\put(76.5,54){\tiny{From}}
\put(76.5,52){\tiny{UV$_2$}}
\put(84,60){\tiny{Bounce}}
\put(84,58){\tiny{$\f_-<0$}}
\put(84,56){\tiny{From}}
\put(84,54){\tiny{UV$_2$}}
\put(91,60){\tiny{Skipping}}
\put(91,58){\tiny{$\f_->0$}}
\put(91,56){\tiny{From}}
\put(91,54){\tiny{UV$_1$}}

\end{overpic}
\caption{Dimensionless curvature $\mathcal{R}$ vs.~flow end point $\f_0$. The leftmost blue region is the IR
  region where non-skipping flows from UV$_1$ can end. The rightmost purple area
  is the IR region for skipping flows from UV$_1$. The light brown region is the IR region for flows starting
  from UV$_2$ with a negative source and no bounces along the flow. The green area is the IR region for
  flows starting from UV$_2$ with a positive source. The red region is the IR
  region for flows from UV$_2$ with a negative source but which exhibit a bounce. The blue line is the critical curve $B(\f) =
  \sqrt{-3V(\f)}$.} \label{phasediagram}
\end{figure}
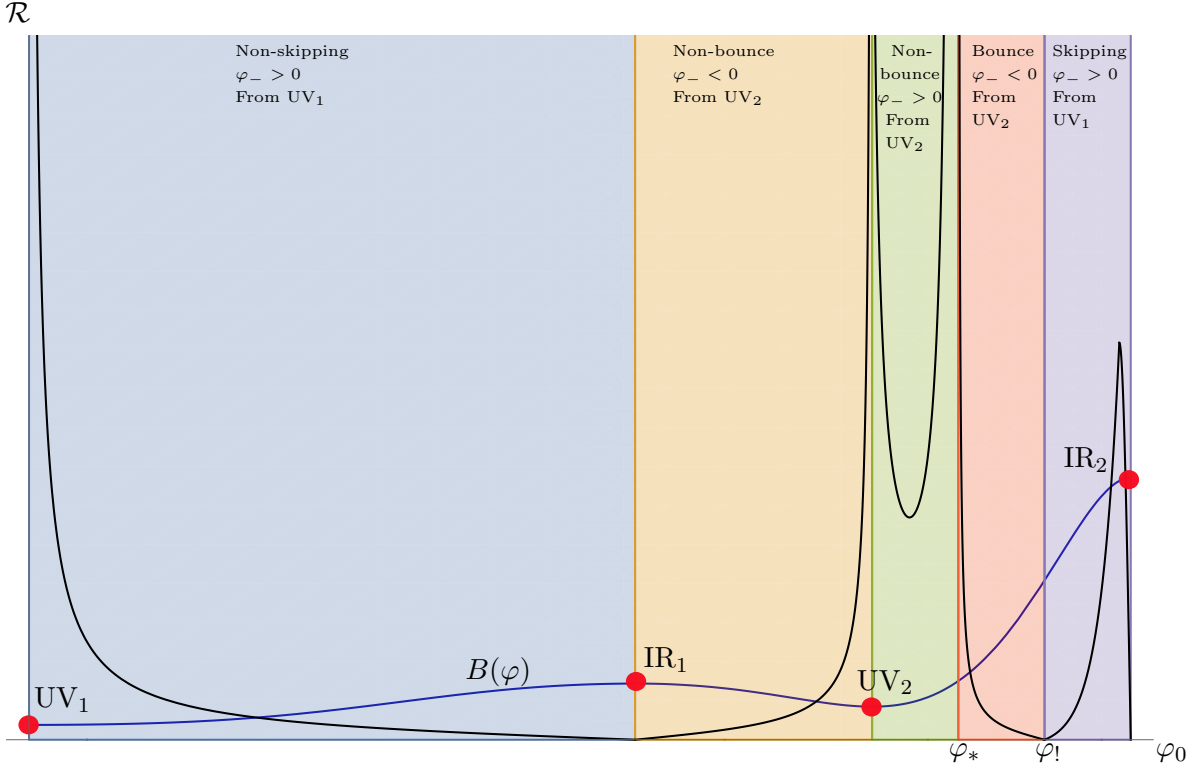

The full space of solutions is represented in Fig.~\ref{phasediagram},
which shows the dimensionless UV curvature parameter $\mathcal{R}$ as
a function of the IR end point $\f_0$ of the flow.  Non-skipping flows
start from UV$_1$ and the end in the leftmost blue region. This type
of solution exists for all values of curvature. On the other hand,
skipping flows from UV$_1$ end up in the rightmost purple region, and
they come in two branches (i.e.~there are two different endpoints in
the right purple region for a given value of  $\mathcal{R}$). These
skipping flows exist up to a maximum value of curvature, where the two
branches merge.

We also display the results for flows starting from
UV$_2$ in Fig.~\ref{phasediagram}. Flows with a negative source (i.e.~going towards the left of
UV$_2$) either end in the light brown region, or bounce
and then end  in the red region. Both these bouncing and non-bouncing
solutions originate as deformations of the ``flat'' flow $W_{21}(\f)$
(see Figures \ref{UVIIneg} and \ref{UVIIbneg}), and exist for any
positive value of  $\mathcal{R}$.

A particularly interesting set of flows from UV$_2$ are those with end points in the green region in Fig.~\ref{phasediagram}. Note that all flows in this region have $\mathcal{R} \geq \mathcal{R}_{\textrm{min}}$ for some minimal value of the dimensionless curvature. This implies that these flows do not arise from a continuous deformation of a flat flow. In fact, there is no flat flow from UV$_2$ with positive source. This is was explained in \cite{exotic} as arising from the fact that there is already a regular flow terminating at IR$_2$ from the left, and therefore no other regular flows can also arrive from the same side. The solutions with end points in the green region in Fig.~\ref{phasediagram} thus correspond to new vacua which do not exist for $\mathcal{R}=0$, but appear when $\mathcal{R}$ becomes large enough.

The fact that there is a regular flow starting at UV$_2$ for negative coupling (towards the left of UV$_2$) but not positive coupling (towards the right of UV$_2$) has analogues in QFT. Both for YM and $\lambda \phi^4$ theory the coupling must have a fixed sign. The theories do not exist for negative $g_{\textrm{YM}}^2$ or negative $\lambda$. Something similar also happens here. The unusual occurence however is that the theory with the `wrong' sign exists for sufficiently large space-time curvature. In a sense positive curvature cures the sickness of the flat theory. It would be interesting to understand this phenomenon better and find examples of a similar behavior in QFT.

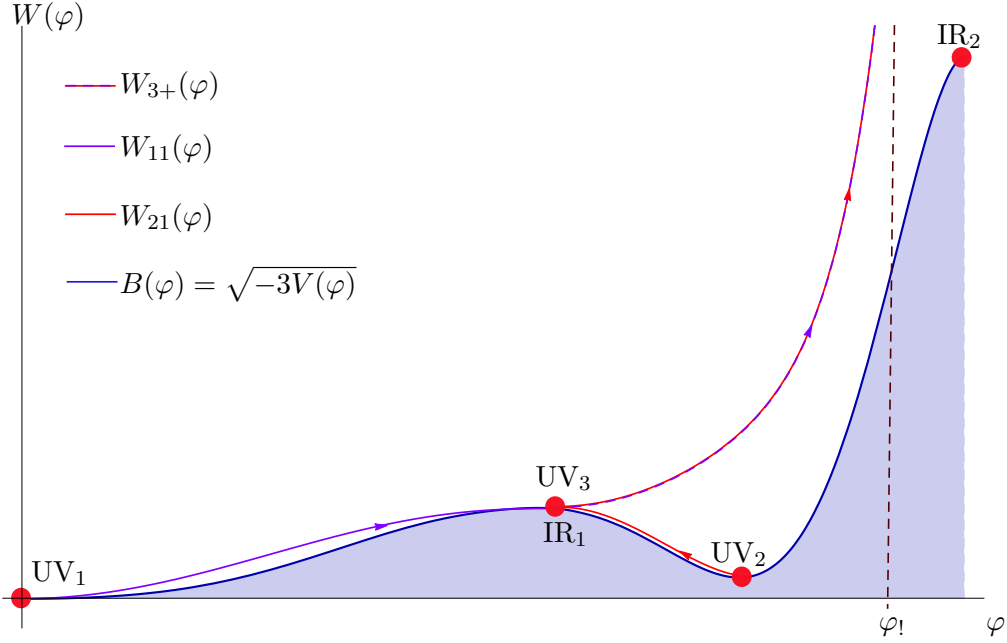
\begin{figure}[t]
\centering
\begin{overpic}
[scale=.4]{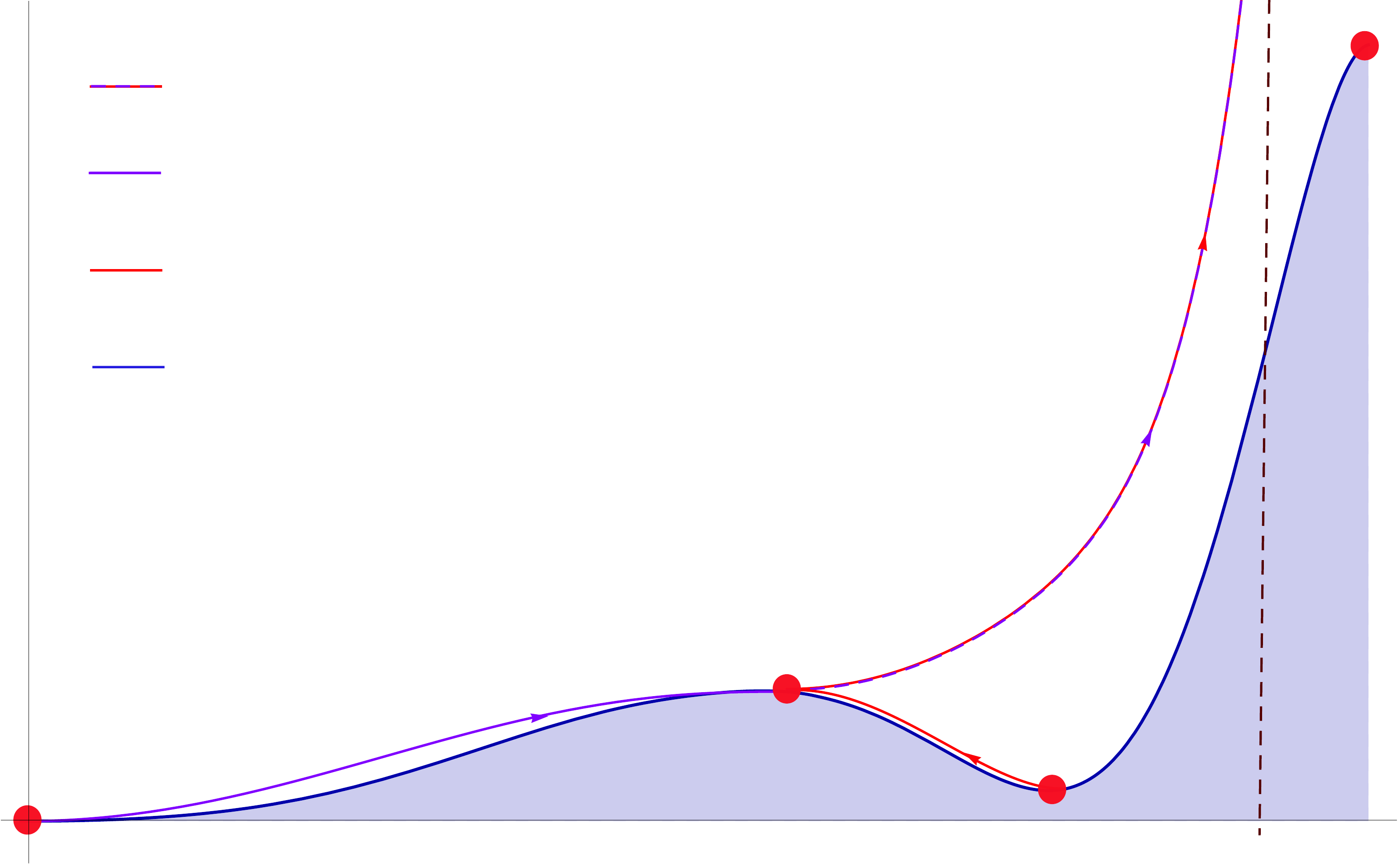}
\put(100,0){$\f$}
\put(89.1,0){$\f_{!}$}
\put(1,62){$W(\f)$}
\put(3,5){UV$_1$}
\put(55,9){IR$_1$}
\put(54.2,14.75){UV$_3$}
\put(72,7){UV$_2$}
\put(95,60){IR$_2$}
\put(12,55){$W_{3+}(\f)$}
\put(12,48.5){$W_{11}(\f)$}
\put(12,41.5){$W_{21}(\f)$}
\put(12,34.5){$ B(\f)=\sqrt{-3V(\f)}$}
\end{overpic}
\caption{RG flows with IR endpoint $\f_0 \rightarrow \f_!$. When the endpoint $\f_0$ approaches $\f_!$ flows from both UV$_1$ and UV$_2$ pass by closely to IR$_1$, passing through IR$_1$ exactly for $\f_0 = \f_!$. This is shown by the purple and red curves. Beyond IR$_1$ both these solutions coincide, which is denoted by the colored dashed curve. These have the following interpretation. The flows from UV$_1$ and UV$_2$ should not be continued beyond IR$_1$, which becomes the IR endpoint for the zero curvature flows $W_{11}$ and $W_{21}$. The remaining branch (the colored dashed curve) is now an independent flow denoted by $W_{3+}$. This is a flow from a UV fixed point at a minimum of the potential (denoted by UV$_3$ above) to $\f_!$ and corresponds to a $W_+$ solution in the notation of section \protect\ref{sec:asymp} with fixed value $\mathcal{R}= R^{\textrm{uv}} |\f_+|^{-2 / \Delta_+} \neq 0$. While flows from UV$_1$ and UV$_2$ can end arbitrarily close to $\f_!$, the endpoint $\f_0 = \f_!$ cannot be reached from UV$_1$ or UV$_2$.}
\label{phiexclamation}
\end{figure}

Notice that there are several values of the endpoint where the
curvature parameter diverges. Two of them correspond to the UV fixed
points UV$_{1}$ and UV$_2$, in line with the generic behavior found
in section \ref{sec:examples}. The third one (namely the point  $\f_*$ in Fig.~\ref{phasediagram} separating
the green and red regions) is more interesting. Across this point the
source changes sign. For the endpoint exactly at $\f_*$  the solution must have zero
source, implying that for finite $R^{\textrm{uv}}$, the
quantity $\mathcal{R}$ diverges (recall the definition in equation
(\ref{r})) as we approach $\f_*$.  This fact has a remarkable
consequence: It implies
that, above some positive threshold value for the curvature, there exist
RG flows with no source (i.e.~of the $W_+$ branch, the ones driven
by a vev) starting at UV$_2$ and ending at $\f_*$. Compare this to the
flat case, where regular vev flows are highly non-generic, and require
a fine-tuned potential.

Note that there seems to exist an endpoint for flows with $\mathcal{R}=0$ that does not coincide with an extremum of the potential. This is the point $\f_{!}$ in Fig.~\ref{phasediagram} separating the purple and red regions. However, one finds that this endpoint cannot be reached by a flow from UV$_1$ or UV$_2$. To illustrate this, let us consider what happens for flows ending in both the red and purple regions of Fig.~\ref{phasediagram} when we let $\f_0 \rightarrow \f_!$. Flows ending in the red region of Fig.~\ref{phasediagram} leave UV$_2$ to the left before bouncing and reversing direction (see Fig.~\ref{UVIIbneg}). Taking $\f_0 \rightarrow \f_!$ the bounce point occurs closer and closer to IR$_1$, coinciding with IR$_1$  for $\f_0 = \f_!$. Once this happens the two branches of the previously bouncing solution become two independent RG flows. For one, there is the flow starting from UV$_2$ and ending at IR$_1$ (denoted by $W_{21}$ in Fig.~\ref{phiexclamation}). This is a zero-curvature flow from extremum to extremum. The 2nd branch of the bouncing flow now becomes a flow starting at the minimum associated with IR$_1$ and ending at $\f_!$. This is shown as $W_{3+}$ in Fig.~\ref{phiexclamation}. The minimum thus plays the role of a UV fixed point, which we label UV$_3$ in Fig.~\ref{phiexclamation}. This corresponds to a $W_+$ solution in the language of section \ref{sec:asymp} with fixed value $\mathcal{R}= R^{\textrm{uv}} |\f_+|^{-2 / \Delta_+} \neq 0$. We can make an analogous observation for flows ending in the purple region, which originate from UV$_1$ and skip IR$_1$. Taking $\f_0 \rightarrow \f_!$ the flows miss IR$_1$ with ever decreasing distance, and pass through IR$_1$ for $\f_0 = \f_!$. Again, in this case we should not continue the flow beyond IR$_1$, which becomes the endpoint. This corresponds to the solution $W_{11}$ in Fig.~\ref{phiexclamation}.

\subsubsection*{A quantum phase transition}
We have observed that there are more than one flows with the  same
dimensionless UV curvature parameter, $\mathcal{R}$, and now we will
focus on  flows which start from UV$_1$ and have the same
source. These correspond to three branches of solutions contained in the blue and purple regions in figure
\ref{phasediagram}. Beyond a certain value of the curvature, only the
non-skipping solution in the blue region  exists, so this solution  it
is necessarily the ground state. On the other hand, as we mentioned at
the beginning of this section,  at zero curvature one of the {\em
  skipping} solutions is the true ground state.  This suggests that
varying   $\mathcal{R}$ we should encounter a phase transition.

\begin{figure}[t]
\centering
\begin{overpic}
[width=.6\textwidth]{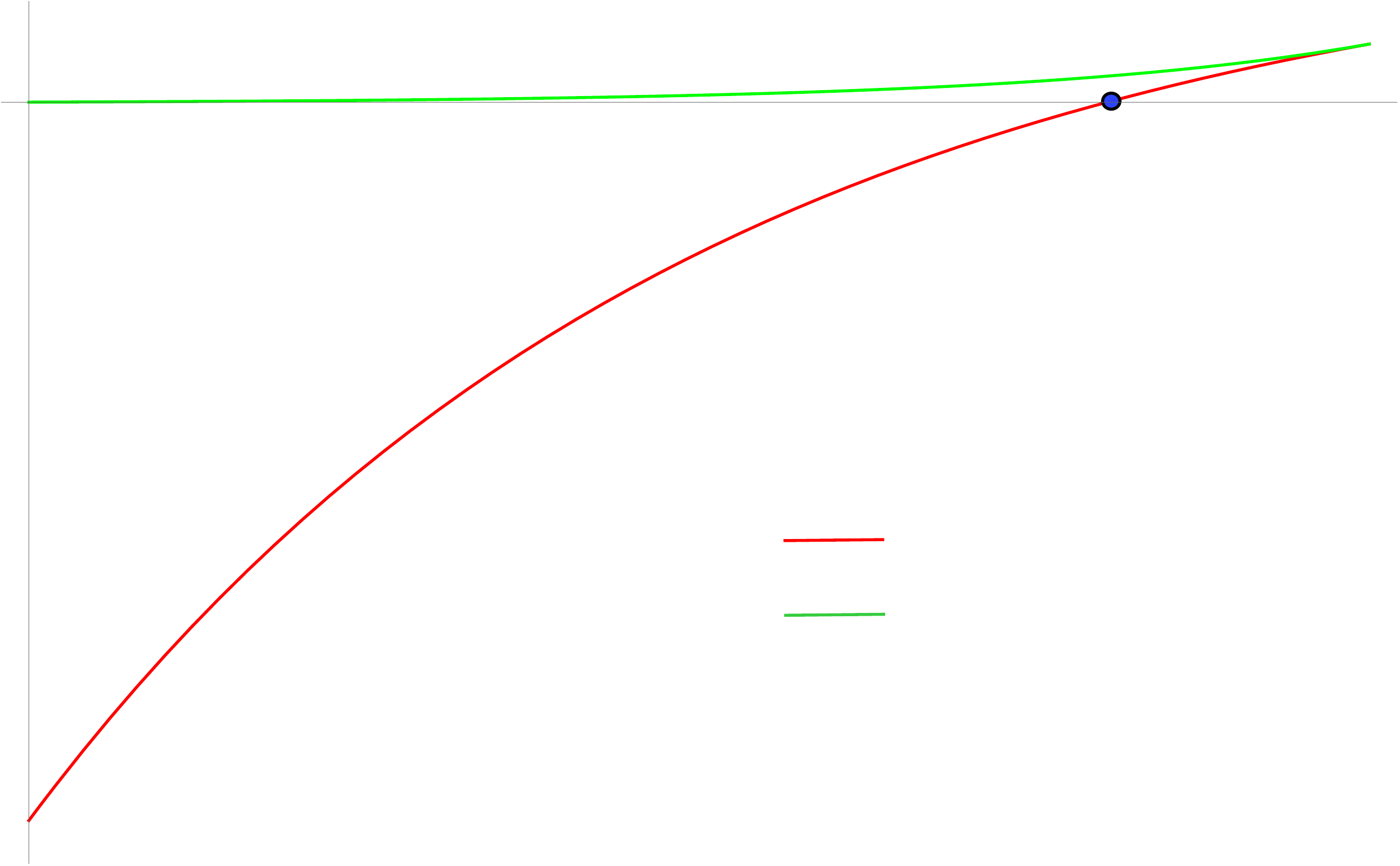}
\put(96.5,49){$\mathcal{R}$}
\put(80,49){$\mathcal{R}_{c}$}
\put(4,58.5){$\Delta F=F_{\text{skip}}-F_{\text{non-skip}}$}
\put(65,22){$ F_{\text{skip},1}-F_{\text{non-skip}}$}
\put(65,17){$F_{\text{skip},2}-F_{\text{non-skip}}$}
\end{overpic}
\caption{Free energy difference between the skipping and the non-skipping solution. For a fixed curvature, there are two skipping solutions  for which the superpotential $W(\f)$ was denoted as $W_{s,1}(\f)$ and $W_{s,2}(\f)$ (they were denoted as red and green curves respectively in figures \protect\ref{skipcurvedsmall} and \protect\ref{skipincreased}). In this figure, the red curve corresponds to the on-shell action difference between the $W_{s,1}(\f)$ solution and the non-skipping solution. The green curve corresponds to the on-shell action difference between the $W_{s,2}(\f)$ solution and the non-skipping solution $W_{ns}(\f)$. }\label{phasetransition}
\end{figure}
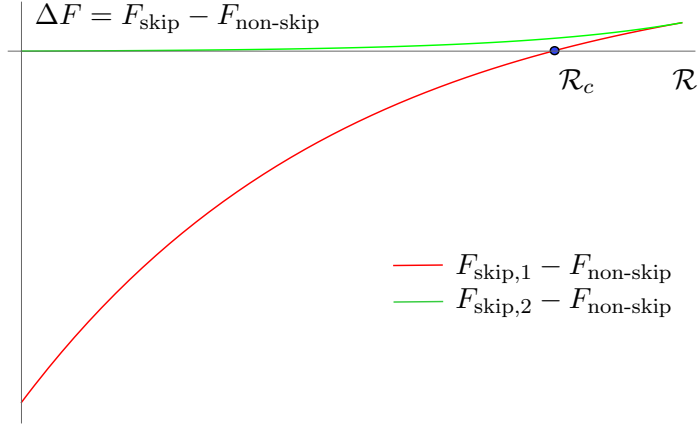

A numerical analysis shows that this expectation is correct, and the
phase transition is first order. To compare the free energies, we
evaluate
expression (\ref{Sonshellu}) numerically, supplemented by the appropriate counter-terms (\ref{ct0}-\ref{ct2}), for each of the
solutions.\footnote{In this case, there is no analytic
  result for all values of curvature because of the bulk integral in expression \eqref{Sonshellu} which is non-trivial to evaluate.} The numerical evaluation of the free energy difference
between the two skipping solutions and the non-skipping solution is
shown in Fig.~\ref{phasetransition}, which clearly shows that we are
in presence of a first order quantum phase transition driven by space-time curvature.

At zero curvature, one of the   skipping solutions dominates, and the
second one becomes degenerate with the non-skipping solution.   As the
curvature is increased, the difference between the free energies
decreases. At a certain value of the dimensionless curvature,
$\mathcal{R}=\mathcal{R}_{c}$, the free energy difference
changes sign. (In our example we find $\mathcal{R}_{c} \approx 80$.) Now, the non-skipping solution is the dominant saddle point in the path integral. Finally, the two skipping solutions merge
at a maximal value $\mathcal{R}$ and disappear for larger values of the
curvature.

\section*{Acknowledgements}\label{ACKNOWL}
\addcontentsline{toc}{section}{Acknowledgements}

We thank Arthur Hebecker, Leandro Silva Pimenta, Kostas Skenderis for
useful discussions
and Igor Klebanov, Dario Martelli and Marika Taylor
for correspondence.

\noindent This work was supported in part  by the Advanced ERC grant SM-grav, No 669288.

\appendix
\renewcommand{\theequation}{\thesection.\arabic{equation}}
\addcontentsline{toc}{section}{Appendix\label{app}}
\section*{Appendix}

\section{Curvature Invariants} \label{curvature inv}
In this appendix we record expressions for various curvature invariants. In particular, for the metric \eqref{eq:metric},
we will compute $R$, $R_{AB}R^{AB}$ and $R_{ABCD}R^{ABCD}$ and we will express them in terms of the functions $W$, $S$ and $T$ defined in \eqref{eq:defWc}--\eqref{eq:defTc}.

\vspace{0.3cm}
\noindent \textbf{Ricci scalar} \newline
The expression for the Ricci scalar is found as,
\begin{equation}
R=-\left[ 2d \ddot{A}(u)+d(d+1)\dot{A}^2(u) \right]+e^{-2A(u)} R^{(\zeta)}
\end{equation}
which in terms of $W(\f)$, $S(\f)$ and $T(\f)$ can be written as,
\begin{equation}
R=\dfrac{d}{d-1}W' S -\dfrac{d(d+1)}{4(d-1)^2}W^2+T={S^2\over 2}+{d+1\over d-1}V \ .
\end{equation}
In the last step we have used equations \eqref{eq:EOM4}--\eqref{eq:EOM6} to simplify.
This shows that the scalar curvature diverges when the potential $V(\f)$ diverges or when $S(\f)$ diverges.

\vspace{0.3cm}
\noindent \textbf{Ricci squared} \newline
The square of the Ricci tensor is given as,
\begin{align}
R_{AB}R^{AB} &=d^2 \left( \ddot{A}(u)+\dot{A}^2(u) \right) ^2+\dfrac{(R^{(\zeta)})^2}{d}e^{-4A(u)}-2R^{(\zeta)}e^{-2A(u)} (\ddot{A}(u) \nonumber \\
&+d \dot{A}^2(u))+d \left( \ddot{A}(u)+d \dot{A}^2(u) \right)^2  .
\end{align}
On shell, this can be written in terms of $V$ and $S$ as
\begin{equation}
R_{AB}R^{AB}=d^2\left(\dfrac{S^2}{2d}+\dfrac{V}{d(d-1)} \right)^2+\frac{d V^2}{(d-1)^2} .
\end{equation}
This also diverges when $V(\f)\to\infty$ or when $S(\f)$ diverges.

\vspace{0.3cm}
\noindent \textbf{Riemann squared} \newline
The so-called Kretschmann scalar is found as,
\begin{equation}
R_{ABCD}R^{ABCD} =4d \left( \ddot{A}(u)+\dot{A}^2(u) \right)^2 +2d(d-1)\left(  \dfrac{e^{-2A(u)}}{\alpha^2 }-\dot{A}^2(u) \right)^2  .
\end{equation}
Using the equations of motion this can be written as,
\begin{equation}
R_{ABCD}R^{ABCD}=\dfrac{4}{d}\left( \dfrac{S^2}{2}+\dfrac{V}{d-1} \right)^2 +\dfrac{1}{2d(d-1)}\left(S^2-2V \right)^2 .
\end{equation}
Again, this diverges whenever the potential $V(\f)$ or the function $S(\f)$ diverge.

\section{Relation to the complex first order formalism} \label{complex}

In \cite{SkenderisTownsend} a first order formalism using a complex (`superpotential') function $Z$ was introduced. In this appendix, we will relate quantities calculated using the complex first order formalism of \cite{SkenderisTownsend} to expressions calculated in the first order formalism employing the real functions $W$, $S$ and $T$ used in this work.

The bulk theory considered in \cite{SkenderisTownsend} is Einstein-dilaton gravity with a dilaton potential, just like in this work. In the conventions of \cite{SkenderisTownsend} the Lagrangian is given by
\begin{equation}
\mathcal{L}=\sqrt{-g}\left[ R-\frac{1}{2}(\partial \sigma)^2-V(\sigma) \right] \, .
\end{equation}
and the following ansatz is used:
\begin{equation}
ds^2=\eta (f(z) e^{\alpha \f(z)})^2 dz^2 +e^{2\beta\f(z)} ds_d^2 \, , \qquad \sigma = \sigma(z) \, .
\end{equation}
To make contact with the approach in this paper, the following map is required:
\begin{align}
 \label{eq:1stformrel1} \eta&=1 \\
 f(z)&=e^{-\alpha \f(z)} \\
 z &\rightarrow u \\
 \beta\f (z) &\rightarrow A(u) \\
\label{eq:1stformrellast}  \sigma(z) &\rightarrow \f (u)
\end{align}
where $\alpha=\sqrt{\frac{d}{2(d-1)}}$ and $\beta=\sqrt{\frac{1}{2 d (d-1)}}$.

In the Hamiltonian formulation, the following Hamiltonian is used in \cite{SkenderisTownsend}:
\begin{equation}
 \mathcal{H}=\frac{1}{2} \left( \pi^2-p^2 \right)+e^{2\alpha\f}(\eta V -\frac{\eta k}{2\beta^2 }e^{-2 \beta \f} ) \, ,
 \end{equation}
 with $k=-1,0,1$ and where the conjugate momenta are defined as
 \begin{align}
 f^{-1}\dot{\f}=\pi , \quad  f^{-1}\dot{\sigma}=-p \ .
 \end{align}
 Using the relations \eqref{eq:1stformrel1}--\eqref{eq:1stformrellast} the Hamiltonian constraint $\mathcal{H}=0$ reduces to
 \begin{equation}
 d((d-1)\dot{A}^2 -\frac{1}{2}\dot{\f}^2+V-R^{(\zeta)}e^{-2A}=0 \, , \label{HC}
 \end{equation}
 with $R^{(\zeta)}=d(d-1)k$. This means that the bulk is sliced by dS$_d$/$S^d$ ($k=1$) or AdS$_d$ ($k=-1$) space-times of unit radius or Minkowski space $\mathcal{M}_d$ ($k=0$). Eq.~(\ref{HC}) is identical to the equation of motion \eqref{eq:EOM2} in our conventions. The other two equations of motion from Hamilton's equations can be shown to reproduce the remaining two equations of motion \eqref{eq:EOM1} and \eqref{eq:EOM3} in our notation.

In the following, we want to make contact between expressions using $W$ and $S$ in our conventions and expressions using $Z$. To this end note that the complex function $Z$ is related to $W$ and $S$ as follows:
\begin{align}
W&=\pm 2 \left[ \frac{Re(\bar{Z}Z')}{|Z'|}\right]  \label{WA} \, , \\
S&=\pm 2 |Z'|  \label{SA} \, .
\end{align}
Writing
\begin{equation}
Z=X+i Y, \quad \bar{Z}=X-iY ,\quad  Z'=X'+iY', \quad \bar{Z}'=X'-iY'
\end{equation}
the expressions \eqref{WA}--\eqref{SA} can be written as
\begin{align}
W&= \pm 2 \frac{X X'+Y Y'}{\sqrt{(X')^2+(Y')^2}} \, , \\
S&= \pm 2 \sqrt{(X')^2+(Y')^2} \, .
\end{align}
Now consider our equation of motion \eqref{eq:EOM6} in the first order fornalism, which can be written as
\begin{align}
& \frac{1}{2} \frac{d}{d\f}S^2 -\frac{d}{2(d-1)}SW= V' \, . \label{KG}
\end{align}
Using \eqref{WA} and \eqref{SA} we can write eq.~(\ref{KG}) as
\begin{align}
\frac{1}{2} \frac{d}{d\f}S^2- 4 \alpha^2 Re(\bar{Z}Z') &=V' \, , \nonumber \\
\Rightarrow \ \ \frac{1}{2} \frac{d}{d\f}S^2- 2 \alpha^2 \frac{d}{d\f}|Z|^2 &=V' \, .
\end{align}
Integrating this equation we find
\begin{align}
V=\frac{1}{2}S^2- 2 \alpha^2 |Z|^2+2\alpha^2 v \label{KG1} \, ,
\end{align}
where $2\alpha^2 v$ is an integration constant. Using $S(\f)= 2 |Z'|$, we can write eq.~(\ref{KG1}) as:
\begin{align}
V=2 |Z'|^2- 2 \alpha^2 (|Z|^2- v) \, .
\end{align}
This is eq.~(2.7) in \cite{SkenderisTownsend}. Overall, we conclude that the first order formalism using a complex function $Z$ and the approach presented in this paper are two equivalent descriptions. The results of this section can be used to translate between the two approaches.

\section{Properties of the functions $W$, $S$ and $T$}\label{properties}

\subsection{Positive curved case ($S^d$ or dS$_d$)}
\begin{enumerate}
\item There are two branches of solutions $W$ and $S$ at a generic point. On a single branch the signs of $S$ and $W'$ coincide.
This can be seen from Eq.~\eqref{eq:EOM4} which we can write as:
\begin{equation}
SW'= S^2  + \frac{2}{d} T \ .
\end{equation}
The RHS is always positive. Therefore the signs of $S$ and $W'$ must be same. The two possible signs for $S$ and $W'$ give rise to the two branches.
\item The absolute value of $W(\f)$ is bounded by the critical curve $B(\f)=\sqrt{-\frac{4(d-1)}{d}V(\f)} $ and on this critical curve the two functions $S(\f)$ and $T(\f)$ go to zero.
This can be shown from Eq.~\eqref{eq:EOM5} which we can write as
\begin{equation}
\frac{d}{2(d-1)} W^2= S^2 +2 T -2V\geq -2V \ .
\end{equation}
 Therefore, we can write,
 \begin{equation}
  W^2(\f)\geq -\frac{4(d-1)}{d}V(\f)=B^2(\f).
  \end{equation}

  On the critical curve $\frac{d}{2(d-1)} W^2(\f)=-2V(\f)$. So Eq.~\eqref{eq:EOM5}  can be written as
  \begin{equation}
  S^2+2T=0 \ .
  \end{equation}
 This is a sum of two positive numbers which can be zero iff the individual contributions are zero. Hence, on the critical curve, the functions $S(\f)$ and $T(\f)$ vanish.
\item
  Bounces cannot happen on the critical curve.
 Let us assume that along a flow originating from a UV fixed point a bounce occurs at $\f=\f_*$, which happens to lie on the critical curve. From property 2, we see that at $\f=\f_*$, $T(\f_*)=0$. But we started from the UV, which is also on the critical curve and consequently had $T(\f_{UV})=0$. So $T(\f)$ is starting from a value of $0$ and ending with a value of  $0$. This cannot happen because $\frac{d T}{du}=\frac{TW}{d-1} \geq 0$ (assuming $W\geq 0$). Instead, a bounce can happen when $S=0$ and
  \begin{equation}
  W^2=\frac{2(d-1)}{d}\left( 2T -2V \right) \, \quad \text{where}\  T>0 \ .
  \end{equation}

 \subsection{Positive and negative curved case}

  \item
  The functions $W(\f)$, $S(\f)$ and $T(\f)$ satisfy the following relation:
  \begin{equation}
  T' S=\frac{TW}{d-1}\ .
  \end{equation}
  This comes from the definition of the functions $W(\f)$, $S(\f)$ and $T(\f)$. We can take a $u$-derivative of $T=e^{-2 A(u)}R^{(\zeta)}$ and then use the definitions \eqref{eq:defTc}--\eqref{eq:defWc} .

  \item
  At the zeros of the function $S(\f)$, the geometry is approximately maximally symmetric.
  We can express the Ricci tensor in terms of the function $S(\f)$. The only non-zero components of the Ricci tensor are:
\begin{align}
R_{uu}&=\left( \frac{S^2}{2}+\frac{V}{d-1}  \right)g_{uu},\\
R_{\mu\nu}&=\frac{V}{d-1}  g_{\mu\nu}.
\end{align}
Let $S$ be zero at a value $\f=\f_*$. Then at this point the Ricci tensor can be written as
\begin{equation}
R_{AB}=\frac{V(\f_*)}{d-1} g_{AB}.
\end{equation}
Therefore the space is maximally symmetric near the zeros of $S$. As the potential is negative, this space is approximately AdS.

\item Symmetry properties. Let $W(\f)$, $S(\f)$ and $T(\f)$ satisfy the equations \eqref{eq:EOM4}--\eqref{eq:EOM6} for a generic potential $V(\f)$. Then
\begin{align}
\bar{S}(\f)&=-S(\f) \ , \\
\bar{W}(\f)&=-W(\f) \ , \\
\bar{T}(\f)&= T(\f) \
\end{align}
are also solutions of the equations \eqref{eq:EOM4}--\eqref{eq:EOM6}. The scale factor and the curvature invariants behave as
\begin{align}
&\bar{A}(u)=A(u) \ , \\
&\bar{R}(\f)=R(\f) \ , \\
&\bar{R}_{AB}\bar{R}^{AB}(\f)= R_{AB}R^{AB}(\f) \ , \\
&\bar{R}_{ABCD}\bar{R}^{ABCD}(\f)= R_{ABCD}R^{ABCD}(\f) \ .
\end{align}
This means that we have two equivalent copies of geometries below and above the $\f$ axis.

If $V(\f)$ is an even function of $\f$, then
\begin{align}
\tilde{S}(\f)&=-S(-\f) \ , \\
\tilde{W}(\f)&=W(-\f) \ , \\
\tilde{T}(\f)&=T(-\f) \ ,
\end{align}
satisfy the equations  \eqref{eq:EOM4}--\eqref{eq:EOM6}. This is because the above relations imply
\begin{align}
\tilde{S}'(\f)&=S'(-\f) \ , \\
\tilde{W}'(\f)&=-W'(-\f) \ ,
\end{align}
and we also have that $V'(\f)=-V'(-\f)$. The scale factor and the curvature invariants behave as
\begin{align}
&\tilde{A}(u)=A(u) \ , \\
&\tilde{R}(\f)=R(-\f) \ , \\
&\tilde{R}_{AB}\tilde{R}^{AB}(\f)= R_{AB}R^{AB}(-\f) \ , \\
&\tilde{R}_{ABCD}\tilde{R}^{ABCD}(\f)= R_{ABCD}R^{ABCD}(-\f) \ .
\end{align}
This shows we have two equivalent copies of geometries on the positive and negative side of $\f$ axis.
\end{enumerate}

\section{Near boundary solution: Small curvature expansion}
\label{app:smallRexp}
In this appendix we will collect analytical expressions for $W(\f)$, $S(\f)$ and $T(\f)$ describing the geometry in the vicinity of the boundary of the asymptotically AdS${}_{d+1}$ space-time.

\subsection{The small curvature expansion}
It will be advantageous to organize the solution as a perturbative expansion in $R^{\textrm{uv}}$, expanding about the result for $R^{\textrm{uv}}=0$. The reason is as follows. We will find that for asymptotically AdS${}_{d+1}$ space-times \emph{a near-boundary expansion of the quantities} $\f$, $A$, $W$, $S$ and $T$ \emph{will automatically give rise to an expansion in} $R^{\textrm{uv}}$. We have already encountered this in \eqref{eq:AdSScaleFactorExpansion} where we expanded the scale factor $A(u)$ for AdS${}_{d+1}$ in the vicinity of the boundary. We found that $R^{\textrm{uv}}$ enters the expansion exclusively in the combination $R^{\textrm{uv}} e^{2 u / \ell}$. Expanding $A(u)$ in powers of $e^{2u / \ell}$ in the vicinity of the boundary at $u \rightarrow - \infty$ is thus is equivalent to expanding in powers of $R^{\textrm{uv}}$. This observation will also hold more generally for asymptotically AdS${}_{d+1}$ space-times.


\label{app:smallRdetails}
Consequently, we will expand $W(\f)$, $S(\f)$ and $T(\f)$ as follows:
\begin{align}
\label{eq:smallRWexp} W(\f) &= \sum_{n=0}^{\infty} W_n (\ell^2 R^{\textrm{uv}})^n  \, , \\
\label{eq:smallRSexp} S(\f) &= \sum_{n=0}^{\infty} S_n (\ell^2 R^{\textrm{uv}})^n  \, , \\
\label{eq:smallRTexp} T(\f) &= \sum_{n=0}^{\infty} T_n (\ell^2 R^{\textrm{uv}})^{n+1}  \, ,
\end{align}
where $\ell$ will be identified with the AdS length scale of the asymptotically AdS${}_{d+1}$ bulk space-time. Note that the expansion of $T(\f) \equiv R^{\textrm{uv}} e^{-2 A(\f)}$ starts at linear order in $R^{\textrm{uv}}$. We proceed by inserting the above expansions into the equations of motion \eqref{eq:EOM4}--\eqref{eq:EOM6} and solve these order by order in $R^{\textrm{uv}}$.

\subsubsection*{Calculating $S_n$}
Inserting the above expansions into \eqref{eq:EOM4} the resulting equations at order $(R^{\textrm{uv}})^0$ and $(R^{\textrm{uv}})^1$ can be rewritten as
\begin{align}
\label{eq:S0} S_0 & = W_0' \, , \\
\label{eq:S1} S_1 & = - \frac{2}{d} \frac{T_0}{W_0'} + W_1' \, ,
\end{align}
where we assumed $S_0 \neq 0$ as we are interested in non-zero flows. In virtue of \eqref{eq:S0} we will freely replace $S_0$ by $W_0'$ whenever convenient. Eq.~\eqref{eq:S1} will allow us to calculate $S_1$ once $T_0$, $W_0$ and $W_1$ have been determined.

\subsubsection*{Calculating $T_n$}
Note that we can write
\begin{align}
\nonumber -2A &= -2A_0 + \frac{1}{d-1} \int_{\varphi_0}^{\varphi} \frac{W}{S} \textrm{d} \varphi \\
&=  -2A_0 + \frac{1}{d-1} \int_{\varphi_0}^{\varphi} \frac{W_0}{S_0} \textrm{d} \varphi - \frac{\ell^2 R^{\textrm{uv}}}{d-1} \int_{\varphi_0}^{\varphi} \frac{S_1 W_0 -S_0 W_1}{S_0^2} \, \textrm{d} \varphi + \ldots \, \\
&=  -2A_0 + \frac{1}{d-1} \int_{\varphi_0}^{\varphi} \frac{W_0}{W_0'} \textrm{d} \varphi - \frac{\ell^2 R^{\textrm{uv}}}{d-1} \int_{\varphi_0}^{\varphi} \frac{S_1 W_0 -W_0' W_1}{(W_0')^2} \, \textrm{d} \varphi + \ldots \, ,
\end{align}
where we defined $A_0 \equiv A(\f_0)$ for some $\varphi_0$. Despite the explicit appearance $\f_0$ above, the result is independent of $\f_0$. Then, from the defining expression $T \equiv R^{\textrm{uv}} e^{-2A}$ we obtain
\begin{align}
\label{eq:T0}T_0 &= \ell^{-2} \, e^{-2A_0} e^{\frac{1}{d-1} \int_{\f_0}^{\f} \frac{W_0}{W_0'} d\varphi} \, , \\
T_1 &= - \ell^{-2} \, e^{-2A_0} e^{\frac{1}{d-1} \int_{\f_0}^{\f} \frac{W_0}{W_0'} d\varphi} \frac{1}{d-1} \int_{\varphi_0}^{\varphi} \frac{S_1 W_0 -W_0' W_1}{(W_0')^2} \, d\varphi \, .
\end{align}
We will calculate $T_0$ explicitly in section \ref{app:T0}.

\subsubsection*{Calculating $W_n$}
Inserting the expansions into \eqref{eq:EOM5} we obtain the following two equations at the orders $(R^{\textrm{uv}})^0$ and $(R^{\textrm{uv}})^1$:
\begin{align}
\label{eq:scalarpot} \frac{d}{4(d-1)} W_0^2 - \frac{1}{2} (W_0')^2 &= -V \, , \\
W_1' - \frac{d}{2(d-1)} \frac{W_0}{W_0'} W_1 + \frac{d-2}{d} \frac{T_0}{W_0'} &=0 \, .
\end{align}
Inserting our expression \eqref{eq:T0} for $T_0$ into the last equation, this can be integrated to give
\begin{align}
\label{eq:W1} W_1 = e^{\frac{d}{2(d-1)}\int_{\f_0}^{\f} \frac{W_0}{W_0'}} \left(\tilde{C}_1 + \frac{2-d}{d} \, e^{-2A_0}  \int_{\f_0}^{\f}  \frac{e^{\frac{2-d}{2(d-1)} \int_{\f_0}^{\f} \frac{W_0}{W_0'}}}{W_0'} \right) \, ,
\end{align}
where $\tilde{C}_{1}$ is an integration constant.

Overall, given a solution for $W_0$ and $W_0'$ we can now determine $W_1$ and $T_0$, which in turn will allow us to calculate $S_1$ and finally $T_1$. While these results will be sufficient for the scope of this paper, the analysis in this appendix can in principle be extended to determine the coefficients $W_n$, $S_n$ and $T_n$ to an arbitrarily high order.

\subsection{Extrema of $V$}
Here we will derive solutions for $W$, $S$ and $T$ in the vicinity of extrema of the potential, using the small curvature expansion introduced in section \ref{app:smallRdetails}.\footnote{This is consistent as long as extrema of the potential
coincide with the boundary of the bulk space-time. We check this explicitly when discussing the resulting geometries in section \ref{sec:asymp}.} To this end, it will be sufficient to consider the potential
\begin{align}
\label{eq:Vextremumapp} V = - \frac{d(d-1)}{\ell^2} - \frac{m^2}{2} \f^2 + \mathcal{O}(\f^3) \, ,
\end{align}
where we will choose $m^2 >0$ for maxima and $m^2 <0$ for minima.

\subsubsection*{Calculating $W_0$}
At order $(R^{\textrm{uv}})^0$ the analysis of our system reduces to a study of holographic RG flows for field theories on flat manifolds. This has been studied extensively in the past and we can hence be brief. For details we will refer readers to e.g.~\cite{exotic}. To be specific, at order $(R^{\textrm{uv}})^0$ we have $W=W_0$, $S=S_0=W_0'$ and $T=0$. Thus  at order $(R^{\textrm{uv}})^0$ the solution is completely determined by $W_0$.

We can determine $W_0$ by solving \eqref{eq:scalarpot}. In the vicinity of the extremum of $V$ at $\f=0$ this can be done by writing $W_0$ as a regular expansion in powers of $\f$. There exist two independent solutions, which we will label by the two subscripts $(\pm)$:
\begin{align}
W_{0, \pm}^{\textrm{reg}} &= \frac{2(d-1)}{\ell} + \frac{\Delta_{\pm}}{2 \ell} \f^2 + \mathcal{O}(\f^3) \, , \\
\nonumber \textrm{with} \quad \Delta_{\pm} &= \frac{1}{2}\left( d \pm \sqrt{d^2-  4 m^2 \ell^2} \right) \, .
\end{align}
Terms in $W_{0, \pm}^{\textrm{reg}}$ of order $\f^3$ and higher will depend on cubic and higher terms in $V$, which we ignore in the vicinity of extrema.

The solution permits a continuous deformation $\delta W_{01}$ as long as the deformation is subleading compared to $W_{0, \pm}^{\textrm{reg}}$. We will make this condition precise at the end of this section. The deformation can be determined by inserting $W_0=W_{0, \pm}^{\textrm{reg}} + \delta W_{01}$ into \eqref{eq:scalarpot} and solving for $\delta W_{01}$:
\begin{align}
\delta W_{01} &= \frac{C}{\ell} \exp \left(\frac{d}{2(d-1)} \int^{\f} \textrm{d} \f \frac{W_{0, \pm}^{\textrm{reg}}}{(W_{0, \pm}^{\textrm{reg}})'} \right) = \frac{C}{\ell} \, |\f|^{\frac{d}{\Delta_\pm}} \left[1+ \mathcal{O}(\f) \right] \, ,
\end{align}
where we introduced the dimensionless integration constant $C$. Putting everything together, we thus obtain:
\begin{align}
\label{eq:W0sol} W_0 &= \frac{2(d-1)}{\ell} + \frac{\Delta_{\pm}}{2 \ell} \f^2 + \mathcal{O}(\f^3) + \frac{C}{\ell} \, |\f|^{\frac{d}{\Delta_\pm}} \left[1+ \mathcal{O}(\f) + \mathcal{O}(C)\right] \, .
\end{align}
From this we also have that
\begin{align}
\label{eq:S0sol} S_0=W_0' &= \frac{\Delta_{\pm}}{\ell} \f + \mathcal{O}(\f^2) + \frac{Cd}{\Delta_\pm \ell} \, |\f|^{\frac{d}{\Delta_\pm}-1} \left[1+ \mathcal{O}(\f) + \mathcal{O}(C)\right] \, .
\end{align}
We can now return to the question under which circumstances the deformation $\propto C$ is permitted. To this end consider eq.~\eqref{eq:EOM9} which involves $S$ only and which any solution for $S$ must satisfy. It is easy to check that the corresponding equation at order $(\mathcal{R}^{\textrm{uv}})^0$ is only satisfied by $S_0$ if the linear term in \eqref{eq:S0sol} dominates over the term involving $C$. This implies that solutions for $S$ and $W$ permit a deformation only if
\begin{align}
\frac{d}{\Delta_{\pm}} > 2 \, .
\end{align}
If this is not the case, no deformation is permitted, which is equivalent to setting $C=0$.

\subsubsection*{Calculating $T_0$:}
\label{app:T0}
Given the expression for $W_0$ in \eqref{eq:W0sol}, we can now proceed to determining $T_0$ from \eqref{eq:T0}. In particular, the exponent in  \eqref{eq:T0} is given by
\begin{align}
\nonumber &-2 A(\f_0) + \frac{1}{(d-1)} \int_{\f_0}^\f \textrm{d} \f \, \frac{W_0}{W_0'} \\
\label{eq:T0int} =& -2 A(\f_0) + \frac{2}{\Delta_\pm} \int_{\f_0}^\f \frac{\textrm{d} \f}{\f} \, \left[1+ \mathcal{O}(\f) + \mathcal{O}(C |\f|^{d/\Delta_{\pm}-2}) \right] \, ,
\end{align}
where we have used \eqref{eq:W0sol} to get to the second line. One unattractive feature is the explicit appearance of the arbitrary parameter $\f_0$. In particular, there is physical information in $A(\f_0)$ which is obscured by this notation. Thus, in the following, we will explain how we can remove $\f_0$ from the expression for $T_0$. The idea is to trade the arbitrary parameter $\f_0$ for $\f_-$ or $\f_{+}$ which are physical parameters of the boundary field theory.

To this end, given our expression \eqref{eq:W0sol} for $W_0$, let us calculate the corresponding solutions for $A(u)$ and $\f(u)$ using $W_0=-2(d-1) \dot{A}$ and $W_0' = \dot{\f}$:
\begin{align}
\f(u) &= \left\{
  \begin{array}{l l}
  \f_+ \, \ell^{\Delta_+} \, e^{\Delta_+ u / \ell} + \mathcal{O}(e^{2 \Delta_- u / \ell}) & \qquad (+)\textrm{-branch,} \\
\f_- \, \ell^{\Delta_-} \, e^{\Delta_- u / \ell} + \mathcal{O}(e^{2 \Delta_- u / \ell}, e^{\Delta_+ u / \ell}) & \qquad (-)\textrm{-branch,} \\
  \end{array} \right. \, \\
A(u)  &= \bar{A} - \frac{u}{\ell} + \mathcal{O}(e^{2 \Delta_{\pm} u / \ell}, e^{d u / \ell}) \, ,
\end{align}
where we introduced $\f_+$, $\f_-$ and $\bar{A}$ as integration constants.\footnote{The leading terms in $\f(u)$ and $A(u)$ will turn out to be universal, i.e.~higher order corrections in $R^{\textrm{uv}}$ to $W$ and $S$ will not affect the leading terms.} We can always set $\bar{A}=0$ as argued in section \ref{sec:AdSfixedpoints} and we will do so in the following. As explained in section \ref{sec:asymp}, $\f_-$ is interpreted as the UV value of source for the operator $\mathcal{O}$ in the boundary field theory. The parameter $\f_+$ is related to the vev of $\mathcal{O}$ as follows: $\langle \mathcal{O} \rangle = (2 \Delta_+-d) \f_+$. Using these expressions we can now invert $\f(u)$ and insert the result into $A(u)$ to arrive at an expression for $A(\f_0)$:
\begin{align}
\nonumber A(\f_0) &= - \frac{1}{\Delta_{\pm}} \ln \left(\frac{\f_0}{\f_{\pm}\ell^{\Delta_\pm}} \right) + \mathcal{O}(\f_0^a) \\
&= - \frac{1}{\Delta_{\pm}} \int_{\f_{\pm} \ell^{\Delta_\pm}}^{\f_0} \frac{\textrm{d} \f}{\f} + \mathcal{O}(\f_0^a) \, .
\end{align}
Here $a$ is an exponent that will depend on the precise values of $\Delta_{\pm}$. The most important fact for this analysis is that $a>0$, as can be verified explicitly. Inserting this expression for $A(\f_0)$ into \eqref{eq:T0int} we find
\begin{align}
\nonumber & -2 A(\f_0) + \frac{2}{\Delta_\pm} \int_{\f_0}^\f \frac{\textrm{d} \f}{\f} \, \left[1+ \mathcal{O}(\f) + \mathcal{O}(C |\f|^{d/\Delta_{\pm}-2}) \right] \\
\nonumber = \ & \frac{2}{\Delta_\pm}  \int_{\f_{\pm} \ell^{\Delta_\pm}}^{\f_0} \frac{\textrm{d} \f}{\f} + \frac{2}{\Delta_\pm} \int_{\f_0}^\f \frac{\textrm{d} \f}{\f} + \int_{\f_0}^\f \textrm{d} \f \left[\mathcal{O}(\f^0) + \mathcal{O}(C \f^{d/\Delta_{\pm}-3}) \right] + \mathcal{O}(\f_0^a) \\
\nonumber  = \ & \frac{2}{\Delta_\pm}  \int_{\f_{\pm} \ell^{\Delta_\pm}}^{\f} \frac{\textrm{d} \f}{\f} + \int_{\f_0}^\f \textrm{d} \f \left[\mathcal{O}(\f^0) + \mathcal{O}(C \f^{d/\Delta_{\pm}-3}) \right] + \mathcal{O}(\f_0^a) \\
\label{eq:T0int2}  = \ & \frac{2}{\Delta_\pm}  \ln \left(\frac{\f}{\f_{\pm}\ell^{\Delta_\pm}} \right) + \int_{\f_0}^\f \textrm{d} \f \left[\mathcal{O}(\f^0) + \mathcal{O}(C \f^{d/\Delta_{\pm}-3}) \right] + \mathcal{O}(\f_0^a) \, .
\end{align}
The main observation is that we can now remove $\f_0$ by setting $\f_0 \rightarrow 0$.\footnote{However, we could in principle also choose some finite value for $\f_0$. It is easy to check that this would have the same effect as a shift in the integration constant $\bar{A}$.} There is no danger of picking up a divergence from the integral. (Recall that the term involving $C$ is only present if $d/ \Delta_{\pm} >2$.) The lower limit of the integral simply becomes $0$ while the term $\mathcal{O}(\f_0^a)$ simply vanishes as $a>0$. We are thus finally in a position to state the result for $T_0$. Using \eqref{eq:T0int2} in \eqref{eq:T0} we obtain:
\begin{align}
\label{eq:T0sol} T_0 = \ell^{-4} \, {\left(\frac{\f}{\f_{\pm}} \right)}^{\frac{2}{\Delta_{\pm}}} \left[ 1+ \mathcal{O}(\f)+ \mathcal{O}(C |\f|^{d/\Delta_{\pm}-2} ) \right] \ .
\end{align}
Thus, in the vicinity of an extremum we find that $T$ has an expansion of the form
\begin{align}
T_{\pm}(\f) =  \ell^{-2} \left(R^{\textrm{uv}} |\f_\pm|^{-2 / \Delta_\pm} \right) \, |\f|^{2 / \Delta_{\pm}} \left[ 1+ \mathcal{O}(\f) + \mathcal{O}(C |\f|^{d/\Delta_{\pm}-2} ) + \mathcal{O}(\ell^2 R^{\textrm{uv}}) \right] \ .
\end{align}
Here we wish to highlight the appearance of the dimensionless combination of quantities $R^{\textrm{uv}} |\f_\pm|^{-2 / \Delta_\pm}$. This is not a coincidence. In fact, we will find that in $W$, $S$ and $T$ the quantity $R^{\textrm{uv}}$ will exclusively appear in the combination $R^{\textrm{uv}} |\f_\pm|^{-2 / \Delta_\pm}$. Thus, it will be useful in assigning a label to this particular combination and we hence define
\begin{align}
\label{eq:Rdefapp} \mathcal{R} = \left\{ \begin{array}{c l}
R^{\textrm{uv}} |\f_+|^{-2/ \Delta_+} & \qquad \textrm{on the }(+)\textrm{-branch}  \\
R^{\textrm{uv}} |\f_-|^{-2/ \Delta_-} & \qquad \textrm{on the }(-)\textrm{-branch} \\
\end{array} \right. \, .
\end{align}
Strictly speaking, we should understand the small curvature expansions in \eqref{eq:smallRWexp}--\eqref{eq:smallRTexp} as expansions in $\mathcal{R}$.
\subsubsection*{Calculating $W_1$ and $S_1$:}
Given our expressions \eqref{eq:W0sol} and \eqref{eq:T0sol} for $W_0$ and $T_0$ we are now in a position to calculate $W_1$ and $S_1$ from \eqref{eq:W1} and \eqref{eq:S1}, respectively. Starting with $W_1$, after some work one finds
\begin{align}
W_1 = \ & \frac{1}{d \ell^3} \, {\left(\frac{\f}{\f_{\pm}} \right)}^{\frac{2}{\Delta_{\pm}}} \left[ 1+ \mathcal{O}(\f) + \mathcal{O}(C |\f|^{d/\Delta_{\pm}-2} ) \right] \\
\nonumber &+ \frac{C_1}{\ell} |\f|^{d / \Delta_{\pm}} \left[ 1+ \mathcal{O}(\f) + \mathcal{O}(C |\f|^{d/\Delta_{\pm}-2} ) \right]
\end{align}
Most importantly, any dependence on the arbitrary parameter $\f_0$ appearing in \eqref{eq:W1} can be absorbed into the integration constant $C_1$. Another observation is that the term $C_1 |\f|^{d / \Delta_{\pm}}$ in $W_1$ combines with the term $C |\f|^{d / \Delta_{\pm}}$ in $W_0$ and thus $C_1$ does not represent an independent integration constant.

We are now in a position to calculate $S_1$. We obtain:
\begin{align}
S_1 = \frac{C_1 d}{\Delta_\pm \ell} |\f|^{\frac{d}{\Delta_\pm}-1} \left[ 1+ \mathcal{O}(\f) + \mathcal{O} \left(C |\f|^{\frac{d}{\Delta_{\pm}}-2} \right)\right] + \frac{1}{\ell} \mathcal{O}\left(|\f|^{\frac{2}{\Delta_\pm}+1} \right) + \frac{1}{\ell} \mathcal{O}\left(C |\f|^{\frac{2+d}{\Delta_\pm}-1} \right) .
\end{align}
Here, the term involving $C_1$ will combine with the corresponding term in $S_0$. We also determined the leading order in $\f$ of the terms not containing $C_1$. One can check that the exact numerical coefficients will depend on the cubic and quartic terms in the potential, and which we hence leave implicit.

\subsection{Summary}
\label{app:smallRsummary}
Putting everything together, we are now in a position to write down expressions for $W$, $S$ and $T$ in the vicinity of an extremum of $V$ and up to order $R^{\textrm{uv}}$ (or $\mathcal{R}$):
\begin{align}
\label{eq:Wgensol} W_{\pm}(\f) & = \frac{1}{\ell} \left[2(d-1) + \frac{\Delta_\pm}{2} \f^2 + \mathcal{O}(\f^3) \right] \\
\nonumber & \hphantom{=} \,  + \frac{\mathcal{R}}{d \ell} \, |\f|^{\frac{2}{\Delta_\pm}} \ [1+ \mathcal{O}(\f) + \mathcal{O}(C) + \mathcal{O}(\mathcal{R})] \\
\nonumber & \hphantom{=} \,  + \frac{C}{\ell} \, |\f|^{\frac{d}{\Delta_\pm}} \ [1+ \mathcal{O}(\f)+ \mathcal{O}(C) + \mathcal{O}(\mathcal{R})] \, , \\
\label{eq:Sgensol} S_{\pm}(\f) & = \frac{\Delta_-}{\ell} \f \ [1+ \mathcal{O}(\f)] + \frac{Cd}{\Delta_- \ell} \, |\f|^{\frac{d}{\Delta_-}-1} \ [1+ \mathcal{O}(\f) + \mathcal{O}(C)] \, , \\
\nonumber & \hphantom{=} \,  + \frac{1}{\ell} \mathcal{O}\left( \mathcal{R} |\f|^{\frac{2}{\Delta_-}+1} \right) + \frac{1}{\ell} \mathcal{O}\left(\mathcal{R} C |\f|^{\frac{2+d}{\Delta_-}-1} \right) \\
\label{eq:Tgensol} T_{\pm}(\f) &= \ell^{-2} \mathcal{R} \, |\f|^{\frac{2}{\Delta_-}} [1+ \mathcal{O}(\f) + \mathcal{O}(C) + \mathcal{O}(\mathcal{R})] \, ,
\end{align}
where $\mathcal{R}$ has been defined in \eqref{eq:Rdefapp}. While we write $\mathcal{O}(C)$ and $\mathcal{O}(\mathcal{R})$ to remove clutter when possible, it is to be understood that $C$ and $\mathcal{R}$ are always accompanied by an appropriate power in $\f$. Also, we absorbed the integration constant $C_1$ into $C$. Further, if $d / \Delta_{\pm} < 2$ we have to set $C=0$, otherwise $S$ is not a solution of \eqref{eq:EOM9}. While this is the most general result for an extremum of $V$, we now look at maxima and minima of $V$ in turn.

\subsubsection*{Maxima of $V$}
At a maximum of $V$ we have:
\begin{align}
\frac{d}{2} < \Delta_+ &< d \, ,  & 0 < \ & \Delta_- < \frac{d}{2} \, , \\
1 < \frac{d}{\Delta_+} &< 2 \, ,  & & \frac{d}{\Delta_-} > 2 \, .
\end{align}
As $d / \Delta_+ < 2$ we have to set $C=0$ in the $(+)$-branch of solutions. This is not required for the $(-)$-branch as we have $d / \Delta_- > 2$. The resulting solutions are shown in discussed in the main body of the paper in section \ref{sec:maximaofV}.

\subsubsection*{Minima of $V$}
At a minimum of $V$ we have:
\begin{align}
 \Delta_+ &< d \, ,  &  \Delta_- &< 0 \, , \\
\frac{d}{\Delta_+} &< 1 \, , &  \frac{d}{\Delta_-}& < 0 \, .
\end{align}
As $d / \Delta_+ < 2$ we have to set $C=0$ in the $(+)$-branch of solutions. However, on the $(-)$-branch we have $\Delta_- <0$ and as a result, any term involving $C$ or $\mathcal{R}$ in \eqref{eq:Wgensol}--\eqref{eq:Tgensol} diverges! To arrive at an acceptable solution, we have to set both $C$ and $R^{\textrm{uv}}$ (and thus $\mathcal{R}$) to zero. We then recover the solution for RG flows for field theories on flat manifolds. A more comprehensive discussion of the solutions on both the $(+)$ and $(-)$-branches can be found in the main text in section \ref{sec:minimaofV}.

\section{Solution in the vicinity of minima of $V$}
\label{app:Aphiminima}
Here we will examine solutions for $A(u)$ and $\f(u)$ in the vicinity of a minimum. In particular, we will be interested in flows starting or ending at the minimum. The potential is given by \eqref{eq:Vextremum} with $m^2 < 0$. As we work in the vicinity of the minimum, we will express solutions as a perturbation about the solution for an AdS fixed point:
\begin{align}
A(u) &= A_0(u) + \epsilon^2 A_2(u) + \mathcal{O}(\epsilon^4) \, , \\
\f(u) &= 0 + \epsilon \f_1(u) + \mathcal{O}(\epsilon^3) \, ,
\end{align}
with
\begin{align}
A_0(u) = \left\{
  \begin{array}{c l}
   \ln \left(- \frac{\ell}{\alpha} \sinh \frac{u+c}{\ell} \right) & \quad S^d \textrm{ or dS}_d \\
   - \frac{u+c}{\ell} & \quad \mathcal{M}_d\\
 \ln \left( \hphantom{-} \frac{\ell}{\alpha} \cosh \frac{u+c}{\ell} \right) & \quad \textrm{AdS}_d \\
  \end{array} \right. \, \ .
\end{align}
With this ansatz we proceed to solving the equations of motion \eqref{eq:EOM1}--\eqref{eq:EOM3} order by order in $\epsilon$. To be specific, at order $\mathcal{O}(\epsilon)$ we obtain:
\begin{align}
\label{eq:minexp1}
\ddot{\f}_1 + d \dot{A}_0 \dot{\f}_1 + m^2 \f_1 =0 \, .
\end{align}

\paragraph*{Minima as UV fixed points}
First, we wish to show that a minimum can indeed give rise to a UV fixed point. At a UV fixed point the scale factor diverges as $A(u) \rightarrow \infty$ which is the case for $u \rightarrow - \infty$. A flow leaving a UV fixed point will thus correspond to a solution to \eqref{eq:minexp1} subject to the boundary conditions $\f_1 (u \rightarrow - \infty) \rightarrow 0$ and $\dot{\f}_1 (u \rightarrow - \infty) \rightarrow 0$.

We can discuss the case of dS, AdS and Minkowski slices in a unified way, due to $A_0(u \rightarrow - \infty) \rightarrow - \tfrac{u}{\ell}$ for all three cases. Thus, for $u \rightarrow - \infty$ \eqref{eq:minexp1} becomes
\begin{align}
\label{eq:minUV}
\ddot{\f}_1 - \frac{d}{\ell} \dot{\f}_1 +m^2 \f_1 =0 \, ,
\end{align}
which is solved by
\begin{align}
\label{eq:minUVsol}
\f_1(u) = c_1 \, e^{\Delta_- u / \ell} + c_2 \, e^{\Delta_+ u / \ell} + \ldots \, ,
\end{align}
where $c_{1,2}$ are integration constants and $\Delta_{\pm}$ have been defined before in \eqref{eq:Deltadef}. This is just the usual asymptotic form of the scalar field in the vicinity of a UV fixed point, with $c_1$ related to the source of the perturbing operator and $c_2$ related to its vev. However, note that for $m^2 <0$ we have $\Delta_- < 0$ while $\Delta_+ > d$. The boundary conditions hence require that we set $c_1=0$. As a result, flows away from a UV fixed point at a minimum of the potential are purely driven by a vev. As the leading behavior of $\f$ is determined by $\Delta_+$, solutions with minima as UV fixed points are associated with the $(+)$-branch of solutions for $W$, $S$ and $T$.

\paragraph*{Minima as IR fixed points}
For the case of Minkowski slicings it is well-known that minima of the potential can be identified as IR fixed points, a fact we will confirm presently. We define an IR endpoint as the locus where the scale factor diverges as $A(u) \rightarrow - \infty$. For the case of flat slicings this occurs for $u \rightarrow \infty$. The relevant equation is again \eqref{eq:minUV}, but the boundary conditions now read $\f_1 (u \rightarrow \infty) \rightarrow 0$ and $\dot{\f}_1 (u \rightarrow \infty) \rightarrow 0$. The general solution is, as before, given by \eqref{eq:minUVsol}, but the boundary conditions now require that $c_2=0$ instead. Thus, for $R^{\textrm{uv}} = 0$ minima of the potential can play the role of IR endpoints of RG flows. When rewritten in terms of $W$, $S$ and $T$, this is a solution on the $(-)$-branch.

In the following we will show that in space-times with a curved foliation RG flows cannot end at a minimum of the potential. For the case of dS and AdS slicings the IR is identified with $u \rightarrow -c_{\textrm{IR}}$ where the bulk geometry asymptotes to AdS${}_{d+1}$:
\begin{align}
A_0(u) \underset{u \rightarrow -c_{\textrm{IR}}}{\rightarrow} \left\{
  \begin{array}{c l}
   \ln \left(- \frac{\ell}{\alpha} \sinh \frac{u+c_{\textrm{IR}}}{\ell_{\textrm{IR}}} \right) & \quad S^d \textrm{ or dS}_d \\
 \ln \left( \hphantom{-} \frac{\ell}{\alpha} \cosh \frac{u+c_{\textrm{IR}}}{\ell_{\textrm{IR}}} \right) & \quad \textrm{AdS}_d\\
  \end{array} \right. \, \ .
\end{align}
For the following analysis it will be useful to introduce the coordinate $w \equiv \tfrac{u+c_{\textrm{IR}}}{\ell_{\textrm{IR}}}$. In the remainder of this subsection we will also define $\dot{}\equiv \tfrac{d}{dw}$. RG flows ending at an IR endpoint at the minimum of the potential will then correspond to solutions to \eqref{eq:minexp1} subject to the boundary conditions $\dot{\f}_1(w=0)=0$ and $\f(w=0)=0$.

In the vicinity of the IR locus equation \eqref{eq:minexp1} becomes
\begin{align}
S^d \textrm{ or dS}_d\textrm{:} \quad & \quad \ddot{\f}_1 + d \left(\frac{1}{w} +\frac{w}{3} + \mathcal{O}(w^3) \right) \, \dot{\f}_1 +m^2 \ell_{\textrm{IR}}^2 \f_1 =0 \, , \\
\textrm{AdS}_d\textrm{:} \quad & \quad \ddot{\f}_1 + d \left(w - \frac{w^3}{3} + \mathcal{O}(w^5) \right) \, \dot{\f}_1 +m^2 \ell_{\textrm{IR}}^2 \f_1 =0 \, .
\end{align}
Solving and implementing the boundary condition $\dot{\f}_1(w=0)=0$ we find
\begin{align}
S^d \textrm{ or dS}_d\textrm{:} \quad & \quad \f_1(w) = c_1 \left( 1- \frac{m^2 \ell_{\textrm{IR}}^2}{2(d+1)} w^2 + \mathcal{O}(w^4) \right) \, , \\
\textrm{AdS}_d\textrm{:} \quad & \quad \f_1(w) = c_1 \left( 1- \frac{m^2 \ell_{\textrm{IR}}^2}{2} w^2 + \mathcal{O}(w^4) \right) \, ,
\end{align}
with $c_1$ the remaining constant of integration. Note that the only way of satisfying the second boundary condition $\f_1(w=0)=0$ is to set $c_1=0$, which causes $\f(w)$ to vanish identically. {\em Hence there are no solutions that smoothly arrive at a minimum of the potential for $S^d$/dS$_d$ and AdS$_d$ slicings.} The only solutions that exist for $R^{\textrm{uv}} \neq 0$ are solutions with $\f=0$, i.e.~conformal fixed points. Most importantly, there are no RG flows for $R^{\textrm{uv}} \neq 0$ that end at a fixed point at a minimum of the potential. This is equivalent to the absence of the $(-)$-branch of solutions for $W$, $S$ and $T$ in the vicinity of a minimum of $V$ for $R^{\textrm{uv}} \neq 0$.

\section{Extremal points of the first order flow equations} \label{App:interior}

We will start by examining the critical power behavior for the first order functions that is allowed by the equations of motion \eqref{eq:EOM4}--\eqref{eq:EOM6}. Then we will proceed to examine the solutions in the regime where the flow stops. To find where the flow of $\f$ stops, we need to find (finite) points $\f_0$ where $S$ vanishes.

We  parametrize $S$ as
\be
S\simeq C_0~ (\f_0-\f)^a\sp \f\to\f_0\sp a>0
\label{a21}\ee
We assume without loss of generality  that $\f$ is approaching to $\f_0$ from the left, i.e.~$\f<\f_0$.  As the flow does not end up at the minimum at the presence of curvature,  $\f_0$ is a generic point and therefore
\be
V(\f)=V_0+V_1(\f_0-\f)+V_2(\f_0-\f)^2+{\cal O}((\f_0-\f)^3)\ .
\label{a22}\ee
Near $\f_0$ the various terms of \eqref{eq:EOM9} behave as follows
\begin{align}
& S^4\sim (\f_0-\f)^{4a},  \quad S^3S'',S^2S'^2\sim (\f_0-\f)^{4a-2},\quad    S^2V\sim (\f_0-\f)^{2a}, \nonumber \\
& SS'V'\sim (\f_0-\f)^{2a-1}, \quad S^2V''\sim (\f_0-\f)^{2a},  \quad V'^2\sim (\f_0-\f)^{0}.
\label{a24}
\end{align}

It is apparent  that near $\f_0$, if $a=1/2$, then to leading order Eq.~\eqref{eq:EOM9} is satisfied. After getting the leading order power law behavior, it is clear that the superpotentials have a square root series expansion near the IR end-point. For brevity, let us write $x=\f_0-\f$. Let us also write the following expansions near $x=0$ :
\begin{align}
S(x)&=\sqrt{x} \left(S_0+S_1 \sqrt{x}+S_2 x +\cdots\right) \ ,\\
W(x)&=\frac{1}{\sqrt{x}}\left(W_0+W_1 \sqrt{x}+W_2 x+\cdots \right) \ ,\\
T(x)&=\frac{1}{x}\left(T_0+T_1\sqrt{x}+T_2 x+\cdots  \right)\ , \\
V(x)&=V_0+V_1x+V_2 x^2 +\cdots\ .
\end{align}
Now we can use Eqs.~\eqref{eq:EOM7} and \eqref{eq:EOM8} to find the various
coefficients. There are two cases which are generic.\\

\noindent \textbf{Case (a): }
\begin{align}
S_0^2&=2 V_1\ , S_1=\frac{-d W_1}{3 (d-1)} \ ,  S_2= \frac{36 (d-1) \left((d-1) V_2+V_0 \right)+d (d+9) W_1^2}{36
   (d-1)^2 S_0} \label{bounceS}\\
W_0&=0\ , \quad W_1= \textrm{arbitrary} \ , \quad W_2= -\frac{4 (d-1) V_0+d W_1^2}{d(d-1)
   S_0} \label{bounceW}\\
   T_0&=0\ , \quad T_1=0\ , T_2=\frac{d W_1^2}{4 (d-1)}+V_0 \ , T_3=-\frac{W_1 \left(4 (d-1) V_0 +d W_1^2\right)}{2  (d-1)^2 S_0} \label{bounceT}
\end{align}
In this case $\ddot{\f}=\frac{1}{2}\frac{d}{dx}S^2=S_0^2=2 V_1 \neq 0$. So the flow does not stop here. We will find that this corresponds to a bouncing solution. Note that $S_0 = \pm \sqrt{2V_1}$, i.e.~there are two branches. These will be the two branches those meet at the bouncing point. As $V_0<0$ we find that $T$ can be both positive or negative, depending on the value of $W_1$. Hence bounces are expected to exist for both positively and negatively curved slices.

\vspace{0.3cm}
\noindent \textbf{Case (b): }
\begin{align}
S_0^2&=\frac{2 V_1}{d+1}\ , S_1=0, S_2= \frac{V_0 + 3 (d-1) V_2}{ (d-1) (d+3)S_0} \\
\ \nonumber \\
W_0&= (d-1) S_0 \ , W_1=0\ , W_2= -\frac{(d+4) V_0 -(d-1) d V_2}{d (d+3)S_0 }\\
\ \nonumber \\
T_0&=\frac{d(d-1)}{4}S_0^2 \ , T_1=0 \ , T_2=\frac{d S_0 W_2}{2}+V_0.
\end{align}
In this case, $\ddot{\f}=\frac{1}{2}\frac{d}{dx}S^2=S_0^2=\frac{2 V_1}{d+1} \neq 0$. Although the flow does not stop here, the space-time ends. Hence this is an IR solution. Again, it seems that there are two possible solutions with $S_0 = \pm \sqrt{\tfrac{2 V_1}{d+1}}$. However, note that $W_0 = (d-1) S_0$. As we have chosen $W>0$, only the solution with $S_0 = + \sqrt{\tfrac{2 V_1}{d+1}}$ is acceptable. Furthermore, here we have tacitly assumed that $V_1 >0$. As $T_0>0$, this solution only occurs for positive curvature (e.g.~a dS slicing).\footnote{If $V_1<0$ we can replace $V_1$ by $|V_1|$ in all expressions. The solution persists and corresponds to a space-time with positively curved $d$-dimensional slices.}\\

\noindent \textbf{Case (c): }
It is worthwhile to highlight another solution, which corresponds to a special case of situation $(a)$ described above. In particular, this is the case with $W=0$ at $x=0$, i.e.~we have $W_0=0=W_1$. The solution is given by
\begin{align}
S_0^2 &= 2 V_1 \, ,  \\
W_0 &=0 \, , & & W_1 =0 \, , & &W_2=\frac{4V_0}{dS_0} \, , \\
T_0 &=0 \, & &T_1=0 \, , & &T_2 = V_0 \, .
\end{align}
As $V_0<0$ we find $T<0$ in this case. Hence this corresponds to the IR solution with negatively curved slices.

\section{The large curvature solution}
\label{sec:largeR}
In sections \ref{sec:asymp} and \ref{IRasymp} we examined solutions of the Einstein-dilaton system in the vicinity of individual points, studying extrema of the potential and generic points in turn. In this section, we will present (approximate) solutions for complete flows from an UV fixed point to an IR endpoint. The flows we will derive will originate from a UV fixed point at a maximum of the potential, which we locate at $\varphi=0$ and end at a generic point, which we will denote by $\varphi_{0}$. Most importantly, in this section we restrict our attention to flows, which terminate close to the maximum at $\f=0$, i.e.~in a regime where the potential is well described by a quadratic approximation:
\begin{align}
V \simeq  -V_0 - \frac{1}{2} m^2 \varphi^2 \, .
\end{align}
Furthermore, we will require that
\begin{align}
\varphi_{0}^2 \ll \frac{|V_0|}{m^2} \, ,
\end{align}
i.e.~the potential is approximately constant along the whole flow from $\f=0$ to $\f=\f_{0}$.
This will have important consequences. For one, the existence of a
small parameter $\f_{0}$ will simplify the task of solving this
system, as we will be able to organize the solution perturbatively in
$\f_{0}$. In addition, we will later find that for small
$\f_{0}$, the value of the endpoint $\f_{0}$ is in one-to-one
correspondence with the dimensionless curvature parameter
$\mathcal{R}$. To be specific, we will find that $\mathcal{R} \sim
\f_{0}^{-2/ \Delta_{-}}$. Hence a flow terminating at a small
value of $\f_{0}$ will describe an RG flow for \emph{large} values
of the curvature $\mathcal{R}$. In turn, our perturbative solution
will thus correspond to a large curvature expansion, with
$\mathcal{R}^{- \Delta_-/2}$ the expansion parameter.

The approximation and the method in this section here are the same
  which were presented in \cite{Taylor} in the case $d=3$, where curved RG flows
  on spheres with endpoints  near the UV boundary were studied. In
  this section we obtain similar results for $d=4$.

\subsection{Perturbative solution}
For the task at hand, it will be most convenient to solve the second order differential equations for $A(u)$ and $\varphi(u)$ directly, rather than working with $W(\f)$, $S(\f)$ and $T(\f)$. Hence, the relevant equations of motion in this section are \eqref{eq:EOM1}--\eqref{eq:EOM3}.

Here we will outline the solution strategy and present the results. To solve the Einstein-dilaton system, we exploit the fact that in the regime of interest, flows remain in the vicinity of the maximum of the potential. This implies that we can organize the solution as an expansion about the solution at a maximum of the potential, which we derived before in sec.~\ref{sec:AdSfixedpoints}.
The solution at a maximum is characterized by $\f=0$, while $A(u)$ takes the form of the AdS scale factor \eqref{eq:AdSScaleFactor}, which we reproduce below for convenience:
\begin{align}
\label{eq:AlargeRdS} S^d / \textrm{dS}_d \textrm{:} \qquad A(u) &= \log \left(- \frac{\ell}{\alpha} \sinh \frac{u+c}{\ell} \right) \, , \\
\label{eq:AlargeRAdS} \textrm{AdS}_d\textrm{:} \qquad A(u) &= \log \left(\frac{\ell}{\alpha} \cosh \frac{u+c}{\ell} \right) \, ,
\end{align}
where we distinguish between space-times with $S^d$/dS$_d$ slicings and AdS$_d$ slicings.
Inserting this ansatz for $A(u)$ into the Klein-Gordon equation \eqref{eq:EOM3} then results in a differential equation for $\f(u)$, which will allow us to determine the leading order contribution to flows $\f(u)$. The resulting equations take the form:
\begin{align}
\label{eq:phi1dS} S^d / \textrm{dS}_d \textrm{:} \qquad & \ddot{\f} + \frac{d}{\ell} \coth \left( \frac{u+c}{\ell} \right) \dot{\f} +m^2 \f =0 \, , \\
\label{eq:phi1AdS} \textrm{AdS}_d\textrm{:} \qquad &\ddot{\f} + \frac{d}{\ell} \tanh \left( \frac{u+c}{\ell} \right) \dot{\f} +m^2 \f =0 \, .
\end{align}
The leading contributions to RG flows are given by regular solutions
to these equations subject to the boundary conditions:
\be\label{eq:bcsmallphi}
\f(-c)=\f_{0}\, , \qquad \dot{\f}(-c)=0.
\ee
These boundary conditions ensure that flows end at $u_{\textrm{IR}}=-c$, the value at which the scale factor $e^{A(u)}$ vanishes.

We can discuss both equations \eqref{eq:phi1dS} and \eqref{eq:phi1AdS} in a unified way if we define a new coordinate $U$ as follows:
\begin{align}
U \equiv \left\{
  \begin{array}{c l}
   - \coth  \tfrac{u+c}{\ell} \, , & \qquad S^d / \textrm{dS}_d \\
 -\tanh  \tfrac{u+c}{\ell} \, ,& \qquad \textrm{AdS}_d \\
  \end{array} \right. \, \ .
\end{align}
Equations \eqref{eq:phi1dS} and \eqref{eq:phi1AdS} then become
\begin{align}
(1-U^2)^2 \f'' + (d-2) U (1-U^2) \f' + \Delta_- (d-\Delta_-) \f =0 \, ,
\end{align}
where we defined ${}^{\prime} \equiv \tfrac{d}{dU}$. The general solution is given by:
\begin{align}
\label{eq:phialld} \f(U) = & \left[ \ \ \tilde{C}_1 \, (U-1)^{\frac{\Delta_-}{2}}(U+1)^{\frac{d-\Delta_-}{2}} \, {}_2F_1 \left(\frac{2-d}{2}, \frac{d}{2}, \frac{2-d+2 \Delta_-}{2}, \frac{1-U}{2} \right) \right. \\
\nonumber & \left. \ - \tilde{C}_2 \, (U-1)^{\frac{\Delta_+}{2}} (U+1)^{\frac{d-\Delta_+}{2}} \, {}_2F_1 \left(\frac{2-d}{2}, \frac{d}{2}, \frac{2-d+2 \Delta_+}{2}, \frac{1-U}{2} \right) \right] ,
\end{align}
where $\tilde{C}_{1,2}$ are integration constants.

So far this discussion is valid for general $d$. In the following, we will find it more instructive to restrict our attention to the case with $d=4$, as this will be most relevant for applications. An additional benefit is that this will make the expressions more manageable. For $d=4$, the general solution \eqref{eq:phialld} reduces to
\begin{align}
S^4/\textrm{dS}_4 \textrm{:} \ \f(u) &= \frac{1}{\sinh^2 \frac{u+c}{\ell}} \left[e^{\delta u / \ell} C_1 \left(\delta - \coth \frac{u+c}{\ell} \right) + e^{-\delta u / \ell} C_2 \left(\delta + \coth \frac{u+c}{\ell} \right) \right] , \\
\textrm{AdS}_4\textrm{:} \ \f(u) &= \frac{1}{\cosh^2 \frac{u+c}{\ell}} \left[e^{\delta u / \ell} C_1 \left(\delta - \tanh \frac{u+c}{\ell} \right) + e^{-\delta u / \ell} C_2 \left(\delta + \tanh \frac{u+c}{\ell} \right) \right] \, ,
\end{align}
where $C_{1,2}$ are integration constants and we have defined $\delta
\equiv \sqrt{4- m^2 \ell^2}$. Implementing the boundary conditions (\ref{eq:bcsmallphi}) we obtain:
\begin{align}
S^4/\textrm{dS}_4 \textrm{:} \ \f(u) &= \frac{3 \f_{0}}{2 \delta (\delta^2-1)} \frac{1}{\sinh^2 \frac{u+c}{\ell}} \left[e^{\delta \frac{u+c}{\ell}} \left(\delta - \coth \frac{u+c}{\ell} \right) + e^{-\delta \frac{u+c}{\ell}} \left(\delta + \coth \frac{u+c}{\ell} \right) \right], \\
\textrm{AdS}_4\textrm{:} \ \f(u) &= \frac{\f_{0}}{2 \delta} \frac{1}{\cosh^2 \frac{u+c}{\ell}} \left[e^{\delta \frac{u+c}{\ell}} \left(\delta - \tanh \frac{u+c}{\ell} \right) + e^{-\delta \frac{u+c}{\ell}} \left(\delta + \tanh \frac{u+c}{\ell} \right) \right] \, .
\end{align}
Having an analytic expression for a complete flow will allow us to relate IR quantities such as $\f_{0}$ to UV data such as $\mathcal{R}$ and $C$, which we will do in the next subsection.

Before embarking on this, we need to check the validity of our results by examining how the existence of a non-vanishing flow backreacts on the metric. To this end we substitute our solution for $\f(u)$ back into \eqref{eq:EOM1} and \eqref{eq:EOM2} and determine how $A(u)$ is corrected. The important observation is that since $\f(u) \sim \mathcal{O}(\f_{0})$, the correction to $A(u)$ arises at $\mathcal{O}(\f_{0}^2)$. Thus, we find that in the regime of interest, i.e.~the regime of small $\f_{0}$, our perturbative ansatz is self-consistent.

We could continue and solve for $A(u)$ and $\f(u)$ to higher orders in $\f_{0}$. The pattern that emerges is that $A(u)$ only contains even powers of $\f_{0}$ while only odd powers of $\f_{0}$ appear in $\f$. To summarize, we collect our results for $d=4$ in the following:
\begin{align}
S^4 / \textrm{dS}_4 \textrm{:} \ \, A(u) &= \log \left(- \frac{\ell}{\alpha} \sinh \frac{u+c}{\ell} \right) + \mathcal{O}(\f_{0}^2) \, , \\
\label{eq:phismallphi0dS}  \f(u) &= \frac{3 \f_{0}}{2 \delta (\delta^2-1)} \frac{1}{\sinh^2 \frac{u+c}{\ell}} \left[e^{\delta \frac{u+c}{\ell}} \left(\delta - \coth \frac{u+c}{\ell} \right) \right. \\
\nonumber & \hphantom{AAAAAAAAAAAAA} \left. + e^{-\delta \frac{u+c}{\ell}} \left(\delta + \coth \frac{u+c}{\ell} \right) \right] + \mathcal{O}(\f_{0}^3) \, , \\
\textrm{AdS}_4\textrm{:} \ \, A(u) &= \log \left(\frac{\ell}{\alpha} \cosh \frac{u+c}{\ell} \right) + \mathcal{O}(\f_{0}^2) \, , \\
\label{eq:phismallphi0AdS} \f(u) &= \frac{\f_{0}}{2 \delta} \frac{1}{\cosh^2 \frac{u+c}{\ell}} \left[e^{\delta \frac{u+c}{\ell}} \left(\delta - \tanh \frac{u+c}{\ell} \right) \right. \\
\nonumber & \left. \hphantom{AAAAAAAAAA} + e^{-\delta \frac{u+c}{\ell}} \left(\delta + \tanh \frac{u+c}{\ell} \right) \right] + \mathcal{O}(\f_{0}^3) \, , \\
\nonumber \textrm{with} \qquad \delta \equiv &\sqrt{4- m^2 \ell^2} \qquad \textrm{and} \qquad \f_{0} \ll \sqrt{\frac{|V_0|}{m^2}} \, .
\end{align}
Before moving on, there is one interesting feature which we wish to point out. One can check that for $S^4$/ dS$_4$ slicings $\f(u)$ grows strictly monotonically along a flow from UV to IR. For the AdS$_4$ case, this is only true as long as $\Delta_- < 1$. For $\Delta_- >1$ we find that $\f(u)$ changes direction along the flow: The flow leaves the UV point at $\f=0$ in one direction, then turns around, before terminating  on the other side of $\f=0$. This is what was referred to as a `bounce'. Interestingly, we find that in the regime of small $\f_{0}$ bounces are generic for $\Delta_- >1$.

\subsection{Relating IR quantities and UV data}
The analytical expressions \eqref{eq:phismallphi0dS} and \eqref{eq:phismallphi0AdS} will allow us to relate the IR quantity $\f_{0}$ to UV data such as $\mathcal{R}$ and $C$. Thus we will be able to show that our perturbative solution in $\f_{0}$ is in fact a large-curvature expansion in $\mathcal{R}^{-\Delta_-/2}$. In addition, we will be able to derive an analytical relation between $C$ and $\mathcal{R}$ valid at large curvatures.

The main observation is that, when approaching the boundary $u \rightarrow -\infty$,  a solution $\f(u)$ corresponding to an RG flow takes the form given by \eqref{eq:phimsol}. This depends on two integration constants $\f_-$ and $C$. In particular, the UV limit of $\f(u)$ in $d=4$ is given by (see \eqref{eq:phimsol})
\begin{align}
\label{eq:phiexpUV} \f(u) &\underset{u \rightarrow - \infty}{\rightarrow} \f_{-} \, \ell^{\Delta_-} \, e^{\Delta_{-}u/\ell} + \frac{4 C \, \f_{-}^{\frac{\Delta_{+}}{\Delta_{-}}}}{\Delta_{-}(4-2 \Delta_{-})} \, \ell^{\Delta_+} \, e^{\Delta_{+}u/\ell} + \ldots \, .
\end{align}
Given our analytical solutions \eqref{eq:phismallphi0dS} and \eqref{eq:phismallphi0AdS}, we can arrive at an alternative expression in the limit $u \rightarrow -\infty$ that will involve $\f_{0}$. Comparing the coefficients of the terms $e^{\Delta_{\pm}u/\ell}$ will then allow us to relate UV to IR data. We will discuss the results for $S^4$/ dS$_4$ slicings and AdS$_4$ slicings separately.

\subsection*{dS$_4$/$S^4$ slicing}
In the following, we focus on the example with $S^4$/ dS$_4$ slices. Expanding \eqref{eq:phismallphi0dS} for $u \rightarrow - \infty$ we find:
\begin{align}
\nonumber \f(u) &\underset{u \rightarrow - \infty}{\rightarrow} \hphantom{+} \frac{6 \, \f_{0} \, e^{\Delta_{-}c/\ell}}{(2-\Delta_{-})(3-\Delta_{-})} e^{\Delta_{-}u/\ell} \ \left(1 + \mathcal{O}(e^{2u/\ell}) \right) \\
& \hphantom{\underset{u \rightarrow - \infty}{\rightarrow}} + \frac{6 \, \f_{0} \, e^{\Delta_{+}c/\ell}}{(2-\Delta_{-})(1-\Delta_{-})} e^{\Delta_{+}u/\ell} \ \left(1 + \mathcal{O}(e^{2u/\ell}) \right) +\mathcal{O}(\f_{0}^3) \, .
\end{align}
Comparing with \eqref{eq:phiexpUV} and matching the coefficients of $e^{\Delta_{-}u/\ell}$ we find
\begin{align}
\label{eq:relationRphi0} \left(\frac{\mathcal{R}}{48}\right)^{- \frac{\Delta_{-}}{2}} = \frac{6}{(2-\Delta_{-})(3-\Delta_{-})} \, \f_{0} \ \left( 1+\mathcal{O}(\f_{0}^2) \right) \, ,
\end{align}
where we used that $c = - \frac{\ell}{2} \log \left( \frac{48}{\ell^2 R^{\textrm{uv}}} \right)$. Similarly, by comparing the coefficients of $e^{\Delta_{+}u/\ell}$ we obtain
\begin{align}
C = \frac{\Delta_{-} (2-\Delta_{-})}{2} \frac{(3-\Delta_{-})}{(1-\Delta_{-})} {\left( \frac{\mathcal{R}}{48} \right)}^{2-\Delta_{-}} \ \left( 1+\mathcal{O}(\mathcal{R}^{-\Delta_-}) \right) \, ,
\end{align}
where we also used \eqref{eq:relationRphi0}. We learn the following from these results, (which will also hold for the AdS$_4$ slicing).
\begin{itemize}
\item We found that  $\f_{0} \sim |\mathcal{R}|^{-\Delta_-/2}$. We hence confirm that our expansion in small $\f_{0}$ indeed corresponds to an expansion in the  large (dimensionless) curvature.
\item Therefore, in the regime of validity of our solution, we find that the larger the (dimensionless) curvature $\mathcal{R}$, the closer to the UV point RG flows end. The numerical examples of section \ref{sec:examples} show that this observation holds more widely, also outside of the regime of large curvature.
\item Last, as the quantity $C$ determines the vev of the operator dual to $\f$, we find
\begin{align}
\langle \mathcal{O} \rangle_{-} &\equiv \frac{4}{\Delta_{-}} C |\f_{-}|^{\frac{\Delta_{+}}{\Delta_{-}}} \\
\nonumber &= \frac{2(2-\Delta_{-}) (3-\Delta_{-})}{(1-\Delta_{-})} {\left( \frac{\mathcal{R}}{48} \right)}^{2-\Delta_{-}} \, |\f_{-}|^{\frac{\Delta_{+}}{\Delta_{-}}} \ \left( 1+\mathcal{O}(\mathcal{R}^{-\Delta_-}) \right) \, .
\end{align}
\end{itemize}

\subsection*{AdS$_4$ slicing}
We can repeat the analysis for the case with AdS$_4$ slicing. While we find the same qualitative behavior as for the $S^4$/ dS$_4$ slicing, there are some quantitative differences.
Expanding the solution \eqref{eq:phismallphi0AdS} for $u \rightarrow - \infty$ we find:
\begin{align}
\nonumber \f(u) &\underset{u \rightarrow - \infty}{\rightarrow} \hphantom{+} \frac{2 \, \f_{0} \, (1-\Delta_-) \, e^{\Delta_{-}c/\ell}}{(2-\Delta_{-})} e^{\Delta_{-}u/\ell} \ \left(1 + \mathcal{O}(e^{2u/\ell}) \right) \\
& \hphantom{\underset{u \rightarrow - \infty}{\rightarrow}} + \frac{2 \, \f_{0} \, (3-\Delta_-) \, e^{\Delta_{+}c/\ell}}{(2-\Delta_{-})} e^{\Delta_{+}u/\ell} \ \left(1 + \mathcal{O}(e^{2u/\ell}) \right) +\mathcal{O}(\f_{0}^3) \, .
\end{align}
Comparing with \eqref{eq:phiexpUV} we hence obtain:
\begin{align}
\left(-\frac{\mathcal{R}}{48}\right)^{- \frac{\Delta_{-}}{2}} = \frac{2(1-\Delta_-)}{(2-\Delta_-)} \ \f_{0} \ \left( 1+\mathcal{O}(\f_{0}^2) \right) \, .
\end{align}
In addition, we find
\begin{align}
C = \frac{\Delta_{-} (2-\Delta_{-})}{2} \frac{(3-\Delta_{-})}{(1-\Delta_{-})} {\left(- \frac{\mathcal{R}}{48} \right)}^{2- \Delta_-}  \ \left( 1+\mathcal{O}(|\mathcal{R}|^{-\Delta_-}) \right) \, .
\end{align}
This results in the following expression for the vev $\langle \mathcal{O} \rangle_{-}$:
\begin{align}
\langle \mathcal{O} \rangle_{-} &\equiv \frac{4}{\Delta_{-}} C |\f_{-}|^{\frac{\Delta_{+}}{\Delta_{-}}} \\
\nonumber &= \frac{2 (2-\Delta_{-}) (3-\Delta_{-})}{(1-\Delta_{-})} {\left(- \frac{\mathcal{R}}{48} \right)}^{2-\Delta_{-}} \, |\f_{-}|^{\frac{\Delta_{+}}{\Delta_{-}}} \ \left( 1+\mathcal{O}(|\mathcal{R}|^{-\Delta_-}) \right) \, .
\end{align}

\addcontentsline{toc}{section}{References}
\bibliography{CurvedRG.bib}

\providecommand{\href}[2]{#2}\begingroup\raggedright\begin{thebibliography}{10}

\bibitem{cw}
C.~G. Callan, Jr. and F.~Wilczek, \emph{{Infrared Behavior At Negative
  Curvature}}, \href{https://doi.org/10.1016/0550-3213(90)90451-I}{\emph{Nucl.
  Phys.} {\bfseries B340} (1990) 366--386}.

\bibitem{kk}
E.~Kiritsis and C.~Kounnas, \emph{{Infrared regularization of superstring
  theory and the one loop calculation of coupling constants}},
  \href{https://doi.org/10.1016/0550-3213(95)00156-M}{\emph{Nucl. Phys.}
  {\bfseries B442} (1995) 472--493},
  [\href{https://arxiv.org/abs/hep-th/9501020}{{\ttfamily hep-th/9501020}}].

\bibitem{Komar}
C.~Closset, T.~T. Dumitrescu, G.~Festuccia and Z.~Komargodski,
  \emph{{Supersymmetric Field Theories on Three-Manifolds}},
  \href{https://doi.org/10.1007/JHEP05(2013)017}{\emph{JHEP} {\bfseries 05}
  (2013) 017}, [\href{https://arxiv.org/abs/1212.3388}{{\ttfamily 1212.3388}}].

\bibitem{Marte}
D.~Martelli and A.~Passias, \emph{{The gravity dual of supersymmetric gauge
  theories on a two-parameter deformed three-sphere}},
  \href{https://doi.org/10.1016/j.nuclphysb.2013.09.012}{\emph{Nucl. Phys.}
  {\bfseries B877} (2013) 51--72},
  [\href{https://arxiv.org/abs/1306.3893}{{\ttfamily 1306.3893}}].

\bibitem{NSVZ}
V.~A. Novikov, M.~A. Shifman, A.~I. Vainshtein and V.~I. Zakharov, \emph{{Exact
  Gell-Mann-Low Function of Supersymmetric Yang-Mills Theories from Instanton
  Calculus}}, \href{https://doi.org/10.1016/0550-3213(83)90338-3}{\emph{Nucl.
  Phys.} {\bfseries B229} (1983) 381--393}.

\bibitem{buchel1}
A.~Buchel and A.~A. Tseytlin, \emph{{Curved space resolution of singularity of
  fractional D3-branes on conifold}},
  \href{https://doi.org/10.1103/PhysRevD.65.085019}{\emph{Phys. Rev.}
  {\bfseries D65} (2002) 085019},
  [\href{https://arxiv.org/abs/hep-th/0111017}{{\ttfamily hep-th/0111017}}].

\bibitem{buchel2}
A.~Buchel, \emph{{Gauge / gravity correspondence in accelerating universe}},
  \href{https://doi.org/10.1103/PhysRevD.65.125015}{\emph{Phys. Rev.}
  {\bfseries D65} (2002) 125015},
  [\href{https://arxiv.org/abs/hep-th/0203041}{{\ttfamily hep-th/0203041}}].

\bibitem{F1}
D.~L. Jafferis, \emph{{The Exact Superconformal R-Symmetry Extremizes Z}},
  \href{https://doi.org/10.1007/JHEP05(2012)159}{\emph{JHEP} {\bfseries 05}
  (2012) 159}, [\href{https://arxiv.org/abs/1012.3210}{{\ttfamily 1012.3210}}].

\bibitem{1103.1181}
D.~L. Jafferis, I.~R. Klebanov, S.~S. Pufu and B.~R. Safdi, \emph{{Towards the
  F-Theorem: N=2 Field Theories on the Three-Sphere}},
  \href{https://doi.org/10.1007/JHEP06(2011)102}{\emph{JHEP} {\bfseries 06}
  (2011) 102}, [\href{https://arxiv.org/abs/1103.1181}{{\ttfamily 1103.1181}}].

\bibitem{1006.1263}
R.~C. Myers and A.~Sinha, \emph{{Seeing a c-theorem with holography}},
  \href{https://doi.org/10.1103/PhysRevD.82.046006}{\emph{Phys. Rev.}
  {\bfseries D82} (2010) 046006},
  [\href{https://arxiv.org/abs/1006.1263}{{\ttfamily 1006.1263}}].

\bibitem{1011.5819}
R.~C. Myers and A.~Sinha, \emph{{Holographic c-theorems in arbitrary
  dimensions}}, \href{https://doi.org/10.1007/JHEP01(2011)125}{\emph{JHEP}
  {\bfseries 01} (2011) 125},
  [\href{https://arxiv.org/abs/1011.5819}{{\ttfamily 1011.5819}}].

\bibitem{ABTY}
O.~Aharony, M.~Berkooz, D.~Tong and S.~Yankielowicz, \emph{{Confinement in
  Anti-de Sitter Space}},
  \href{https://doi.org/10.1007/JHEP02(2013)076}{\emph{JHEP} {\bfseries 02}
  (2013) 076}, [\href{https://arxiv.org/abs/1210.5195}{{\ttfamily 1210.5195}}].

\bibitem{Tsamis:1992xa}
N.~C. Tsamis and R.~P. Woodard, \emph{{The Structure of perturbative quantum
  gravity on a De Sitter background}},
  \href{https://doi.org/10.1007/BF02102015}{\emph{Commun. Math. Phys.}
  {\bfseries 162} (1994) 217--248}.

\bibitem{Tsamis:1994ca}
N.~C. Tsamis and R.~P. Woodard, \emph{{Strong infrared effects in quantum
  gravity}}, \href{https://doi.org/10.1006/aphy.1995.1015}{\emph{Annals Phys.}
  {\bfseries 238} (1995) 1--82}.

\bibitem{Tsamis:1996qm}
N.~C. Tsamis and R.~P. Woodard, \emph{{The Quantum gravitational back reaction
  on inflation}}, \href{https://doi.org/10.1006/aphy.1997.5613}{\emph{Annals
  Phys.} {\bfseries 253} (1997) 1--54},
  [\href{https://arxiv.org/abs/hep-ph/9602316}{{\ttfamily hep-ph/9602316}}].

\bibitem{Ramsey:1997qc}
S.~A. Ramsey and B.~L. Hu, \emph{{O(N) quantum fields in curved space-time}},
  \href{https://doi.org/10.1103/PhysRevD.56.661}{\emph{Phys. Rev.} {\bfseries
  D56} (1997) 661--677}, [\href{https://arxiv.org/abs/gr-qc/9706001}{{\ttfamily
  gr-qc/9706001}}].

\bibitem{Burgess:2009bs}
C.~P. Burgess, L.~Leblond, R.~Holman and S.~Shandera, \emph{{Super-Hubble de
  Sitter Fluctuations and the Dynamical RG}},
  \href{https://doi.org/10.1088/1475-7516/2010/03/033}{\emph{JCAP} {\bfseries
  1003} (2010) 033}, [\href{https://arxiv.org/abs/0912.1608}{{\ttfamily
  0912.1608}}].

\bibitem{1105.4539}
J.~Serreau, \emph{{Effective potential for quantum scalar fields on a de Sitter
  geometry}}, \href{https://doi.org/10.1103/PhysRevLett.107.191103}{\emph{Phys.
  Rev. Lett.} {\bfseries 107} (2011) 191103},
  [\href{https://arxiv.org/abs/1105.4539}{{\ttfamily 1105.4539}}].

\bibitem{1306.3846}
J.~Serreau, \emph{{Renormalization group flow and symmetry restoration in de
  Sitter space}},
  \href{https://doi.org/10.1016/j.physletb.2014.01.058}{\emph{Phys. Lett.}
  {\bfseries B730} (2014) 271--274},
  [\href{https://arxiv.org/abs/1306.3846}{{\ttfamily 1306.3846}}].

\bibitem{HSke}
M.~Henningson and K.~Skenderis, \emph{{The Holographic Weyl anomaly}},
  \href{https://doi.org/10.1088/1126-6708/1998/07/023}{\emph{JHEP} {\bfseries
  07} (1998) 023}, [\href{https://arxiv.org/abs/hep-th/9806087}{{\ttfamily
  hep-th/9806087}}].

\bibitem{skenderislec}
K.~Skenderis, \emph{{Lecture notes on holographic renormalization}},
  \href{https://doi.org/10.1088/0264-9381/19/22/306}{\emph{Class. Quant. Grav.}
  {\bfseries 19} (2002) 5849--5876},
  [\href{https://arxiv.org/abs/hep-th/0209067}{{\ttfamily hep-th/0209067}}].

\bibitem{skenderisham}
I.~Papadimitriou and K.~Skenderis, \emph{{Correlation functions in holographic
  RG flows}}, \href{https://doi.org/10.1088/1126-6708/2004/10/075}{\emph{JHEP}
  {\bfseries 10} (2004) 075},
  [\href{https://arxiv.org/abs/hep-th/0407071}{{\ttfamily hep-th/0407071}}].

\bibitem{Duff}
M.~J. Duff, \emph{{Twenty years of the Weyl anomaly}},
  \href{https://doi.org/10.1088/0264-9381/11/6/004}{\emph{Class. Quant. Grav.}
  {\bfseries 11} (1994) 1387--1404},
  [\href{https://arxiv.org/abs/hep-th/9308075}{{\ttfamily hep-th/9308075}}].

\bibitem{Osborn}
H.~Osborn, \emph{{Weyl consistency conditions and a local renormalization group
  equation for general renormalizable field theories}},
  \href{https://doi.org/10.1016/0550-3213(91)80030-P}{\emph{Nucl. Phys.}
  {\bfseries B363} (1991) 486--526}.

\bibitem{9711200}
J.~M. Maldacena, \emph{{The Large N limit of superconformal field theories and
  supergravity}}, \href{https://doi.org/10.1023/A:1026654312961}{\emph{Int. J.
  Theor. Phys.} {\bfseries 38} (1999) 1113--1133},
  [\href{https://arxiv.org/abs/hep-th/9711200}{{\ttfamily hep-th/9711200}}].

\bibitem{9802109}
S.~S. Gubser, I.~R. Klebanov and A.~M. Polyakov, \emph{{Gauge theory
  correlators from noncritical string theory}},
  \href{https://doi.org/10.1016/S0370-2693(98)00377-3}{\emph{Phys. Lett.}
  {\bfseries B428} (1998) 105--114},
  [\href{https://arxiv.org/abs/hep-th/9802109}{{\ttfamily hep-th/9802109}}].

\bibitem{9802150}
E.~Witten, \emph{{Anti-de Sitter space and holography}}, {\emph{Adv. Theor.
  Math. Phys.} {\bfseries 2} (1998) 253--291},
  [\href{https://arxiv.org/abs/hep-th/9802150}{{\ttfamily hep-th/9802150}}].

\bibitem{9810126}
L.~Girardello, M.~Petrini, M.~Porrati and A.~Zaffaroni, \emph{{Novel local CFT
  and exact results on perturbations of N=4 superYang Mills from AdS
  dynamics}}, \href{https://doi.org/10.1088/1126-6708/1998/12/022}{\emph{JHEP}
  {\bfseries 12} (1998) 022},
  [\href{https://arxiv.org/abs/hep-th/9810126}{{\ttfamily hep-th/9810126}}].

\bibitem{9903190}
V.~Balasubramanian and P.~Kraus, \emph{{Space-time and the holographic
  renormalization group}},
  \href{https://doi.org/10.1103/PhysRevLett.83.3605}{\emph{Phys. Rev. Lett.}
  {\bfseries 83} (1999) 3605--3608},
  [\href{https://arxiv.org/abs/hep-th/9903190}{{\ttfamily hep-th/9903190}}].

\bibitem{9904017}
D.~Z. Freedman, S.~S. Gubser, K.~Pilch and N.~P. Warner, \emph{{Renormalization
  group flows from holography supersymmetry and a c theorem}}, {\emph{Adv.
  Theor. Math. Phys.} {\bfseries 3} (1999) 363--417},
  [\href{https://arxiv.org/abs/hep-th/9904017}{{\ttfamily hep-th/9904017}}].

\bibitem{9912012}
J.~de~Boer, E.~P. Verlinde and H.~L. Verlinde, \emph{{On the holographic
  renormalization group}},
  \href{https://doi.org/10.1088/1126-6708/2000/08/003}{\emph{JHEP} {\bfseries
  08} (2000) 003}, [\href{https://arxiv.org/abs/hep-th/9912012}{{\ttfamily
  hep-th/9912012}}].

\bibitem{0105276}
M.~Bianchi, D.~Z. Freedman and K.~Skenderis, \emph{{How to go with an RG
  flow}}, \href{https://doi.org/10.1088/1126-6708/2001/08/041}{\emph{JHEP}
  {\bfseries 08} (2001) 041},
  [\href{https://arxiv.org/abs/hep-th/0105276}{{\ttfamily hep-th/0105276}}].

\bibitem{0404176}
I.~Papadimitriou and K.~Skenderis, \emph{{AdS / CFT correspondence and
  geometry}}, \href{https://doi.org/10.4171/013-1/4}{\emph{IRMA Lect. Math.
  Theor. Phys.} {\bfseries 8} (2005) 73--101},
  [\href{https://arxiv.org/abs/hep-th/0404176}{{\ttfamily hep-th/0404176}}].

\bibitem{0702088}
A.~Ceresole and G.~Dall'Agata, \emph{{Flow Equations for Non-BPS Extremal Black
  Holes}}, \href{https://doi.org/10.1088/1126-6708/2007/03/110}{\emph{JHEP}
  {\bfseries 03} (2007) 110},
  [\href{https://arxiv.org/abs/hep-th/0702088}{{\ttfamily hep-th/0702088}}].

\bibitem{Papadimitriou:2007sj}
I.~Papadimitriou, \emph{{Multi-Trace Deformations in AdS/CFT: Exploring the
  Vacuum Structure of the Deformed CFT}},
  \href{https://doi.org/10.1088/1126-6708/2007/05/075}{\emph{JHEP} {\bfseries
  05} (2007) 075}, [\href{https://arxiv.org/abs/hep-th/0703152}{{\ttfamily
  hep-th/0703152}}].

\bibitem{0707.1324}
U.~Gursoy and E.~Kiritsis, \emph{{Exploring improved holographic theories for
  QCD: Part I}},
  \href{https://doi.org/10.1088/1126-6708/2008/02/032}{\emph{JHEP} {\bfseries
  02} (2008) 032}, [\href{https://arxiv.org/abs/0707.1324}{{\ttfamily
  0707.1324}}].

\bibitem{0707.1349}
U.~Gursoy, E.~Kiritsis and F.~Nitti, \emph{{Exploring improved holographic
  theories for QCD: Part II}},
  \href{https://doi.org/10.1088/1126-6708/2008/02/019}{\emph{JHEP} {\bfseries
  02} (2008) 019}, [\href{https://arxiv.org/abs/0707.1349}{{\ttfamily
  0707.1349}}].

\bibitem{1010.1264}
I.~Heemskerk and J.~Polchinski, \emph{{Holographic and Wilsonian
  Renormalization Groups}},
  \href{https://doi.org/10.1007/JHEP06(2011)031}{\emph{JHEP} {\bfseries 06}
  (2011) 031}, [\href{https://arxiv.org/abs/1010.1264}{{\ttfamily 1010.1264}}].

\bibitem{1010.4036}
T.~Faulkner, H.~Liu and M.~Rangamani, \emph{{Integrating out geometry:
  Holographic Wilsonian RG and the membrane paradigm}},
  \href{https://doi.org/10.1007/JHEP08(2011)051}{\emph{JHEP} {\bfseries 08}
  (2011) 051}, [\href{https://arxiv.org/abs/1010.4036}{{\ttfamily 1010.4036}}].

\bibitem{1106.4826}
I.~Papadimitriou, \emph{{Holographic Renormalization of general dilaton-axion
  gravity}}, \href{https://doi.org/10.1007/JHEP08(2011)119}{\emph{JHEP}
  {\bfseries 08} (2011) 119},
  [\href{https://arxiv.org/abs/1106.4826}{{\ttfamily 1106.4826}}].

\bibitem{1112.3356}
S.~Grozdanov, \emph{{Wilsonian Renormalisation and the Exact Cut-Off Scale from
  Holographic Duality}},
  \href{https://doi.org/10.1007/JHEP06(2012)079}{\emph{JHEP} {\bfseries 06}
  (2012) 079}, [\href{https://arxiv.org/abs/1112.3356}{{\ttfamily 1112.3356}}].

\bibitem{1205.6205}
E.~Kiritsis and V.~Niarchos, \emph{{The holographic quantum effective potential
  at finite temperature and density}},
  \href{https://doi.org/10.1007/JHEP08(2012)164}{\emph{JHEP} {\bfseries 08}
  (2012) 164}, [\href{https://arxiv.org/abs/1205.6205}{{\ttfamily 1205.6205}}].

\bibitem{1310.0858}
J.~Bourdier and E.~Kiritsis, \emph{{Holographic RG flows and nearly-marginal
  operators}},
  \href{https://doi.org/10.1088/0264-9381/31/3/035011}{\emph{Class. Quant.
  Grav.} {\bfseries 31} (2014) 035011},
  [\href{https://arxiv.org/abs/1310.0858}{{\ttfamily 1310.0858}}].

\bibitem{1401.0888}
E.~Kiritsis, W.~Li and F.~Nitti, \emph{{Holographic RG flow and the Quantum
  Effective Action}},
  \href{https://doi.org/10.1002/prop.201400007}{\emph{Fortsch. Phys.}
  {\bfseries 62} (2014) 389--454},
  [\href{https://arxiv.org/abs/1401.0888}{{\ttfamily 1401.0888}}].

\bibitem{exotic}
E.~Kiritsis, F.~Nitti and L.~S. Pimenta, \emph{{Exotic RG Flows from
  Holography}}, \href{https://doi.org/10.1002/prop.201600120}{\emph{Fortsch.
  Phys.} {\bfseries 65} (2017) 1600120},
  [\href{https://arxiv.org/abs/1611.05493}{{\ttfamily 1611.05493}}].

\bibitem{1108.6070}
A.~Buchel, \emph{{Quantum phase transitions in cascading gauge theory}},
  \href{https://doi.org/10.1016/j.nuclphysb.2011.11.007}{\emph{Nucl. Phys.}
  {\bfseries B856} (2012) 278--327},
  [\href{https://arxiv.org/abs/1108.6070}{{\ttfamily 1108.6070}}].

\bibitem{1007.3996}
D.~Marolf, M.~Rangamani and M.~Van~Raamsdonk, \emph{{Holographic models of de
  Sitter QFTs}},
  \href{https://doi.org/10.1088/0264-9381/28/10/105015}{\emph{Class. Quant.
  Grav.} {\bfseries 28} (2011) 105015},
  [\href{https://arxiv.org/abs/1007.3996}{{\ttfamily 1007.3996}}].

\bibitem{staro}
A.~A. Starobinsky, \emph{{A New Type of Isotropic Cosmological Models Without
  Singularity}},
  \href{https://doi.org/10.1016/0370-2693(80)90670-X}{\emph{Phys. Lett.}
  {\bfseries 91B} (1980) 99--102}.

\bibitem{kcosmo}
E.~Kiritsis, \emph{{Holography and brane-bulk energy exchange}},
  \href{https://doi.org/10.1088/1475-7516/2005/10/014}{\emph{JCAP} {\bfseries
  0510} (2005) 014}, [\href{https://arxiv.org/abs/hep-th/0504219}{{\ttfamily
  hep-th/0504219}}].

\bibitem{Ftheorem}
J.~K. Ghosh, E.~Kiritsis, F.~Nitti and L.~T. Witkowski, \emph{{to appear}}, .

\bibitem{HH}
J.~B. Hartle and S.~W. Hawking, \emph{{Wave Function of the Universe}},
  \href{https://doi.org/10.1103/PhysRevD.28.2960}{\emph{Phys. Rev.} {\bfseries
  D28} (1983) 2960--2975}.

\bibitem{HH2}
T.~Hertog and J.~Hartle, \emph{{Holographic No-Boundary Measure}},
  \href{https://doi.org/10.1007/JHEP05(2012)095}{\emph{JHEP} {\bfseries 05}
  (2012) 095}, [\href{https://arxiv.org/abs/1111.6090}{{\ttfamily 1111.6090}}].

\bibitem{HHH}
J.~B. Hartle, S.~W. Hawking and T.~Hertog, \emph{{Quantum Probabilities for
  Inflation from Holography}},
  \href{https://doi.org/10.1088/1475-7516/2014/01/015}{\emph{JCAP} {\bfseries
  1401} (2014) 015}, [\href{https://arxiv.org/abs/1207.6653}{{\ttfamily
  1207.6653}}].

\bibitem{self}
C.~Charmousis, E.~Kiritsis and F.~Nitti, \emph{{Holographic self-tuning of the
  cosmological constant}},
  \href{https://doi.org/10.1007/JHEP09(2017)031}{\emph{JHEP} {\bfseries 09}
  (2017) 031}, [\href{https://arxiv.org/abs/1704.05075}{{\ttfamily
  1704.05075}}].

\bibitem{SkenderisTownsend}
K.~Skenderis and P.~K. Townsend, \emph{{Hamilton-Jacobi method for curved
  domain walls and cosmologies}},
  \href{https://doi.org/10.1103/PhysRevD.74.125008}{\emph{Phys. Rev.}
  {\bfseries D74} (2006) 125008},
  [\href{https://arxiv.org/abs/hep-th/0609056}{{\ttfamily hep-th/0609056}}].

\bibitem{maldamaoz}
J.~M. Maldacena and L.~Maoz, \emph{{Wormholes in AdS}},
  \href{https://doi.org/10.1088/1126-6708/2004/02/053}{\emph{JHEP} {\bfseries
  02} (2004) 053}, [\href{https://arxiv.org/abs/hep-th/0401024}{{\ttfamily
  hep-th/0401024}}].

\bibitem{witten-thermo}
E.~Witten, \emph{{Anti-de Sitter space, thermal phase transition, and
  confinement in gauge theories}},
  \href{https://doi.org/10.4310/ATMP.1998.v2.n3.a3}{\emph{Adv. Theor. Math.
  Phys.} {\bfseries 2} (1998) 505--532},
  [\href{https://arxiv.org/abs/hep-th/9803131}{{\ttfamily hep-th/9803131}}].

\bibitem{1202.2070}
H.~Liu and M.~Mezei, \emph{{A Refinement of entanglement entropy and the number
  of degrees of freedom}},
  \href{https://doi.org/10.1007/JHEP04(2013)162}{\emph{JHEP} {\bfseries 04}
  (2013) 162}, [\href{https://arxiv.org/abs/1202.2070}{{\ttfamily 1202.2070}}].

\bibitem{1202.5650}
H.~Casini and M.~Huerta, \emph{{On the RG running of the entanglement entropy
  of a circle}}, \href{https://doi.org/10.1103/PhysRevD.85.125016}{\emph{Phys.
  Rev.} {\bfseries D85} (2012) 125016},
  [\href{https://arxiv.org/abs/1202.5650}{{\ttfamily 1202.5650}}].

\bibitem{Taylor}
M.~Taylor and W.~Woodhead, \emph{{The holographic F theorem}},
  \href{https://arxiv.org/abs/1604.06809}{{\ttfamily 1604.06809}}.

\bibitem{KarchRandall}
A.~Karch and L.~Randall, \emph{{Locally localized gravity}},
  \href{https://doi.org/10.1088/1126-6708/2001/05/008}{\emph{JHEP} {\bfseries
  05} (2001) 008}, [\href{https://arxiv.org/abs/hep-th/0011156}{{\ttfamily
  hep-th/0011156}}].

\bibitem{Bayona}
C.~A. Bayona and N.~R.~F. Braga, \emph{{Anti-de Sitter boundary in Poincare
  coordinates}}, \href{https://doi.org/10.1007/s10714-007-0446-y}{\emph{Gen.
  Rel. Grav.} {\bfseries 39} (2007) 1367--1379},
  [\href{https://arxiv.org/abs/hep-th/0512182}{{\ttfamily hep-th/0512182}}].

\bibitem{higuchi}
A.~Higuchi, \emph{{Forbidden Mass Range for Spin-2 Field Theory in De Sitter
  Space-time}}, \href{https://doi.org/10.1016/0550-3213(87)90691-2}{\emph{Nucl.
  Phys.} {\bfseries B282} (1987) 397--436}.

\bibitem{derham}
C.~de~Rham and S.~Renaux-Petel, \emph{{Massive Gravity on de Sitter and Unique
  Candidate for Partially Massless Gravity}},
  \href{https://doi.org/10.1088/1475-7516/2013/01/035}{\emph{JCAP} {\bfseries
  1301} (2013) 035}, [\href{https://arxiv.org/abs/1206.3482}{{\ttfamily
  1206.3482}}].

\bibitem{Myers}
R.~Emparan, C.~V. Johnson and R.~C. Myers, \emph{{Surface terms as counterterms
  in the AdS / CFT correspondence}},
  \href{https://doi.org/10.1103/PhysRevD.60.104001}{\emph{Phys. Rev.}
  {\bfseries D60} (1999) 104001},
  [\href{https://arxiv.org/abs/hep-th/9903238}{{\ttfamily hep-th/9903238}}].

\bibitem{Kraus}
P.~Kraus, F.~Larsen and R.~Siebelink, \emph{{The gravitational action in
  asymptotically AdS and flat space-times}},
  \href{https://doi.org/10.1016/S0550-3213(99)00549-0}{\emph{Nucl. Phys.}
  {\bfseries B563} (1999) 259--278},
  [\href{https://arxiv.org/abs/hep-th/9906127}{{\ttfamily hep-th/9906127}}].

\bibitem{HaroSkenderisSolodukhin}
S.~de~Haro, S.~N. Solodukhin and K.~Skenderis, \emph{{Holographic
  reconstruction of space-time and renormalization in the AdS / CFT
  correspondence}}, \href{https://doi.org/10.1007/s002200100381}{\emph{Commun.
  Math. Phys.} {\bfseries 217} (2001) 595--622},
  [\href{https://arxiv.org/abs/hep-th/0002230}{{\ttfamily hep-th/0002230}}].

\end{thebibliography}\endgroup
\bibliographystyle{JHEP}

\end{document}